%% file: main.tex
\documentclass[]{aa}
\usepackage{morefloats}
\usepackage{amsmath}
\usepackage{color}
\usepackage{textcomp}
\usepackage{commath}
\usepackage{array}
\usepackage{siunitx}
\usepackage{subcaption}
\usepackage{caption}
\usepackage[capitalise]{cleveref} 

\DeclareCaptionFormat{cont}{#1 (cont.)#2#3\par}

\bibpunct{(}{)}{;}{a}{}{,} 

\newcolumntype{C}[1]{>{\centering\arraybackslash}m{#1}}

\begin{document}

\title{Implications of three-dimensional chemical transport in hot Jupiter atmospheres: results from a consistently coupled chemistry-radiation-hydrodynamics model}

\titlerunning{Chemical transport in hot Jupiter atmospheres}
\authorrunning{Drummond et al.}
\author{Benjamin Drummond\inst{\ref{inst1},\ref{inst2}}
\and Eric H\'ebrard\inst{\ref{inst1}}
\and Nathan J. Mayne\inst{\ref{inst1}}
\and Olivia Venot\inst{\ref{inst3}}
\and Robert J. Ridgway\inst{\ref{inst1}}
\and Quentin Changeat\inst{\ref{inst4}}
\and Shang-Min Tsai\inst{\ref{inst5}}
\and James Manners\inst{\ref{inst2},\ref{inst6}}
\and Pascal Tremblin\inst{\ref{inst7}}
\and Nathan Luke Abraham\inst{\ref{inst8},\ref{inst9}}
\and David Sing\inst{\ref{inst10}}
\and Krisztian Kohary\inst{\ref{inst1}}
}

\institute{Astrophysics Group, University of Exeter, EX4 4QL, Exeter, UK\label{inst1} \\
\email{b.drummond@exeter.ac.uk}
\and Met Office, Fitzroy Road, Exeter, EX1 3PB, UK\label{inst2}
\and Laboratoire Interuniversitaire des Syst\`{e}mes Atmosph\'{e}riques (LISA), UMR CNRS 7583, Universit\'{e} Paris-Est-Cr\'eteil, Universit\'e de Paris, Institut Pierre Simon Laplace, Cr\'{e}teil, France\label{inst3}
\and Department of Physics and Astronomy, University College London, Gower Street, London, WC1E 6BT, UK\label{inst4}
\and Atmospheric, Oceanic, and Planetary Physics Department, Clarendon Laboratory, University of Oxford, Sherrington Road, Oxford OX1 3PU, UK\label{inst5}
\and Global Systems Institute, University of Exeter, Exeter, UK\label{inst6}
\and Maison de la simulation, CEA, CNRS, Univ. Paris-Sud, UVSQ, Université Paris-Saclay, 91191 Gif-Sur-Yvette, France\label{inst7}
\and National Centre for Atmospheric Science, UK\label{inst8}
\and Department of Chemistry, University of Cambridge, Lensfield Road, Cambridge, CB2 1EW, UK\label{inst9}
\and Physics and Astronomy, Johns Hopkins University, Baltimore, MD, United States\label{inst10}
}

\date{Received /
	Accepted }


\abstract {
We present results from a set of simulations using a fully coupled three-dimensional (3D) chemistry-radiation-hydrodynamics model and investigate the effect of transport of chemical species by the large-scale atmospheric flow in hot Jupiter atmospheres. We coupled a flexible chemical kinetics scheme to the Met Office Unified Model, which enables the study of the interaction of chemistry, radiative transfer, and fluid dynamics. We used a newly-released `reduced' chemical network, comprising 30 chemical species, that was specifically developed for its application in 3D atmosphere models. We simulated the atmospheres of the well-studied hot Jupiters HD~209458b and HD~189733b which both have dayside--nightside temperature contrasts of several hundred Kelvin and superrotating equatorial jets. We find qualitatively quite different chemical structures between the two planets, particularly for methane (CH$_4$), when advection of chemical species is included. Our results show that consideration of 3D chemical transport is vital in understanding the chemical composition of hot Jupiter atmospheres. Three-dimensional mixing leads to significant changes in the abundances of absorbing gas-phase species compared with what would be expected by assuming local chemical equilibrium, or from models including 1D - and even 2D - chemical mixing. We find that CH$_4$, carbon dioxide (CO$_2$), and ammonia (NH$_3$) are particularly interesting as 3D mixing of these species leads to prominent signatures of out-of-equilibrium chemistry in the transmission and emission spectra, which are detectable with near-future instruments.
}

\maketitle

\input{intro}
%
\input{model_description}
%
\input{results}
%
\input{discussion}
%
\input{conclusions}

\begin{acknowledgements}
      The authors thank the anonymous referee for a very useful and constructive report that led to improvements to the manuscript. BD, EH and NJM acknowledge support from a Science and Technology Facilities Council Consolidated Grant (ST/R000395/1). NJM was part funded by a Leverhulme Trust Research Project Grant during this project. This work also benefited from the 2018 Exoplanet Summer Program in the Other Worlds Laboratory (OWL) at the University of California, Santa Cruz, a program funded by the Heising-Simons Foundation. O.V. thanks the CNRS/INSU Programme National de Plan\'etologie (PNP) and the Centre National d'\'Etudes Spatiales (CNES) for funding support. QC acknowledges funding from the European Research Council (ERC) under the European Union's Horizon 2020 research and innovation programme (grant agreement No 758892, ExoAI), under the European Union's Seventh Framework Programme (FP7/2007-2013)/ ERC grant agreement numbers 617119 (ExoLights) and the European Union's Horizon 2020 COMPET programme (grant agreement No 776403, ExoplANETS A) and also by the Science and Technology Funding Council (STFC) grants: ST/K502406/1, ST/P000282/1, ST/P002153/1 and ST/S002634/1. JM acknowledges the support of a Met Office Academic Partnership secondment. PT acknowledges supports by the European Research Council under Grant Agreement ATMO 757858. This work was performed using the DiRAC Data Intensive service at Leicester, operated by the University of Leicester IT Services, which forms part of the STFC DiRAC HPC Facility (www.dirac.ac.uk). The equipment was funded by BEIS capital funding via STFC capital grants ST/K000373/1 and ST/R002363/1 and STFC DiRAC Operations grant ST/R001014/1. DiRAC is part of the National e-Infrastructure. This research also made use of the ISCA High Performance Computing Service at the University of Exeter. Additionally, we acknowledge use of the Monsoon system, a collaborative facility supplied under the Joint Weather and Climate Research Programme, a strategic partnership between the Met Office and the Natural Environment Research Council. Material produced using Met Office Software. The research data supporting this publication are available upon request to the authors.
\end{acknowledgements}

\bibliographystyle{aa} 
\bibliography{referencelist.bib} 

\begin{appendix}
\input{appendix}

\end{appendix}

\end{document}

%% file: intro.tex

\section{Introduction}

Observations of exoplanets, particularly of transiting, tidally--locked hot Jupiters, have yielded the detection of several atmospheric species \citep[e.g.][]{Snellen2008,Wilson2015,Sing2015,KreLB15}, alongside evidence for aerosols \citep[e.g. photochemical hazes or condensate clouds, see for example,][]{SinFN16}. Observations of some hot hydrogen-dominated atmospheres show clear absorption or emission features of absorbing gas-phase species \citep[e.g.][]{NikSF18}; while for other planets, these features appear to be obscured, possibly by the presence of absorbing or scattering clouds and hazes \citep[e.g.,][]{deming_2013}. 

Theoretical modelling of hot Jupiter atmospheres has recently shifted its focus to chemical composition, in terms of both the gas-phase chemistry and cloud formation. Dynamical mixing can alter the local abundances of chemical species in an atmosphere. If the transport of material occurs on a timescale that is faster than the timescale for the atmosphere to reach a state of chemical equilibrium, then advection of chemical species has a strong influence on the local chemical composition. This process of `quenching' was inferred several decades ago in the atmosphere of Jupiter since observed abundances of carbon monoxide (CO) did not agree with predictions based on local chemical equilibrium \citep[e.g.][]{Prinn1977}. Vertical quenching, due to vertical mixing, has also been inferred from observations of young Jupiter analogues \citep[e.g.][]{skemer_2014} and brown dwarfs \citep[see][and references therein]{saumon_2003,Zahnle_2014}.

Until recently, studies of the effect of dynamical mixing on the chemical composition have mainly been restricted to the use of vertical diffusion in one-dimensional (1D) atmosphere models \citep[e.g.][]{ZahMF09,Moses2011,Venot2012,RimH16,DruTB16,TsaLG2017}, primarily because of the high computational expense of the chemistry calculations, making higher-dimensional simulations challenging. However, for short--period, tidally--locked exoplanets material is also expected to be mixed horizontally between a permanently irradiated dayside and an unirradiated nightside. For hot Jupiters, fast horizontal winds \citep[e.g.][]{ShoG02,PerS13} acting to redistribute heat from the dayside to nightside have been inferred from day--night temperature contrasts and shifting of the thermal emission, or `hot spot', from the closest point to the star \citep[e.g.][]{knutson_2007,KnuLF12,Zellem2014,WonKK16}. These winds have also been observed more directly using high--resolution spectroscopy \citep[e.g.][]{Snedd10}, indicating `diverging' flows from the day to night side \citep{Louden2015}. Observations of the emission as a function of orbital phase, or phase curves, have also been shown to depart from model predictions for the nightside, which has been proposed to be caused by horizontal mixing of chemical species \citep[e.g.][]{stevenson_2010,knutson_2012,Zellem2014}.

Future facilities, such as the James Webb Space Telescope (JWST) \citep[e.g.][]{BeaSB18} and the Atmospheric Remote-sensing Infrared Exoplanet Large-survey (ARIEL) \citep[e.g.][]{TinDE18}, will provide more accurate phase curves, for an increasing number of targets, alongside the potential for more accurate measurements of chemical abundances (e.g. methane (CH$_4$)) from emission and transmission spectra. Combining these observations with a more complete understanding of the mixing processes controlling the observed abundances, and phase curves, will provide a unique `window' into the atmospheric dynamics of tidally--locked exoplanets.

\citet{CooS06} were the first to investigate the effect of three-dimensional (3D) atmospheric mixing on the composition of hot Jupiter atmospheres using a simplified chemistry scheme (chemical relaxation) and temperature forcing within a 3D hydrodynamics code. Focusing on HD~209458b, the abundance of CO was `relaxed' back to an equilibrium abundance over a timescale chosen to be indicative of the relevant chemical pathways. The results of \citet{CooS06} indicated that vertical mixing was actually dominant, in terms of controlling the composition in the observable region of the atmosphere, compared to the horizontal mixing. \citet{AguPV14} also investigated the role of vertical and horizontal mixing using a more complete chemistry scheme but a highly idealised approach to the dynamics. Their pseudo two-dimensional (2D) model approximated horizontal mixing with a time--varying pressure--temperature profile within a 1D atmosphere model, intended to represent a column of atmosphere rotating around the equator within the equatorial zonal jet. \citet{AguPV14} concluded that horizontal mixing was also important in controlling the local chemical composition, finding that the nightside can be `contaminated' by material transported from the hotter dayside.

Recently, \citet{DruMM18a} and \citet{DruMM18b} revisited the chemical relaxation scheme developed by \citet{CooS06} and implemented it into a different 3D atmosphere model. The main improvement over the earlier work of \citet{CooS06} was the coupling of the chemistry scheme to a sophisticated radiative transfer scheme \citep{AmuBT14,AmuMB16}. This improvement allows changes in the local composition, due to advection, for instance, to be consistently fed back to the heating rates and the thermal evolution of the atmosphere, as well as the calculation of synthetic observations \citep[e.g.][]{LinMM18}. \citet{DruMM18a} and \citet{DruMM18b} found that both horizontal and vertical quenching play important roles in determining the abundance of CH$_4$ in the observable region of the atmosphere for both HD~209458b and HD~189733b. \citet{MenTM18} recently included an updated chemical relaxation scheme, developed in \citet{TsaKL18}, in a 3D atmosphere model for the case of WASP--43b. In contrast to the results of \citet{DruMM18a,DruMM18b}, who found that chemical gradients are very efficiently removed due to advection, they found that significant horizontal structure in the chemical composition remains. However, a fair comparison cannot be made since there are many differences in the implementation of the model and the physics schemes, as well as the planetary parameters assumed, between our earlier work and that of \citet{MenTM18}.

\citet{BorBO18} considered the quenching of chemical species by modelling the (2D and 3D) fluid dynamics of the convective region of substellar atmospheres including a reactive chemical tracer. Their results suggest that the length scale associated with the quenching is dependent on the chemical properties of the species of interest, in agreement with the earlier 1D work of \citet{Smi98}. \citet{ZhaS18,ZhaS18b} investigated the eddy diffusion coefficient ($K_{zz}$) using results from 3D simulations including a transported reactive chemical species and found that $K_{zz}$ should in principle be different for different chemical species. They derived also an analytical theory for the quench point of chemical species in substellar atmospheres.

As well as the gas-phase chemistry, the study of the formation of clouds and their effects on the atmosphere have recently been a significant focus. In the framework of 3D modelling there are two main approaches to treating cloud formation. One approach is to include cloud formation as a post-processing step, using output from a cloud-free radiation-hydrodynamics code, effectively decoupling the dynamic and thermodynamic evolution of the atmosphere from the cloud treatment \citep[e.g.][]{ParFS16,Helling2016,Powell2019}. Another option is to take a more complete approach by coupling treatments of the cloud formation and their radiative properties directly to an atmosphere model, allowing for cloud-radiative feedback on the atmosphere \citep[e.g.][]{LeeDH16,LinMB18,LinMM18,LinMM19}. In this work we choose to ignore the presence of clouds and focus on the gas-phase chemical composition. Future works will need to treat the kinetics of both gas-phase and cloud chemistry to fully understand the composition of hot Jupiter atmospheres. However, coupling these two processes consistently in a numerical model is a very significant scientific and technical challenge.

We present a significant step towards a better understanding of the chemical composition of hot exoplanet atmospheres by focusing in detail on the gas-phase chemistry. We couple our 3D model, including sophisticated treatments of the radiative transfer and dynamics \citep[as used in, e.g.][]{AmuMB16,DruMB18,MayDD19}, to a chemical kinetics scheme using a chemical network specifically developed for application in 3D atmosphere models \citep{VenBD19}. This model allows for the interactions and feedback between chemistry, radiation, and dynamics. 
 
 In \cref{section:methods} we introduce the model development to this point, focusing on the chemistry (\cref{section:methods_chem}), and detail the parameters used in the simulations that we present (\cref{section:methods_param}). We then move to our results in \cref{section:results} demonstrating that including 3D transport is crucial in understanding the atmospheric chemical structure of hot Jupiters, and that the effects of chemical transport have significant consequences for their predicted spectra. We discuss our results in the context of previous studies in \cref{section:discussion} and finally, we state our conclusions in \cref{section:conclusions}.  

%% file: model_description.tex

\section{Methods and model description}
\label{section:methods}

We use an idealised configuration of the Met Office Unified Model (UM) to simulate the atmospheres of highly-irradiated gas-giant planets, namely HD~209458b and HD~189733b.  The UM, originally developed for Earth atmosphere applications, has previously been used to simulate the atmospheres of hot Jupiters and warm Neptunes \citep{MayBA14,MayDB17,MayDD19,AmuMB16,DruMB18,DruMM18a,DruMM18b,LinMM18,LinMB18,LinMM19}, as well as terrestrial exoplanets \citep{MayBA14b,BouMD17,LewLB18,YatPM19}. Our modifications to the model for this work were made in a development branch using the UM at version 11.4 (released June 2019) as its base. 

The dynamical core of the UM, called ENDGame \citep[][]{WooSW14}, has been validated for exoplanet applications by reproducing long-term, large-scale flows of idealised standard tests \citep{MayBA14b} and tested for the typical flows expected in hot Jupiter like atmospheres \citep{MayBA14,MayDB17}. ENDGame has the unique capability to seamlessly transition between the simplified `primitive' equations of motion and the `full' equations of motion, the latter not adopting a series of assumptions (e.g. hydrostatic balance, shallow-atmosphere) that are used to simplify the equation set \citep[see][for discussion]{MayBA14}. Using this, \citet{MayDD19} demonstrated that the primitive equations produce a significantly different circulation, and hence also thermal structure, compared with the full equation set for warm small Neptune and super Earth atmospheres. For typical hot Jupiter atmospheres, however, the differences in the resulting flow are negligible when using the primitive or full equations \citep{MayBA14,MayDD19}. In this work we use the full equations of motion.

The UM uses a flexible and sophisticated radiative transfer scheme, the open-source Suite Of Community RAdiative Transfer codes based on Edwards and Slingo (SOCRATES) scheme\footnote{https://code.metoffice.gov.uk/trac/socrates} \citep{EdwS96,Edw96}, to calculate the radiative heating rates that, in part, drive the thermal evolution of the atmosphere. SOCRATES has been updated and tested for the high-temperatures and compositions relevant for hot Jupiter atmospheres \citep{AmuBT14,AmuMB16,AmuTM17} and is now being used in other models in the community \citep{WayAA17,WayDA18,ThoV19}. 

SOCRATES adopts the two-stream approximation and treats opacities using the correlated-$k$ approximation, the latter significantly improving computational efficiency while maintaining accuracy, compared with alternative methods \citep{Lacis1991,AmuBT14}. The $k$-coefficients of individual gases are combined on-the-fly using the method of equivalent extinction \citep{Edw96,AmuTM17}, meaning that the composition of the gas does not need to be known a priori, unlike the commonly used alternative method of `pre-mixed' $k$-coefficient tables \citep[e.g.][]{Showman2009}. In the case of pre-mixed $k$-coefficients a chemical equilibrium composition is typically assumed, with abundances tabulated as a function of pressure and temperature. When including 3D chemical mixing combining opacities of individual gases on-the-fly is extremely important as the abundances in a grid cell are not necessarily dependent on the local pressure and temperature, if the chemistry is not in equilibrium.

The model includes absorption opacity due to H$_2$O, CO, CO$_2$, CH$_4$, NH$_3$, HCN, Na, K, Li, Cs, and Rb, as well as collision-induced absorption (CIA) due to H$_2$-H$_2$ and H$_2$-He. We also include Rayleigh scattering due to H$_2$ and He. We note that CO$_2$ and HCN have been added to the model for this work and were not included in our previous modelling studies of hot Jupiters. The implementation of opacities in the model is described in detail in \citet{AmuBT14,AmuMB16,AmuTM17} and we refer the reader to \citet{GoyMS18} for the most up-to-date information on the sources of line-lists used in the model. Where available we use line lists from the ExoMol project\footnote{exomol.com}.

The treatment of chemistry in the exoplanet configuration of the UM has been incrementally improved and presented in a series of recent papers. Initially, \citet{AmuMB16} used a simple and extremely efficient analytical solution to equilibrium chemistry \citep{Burrows1999} for the species H$_2$O, CO, CH$_4$, and NH$_3$, combined with a simple parameterisation for the abundances of alkali species (Na, K, Li, Cs, and Rb) based on their condensation pressure-temperature curves. \citet{DruMB18} coupled a Gibbs energy minimisation scheme to the UM, also a local chemical equilibrium scheme, increasing the flexibility of the chemistry by allowing the numerical calculation of local chemical equilibrium abundances of a very large number of chemical species over a wide range of pressure, temperature, and element abundances.

Most recently, \citet{DruMM18a,DruMM18b} implemented a chemical relaxation scheme, based on the earlier work of \citet{CooS06}, which considers the time-dependent evolution of CO, CH$_4$, and H$_2$O and allows therefore the assumption of local chemical equilibrium to be relaxed for these species. This enables effects such as 3D chemical transport to be included, which is not possible when using a local chemical equilibrium scheme. Using a chemical relaxation scheme involves the estimation of a chemical timescale that represents either the time for interconversion to take place between two chemical species \citep[e.g. for CO and CH$_4$,][]{CooS06} or the time taken for a species to evolve back to its local chemical equilibrium, following a perturbation \citep[e.g.][]{TsaKL18}. In \citet{CooS06} the timescale for CO--CH$_4$ interconversion was derived using the rate-limiting step in the series of elementary reactions that link the two species. This approach was questioned in recent work by \citet{TsaKL18}.

In this work, we further improve the treatment of chemistry in the UM by coupling a flexible chemical kinetics scheme that was previously developed within the framework of the 1D atmosphere code ATMO \citep{Tremblin2015,DruTB16}. This improves on our earlier work \citep{DruMM18a,DruMM18b} by allowing for the inclusion of more chemical species, most importantly those that are included as opacities in the model. In addition, the chemical network that we use \citep{VenBD19} is derived from a larger network that has been experimentally validated across a wide-range of pressure and temperature \citep{Venot2012}, likely leading to a more accurate chemical evolution compared with the \citet{CooS06} chemical relaxation formulation.

In \cref{section:methods_chem} we describe the coupling of the chemical kinetics scheme to the UM. Then, in \cref{section:methods_param}, we describe the setup of the model including the various planet-star specific parameters for each simulation.

\subsection{Chemical kinetics in a 3D atmosphere model}
\label{section:methods_chem}
Implementing a chemical kinetics scheme in a 3D model such as the UM requires its coupling with the dynamics and radiative transfer schemes, and the selection of a suitable chemical network. 

\subsubsection{Coupling the chemical kinetics scheme to the model}

The chemical kinetics scheme that we couple to the UM was initially developed within the framework of the 1D atmosphere code ATMO. The scheme has previously been described in \citet{DruTB16}. The methods are similar to those taken in other chemical kinetics codes applied to hot exoplanet atmospheres presented in the literature \citep[e.g.][]{Moses2011,Kopparapu2012,Venot2012,AguVS2014,TsaLG2017}. We give a brief overview here. 

A chemical kinetics scheme requires the solution of a system of coupled ordinary differential equations (ODEs) that describe the time evolution of a set of chemical species. A chemical equilibrium scheme, in contrast, requires only the solution of algebraic equations. The terms in the ODEs correspond to chemical production and loss terms that are calculated based on a chosen chemical network, which describes the chemical reactions that link each of the chemical species. One of the fundamental choices to make when developing a chemical kinetics scheme is how to solve this system of ODEs.

We use the publicly available Fortran library {ODEPACK} \citep{Hindmarsh1983} to solve the system of ODEs. Specifically, we use the DLSODES routine which is the variant of the package suitable for stiff systems, which uses an implicit Backward Differentiation Formulae (BDF) method \citep{Radhakrishnan1993}. Several user-defined options are required inputs for DLSODES. We use a relative tolerance 1$\times10^{-3}$ and we use an order of 2 in the BDF method (i.e. the results from two previous iterations are used in the calculation for the current iteration). The step size of each iteration is internally generated by DLSODES, though we impose minimum and maximum timesteps of 1$\times10^{-10}$~s and 1$\times10^{10}$~s, respectively. The user must also set a method flag (MF) to control the operation of DLSODES. Here we use MF=222 which is the option for stiff problems using an internally generated (numerically derived) Jacobian, as opposed to user-supplied analytical Jacobian.

Tracer advection is handled by the well-tested semi-implicit, semi-Lagrangian advection scheme of the ENDGame dynamical core \citep{WooSW14}. The tracers (one for each chemical species, totalling 30 in this case) are initialised to chemical equilibrium abundances at the start of the model integration, calculated using the coupled Gibbs energy minimisation scheme \citep{DruMB18}. The quantity that is advected is the mass mixing ratio.

We call the chemical kinetics scheme, that handles the chemical evolution, individually for each cell of the 3D model grid on a frequency that corresponds to a chemical timestep. The chemical timestep is longer than the dynamical timestep meaning that the chemistry is called less frequently than every dynamical timestep, motivated by improving the computational efficiency. A similar approach is taken for the radiative transfer calculations \citep{AmuMB16}. Details of the dynamical, radiative, and chemical timesteps are given in \cref{section:methods_param}, and these were selected to provide accuracy and efficiency. Since the chemistry calculations are done separately from the advection step (i.e. operator splitting), the process could be thought of as solving a series of 0D box models, one for each cell of the 3D grid.

The initial values for chemical integration are the current values of the advected tracers converted from mass mixing ratio into number densities using the total gas number of that grid cell. The time-stepping in the chemical kinetics iterations is done using an adaptive timestep, starting with an initial small timestep of 1$\times10^{-10}$~s. Subsequent timesteps are controlled by the output of the DLSODES routine, as described above. Chemical iterations continue until either the integration time is equal to the set chemical timestep or the chemical system has reached a steady-state. Steady-state is assessed via the maximum relative change in the number densities of the individual species in an approach similar to \citet{TsaLG2017}. We find that continuing iterations significantly beyond the point at which the system has reached a steady-state can lead to numerical errors. This means that for grid cells in the deep high-pressure, high-temperature region where the chemistry reaches a steady-state relatively quickly (as the chemical timescale is short) we may integrate the chemistry for a time less than the chemical timestep. For cells at lower pressures and temperatures (where the chemical timescale is long) the chemistry iterations are more likely to be stopped by reaching an integration time equivalent to the chemical timestep.

\subsubsection{Chemical network}

There are now a large number of chemical networks that have been used to model the atmospheres of hot exoplanets \citep[e.g.][]{Liang2003,ZahMF09,Moses2011,Venot2012,RimH16,TsaLG2017}. Since the early work of \citet{Liang2003} the trend has been for such networks to grow in size and complexity. However, more recently the focus has shifted to the development of smaller chemical networks \citep{TsaLG2017,VenBD19} that are less computationally demanding, making them feasible for use in a 3D atmosphere model.

We use the `reduced' chemical network recently developed by \citet{VenBD19}. The network comprises 30 chemical species and 181 reversible reactions. This network was constructed by reducing the larger (105 species, $\sim$1000 reversible reactions) network of \citet{Venot2012} using commercial software. The original network of \citet{Venot2012} was experimentally validated for a wide-range of pressure and temperature relevant to hot Jupiter atmospheres. The \citet{Venot2012} network, excluding photodissociations, was reduced by eliminating chemical species and reactions while maintaining a set level of accuracy for six target species: H$_2$O, CH$_4$, CO, CO$_2$, NH$_3$, and HCN. The reduced network was compared against the original network within a 1D chemical-diffusion model and results for the six target species agreed extremely well across a remarkably large parameter space (pressure, temperature, and element abundances). The network is publicly available via the KIDA database\footnote{\url{http://kida.obs.u-bordeaux1.fr/}} \citep{WakHL12}.

Recently, \citet{VenCB19} released updates to their original chemical network \citep{Venot2012} with a focus on improvements to the chemistry of methanol. They also re-derived a reduced chemical network from the updated network. Negligible differences were found in the quench points of several chemical species, using a 1D photochemical model, between the original and updated networks for hot Jupiters (specifically HD~209458b and HD~189733b were tested). On the other hand, significant differences were found for cooler atmospheres. As we only consider hot Jupiter atmospheres in this study, our use of the \citet{VenBD19} chemical network, based on the original \citet{Venot2012} network without updates to the methanol chemistry, does not affect our results.

We note that we only consider thermochemical kinetics and transport-induced quenching in this work and do not include photochemistry (photodissociations or photoionisations). Nonetheless the model developments presented in this work represent a significant step forwards in exoplanet climate modelling. Inclusion of photochemistry represents a significant further model development.

\subsection{Model setup and simulation parameters}
\label{section:methods_param}

The basic model setup is similar to that used for previous hot, hydrogen-dominated atmosphere simulations with the UM \citep[e.g.][]{AmuMB16}. We use 144 longitude points and 90 latitude points giving a horizontal spatial resolution of 2.5$^{\circ}$ and 2$^{\circ}$ in longitude and latitude, respectively. The UM uses a geometric height-based grid, for which we use 66 levels, as opposed to a pressure-based grid \citep[see Section 2.5 of ][for details]{MayBA14}. For each planet setup we adjust the height at the top-of-atmosphere ($z_{\rm toa}$) to reach approximately the same minimum pressure range ($\sim1$~Pa) for the initial profile. The bottom boundary of the model has a pressure of $\sim2\times10^7$~Pa. The model includes a vertical damping ``sponge'' layer to reduce vertically propagating waves resulting from the rigid boundaries of the model, and we use the default setup for this as described in \citet{MayBA14} and \citet{AmuMB16}. 

For the planetary and stellar parameters we use the values from the TEPCat database\footnote{\url{https://www.astro.keele.ac.uk/jkt/tepcat/}}, which are given in \cref{table:params}. For the stellar spectra we use the Kurucz spectra for HD~209458 and HD~189733\footnote{\url{http://kurucz.harvard.edu/stars.html}}.

\begin{table*}
\centering
\setlength\extrarowheight{3pt}
\begin{tabular}{l c c c}
\hline\hline
Parameter  & Unit & HD 209458b & HD 189733b \\
\hline
Mass, $M_{\rm P}$ & kg & $1.36\times10^{27}$ & $2.18\times10^{27}$ \\
Planet radius, $R_{\rm P}$ & m &  $9.65\times10^{7}$ & $8.05\times10^7$  \\
Star radius, $R_{\rm S}$ & m & $8.08\times10^8$ & $5.23\times10^8$  \\
Semi major axis, $a$ & AU & 0.04747 & 0.03142  \\
Surface gravity, $g_{\rm surf}$ & ms$^{-2}$ & 9.3 & 21.5  \\
Intrinsic temperature, $T_{\rm int}$ & K & 100 & 100  \\
Lower boundary pressure, $P_{\rm lower}$ & Pa & $2\times10^7$ & $2\times10^7$  \\
Top-of-atmosphere height, $z_{\rm TOA}$ & m & $9.0\times10^6$ & $3.2\times10^6$  \\
Rotation rate, $\Omega$ & s$^{-1}$ &  $2.06\times10^{-5}$ & $3.23\times10^{-5}$  \\
Specific heat capacity, $c_P$ & ${\rm J}~{\rm kg}^{-1}~{\rm K}^{-1}$ &  $1.28\times10^4$ & $1.25\times10^4$  \\
Specific gas constant, $R$ & ${\rm J}~{\rm kg}^{-1}~{\rm K}^{-1}$ &  3516.6 & 3516.1  \\
\hline
\end{tabular}
\caption{Planetary and stellar parameters for each simulation.}
\label{table:params}
\end{table*}

For the element abundances we use a solar composition with values taken from \citet{Caffau2011}. The model requires global values of the specific heat capacity ($c_{P}$) and the specific gas constant ($R=\bar{R}/\mu$, where $\bar{R}=8.314~{\rm J}~{\rm K}^{-1}~{\rm mol}^{-1}$ is the molar gas constant and $\mu$ is the mean molecular weight). We calculate vertical profiles of $c_P$ and $R$ for the initial 1D thermal profiles and use the straight average of these profiles for the globally uniform values required as model inputs, as previously described in \citet{DruMB18}.

We use a main model (dynamical) timestep of 30 s which we find is a good compromise between efficiency and stability \citep{AmuMB16}. The radiative heating rates are updated every 5 dynamical timesteps, giving a radiative transfer timestep of 150 s. The chemistry scheme is called every 125 dynamical timesteps giving a chemical timestep of 3750 s or approximately 1 hour. Using a more efficient model setup (replacing the radiative transfer with a Newtonian cooling approach) we found little difference in the model result after 1000 (Earth) days of simulation between a model using a chemical timestep of 10 dynamical timesteps and 125 dynamical timesteps. We therefore choose to use a longer chemical timestep here simply for improved computational efficiency.

The model is initialised at rest (i.e. zero wind velocities) and with a horizontally-uniform thermal structure. The initial vertical thermal profile is calculated using the 1D radiative-convective equilibrium model ATMO, using the same planetary and stellar parameters \cref{table:params}. We use a zenith angle of $\cos\theta = 0.5$ for the ATMO calculation of the initial thermal profile, representative of a dayside average \citep[e.g.][]{ForM07}. For the chemical kinetics simulations the chemical tracers are initialised to the chemical equilibrium values in each grid cell. Since the initial thermal field is horizontally uniform, this entails that the initial chemical tracer values are also horizontally uniform.

We integrate the model for 1000 Earth days (i.e. 8.64$\times10^7$~s) by which point the maximum wind velocities and chemical abundances have approximately ceased to evolve (see \cref{section:appendix_evo}). A common problem with simulations of gas-giant planet atmospheres is the slow evolution of the deep atmosphere. Here, where the atmosphere is initialised with a ``cold'' initial thermal profile, the radiative and dynamical timescales are too long to feasibly reach a steady-state due to computational limitations \citep{RauscherMenou2012,MayBA14,AmuMB16,sainsbury_martinez_2019}. For HD~209458b we find no significant differences when initialising with a hotter thermal profile (\cref{section:hot_initial}). Conservation of several quantities is presented in \cref{section:appendix_con}

Chemical kinetics calculations are typically computationally demanding because they require the solution of a system of stiff ODEs. In our simulations, including the chemical kinetics calculations increases the computational cost by at least a factor of two, compared with an otherwise similar setup that uses the Gibbs energy minimisation (chemical equilibrium) scheme. However, the simulations remain computationally feasible and required approximately one-week of walltime using $\sim$200 cores on the DiRAC DIaL supercomputing facility\footnote{\url{https://www2.le.ac.uk/offices/itservices/ithelp/services/hpc/dirac/about}}, as an example. Using a larger chemical network would increase the computational cost because (1) chemistry calculations would become more expensive and, (2) more advected tracers would be required.

In \cref{section:results} we present results from four simulations in total. We model the atmospheres of HD~209458b and HD~189733b both under the assumption of local chemical equilibrium and including the effect of 3D transport. For the former we use the coupled Gibbs energy minimisation scheme \citep{DruMB18} to calculate the chemical abundances and we refer to these simulations throughout simply as the ``equilibrium'' simulations. For the latter we use the newly coupled chemical kinetics scheme and include tracer advection and we refer to these simulations as the ``kinetics'' simulations.

%% file: results.tex

\section{Results}
\label{section:results}

We first focus on the overall circulation and thermal structure resulting from our simulations, before moving on to describe the chemical structure, the chemical-radiative feedback, and the impacts on the synthetic observations.

\subsection{Circulation and thermal structure}
\label{section:circ_temp}

\begin{figure}
  \begin{center}
\includegraphics[width=0.45\textwidth]{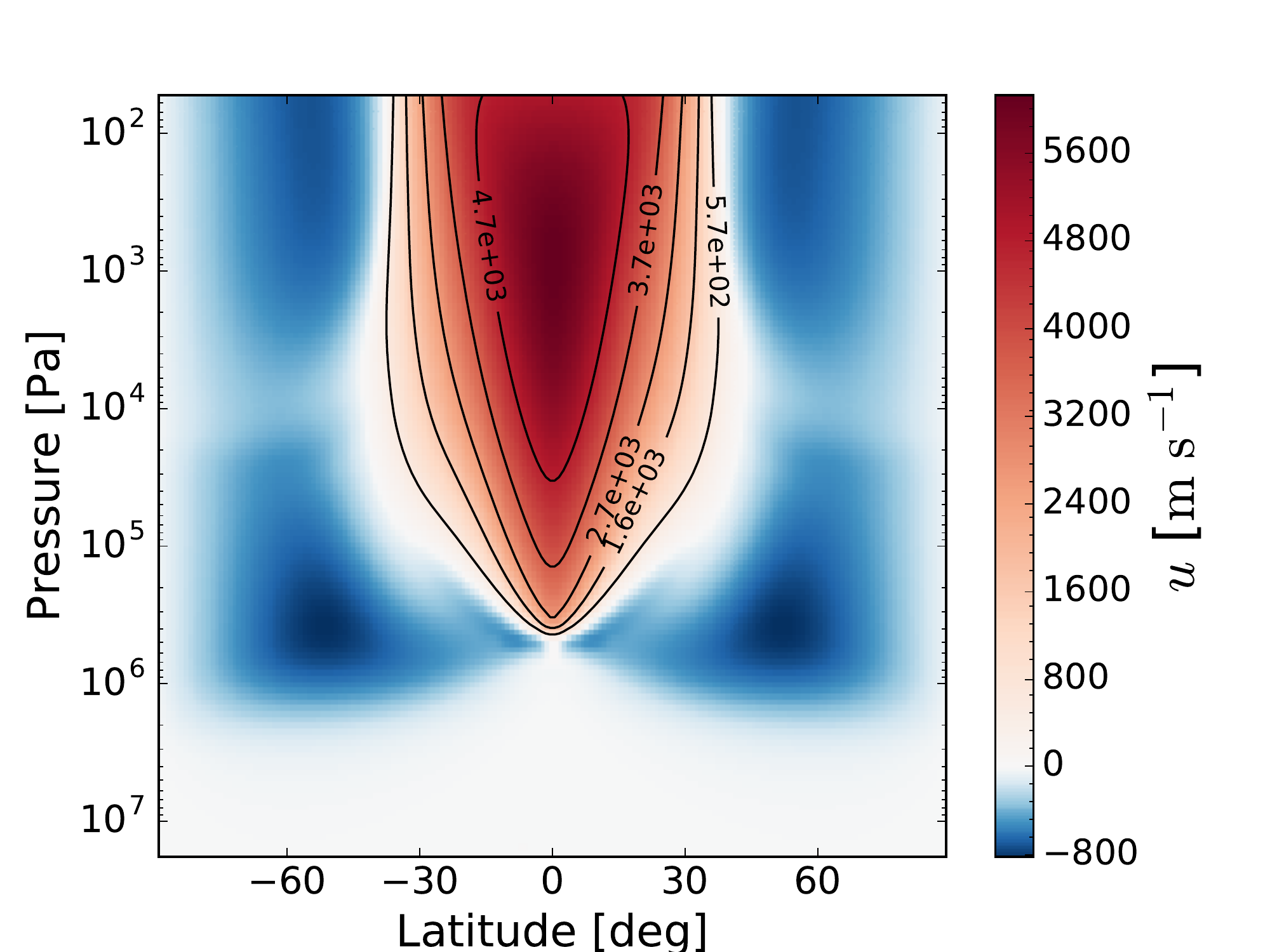} \\
    \includegraphics[width=0.45\textwidth]{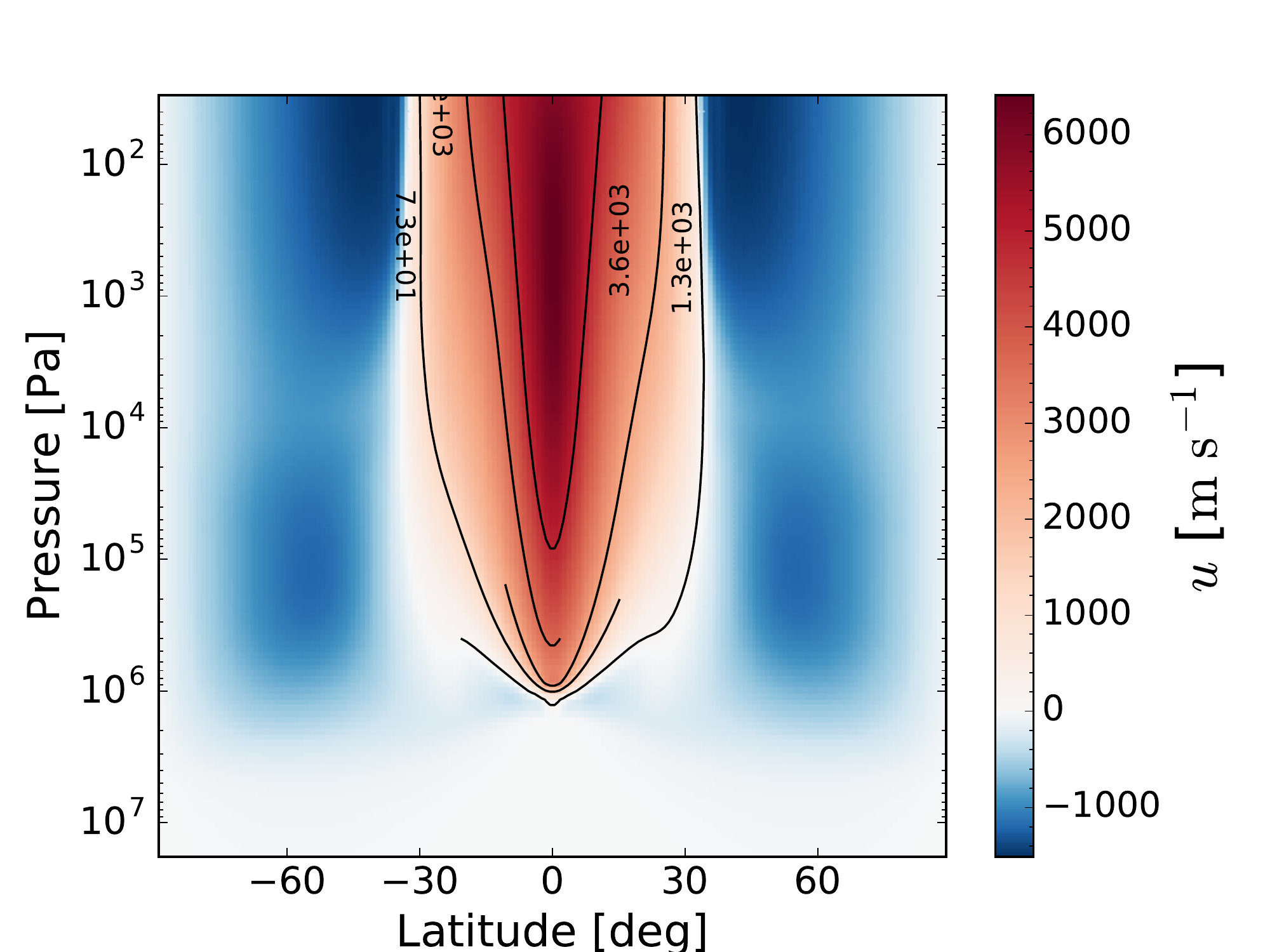} 
  \end{center}
\caption{Zonal-mean zonal wind velocity for the equilibrium simulations of HD~209458b (top) and HD~189733b (bottom).}
\label{figure:zonal_wind}
\end{figure}

\begin{figure}
  \begin{center}
    \includegraphics[width=0.5\textwidth]{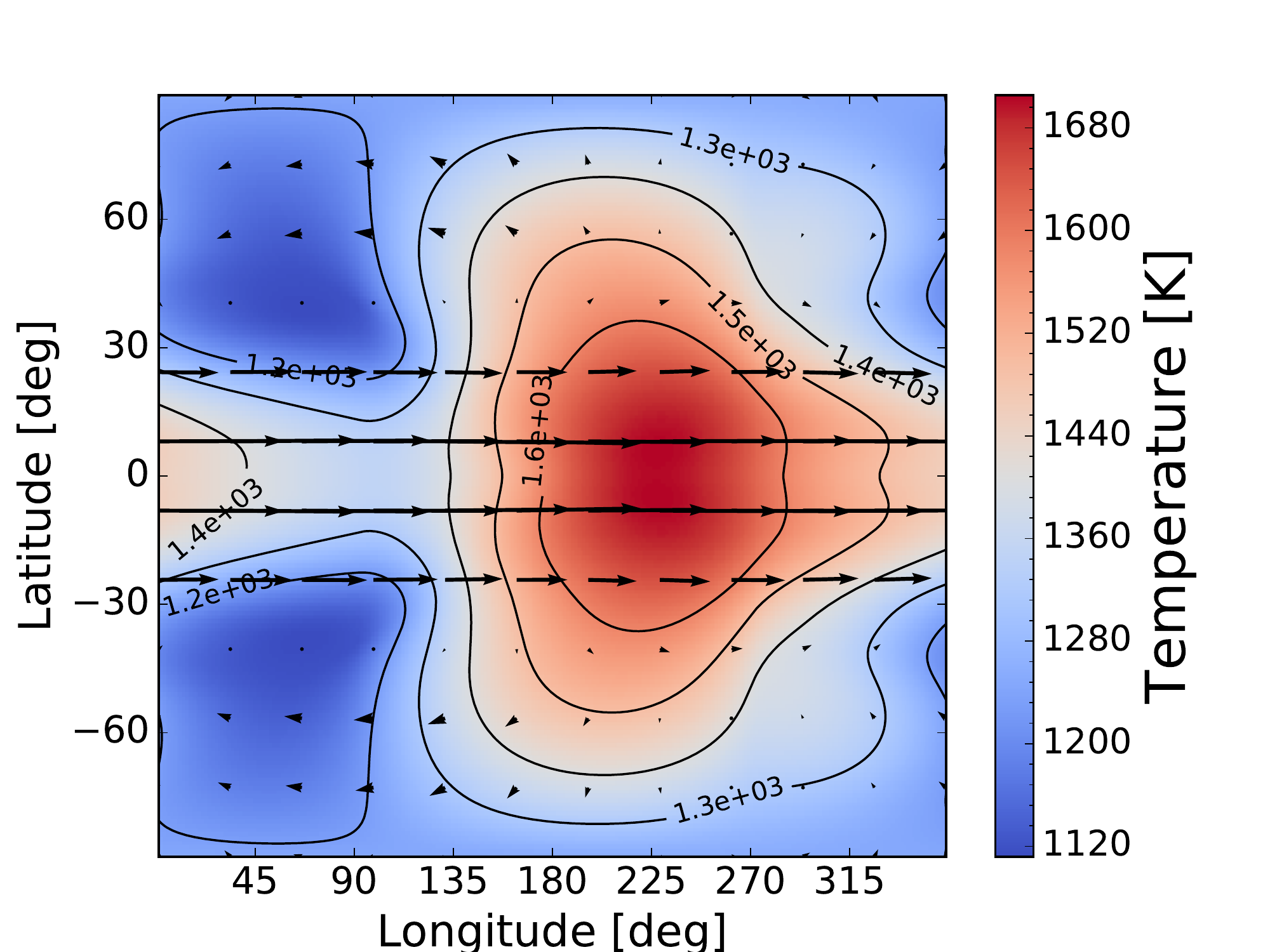} \\
    \includegraphics[width=0.5\textwidth]{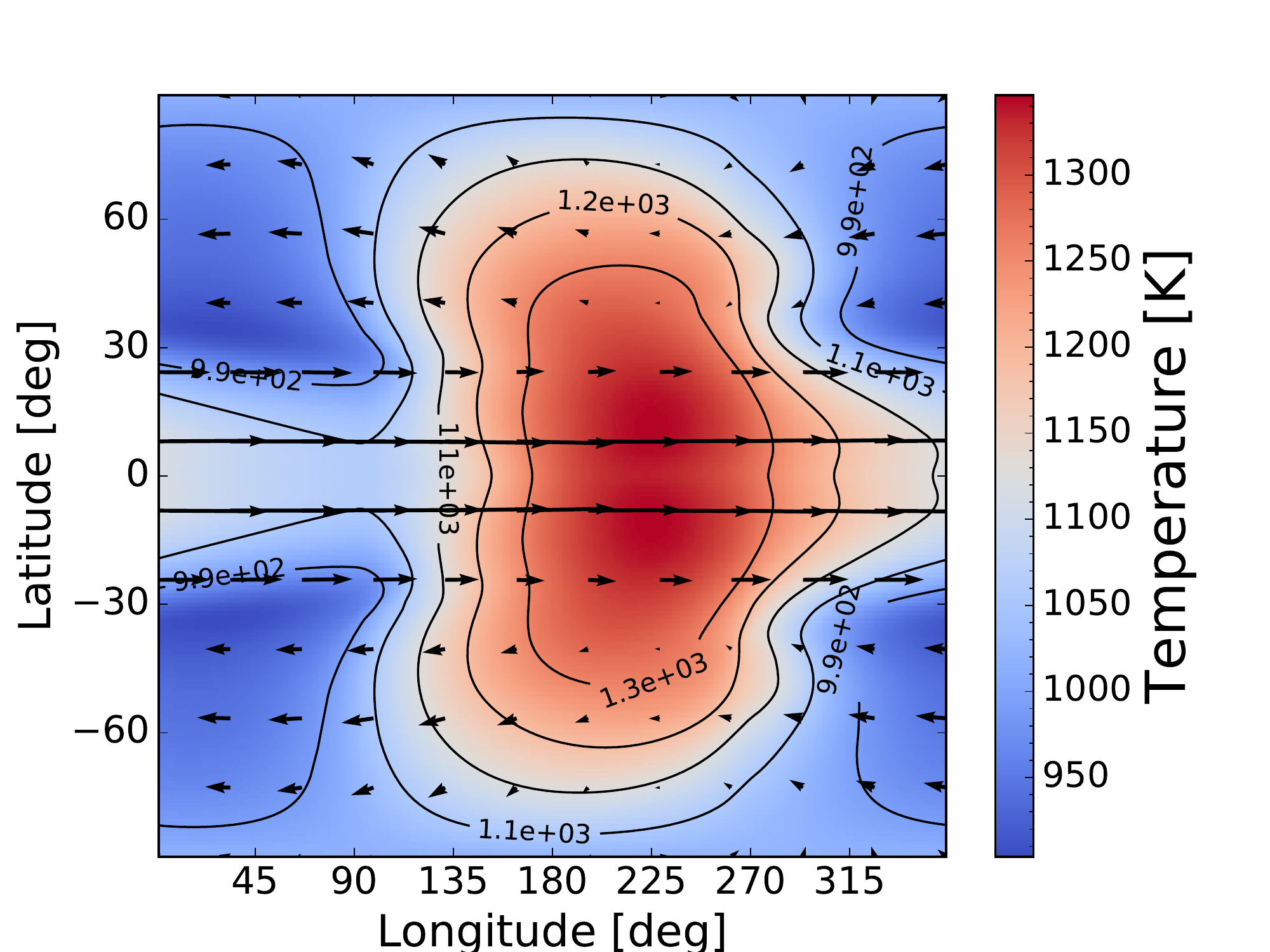} 
  \end{center}
\caption{Atmospheric temperature (colour scale and black contours) on the $1\times10^4$ Pa (0.1 bar) isobar with horizontal wind velocity vectors for the equilibrium simulations of HD~209458b (top) and HD~189733b (bottom).}
\label{figure:temp_p1e4}
\end{figure}

\begin{figure}
  \begin{center}
    \includegraphics[width=0.5\textwidth]{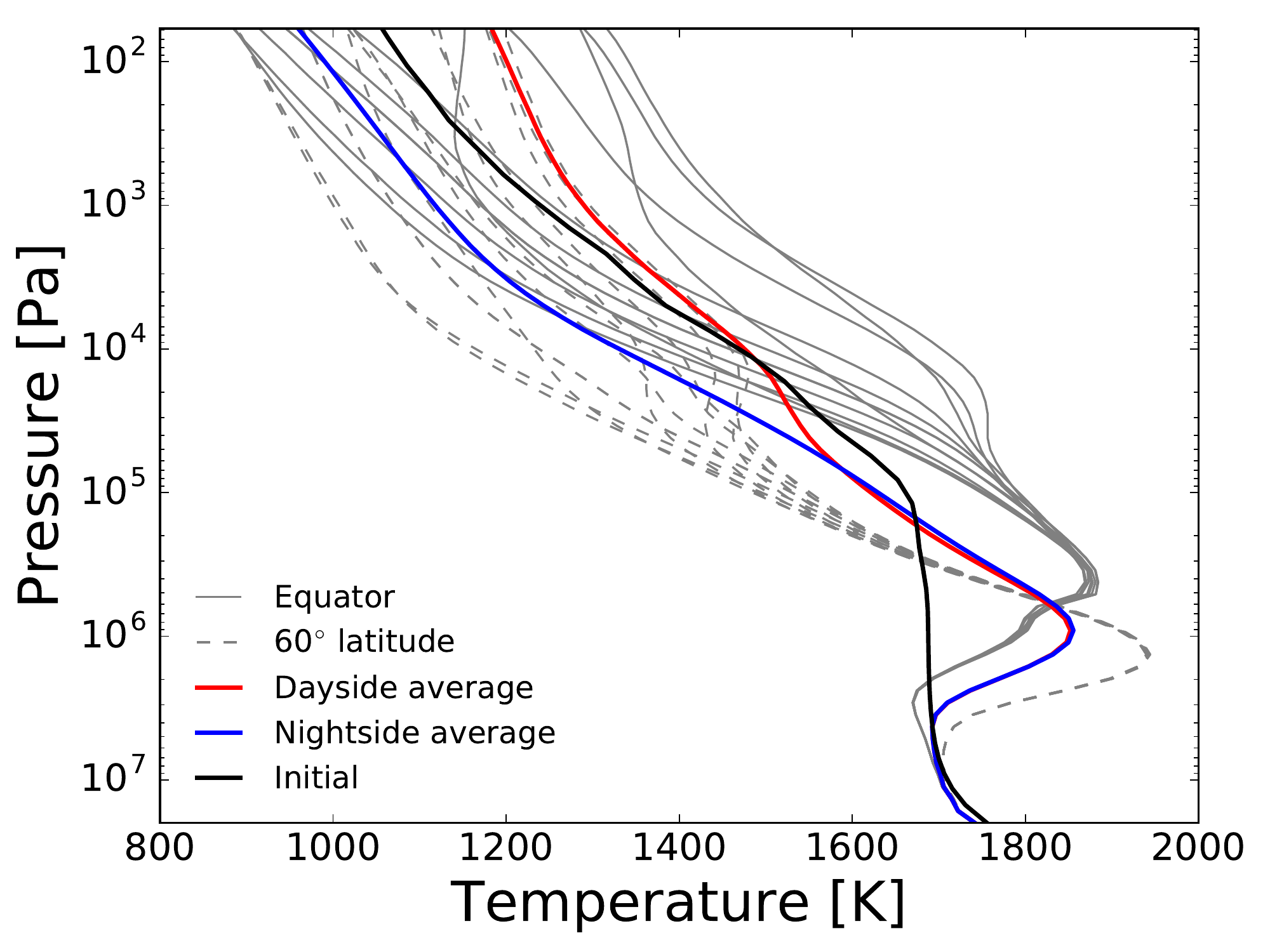} \\
    \includegraphics[width=0.5\textwidth]{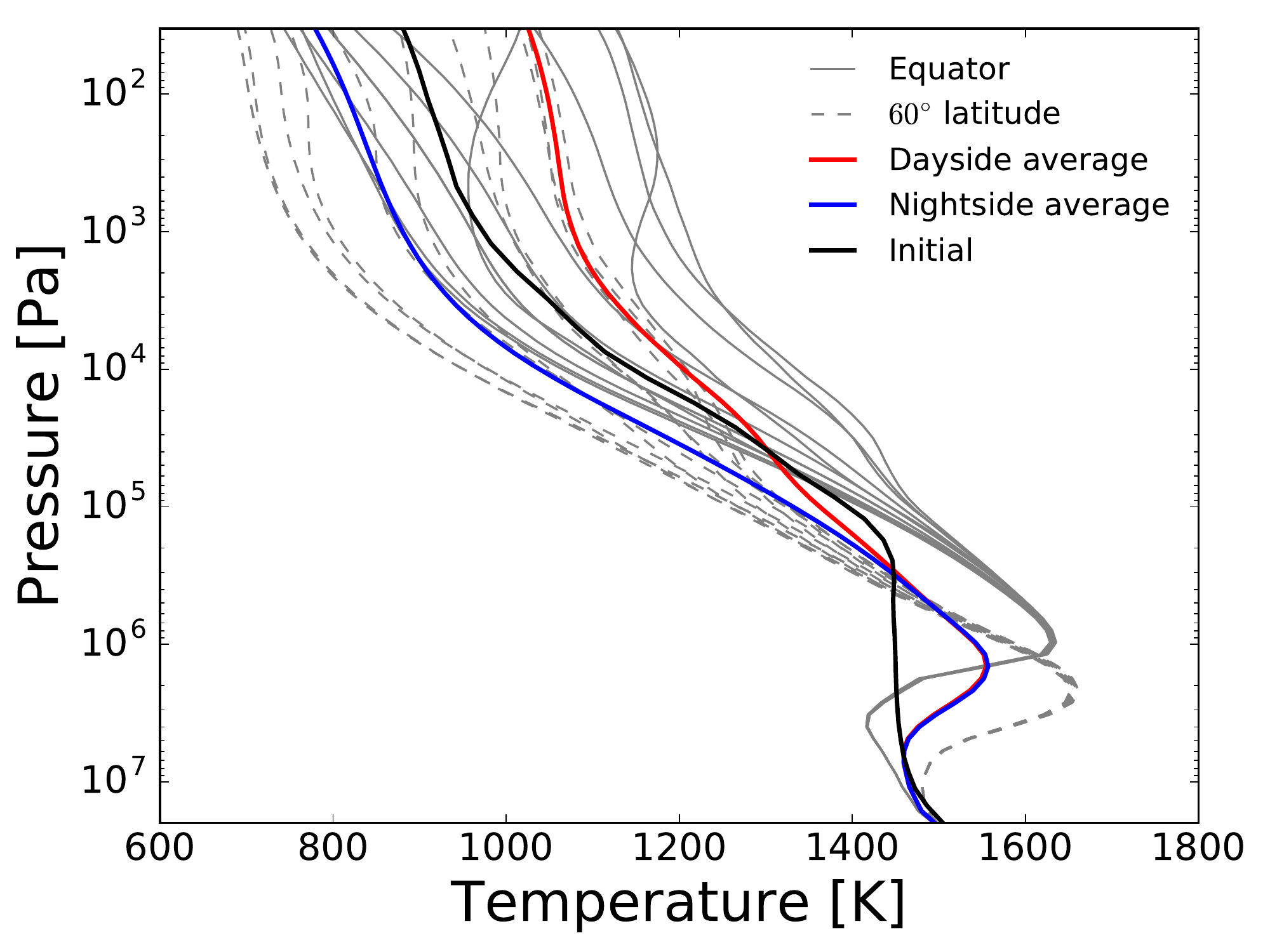} 
  \end{center}
\caption{Pressure-temperature profiles extracted from the 3D grid of equilibrium simulations of HD~209458b (top) and HD~189733b (bottom). Solid grey lines show profiles at various longitude points around the equator ($0^{\circ}$ latitude) while dashed grey lines show profiles at various longitude points at a latitude of $60^{\circ}$. Area-weighted dayside average (red) and nightside average (blue) profiles are shown, along with the 1D radiative-convective equilibrium model profile used to initialise the 3D model (black).}
\label{figure:temp_prof}
\end{figure}

The temporal-mean (800--1000~days) zonal-mean zonal wind for the equilibrium simulations of HD~209458b and HD~189733b are shown in \cref{figure:zonal_wind}. For both planets, the circulation is dominated by a superrotating equatorial zonal jet, which is typical of close-in, tidally-locked atmospheres and is shown to be present across a wide parameter-space  \citep[e.g.][]{KatSL16}. The maximum zonal-mean zonal wind velocity is $\sim6$~km~s$^{-1}$ in both cases. At higher latitudes, and also at higher pressures, the flow is westward (retrograde) with maximum wind velocities of $\sim1$~km~s$^{-1}$.

The maximum zonal-mean zonal wind for HD~209458b in previous results in the literature varies between 5 km~s$^{-1}$ and 6 km~s$^{-1}$ \citep{Showman2009,KatSF14} while for HD~189733b it varies between 3.5 km~s$^{-1}$ and 6 km~s$^{-1}$ \citep{Showman2009,KatSF14,LeeDH16,FloBR19}. The jet speed depends on the magnitude of dissipation implemented in the model (required for numerical stability) and is currently unconstrained by observation \citep{heng_2011,MayBA14}.

The atmospheric temperature on the $10^4$~Pa isobar for the equilibrium simulations of both planets is shown in \cref{figure:temp_p1e4}. The thermal structure is qualitatively quite similar between the two planets, with a hot dayside and relatively much cooler nightside and a hotspot that is shifted eastward of the substellar point. HD~209458b is generally a few hundred Kelvin warmer than HD~189733b.

\cref{figure:temp_prof} shows pressure-temperature profiles extracted from the 3D grid of equilibrium simulations of HD~209458b and HD~189733b, as well as dayside and nightside average profiles. The 1D radiative-convective equilibrium pressure-temperature profile used to initialise each 3D simulation is also shown. Both atmospheres have significant zonal temperature gradients for $P\lesssim10^5$~Pa with maximum day-night temperature contrasts of $\sim500$~K. There is also a significant latitudinal thermal gradient, at a given longitude and pressure, of several hundred Kelvin between the equator and mid-latitudes.

A high-pressure thermal inversion at $P\sim10^6$~Pa is present in both equilibrium simulations. This is a result of the deep atmosphere slowly converging towards a temperature profile that is hotter than the 1D radiative-convective equilibrium profile used to initialise the model (shown in black in \cref{figure:temp_prof}). Using a 2D steady-state circulation model, \citet{TreCM17} showed that the steady--state of the deep atmosphere is actually significantly hotter than predicted by 1D radiative-convective equilibrium models, likely due to the advection of potential temperature. It has also been noted that 3D simulations appear to slowly evolve towards hotter profiles in the deep atmosphere \citep[][]{AmuMB16}, however very long timescale simulations are required to actually reach the end state \citep[as performed in][]{sainsbury_martinez_2019}. Therefore, the results for pressures greater than $\sim10^6$~Pa are likely to be strongly effected by the initial condition.

The atmospheric circulation and thermal structure from the equilibrium simulations of HD~209458b and HD~189733b are almost identical to results presented in our previous work \citep{DruMM18a,DruMM18b}. Minor differences result from a number of changes in the model setup and chosen parameters, including, in no particular order: use of a different chemical equilibrium scheme (Gibbs energy minimisation versus analytical chemical equilibrium formulae), slight adjustments to some planetary and stellar parameters \citep[compare the parameters for HD~209458b and HD~189733b in \cref{table:params} with those presented in][respectively]{DruMM18a,DruMM18b}, and the addition CO$_2$ and HCN as opacity sources which were not previously included. Differences in the wind velocities and temperatures between the equilibrium and kinetics simulations due to chemical-radiative feedback are discussed later in \cref{subsection:Chemical-radiative-feedback}.

\subsection{Chemical structure}
To compare the differing chemical structure resulting from either the equilibrium or kinetics approach, we focus on each planet separately.  

\subsubsection{HD~209458b}
\label{results:209}
\begin{figure*}
  \begin{center}
    \includegraphics[width=0.45\textwidth]{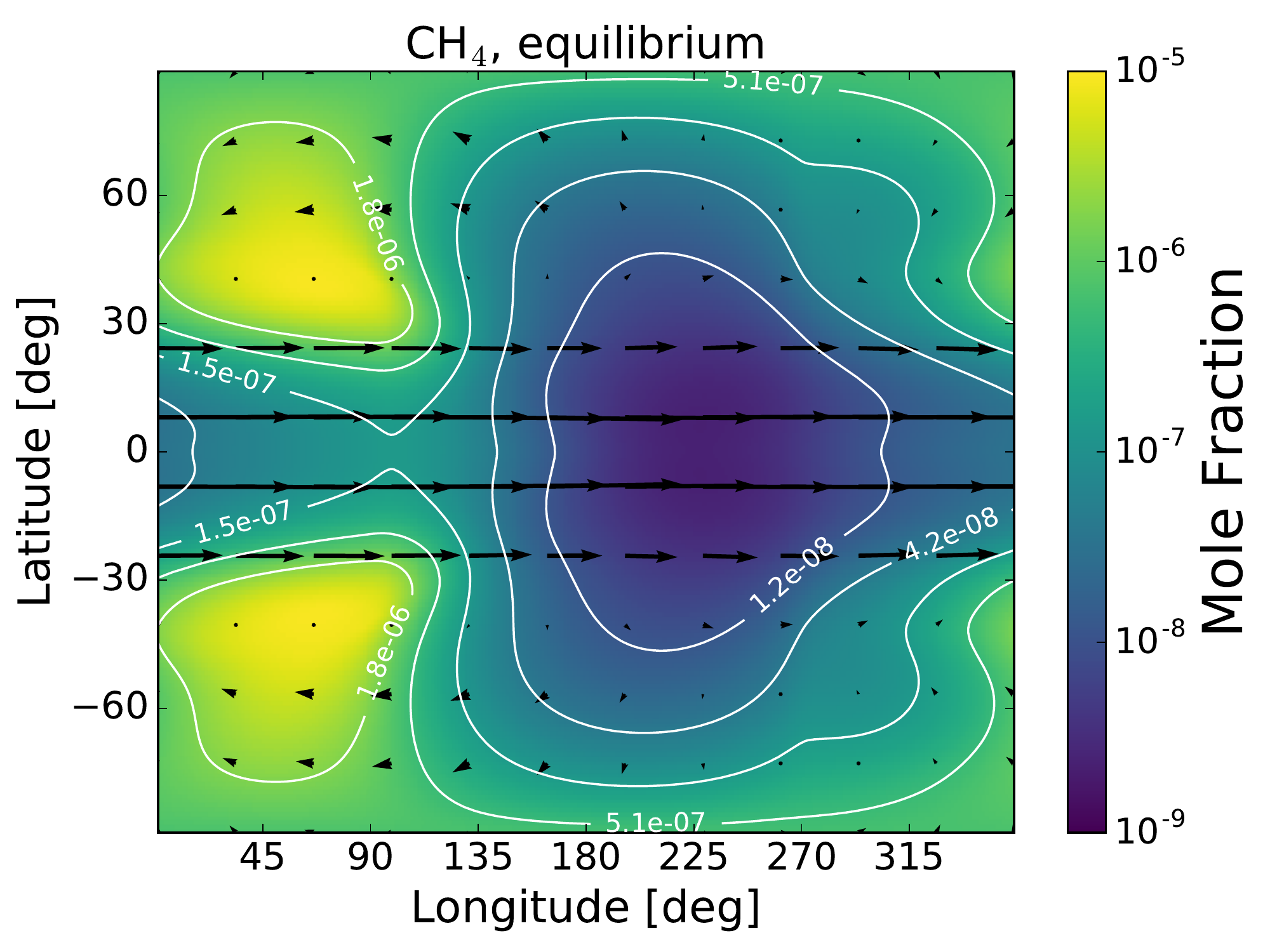}
    \includegraphics[width=0.45\textwidth]{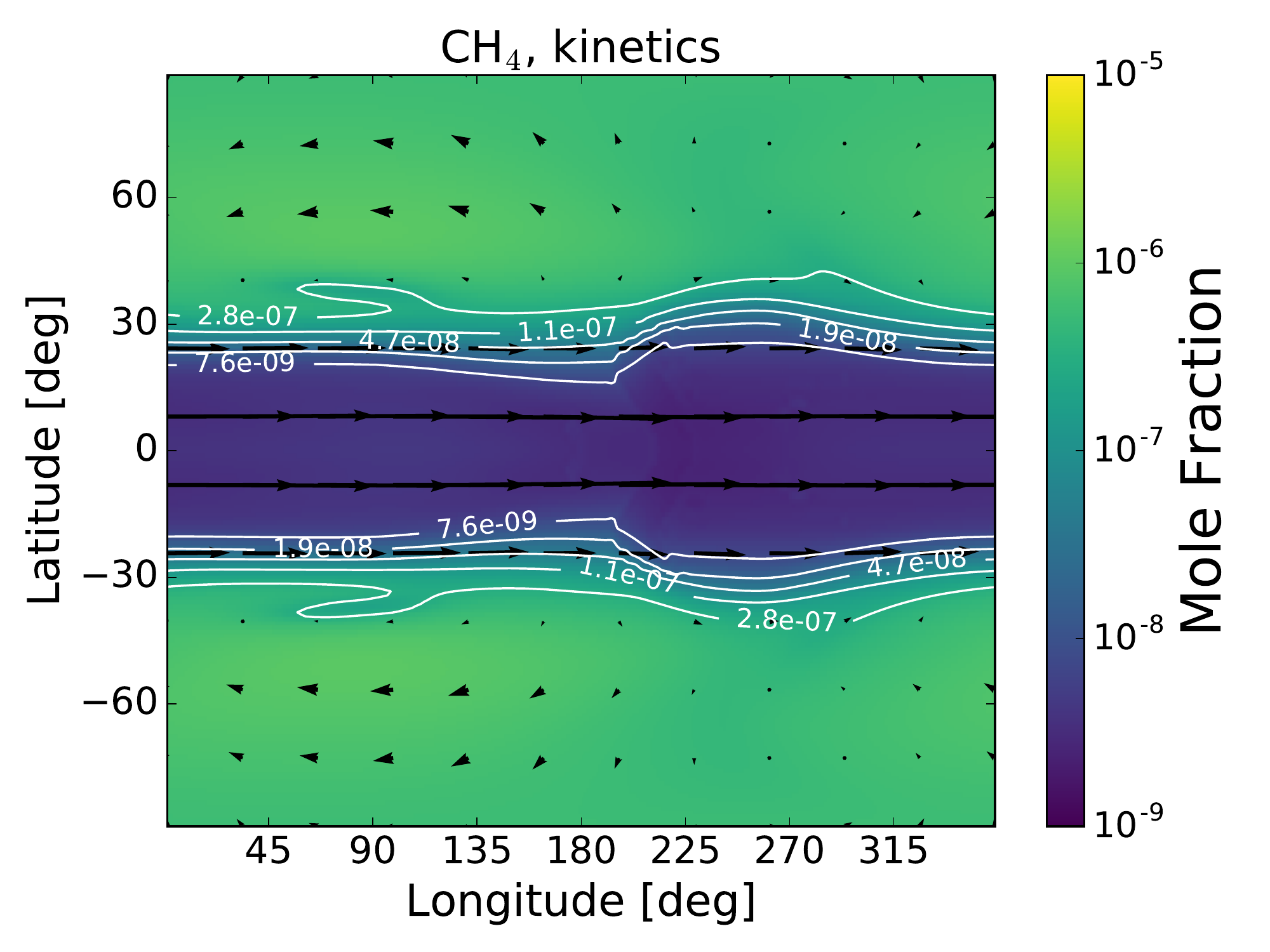} \\
     \includegraphics[width=0.45\textwidth]{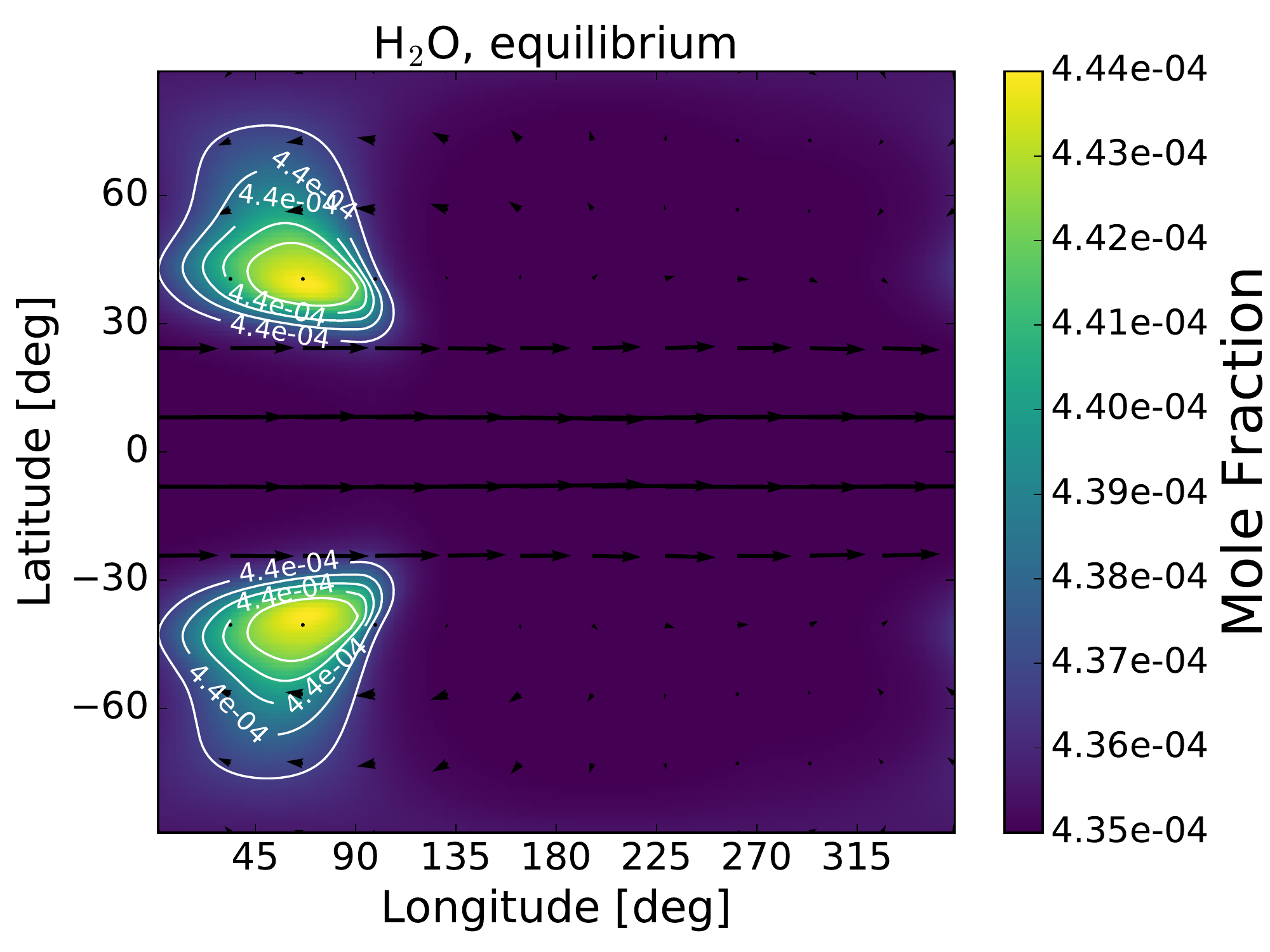}
    \includegraphics[width=0.45\textwidth]{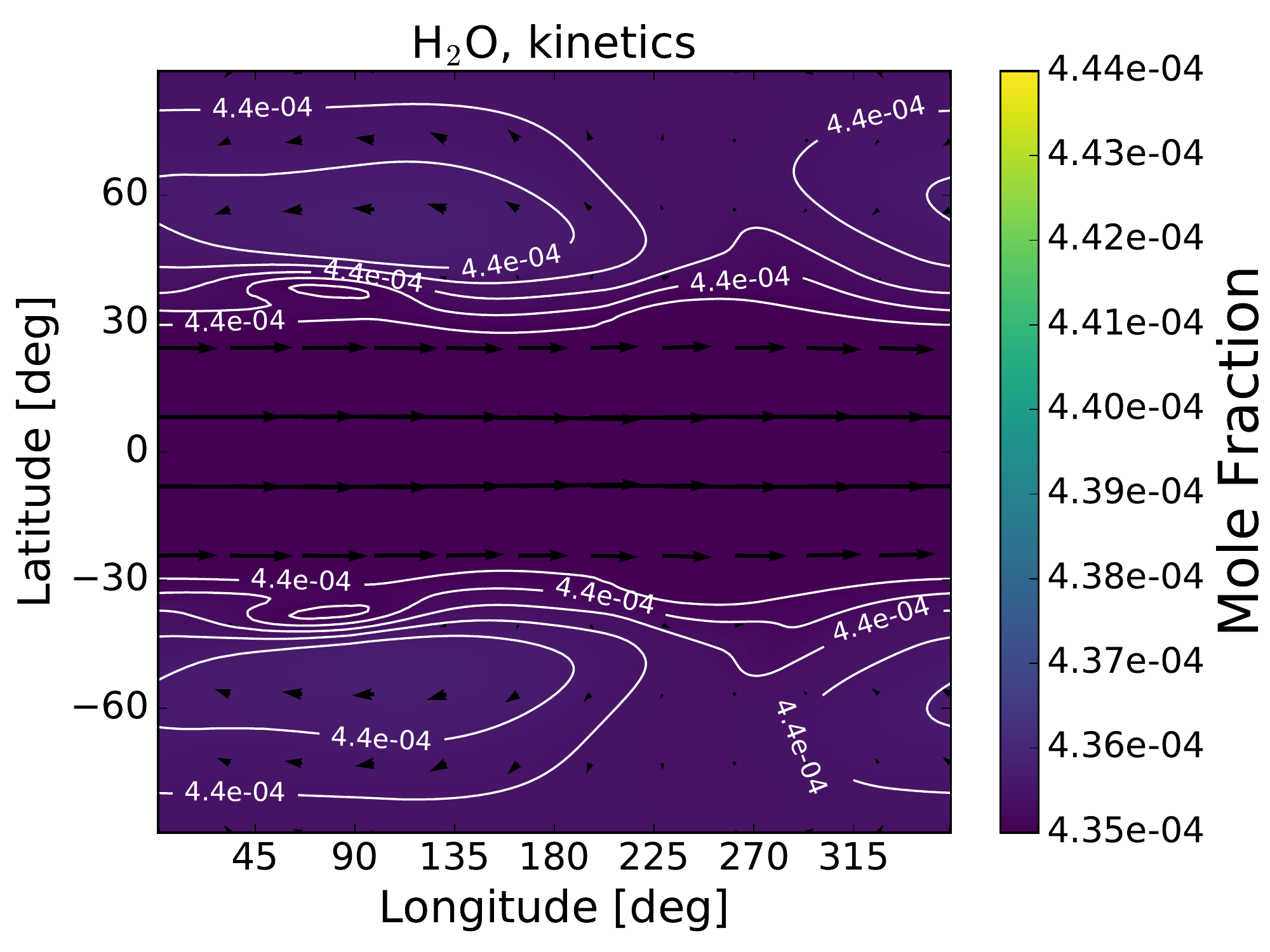} \\
     \includegraphics[width=0.45\textwidth]{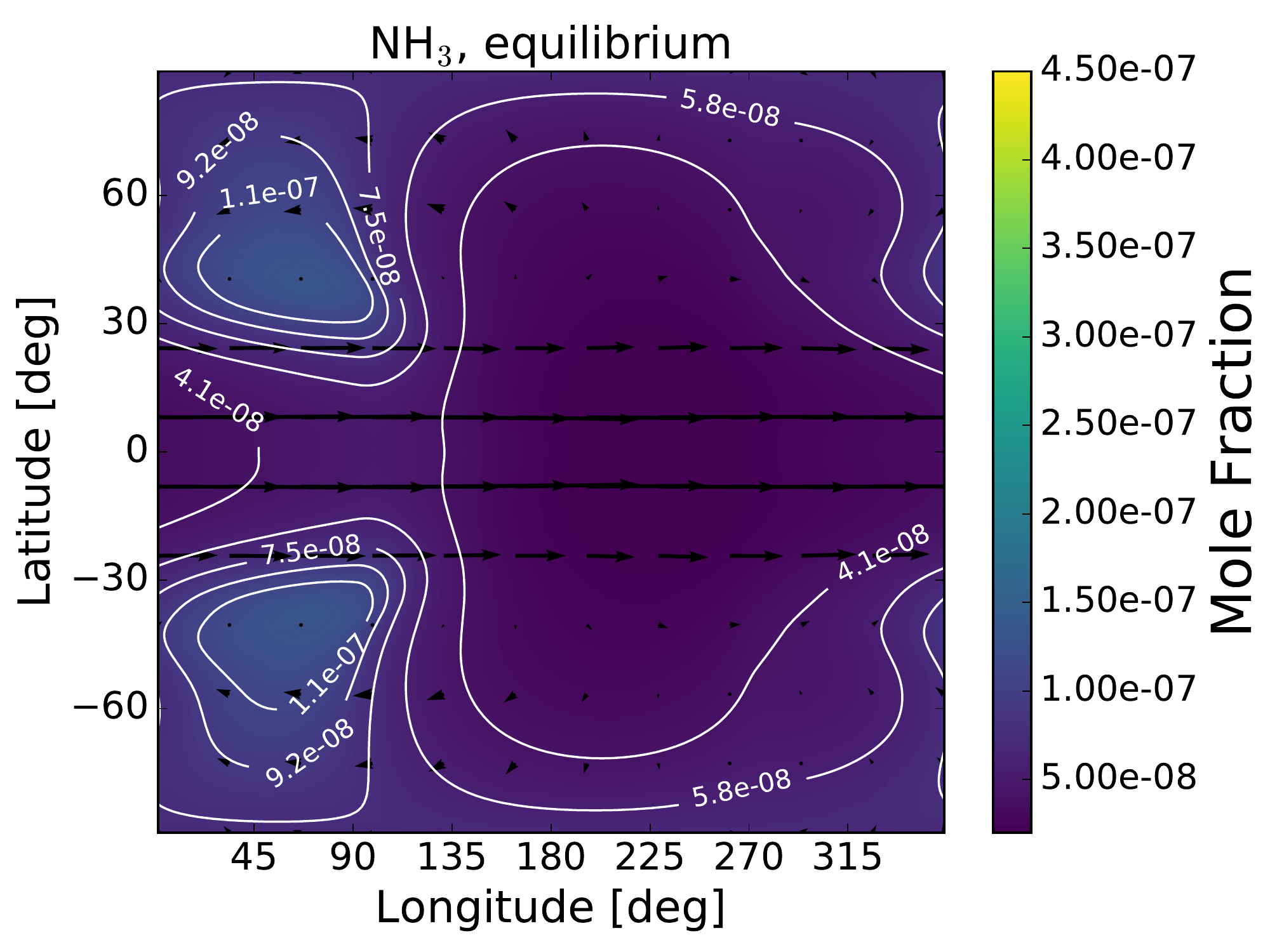}
    \includegraphics[width=0.45\textwidth]{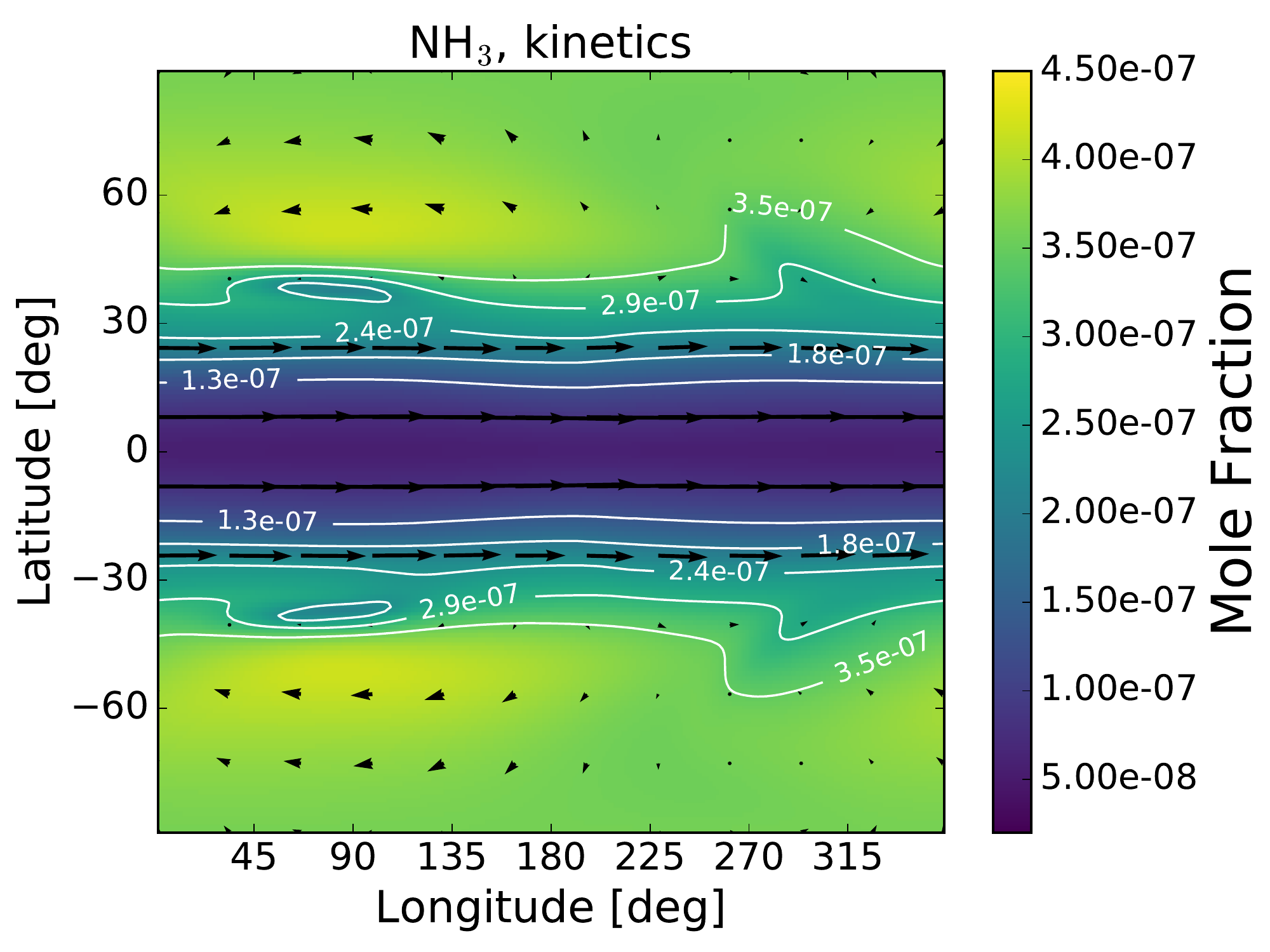} \\
  \end{center}
\caption{Mole fractions (colour scale and white contours) of CH$_4$, H$_2$O, NH$_3$, CO, CO$_2$ and HCN and wind vectors (arrows) on the $1\times10^4$ Pa (0.1 bar) isobar for the equilibrium simulation (left column) and the kinetics simulation (right column) of HD~209458b.}
\label{figure:hd209_mf_eq}
\end{figure*}

\begin{figure*}
  \ContinuedFloat
    \captionsetup{list=off,format=cont}
  \begin{center}
    \includegraphics[width=0.45\textwidth]{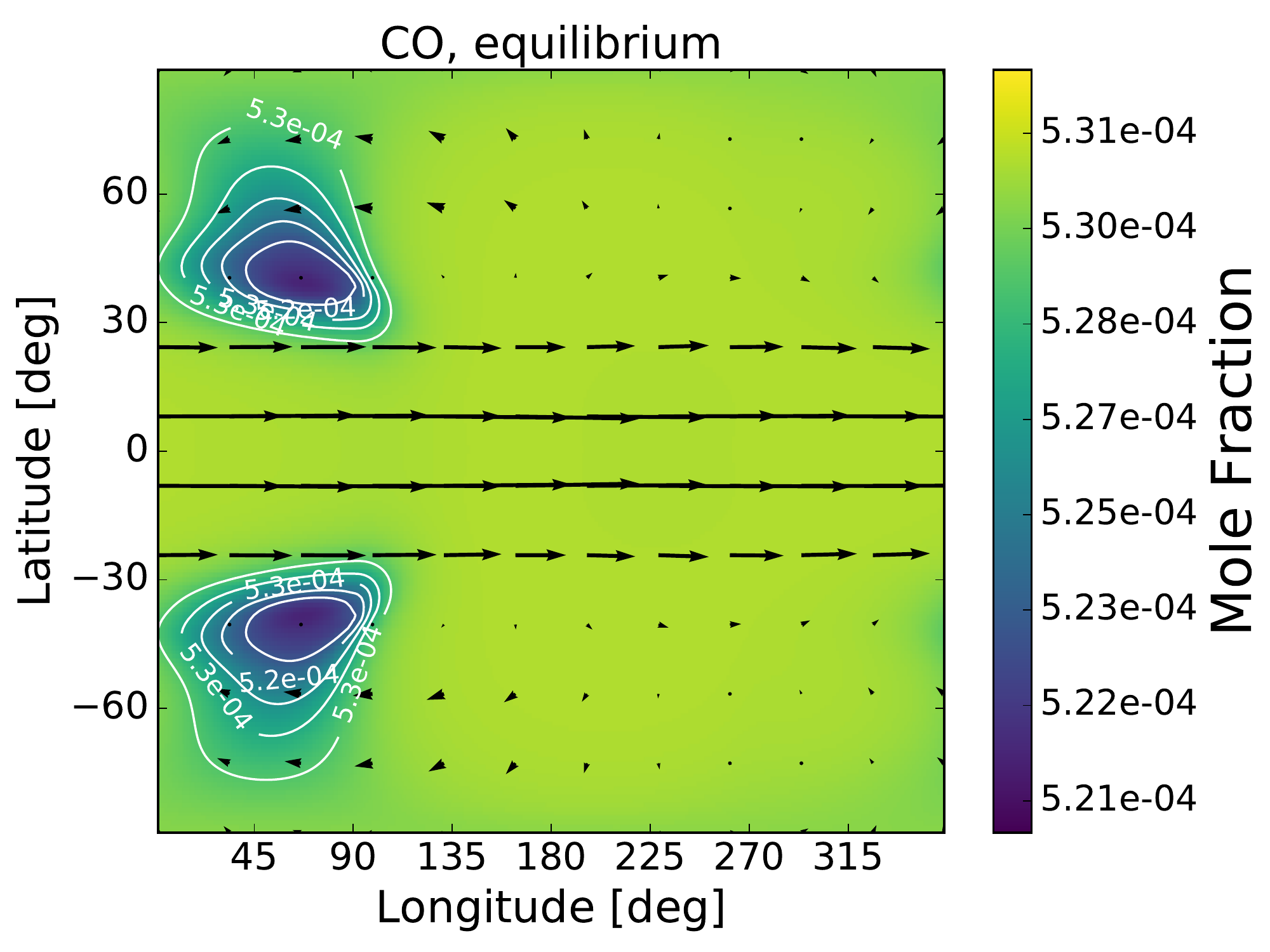}
    \includegraphics[width=0.45\textwidth]{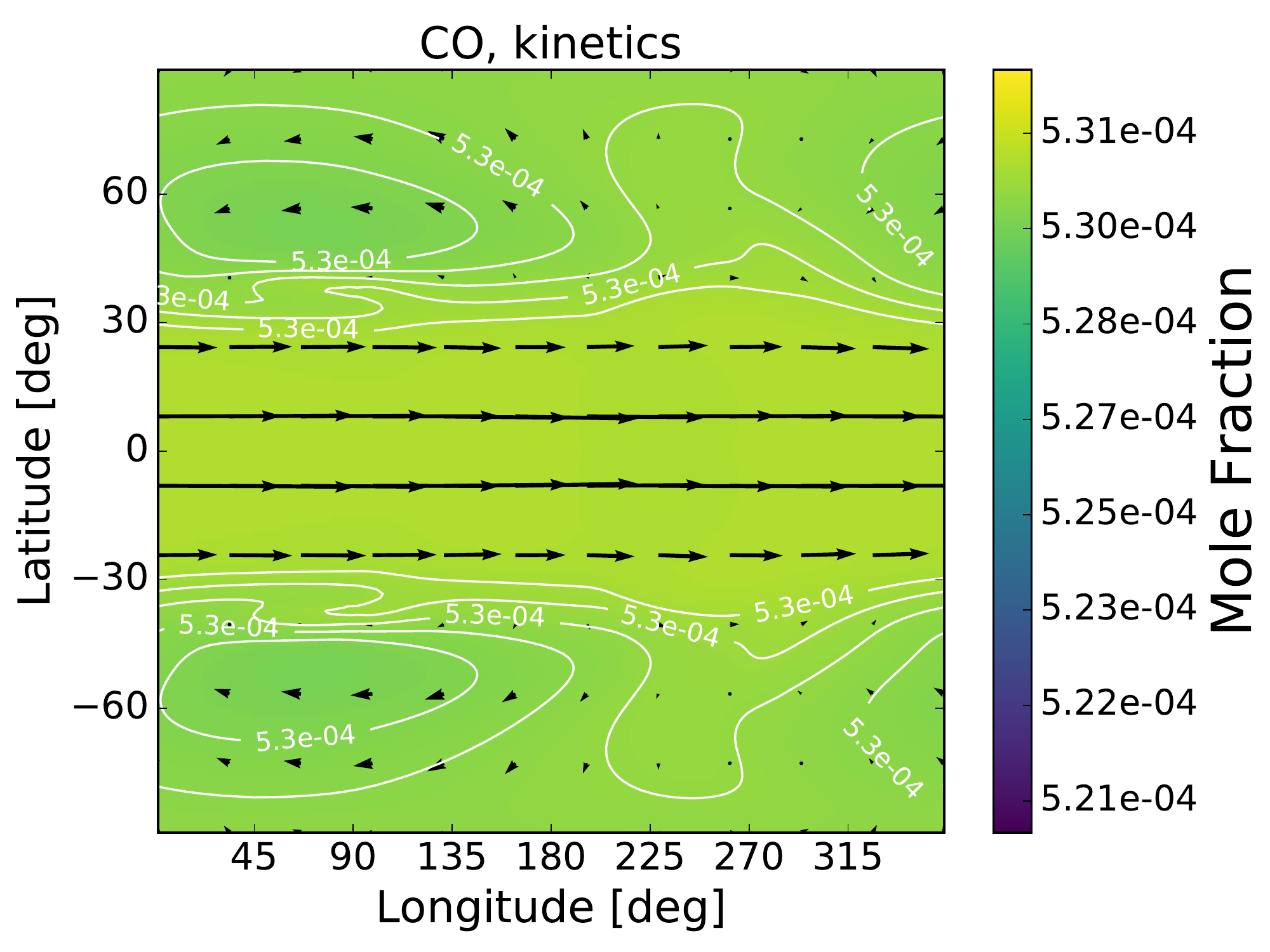} \\
     \includegraphics[width=0.45\textwidth]{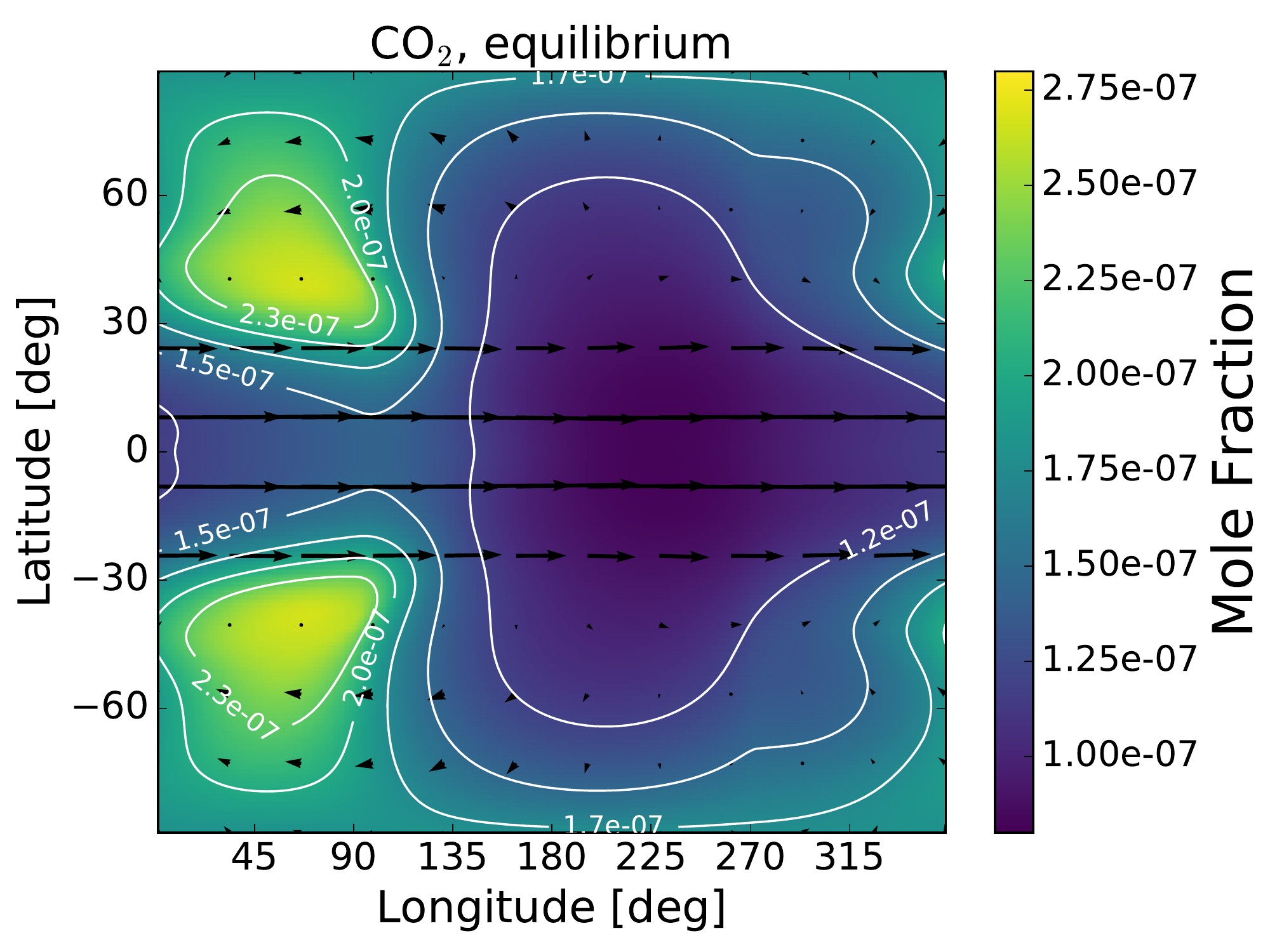}
    \includegraphics[width=0.45\textwidth]{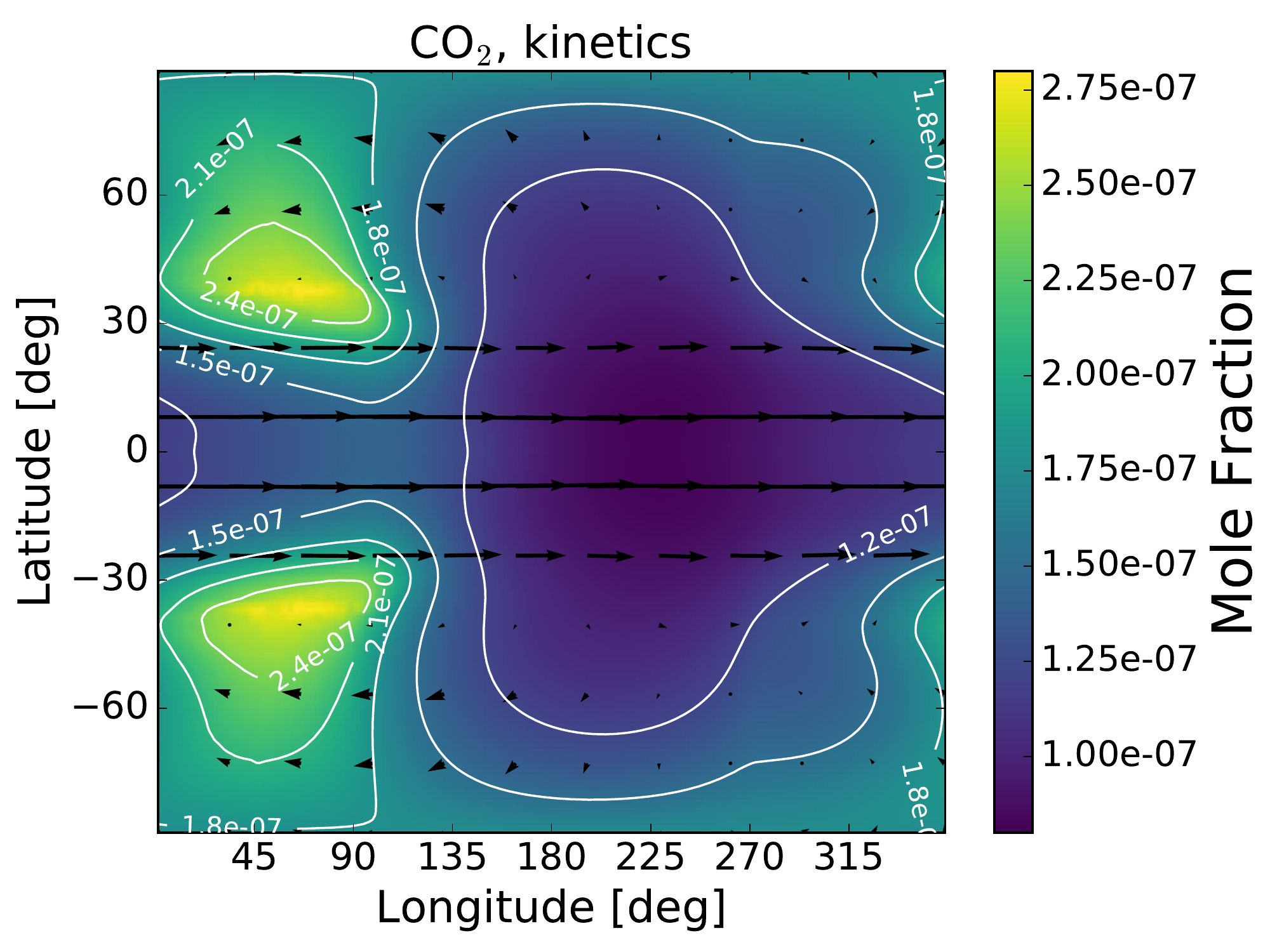} \\
     \includegraphics[width=0.45\textwidth]{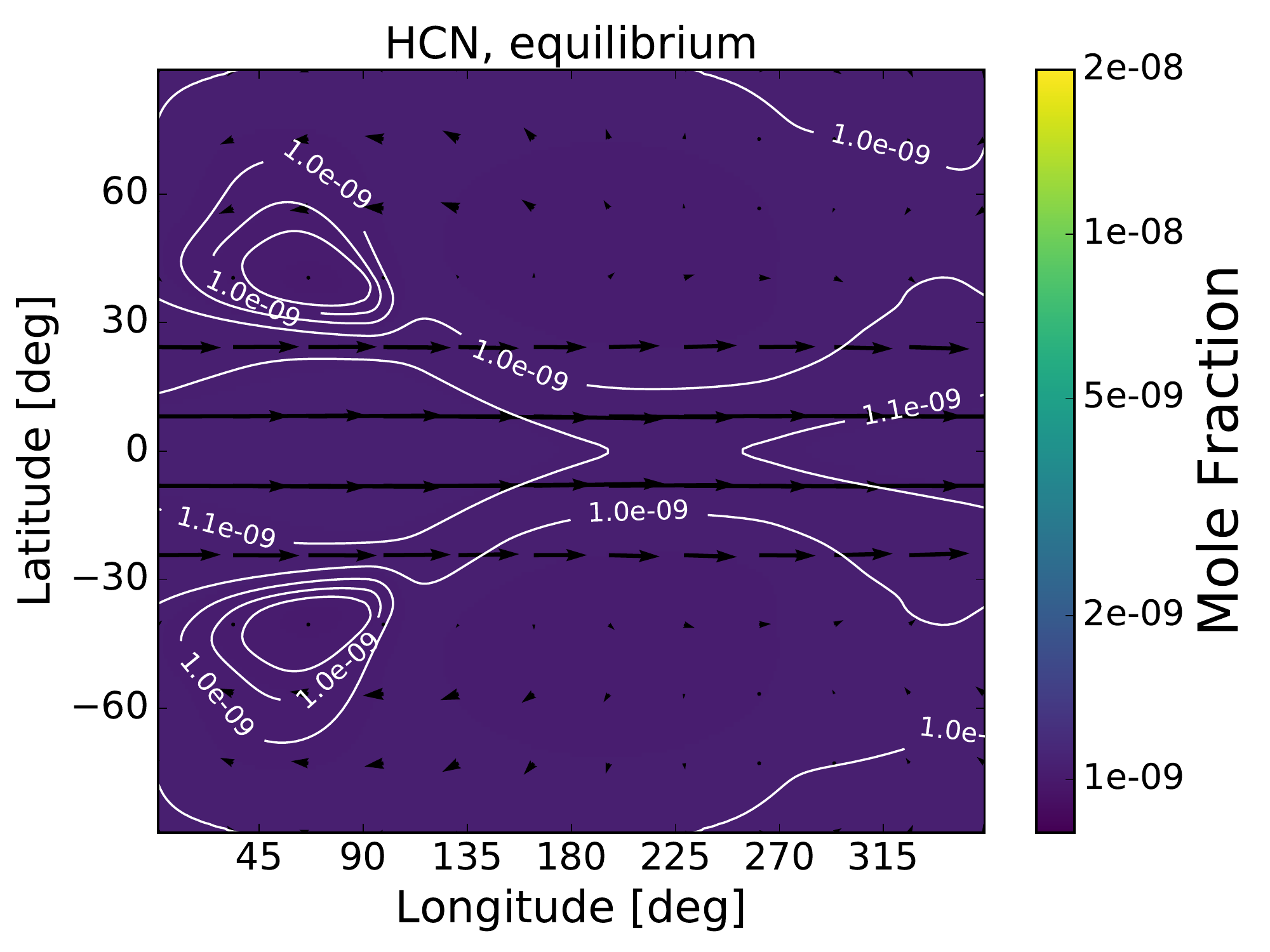}
    \includegraphics[width=0.45\textwidth]{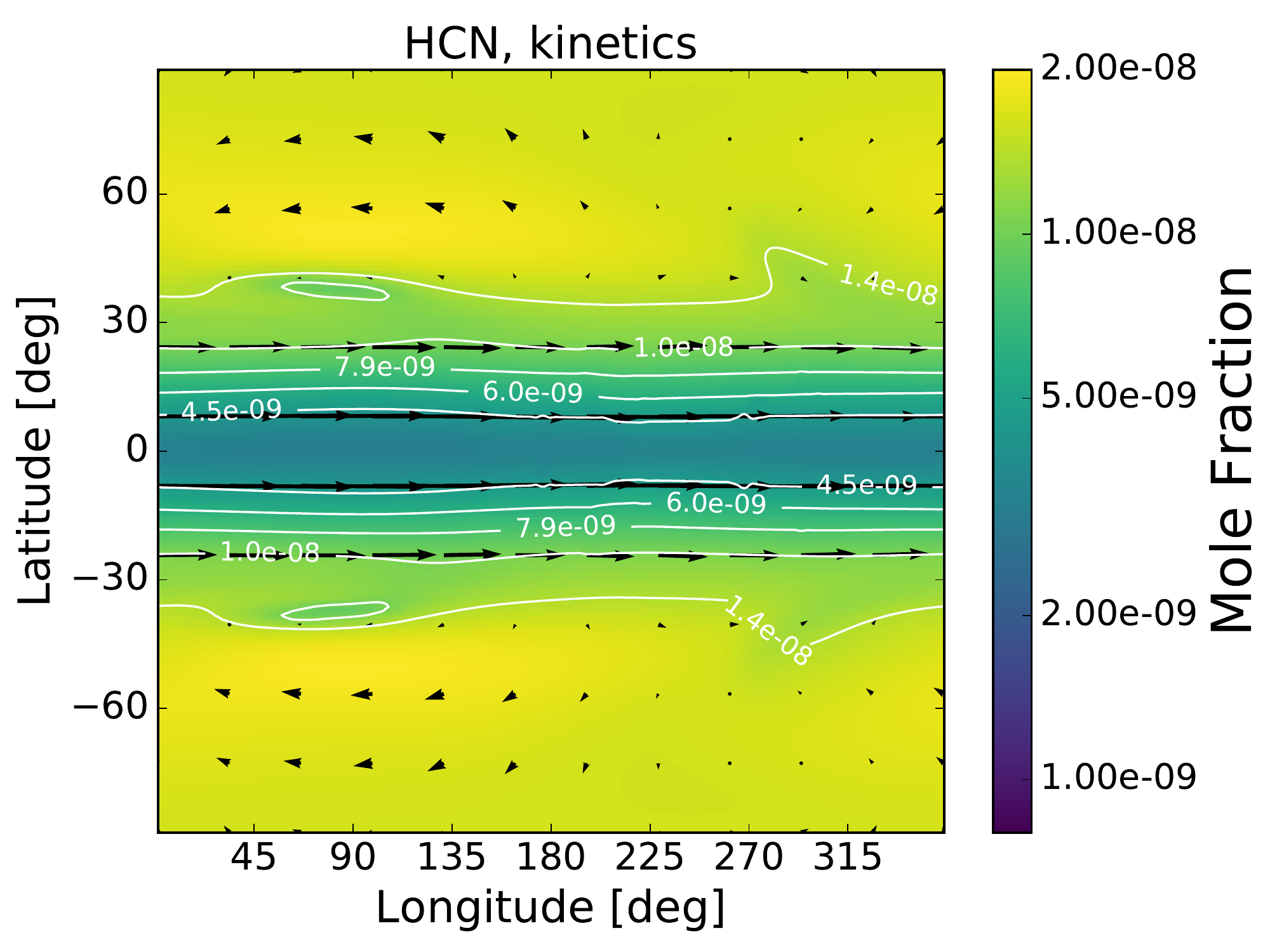} \\
  \end{center}
\caption{}
\end{figure*}

\begin{figure*}
  \begin{center}
    \includegraphics[width=0.45\textwidth]{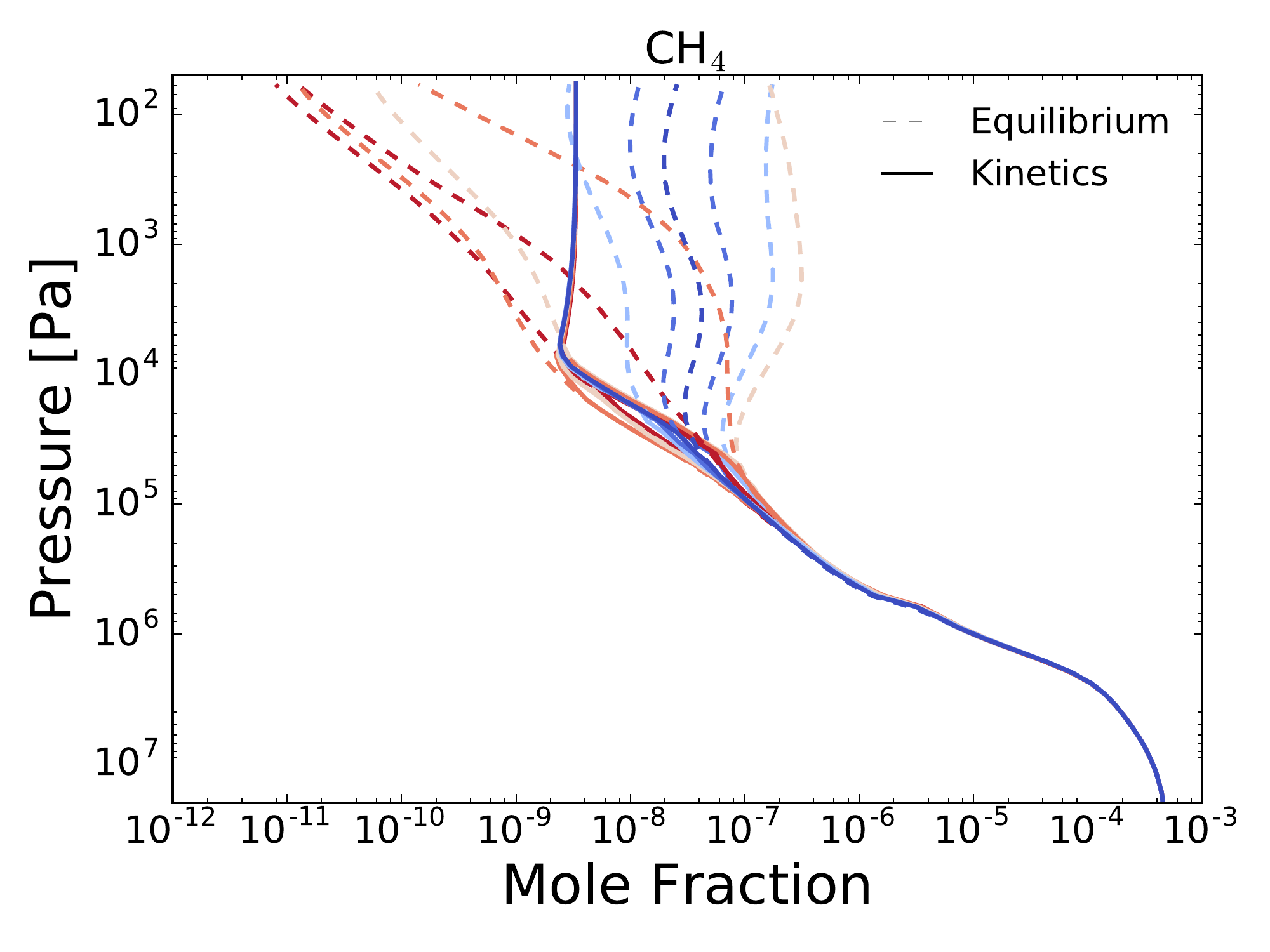}
     \includegraphics[width=0.45\textwidth]{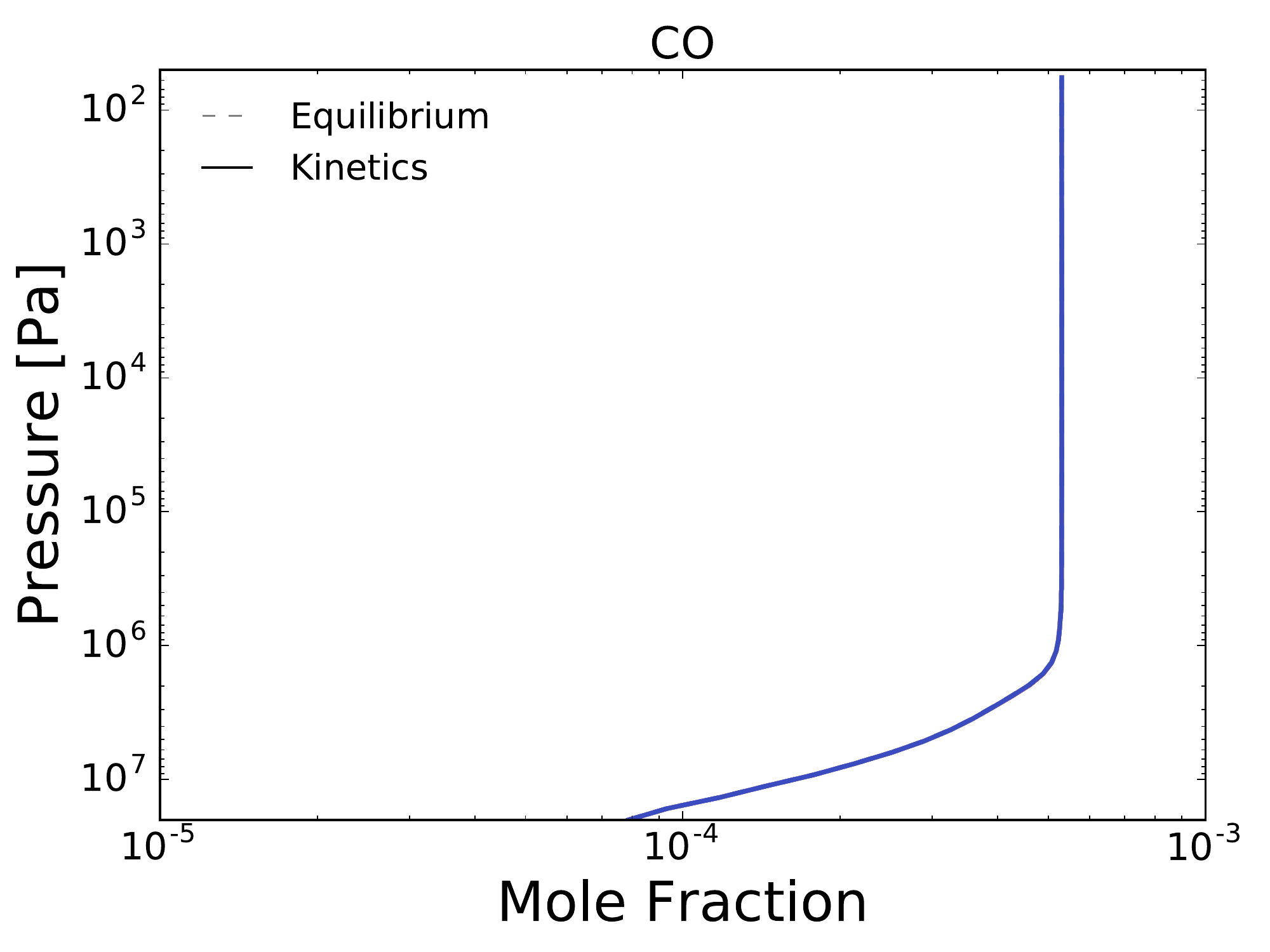} \\
     \includegraphics[width=0.45\textwidth]{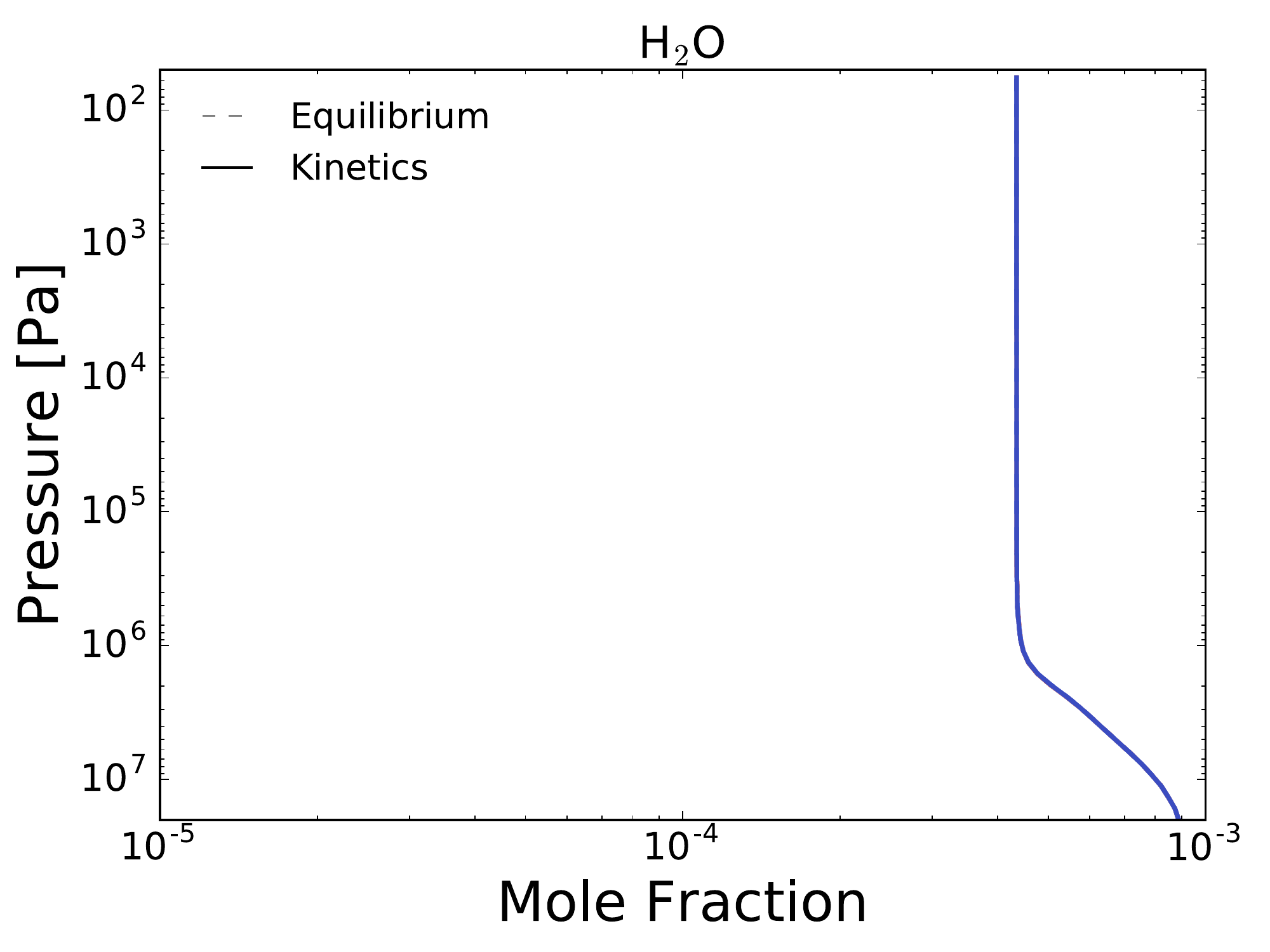}
    \includegraphics[width=0.45\textwidth]{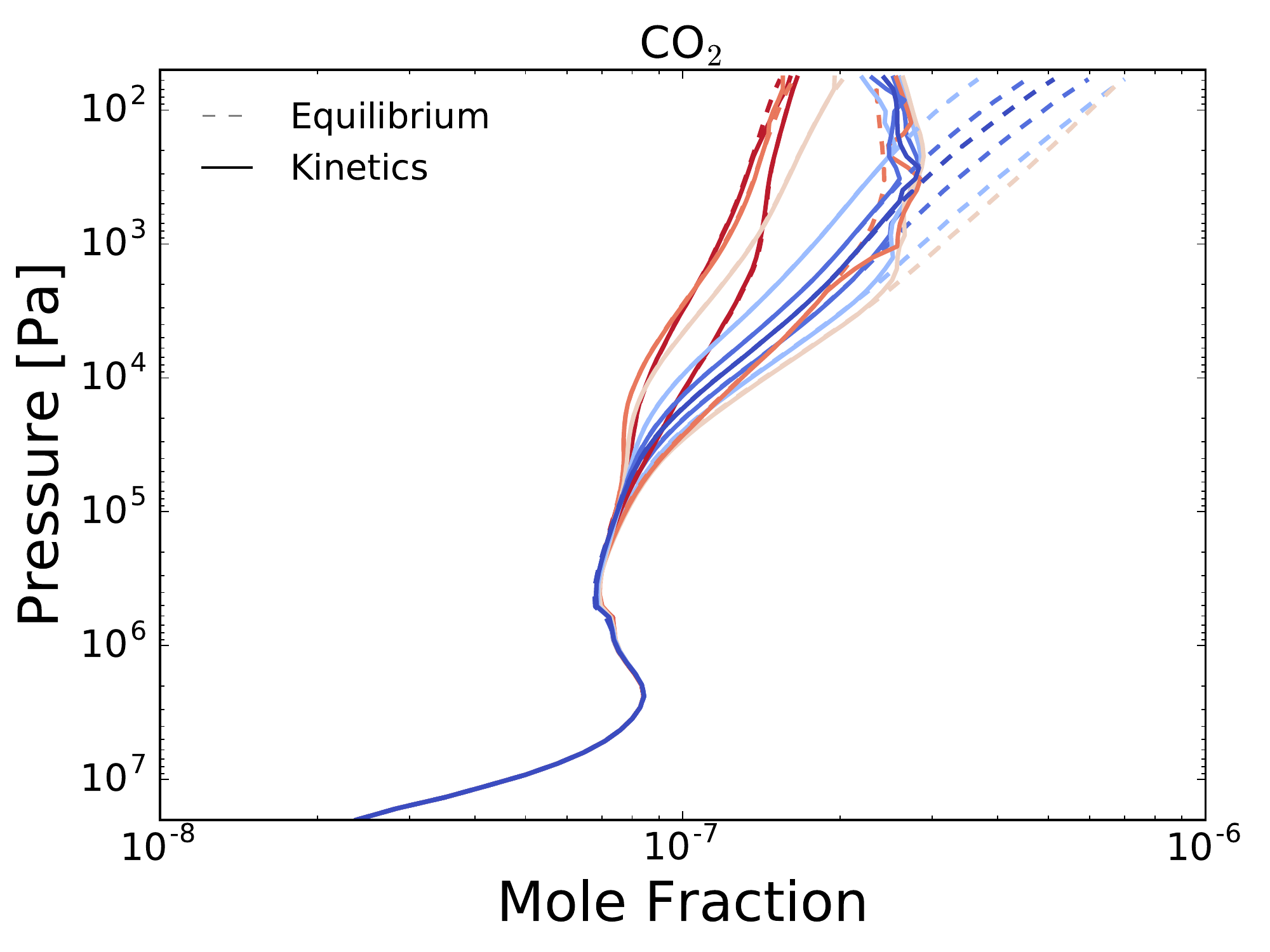}\\
     \includegraphics[width=0.45\textwidth]{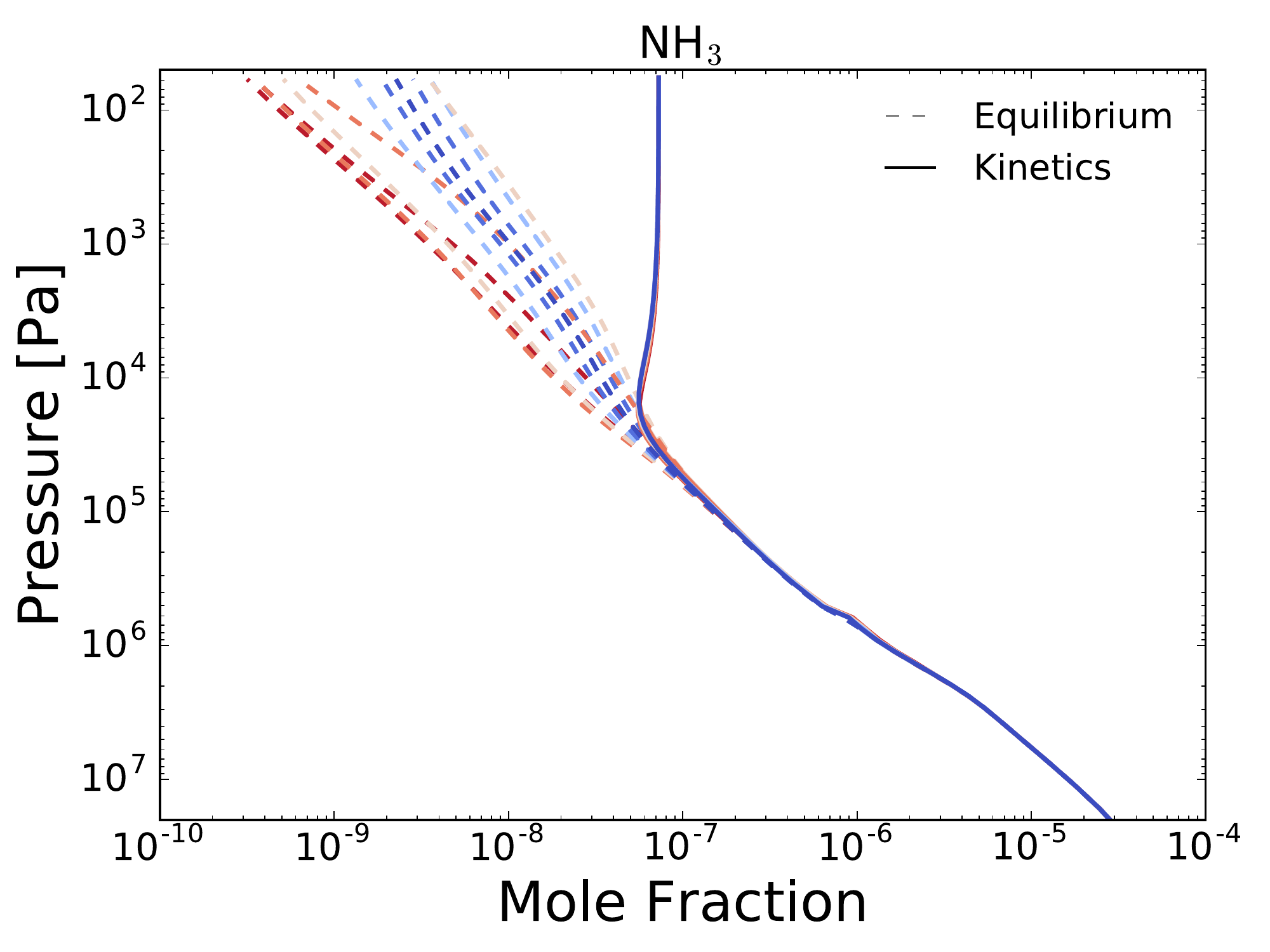}
 \includegraphics[width=0.45\textwidth]{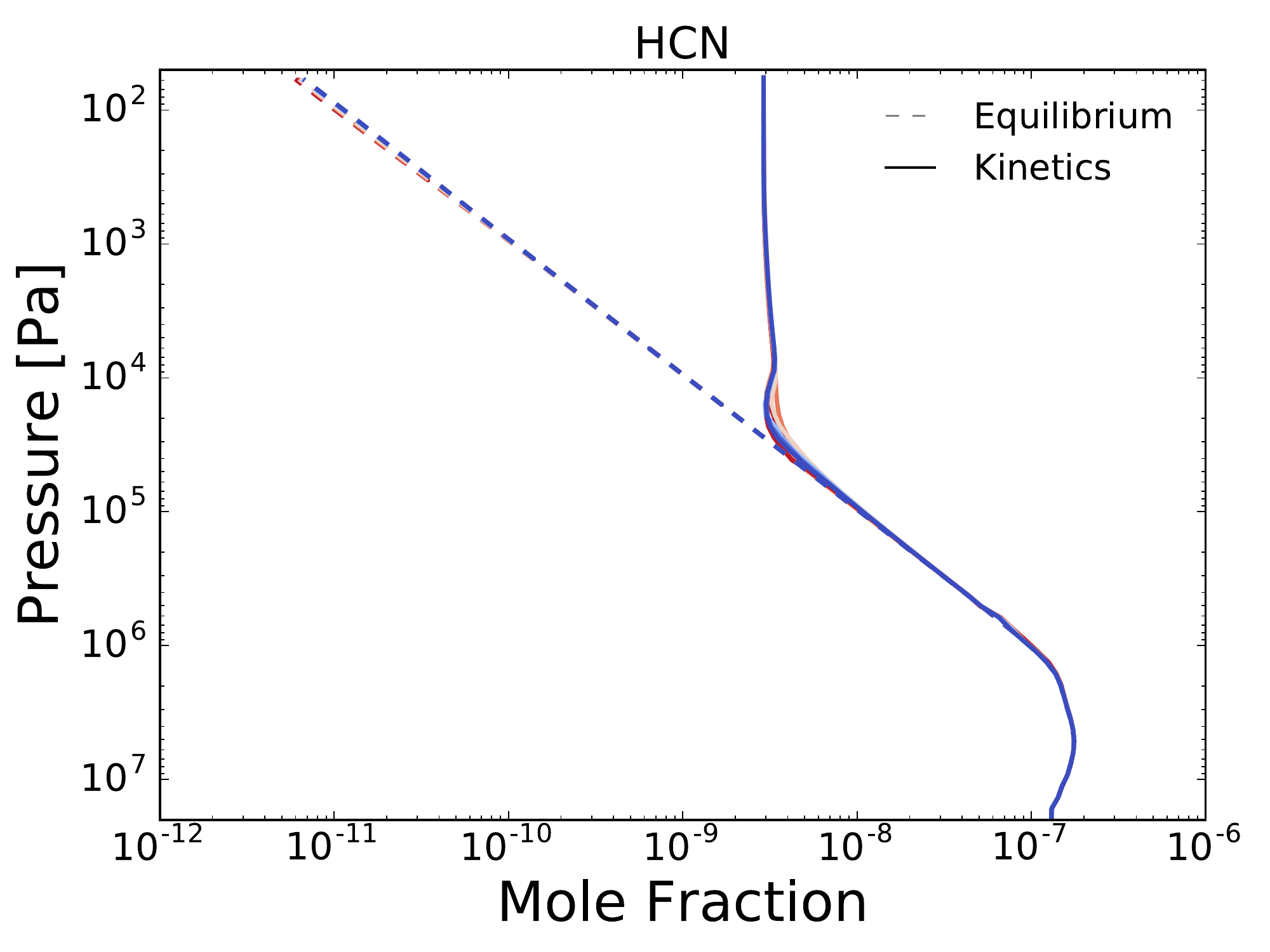}\\
  \end{center}
\caption{Vertical mole fraction profiles for different longitude points around the equator, for the equilibrium simulation (dashed lines) and kinetics simulation (solid lines) of HD~209458b. Warmer colours indicate a profile that is closer to the substellar point, cooler colours are closer to the antistellar point.}
\label{figure:hd209_mf_prof}
\end{figure*}

\cref{figure:hd209_mf_eq} shows the mole fractions of a subset of chemical species (CH$_4$, CO, H$_2$O, CO$_2$, NH$_3$, and HCN) on the $P=10^4$~Pa isobar for the equilibrium and kinetics simulations of HD~209458b. \cref{figure:hd209_mf_prof} shows both their equilibrium and kinetics abundances as vertical (pressure) mole fraction profiles at a series of equally spaced longitude points around the equator (i.e. at a latitude of $\theta=0$). We also show the vertical abundance profiles for a latitude of 45$^{\circ}$ in \cref{section:midlat}. We show this particular subset of six molecules (out of a total of 30 species included in the chemical network) since each of these species are included as opacity sources in the model and they are also the six `target species' for which the reduced chemical network was constructed \citep{VenBD19}.

The equilibrium abundances (left column of \cref{figure:hd209_mf_eq}) clearly trace the temperature structure (\cref{figure:temp_p1e4}), as they necessarily should since the chemical equilibrium composition only depends on local pressure, temperature, and element abundances. The abundances of CH$_4$, CO$_2$, and NH$_3$ are lower on the warm dayside compared with the cooler nightside. The equilibrium abundance of CH$_4$ is large enough in the coolest regions at this pressure level to capture a significant number of carbon atoms, thereby reducing the equilibrium abundance of CO, though CO remains the dominant carbon species. The decrease in CO leads to a small increase in the equilibrium abundance of H$_2$O as more oxygen atoms are available. HCN shows a very small horizontal variation, since its abundance does not strongly depend on temperature for $T>1000$~K \citep[e.g.][]{HenT16}. However, this molecule does show a significant vertical abundance gradient (\cref{figure:hd209_mf_prof}). These trends are simply due to the temperature and pressure dependence of the Gibbs free energy of each chemical species. The equilibrium abundances of CH$_4$, CO, and H$_2$O are almost identical to those presented in \citet{DruMM18a}, since the temperature structure is approximately the same.

We now focus on the differences in the chemical abundances between the equilibrium and kinetics simulations to investigate the effect of advection on the atmospheric composition. On the $P=10^4$~Pa isobar the abundance fields from the kinetics simulation (right column of \cref{figure:hd209_mf_eq}) differ from the equilibrium simulation for all of the presented species. The kinetics abundances of CO and H$_2$O become horizontally more uniform compared with their equilibrium abundances, losing the variations across the cold, nightside mid-latitudes. The quantitative differences between the two simulations, however, are small and CO and H$_2$O are the dominant carbon and oxygen species throughout the atmosphere, as in the equilibrium case. CO$_2$ is close to its chemical equilibrium abundance for most of the domain, on this pressure level, except in the coldest region (nightside mid-latitudes) where there is a small difference between the equilibrium and kinetics simulations. Here CO$_2$ is maintaining a `pseudo-equilibrium' with CO and H$_2$O, since the cycling between oxidised forms of carbon remains efficient \citep[e.g.][]{Moses2011,TsaKL18}. Therefore, CO$_2$ only shows an obvious difference with its chemical equilibrium value in the region where CO and H$_2$O change, at the nightside mid-latitudes. In other regions at this pressure level, the abundances of CO and H$_2$O do not change significantly, and therefore neither does the abundance of CO$_2$.

The effect of advection on the horizontal abundance distribution of CH$_4$ at $P=10^4$~Pa is more complex. The zonal (longitudinal) abundance gradient is decreased compared with the equilibrium simulation but there is a large latitudinal abundance gradient. The abundance of CH$_4$ from the kinetics simulation is around an order of magnitude smaller in the equatorial region compared with at higher latitudes. A similar qualitative structure is found for NH$_3$ and HCN, with lower abundances in the equatorial region compared with higher latitudes.

From the vertical (pressure) profiles of the equatorial abundances of the six chemical species that we focus on here (\cref{figure:hd209_mf_prof}), it is clear that H$_2$O and CO are the most abundant trace species (i.e. after H$_2$ and He), showing negligible abundance variations with both pressure and longitude (for $P>10^6$~Pa). We note that the small variations in the abundances between the equilibrium and kinetics simulations that are apparent in \cref{figure:hd209_mf_eq} are mainly present at mid-latitudes and relatively small differences are also less apparent on the log-scale of this figure (\cref{figure:hd209_mf_prof}). Both the kinetics abundances of NH$_3$ and HCN diverge from their respective equilibrium abundances between $10^5$ and $10^4$~Pa, with their abundances becoming zonally and vertically well mixed towards lower pressures.

The vertical profiles of the equatorial abundance of CH$_4$ are the most interesting. In the pressure range $10^4-10^5$~Pa the kinetics abundances of CH$_4$ tend towards its equilibrium abundance in the hottest part of the atmosphere. This has the effect of significantly reducing the zonal abundance gradient, in this pressure range, by decreasing the nightside abundance. However, there remains a significant vertical abundance gradient in the same pressure range. This structure can be explained by CH$_4$ undergoing efficient zonal mixing in the pressure range $10^4-10^5$~Pa, while not being affected by vertical mixing. As a result, zonal abundance gradients are homogenised but vertical abundance gradients persist. For pressures less than $10^4$~Pa the equatorial abundance of CH$_4$ becomes approximately vertically uniform.

CO$_2$ is an interesting molecule in that it stays close to its chemical equilibrium abundance for much of the equatorial region, except for the coolest nightside profiles at low pressures. As previously discussed, CO$_2$ maintains a pseudo-equilibrium with CO and H$_2$O, even once those molecules themselves have quenched \citep{Moses2011,TsaKL18}. We attribute the departure of the abundances of CO$_2$ from chemical equilibrium as being due to its pseudo-equilibrium with CO and H$_2$O, rather than to quenching of CO$_2$ itself.

To understand the differences between the equilibrium and kinetics simulations, we compare the advection and chemical timescales. Where the advection timescale is much smaller than the chemical timescale ($\tau_{\rm adv}<<\tau_{\rm chem}$) the composition should be well--mixed. Conversely, where the chemical timescale is much smaller than the advection timescale ($\tau_{\rm adv}>>\tau_{\rm chem}$), local chemical equilibrium should hold. Where the two timescales are comparable ($\tau_{\rm adv}\approx\tau_{\rm chem}$) both processes are important to consider and the situation is more complex.

We estimate the zonal, meridional and vertical advection timescales as $\tau_{\rm adv}^u\approx \frac{2\pi R_{\rm P}}{\left|u\right|}$, $\tau_{\rm adv}^v\approx \frac{\pi R_{\rm P}}{2\left|v\right|}$, and $\tau_{\rm adv}^w\approx \frac{H}{\left|w\right|}$,
respectively, where $R_{\rm P}$ is the planet radius, $H$ is the vertical pressure scale height and $u$, $v$, and $w$ are the zonal, meridional, and vertical components of the wind velocity. We sample the wind velocities directly from the 3D grid to capture the spatial variation of the advection timescale. We note that these can be considered only as zeroth-order estimates.

We estimate the chemical timescale using the methods described in \citet{TsaKL18}. Since CH$_4$ shows a particularly interesting effect from advection, and because it is an important absorbing species, we focus on that molecule for this timescale comparison. However, we note that the chemical timescale, for a given pressure and temperature, varies significantly for different molecules \citep{TsaKL18}. The chemical timescale for CH$_4$ that we use in this analysis has been specifically calculated for the \citet{VenBD19} chemical network. In practice, we calculated the CH$_4$ chemical timescale for a grid of pressures and temperatures. We then interpolated these timescales on to the model 3D grid using the pressure and temperature of each grid cell.

\begin{figure}
  \begin{center}
    \includegraphics[width=0.45\textwidth]{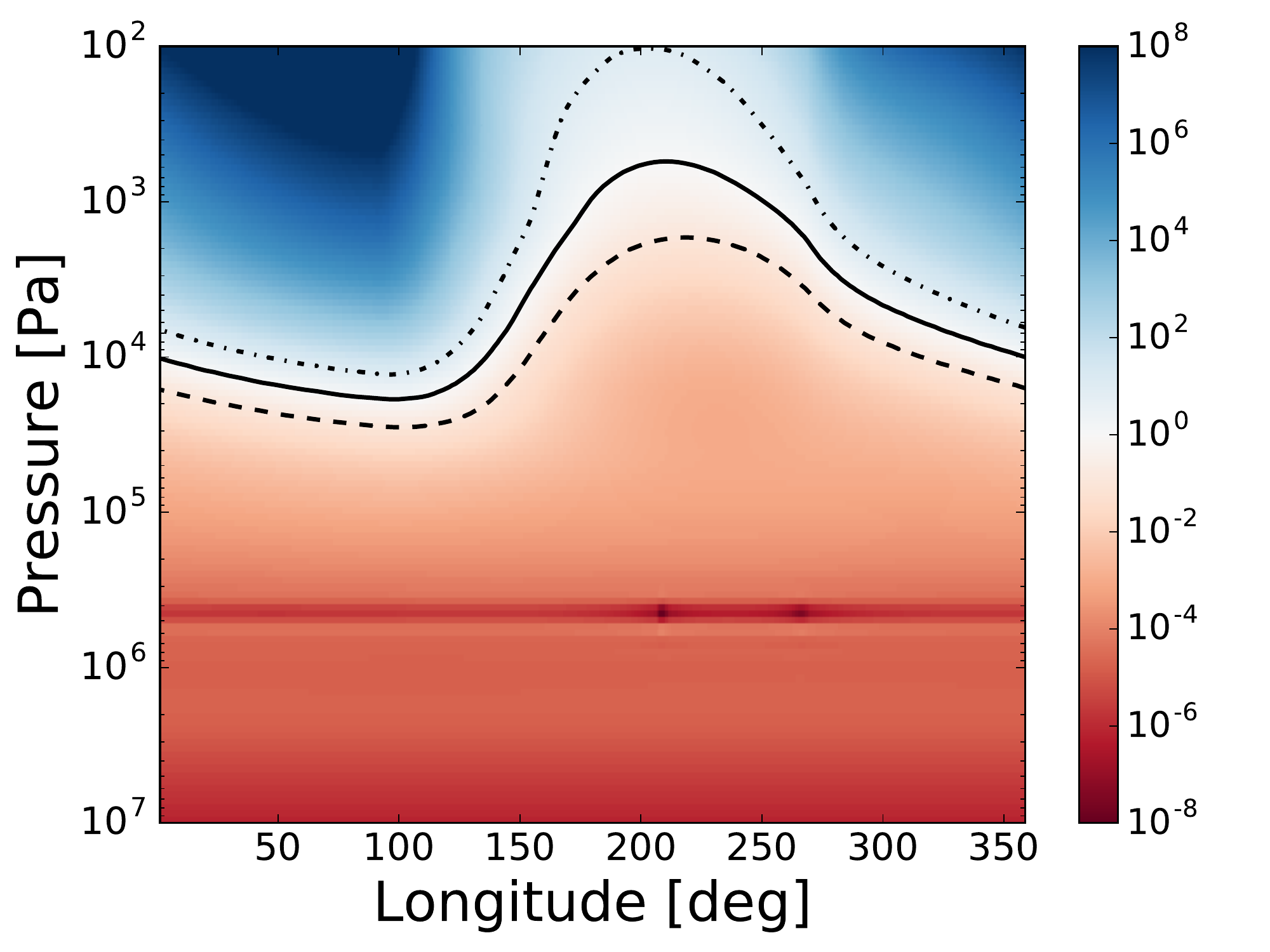} \\
    \includegraphics[width=0.45\textwidth]{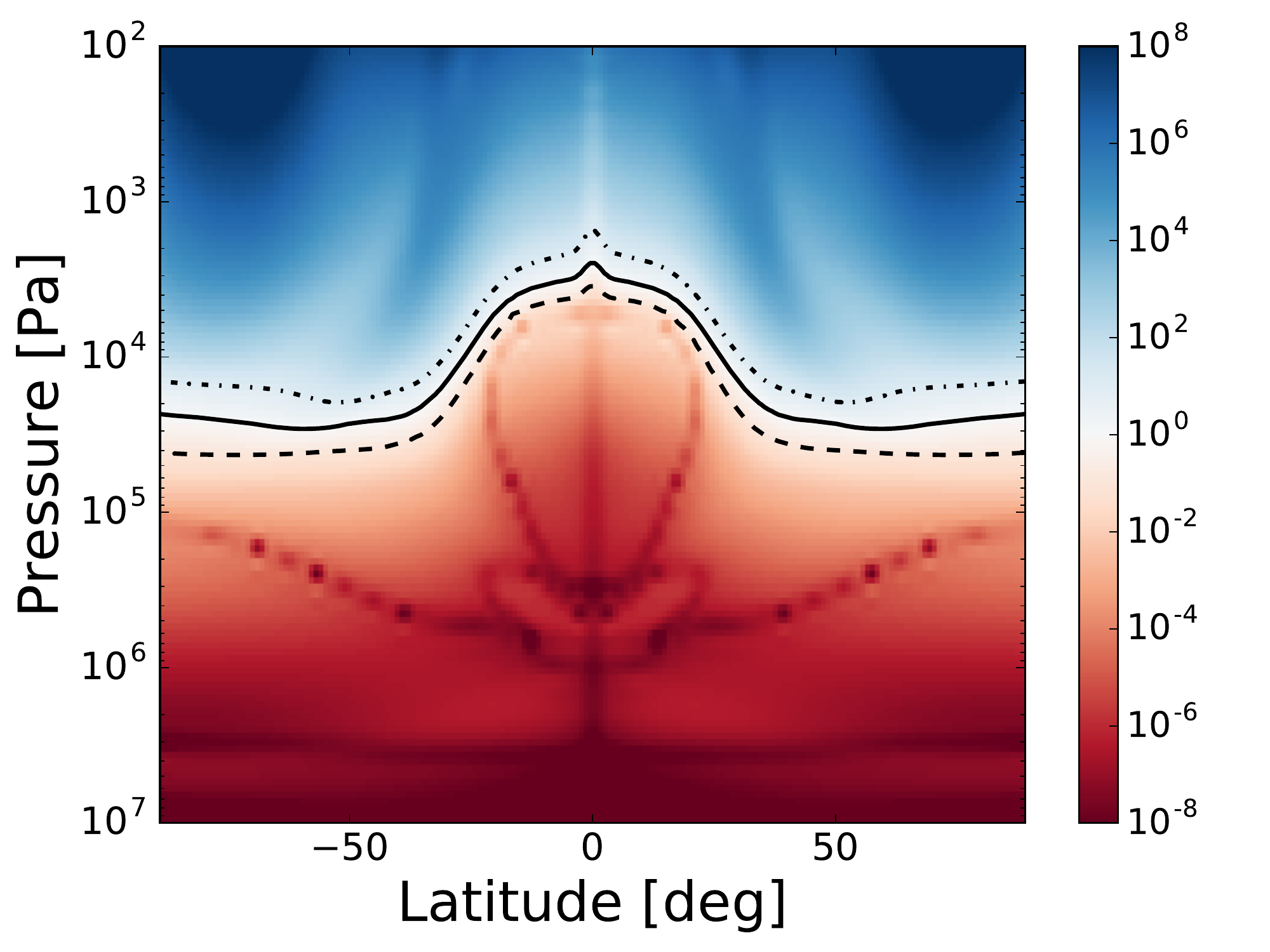} \\
    \includegraphics[width=0.45\textwidth]{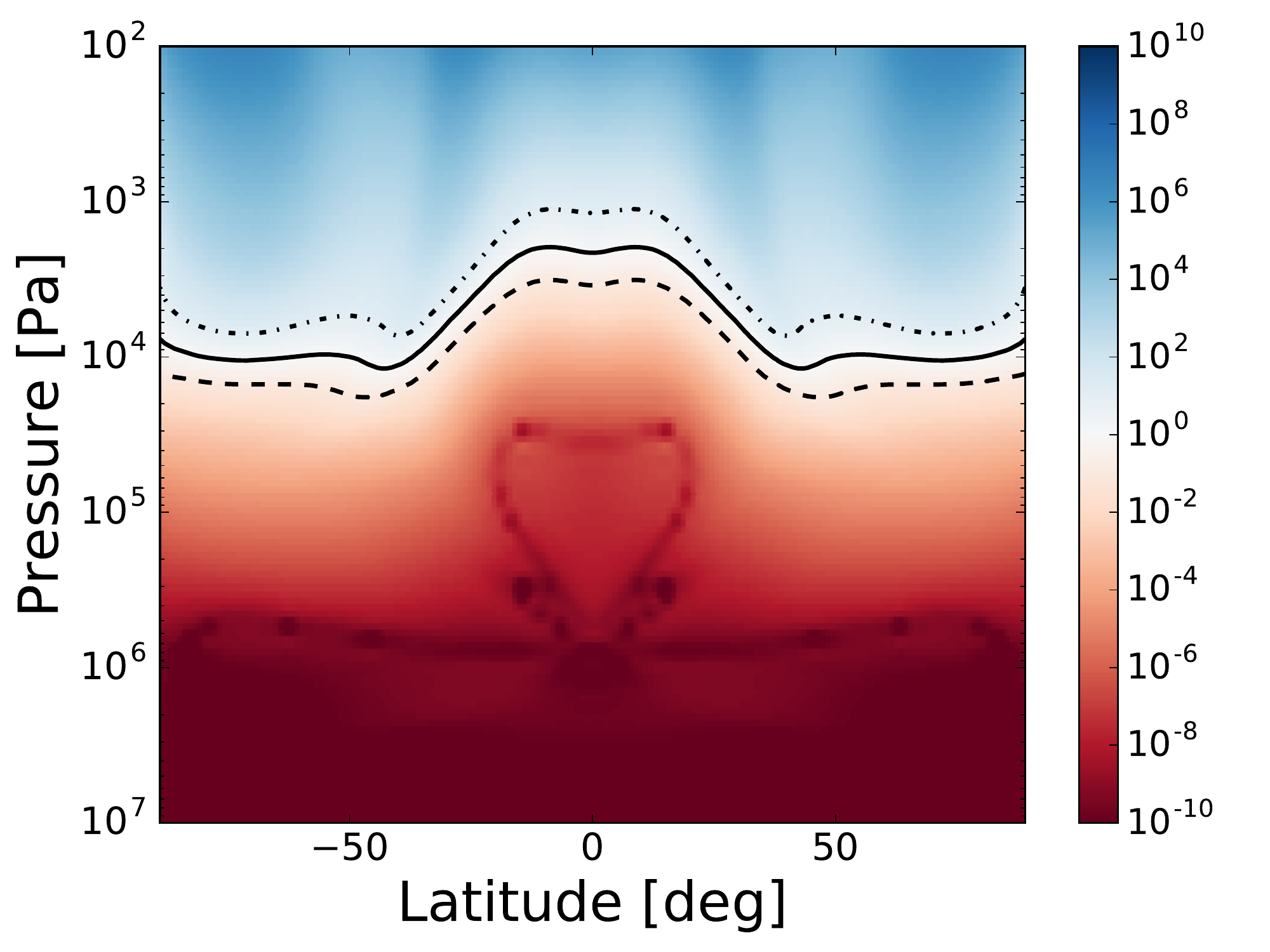} \\
  \caption{Ratios of the chemical to advection timescales for CH$_4$ ($\alpha=\tau_{\rm chem}/\tau_{\rm adv}$), for the zonal (top), meridional (middle), and vertical (bottom) components of the wind for the kinetics simulation of HD~209458b. We show a meridional-mean between $\pm30^{\circ}$ latitude for the zonal component, and slices at a longitude of 0$^{\circ}$ (i.e. the antistellar point) for the meridional and vertical components. The colour scale shows $\alpha$, with blue colours corresponding to $\alpha>1$ and red colours corresponding to $\alpha<1$. The black contours show the $\alpha=0.1,1,10$ values. The fine structure shown in the plots is a reflection of regions where the wind velocity changes signs (e.g. eastward to westward) and the wind velocity approaches a value of zero over a very small region at the transition.}
  \label{figure:hd209_ts}
  \end{center}
\end{figure}

\cref{figure:hd209_ts} shows the ratio of the chemical to advection timescales for CH$_4$ ($\alpha=\tau_{\rm chem}/\tau_{\rm adv}$) for each of the three components of the 3D wind (zonal, meridional, and vertical) for the kinetics simulation of HD~209458b. For the zonal component we show the timescales ratio as a function of pressure and longitude for a meridional mean between $\pm30^{\circ}$, whereas for the meridional and vertical components we show the timescales ratio as a function of pressure and latitude for a pole-to-pole slice centred on the anti-stellar point (i.e. a longitude of $\phi=0^{\circ}$). Where $\alpha<1$ (red colours in \cref{figure:hd209_ts}) the chemical timescale is smaller (faster) than the advection timescale and the chemical equilibrium is expected to hold. However, where $\alpha>1$ (blue colours in \cref{figure:hd209_ts}) the chemical timescale is larger (slower) than the advection timescale and the atmosphere is expected to be well-mixed (in the direction of the considered wind velocity). We reiterate that both the chemical and advection timescales are only estimates, which conservatively should be seen as zeroth-order estimates.

For the zonal component of the timescale ratio (top panel, \cref{figure:hd209_ts}), $\alpha$ decreases with increasing pressure, which is primarily due to the pressure and temperature dependence of the chemical timescale. Following the contour of $\alpha=1$, where the timescales are comparable, it is clear that there is a significant dependence of $\alpha$ on longitude. For a line of constant pressure, in between $\sim10^3$ and $\sim10^4$~Pa, $\alpha>1$ on the nightside whereas $\alpha<1$ for a significant portion of the dayside, largely due to the dependence of the chemical timescale on the local atmospheric temperature. This has the consequence that the atmosphere is expected to hold local chemical equilibrium on the dayside, but not on the nightside in the presence of zonal advection. This relates to the idea of `contamination' of the nightside by the chemistry of the dayside as discussed by previous authors \citep[e.g.][]{AguPV14}. The dayside atmosphere is hot enough to remain in chemical equilibrium and the fast zonal winds can transport this material to the cooler nightside that is not hot enough to maintain chemical equilibrium.

This structure of the timescale ratios helps to explain vertical profiles of the equatorial abundance of CH$_4$ in \cref{figure:hd209_mf_prof} (top left panel) where we see its nightside abundances departing from chemical equilibrium and tending towards its equilibrium abundance in the hottest part of the dayside. Essentially this is horizontal quenching of CH$_4$ \citep{AguPV14}. We note that the pressure range of this feature does not exactly match what we would expect from the timescale comparison shown in \cref{figure:hd209_ts}. We attribute this simply to the fact that these timescale estimates are only zeroth-order estimates that we do not expect to precisely match the results of the full numerical simulations. We use the estimated timescales only as an aid to understand the complex 3D chemical structure that results from 3D advection.

For the meridional component of the timescale ratio (middle panel, \cref{figure:hd209_ts}), the contour of $\alpha=1$ shows a significant latitudinal dependence and lies at lower pressures nearer to the equator. Again, this is primarily due to the temperature dependence of the chemical timescale. Near the equator the atmosphere is warmer, which results in a faster chemical timescale. The spatial variation of the meridional wind velocity also plays a role. From this figure we can see that for some pressure range the atmosphere is expected to be in local chemical equilibrium in the equatorial region, but not for higher-latitudes. This has the important consequence that, for some pressure levels, the equatorial region can become chemically isolated from the mid-latitudes and polar regions. This can explain why our results show different compositions in the equatorial region compared with higher latitudes on the $P=10^4$~Pa isobar for CH$_4$, and also for NH$_3$, and HCN, as shown in \cref{figure:hd209_mf_eq}.

The vertical component of the timescale ratio (bottom panel, \cref{figure:hd209_ts}) is shown as a latitudinal slice at a longitude of 0$^{\circ}$. The structure of the vertical component of the timescale ratio is very similar to that of the meridional component. The contour of $\alpha=1$ lies at lower pressures in the equatorial region, where temperatures are higher, compared with higher latitudes. The transition lies at much lower pressures on the warmer dayside, in fact almost at the upper model boundary (see \cref{figure:hd209_ts_profiles}). Because of this zonal variation in the ratio of the chemical to vertical advection timescale, we only expect vertical mixing to be important for the cooler nightside region. However, because the zonal advection timescale becomes faster than the chemical timescale at higher pressures, once a species becomes vertically homogenised on the nightside, that vertically quenched abundance can be subsequently transported zonally, leading to both horizontal and vertical homogenisation all around the planet. This explains the vertically uniform equatorial abundance of CH$_4$ shown in \cref{figure:hd209_mf_prof}.

\subsubsection{HD~189733b}

We now consider the simulations of the atmosphere of HD~189733b which, though overall cooler, shows a similar qualitative thermal structure and circulation to HD~209458b. \cref{figure:hd189_mf_eq} (left column) shows the mole fractions of CH$_4$, CO, H$_2$O, CO$_2$, NH$_3$, and HCN on the $P=10^4$~Pa isobar for the equilibrium simulation of HD~189733b. The distribution of abundances are generally very similar to those for the warmer HD~209458b (\cref{figure:hd189_mf_eq}) but with quantitative differences due to differences in the temperature (\cref{figure:temp_p1e4}). Since the atmosphere is cooler than HD~209458b there is an overall increased relative abundance of CH$_4$ by one to two orders of magnitude, which is favoured thermodynamically for cooler temperatures. In particular, in the nightside mid-latitude regions the temperature is cold enough at $P=10^4$~Pa for CH$_4$ to become the most abundant carbon species, replacing CO.

\begin{figure*}
  \begin{center}
    \includegraphics[width=0.45\textwidth]{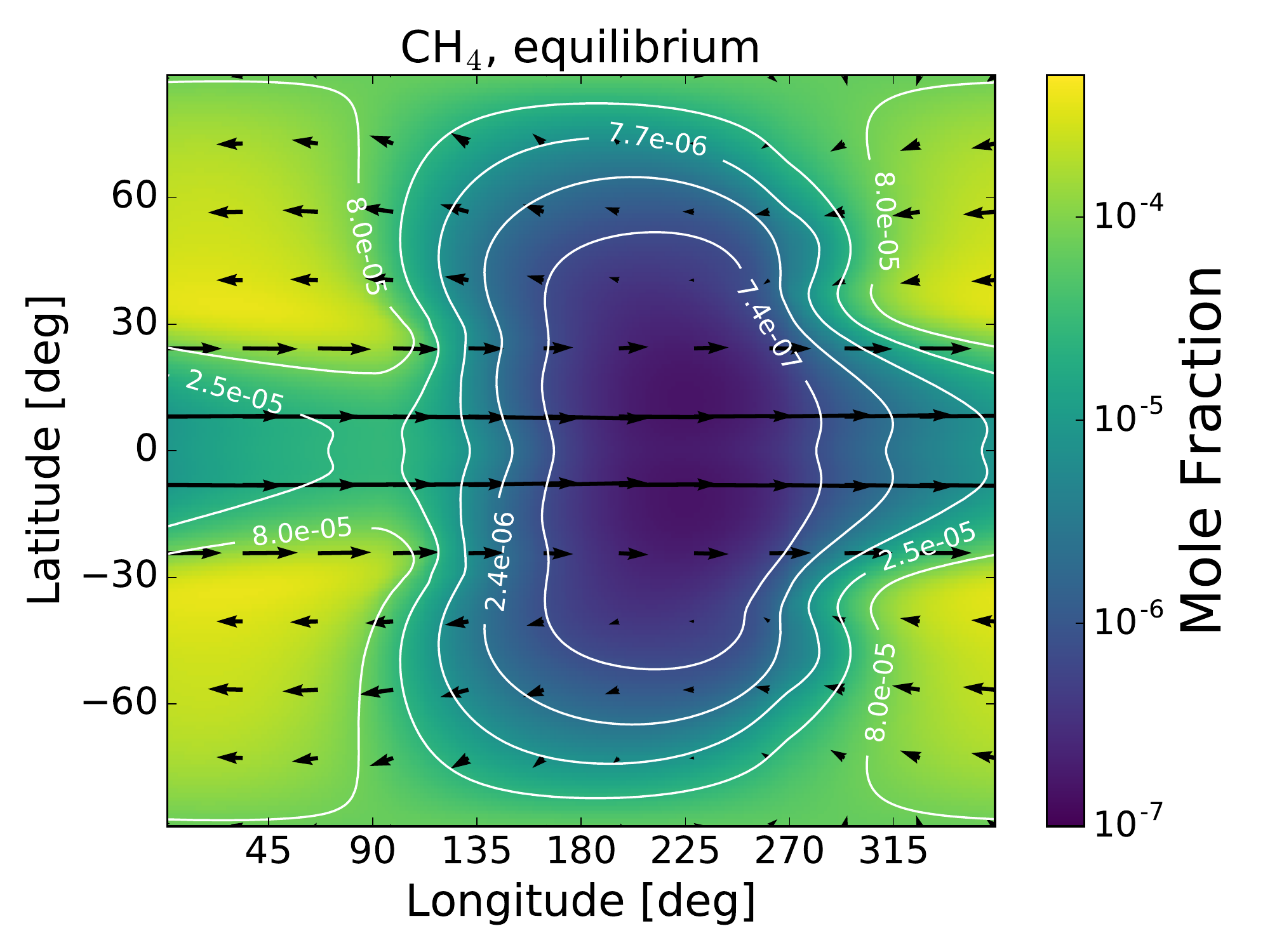}
    \includegraphics[width=0.45\textwidth]{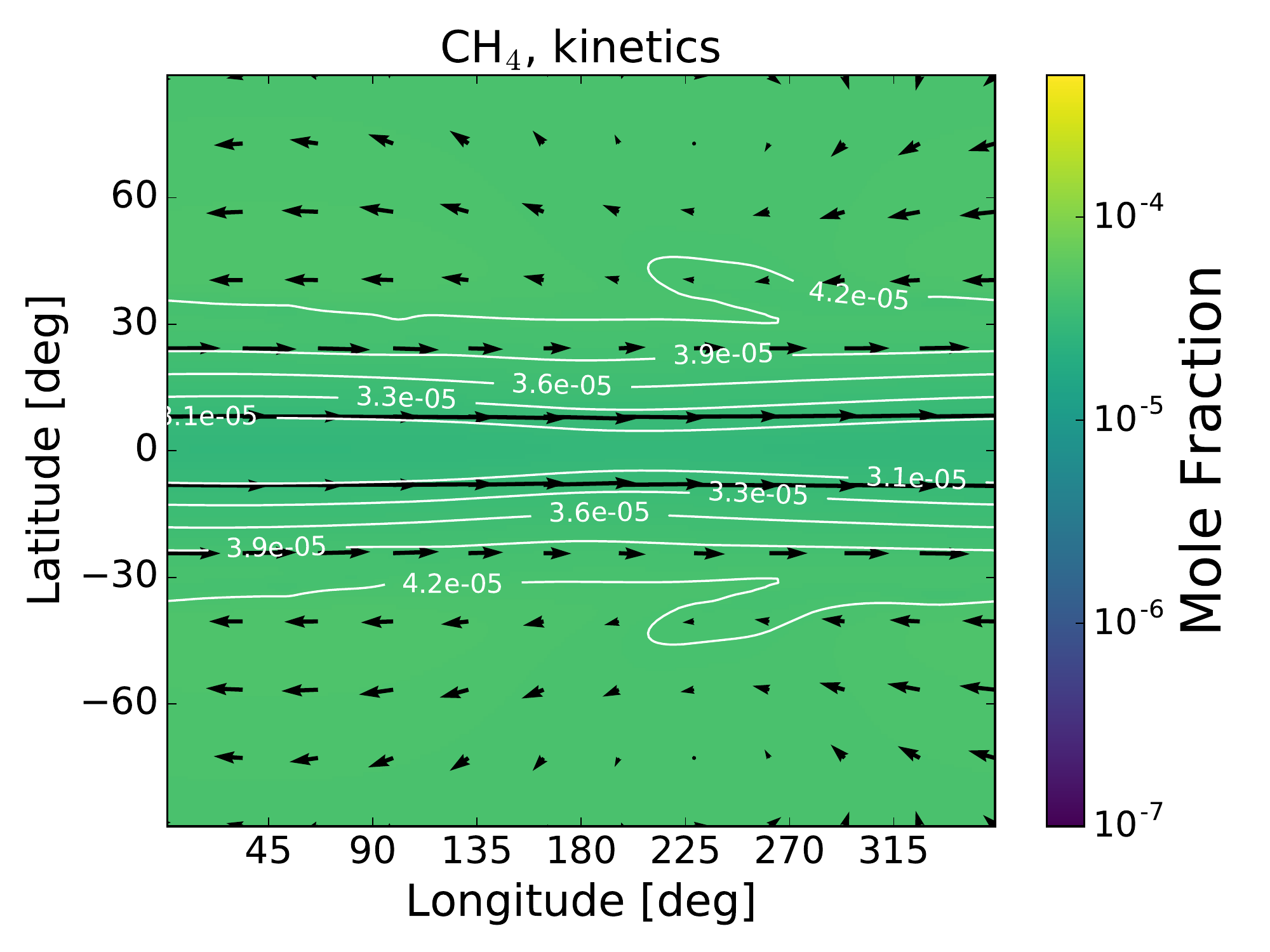} \\
     \includegraphics[width=0.45\textwidth]{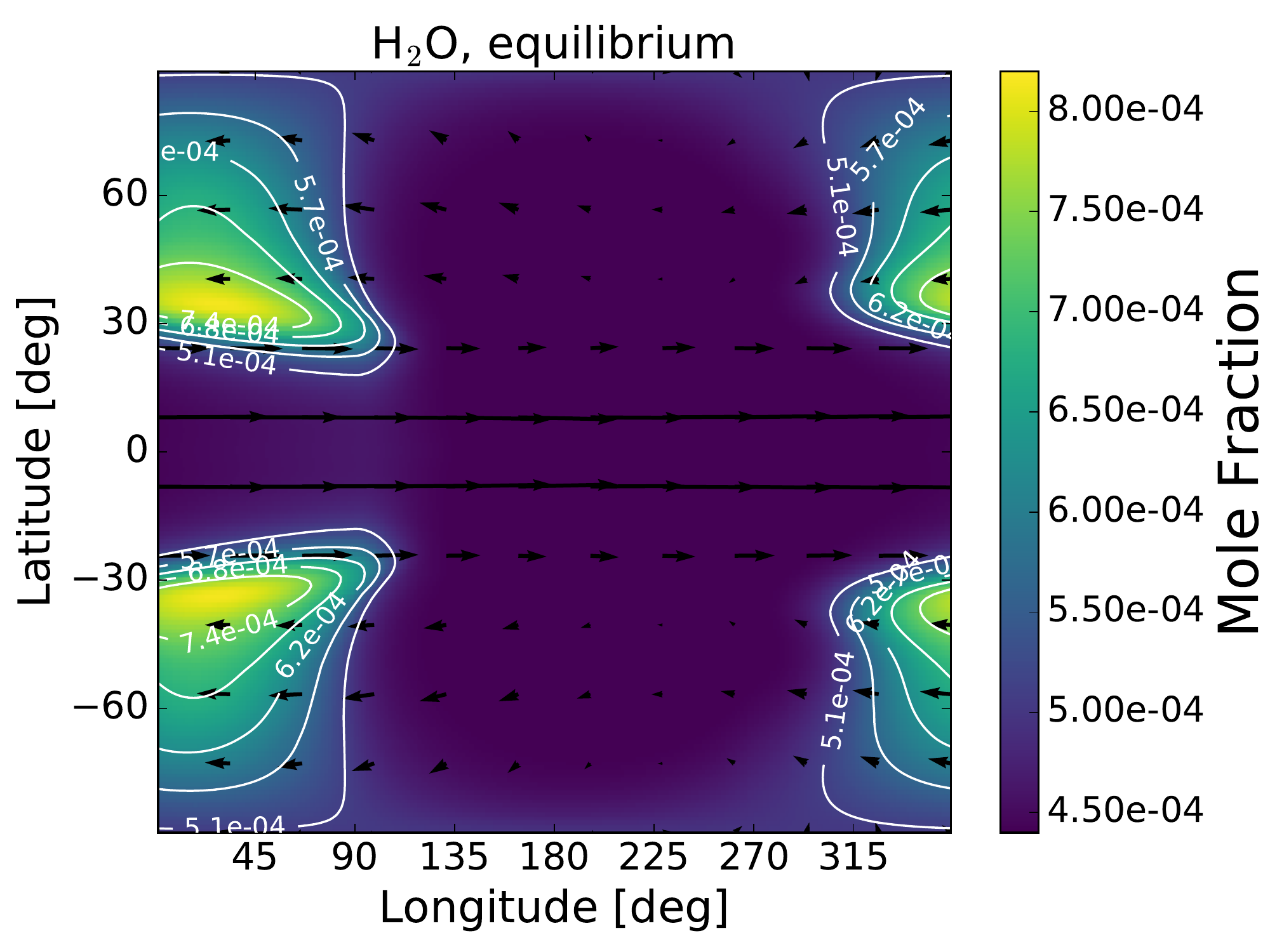}
    \includegraphics[width=0.45\textwidth]{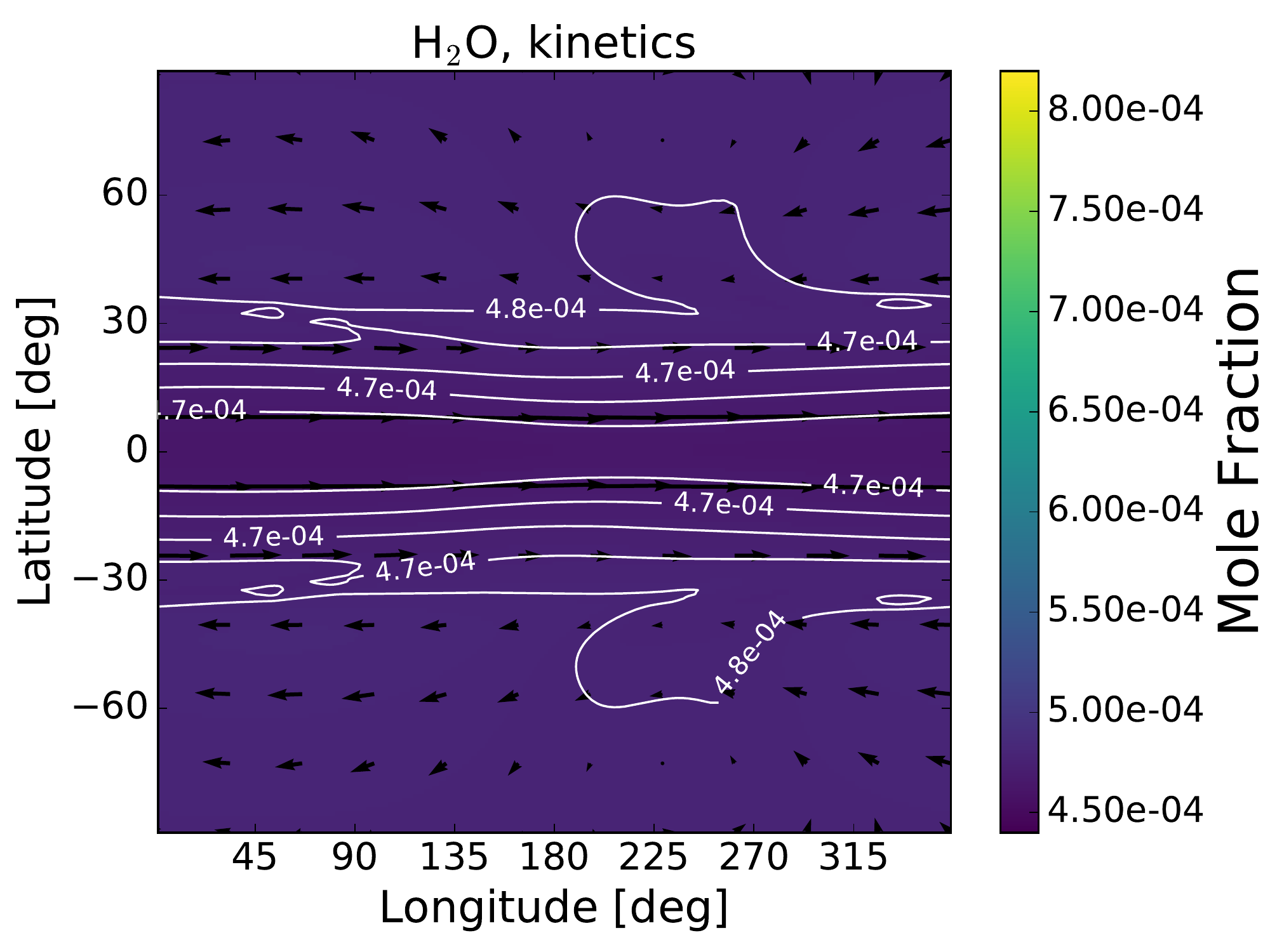} \\
     \includegraphics[width=0.45\textwidth]{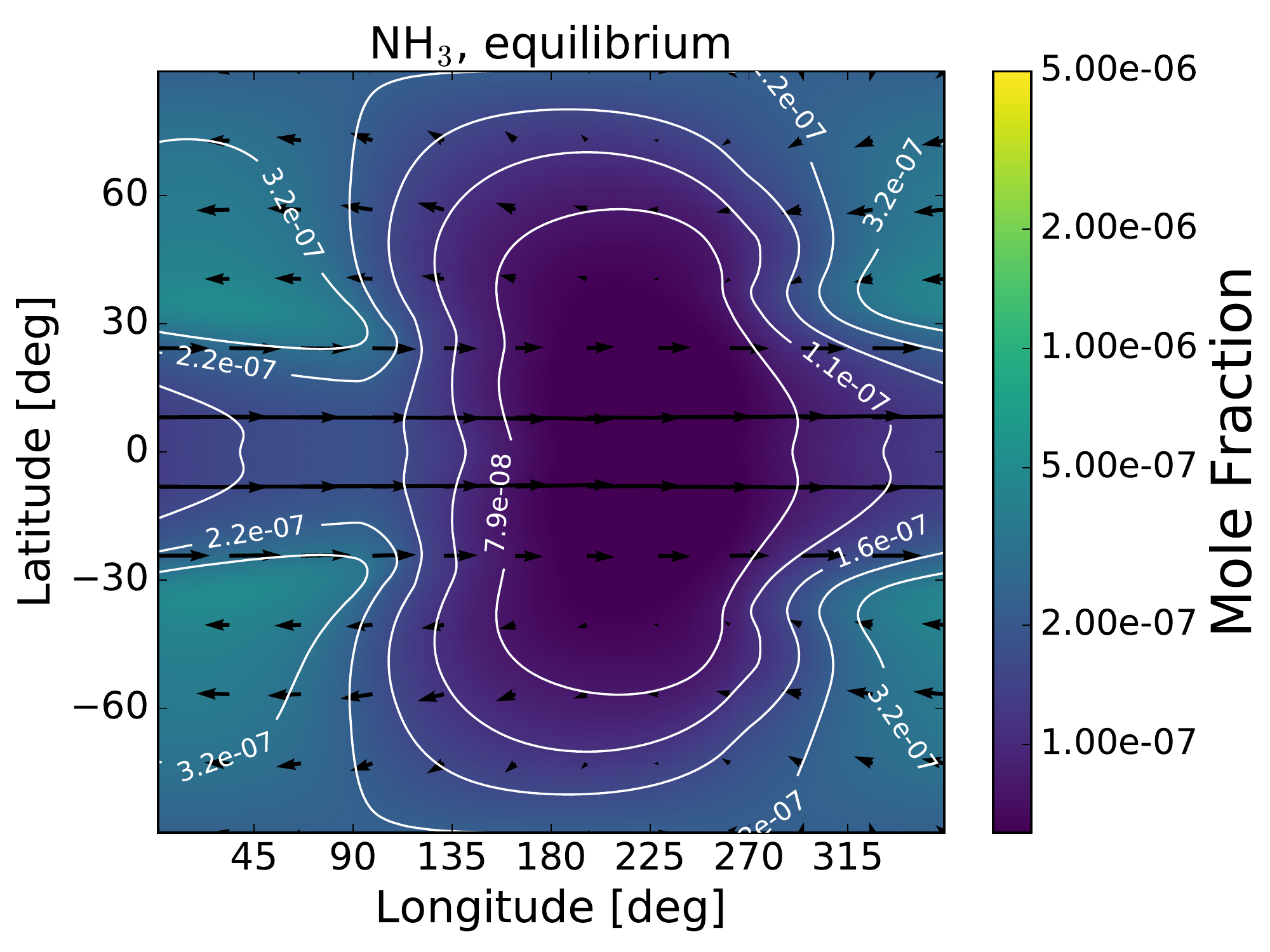}
    \includegraphics[width=0.45\textwidth]{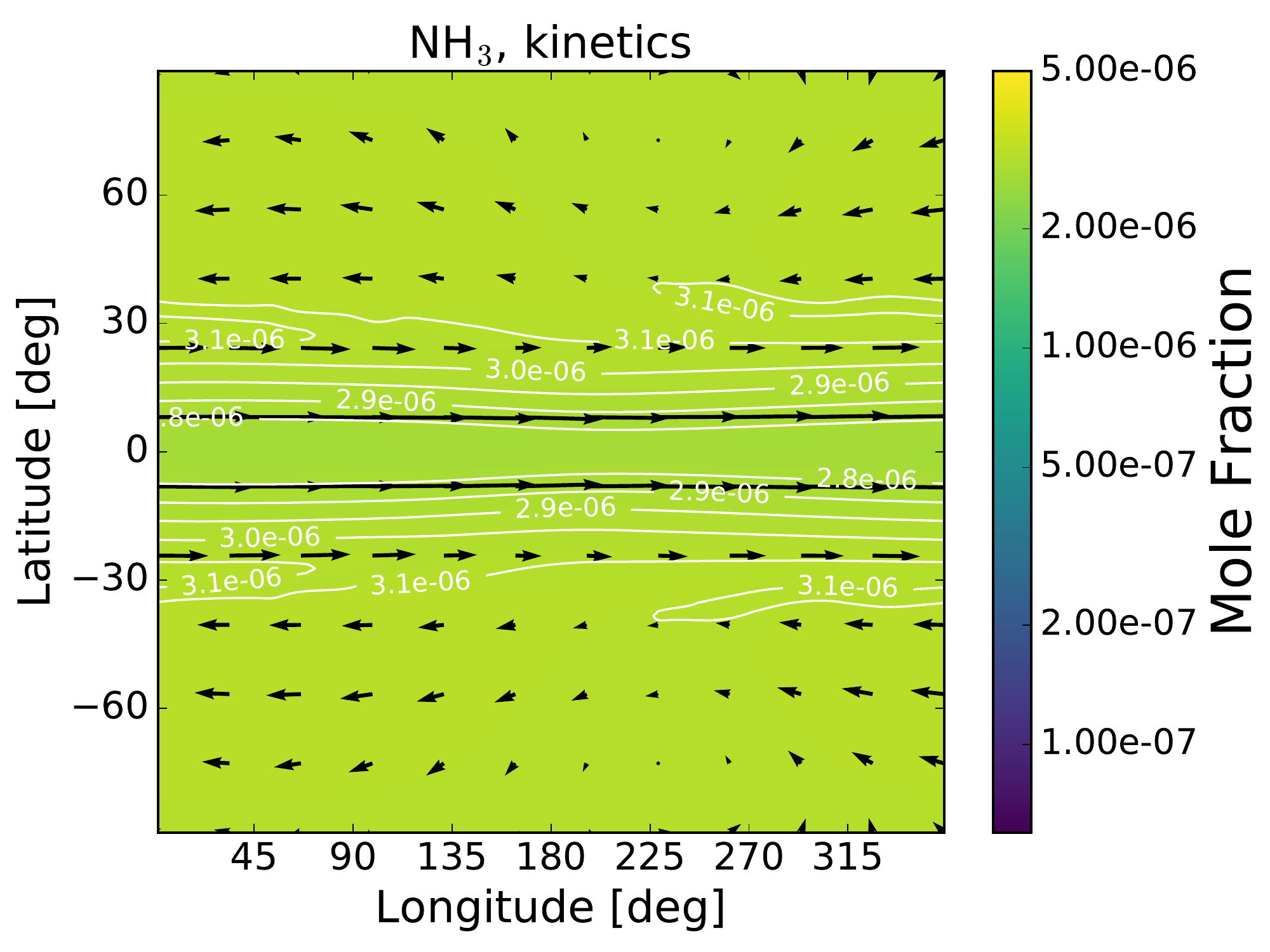} \\
  \end{center}
\caption{Mole fractions (colour scale and white contours) of CH$_4$, H$_2$O, NH$_3$, CO, CO$_2$, and HCN and wind vectors (arrows) on the $1\times10^4$ Pa (0.1 bar) isobar for the equilibrium simulation (left column) and the kinetics simulation (right column) of HD~189733b.}
\label{figure:hd189_mf_eq}
\end{figure*}

\begin{figure*}
  \ContinuedFloat
    \captionsetup{list=off,format=cont}
  \begin{center}
    \includegraphics[width=0.45\textwidth]{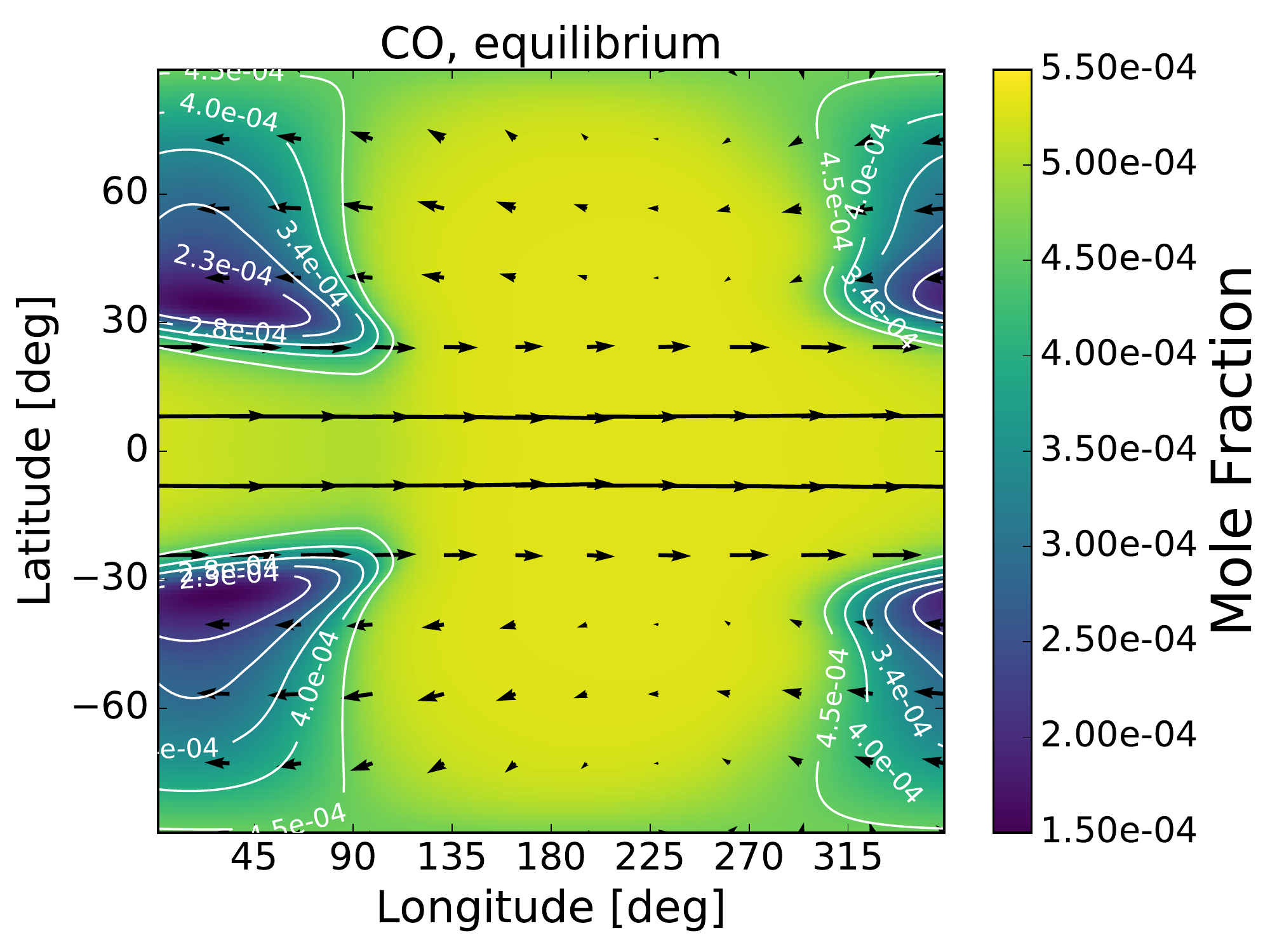}
    \includegraphics[width=0.45\textwidth]{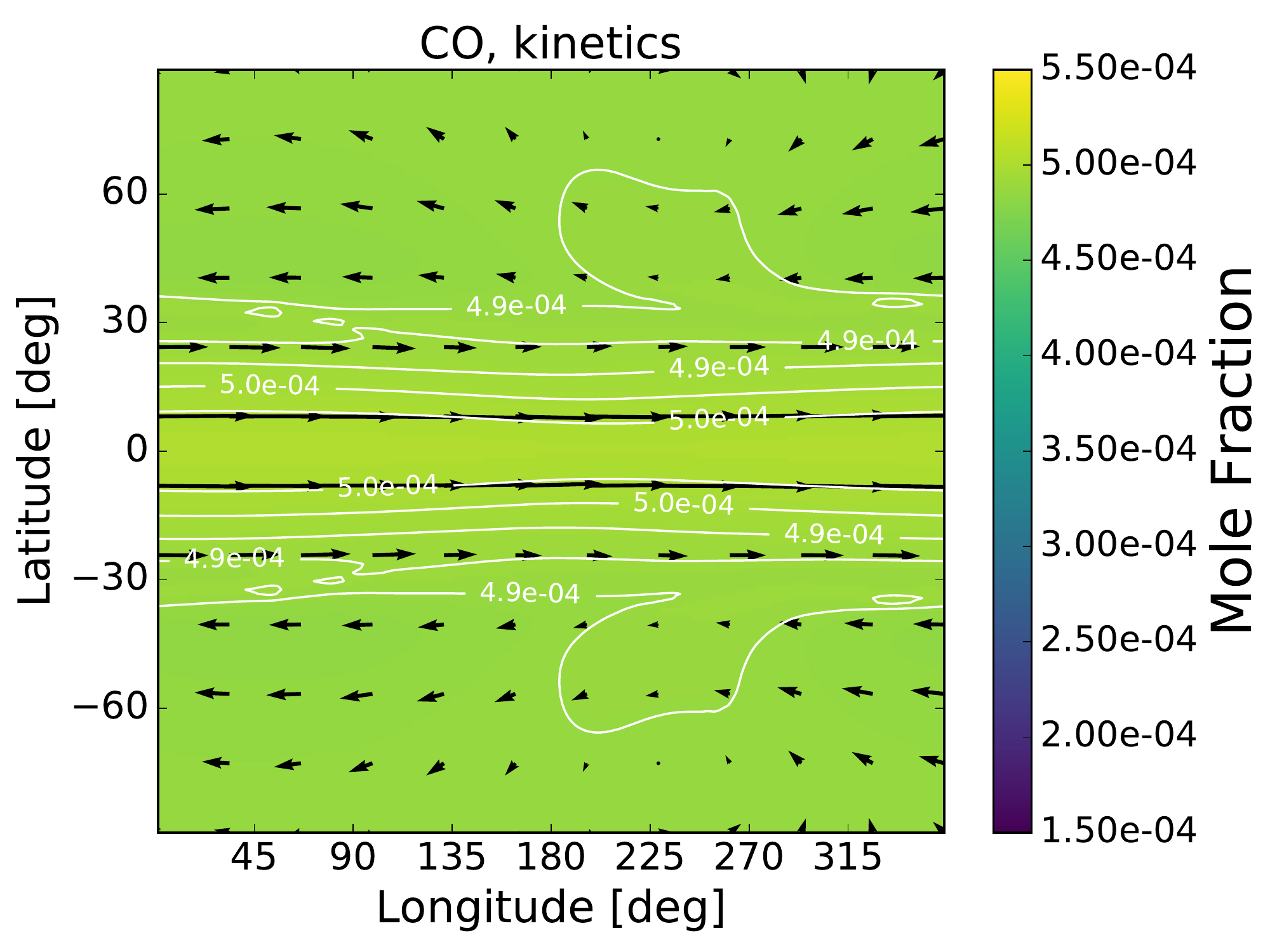} \\
     \includegraphics[width=0.45\textwidth]{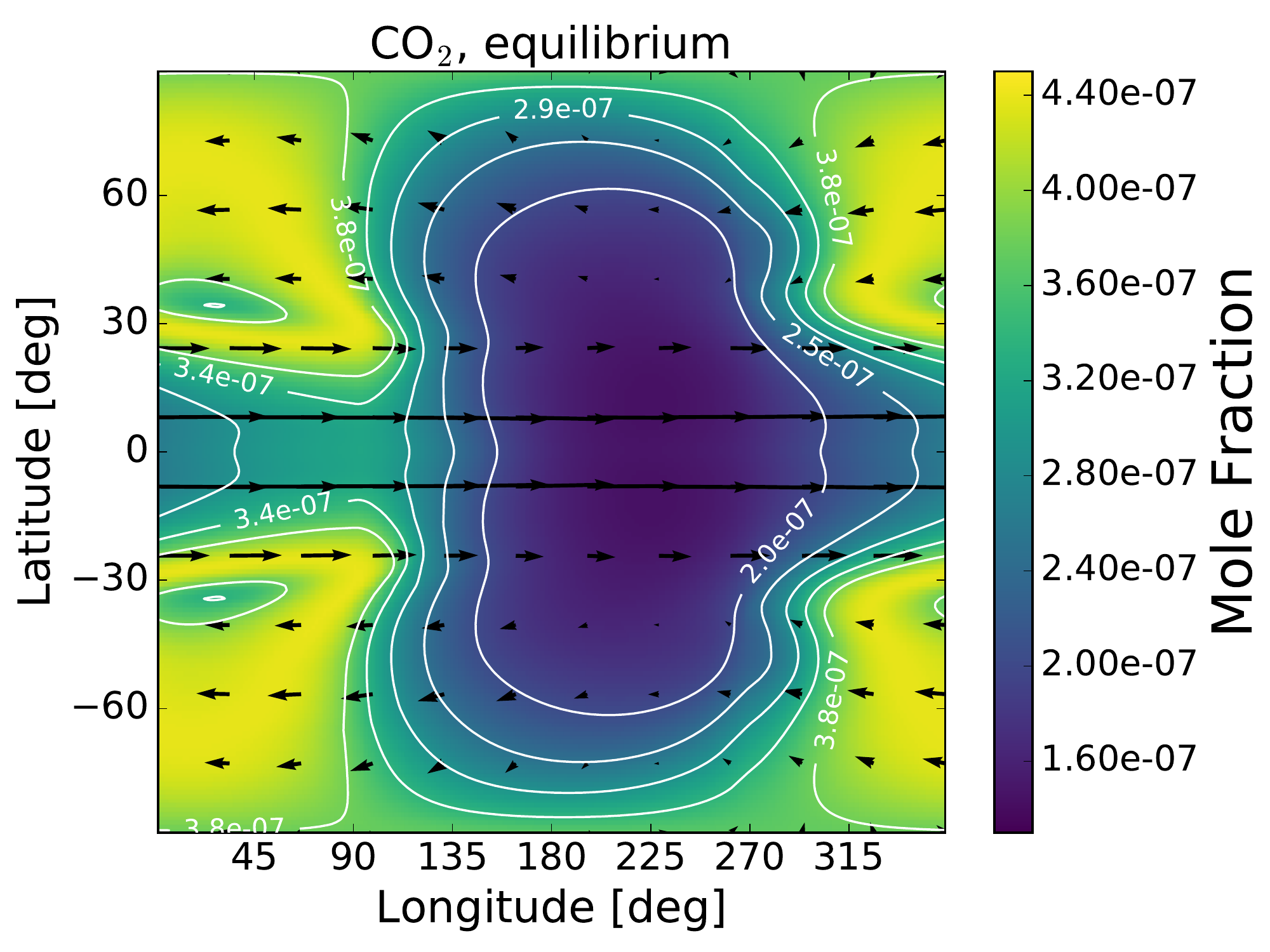}
    \includegraphics[width=0.45\textwidth]{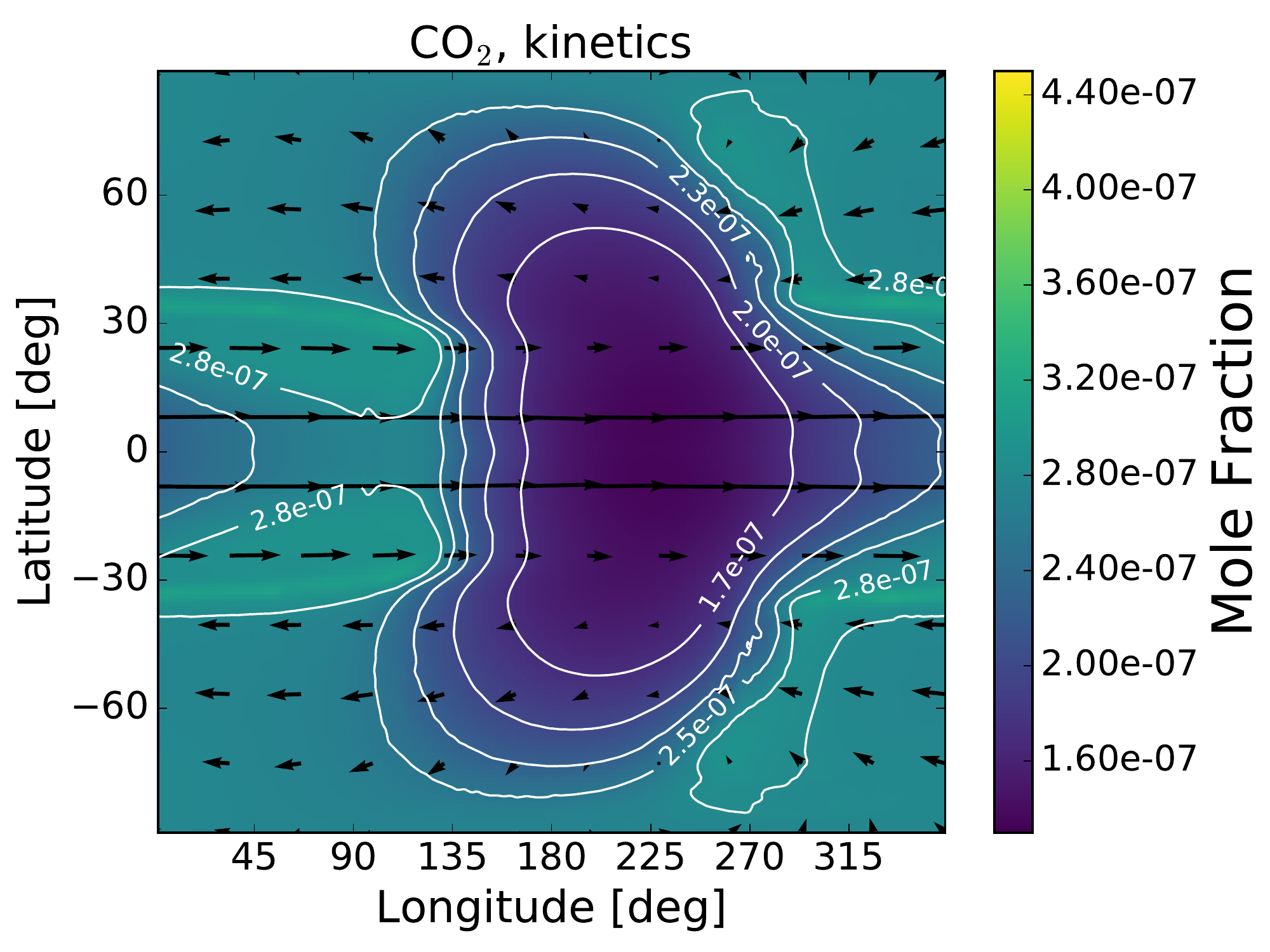} \\
     \includegraphics[width=0.45\textwidth]{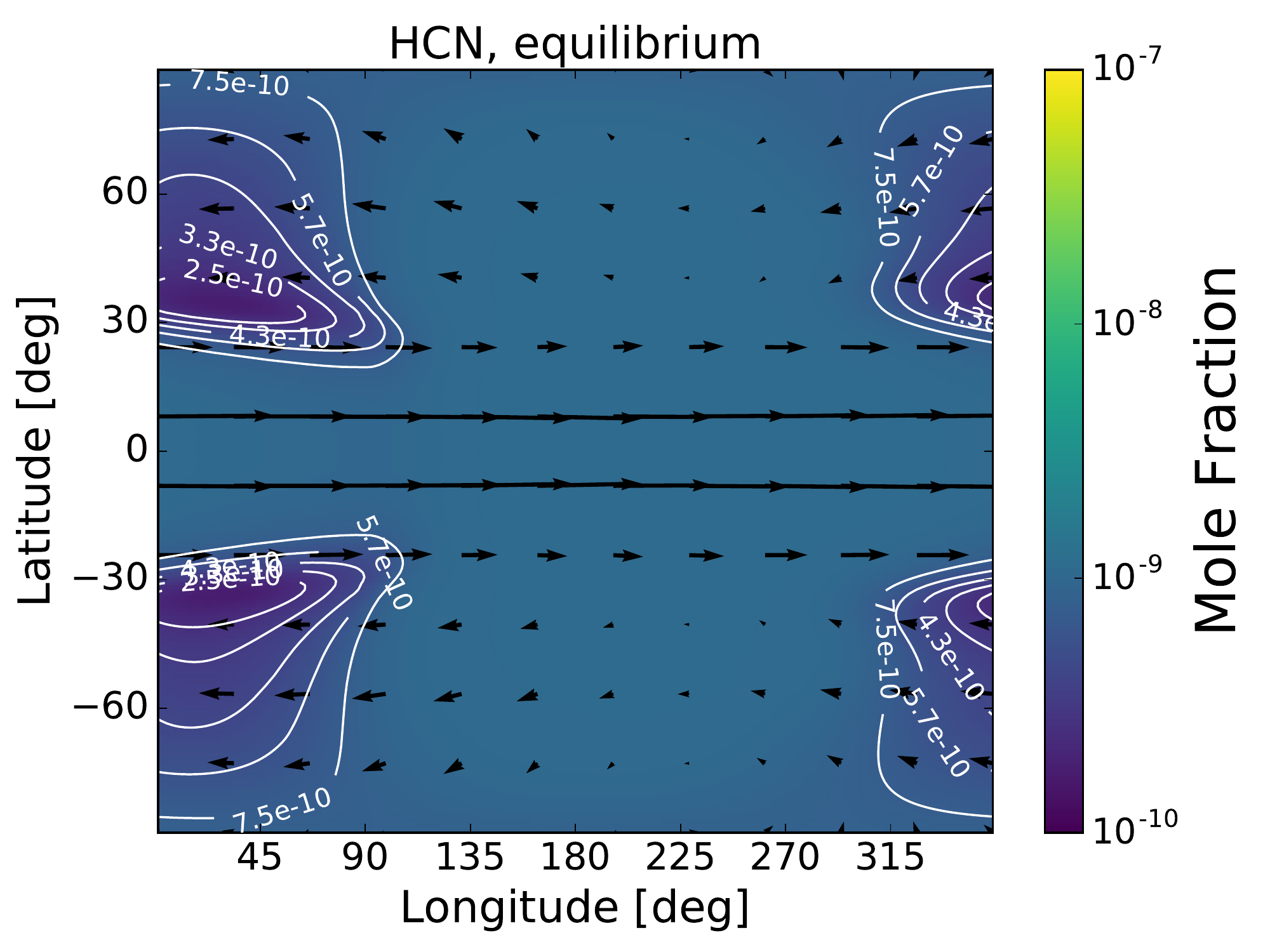}
    \includegraphics[width=0.45\textwidth]{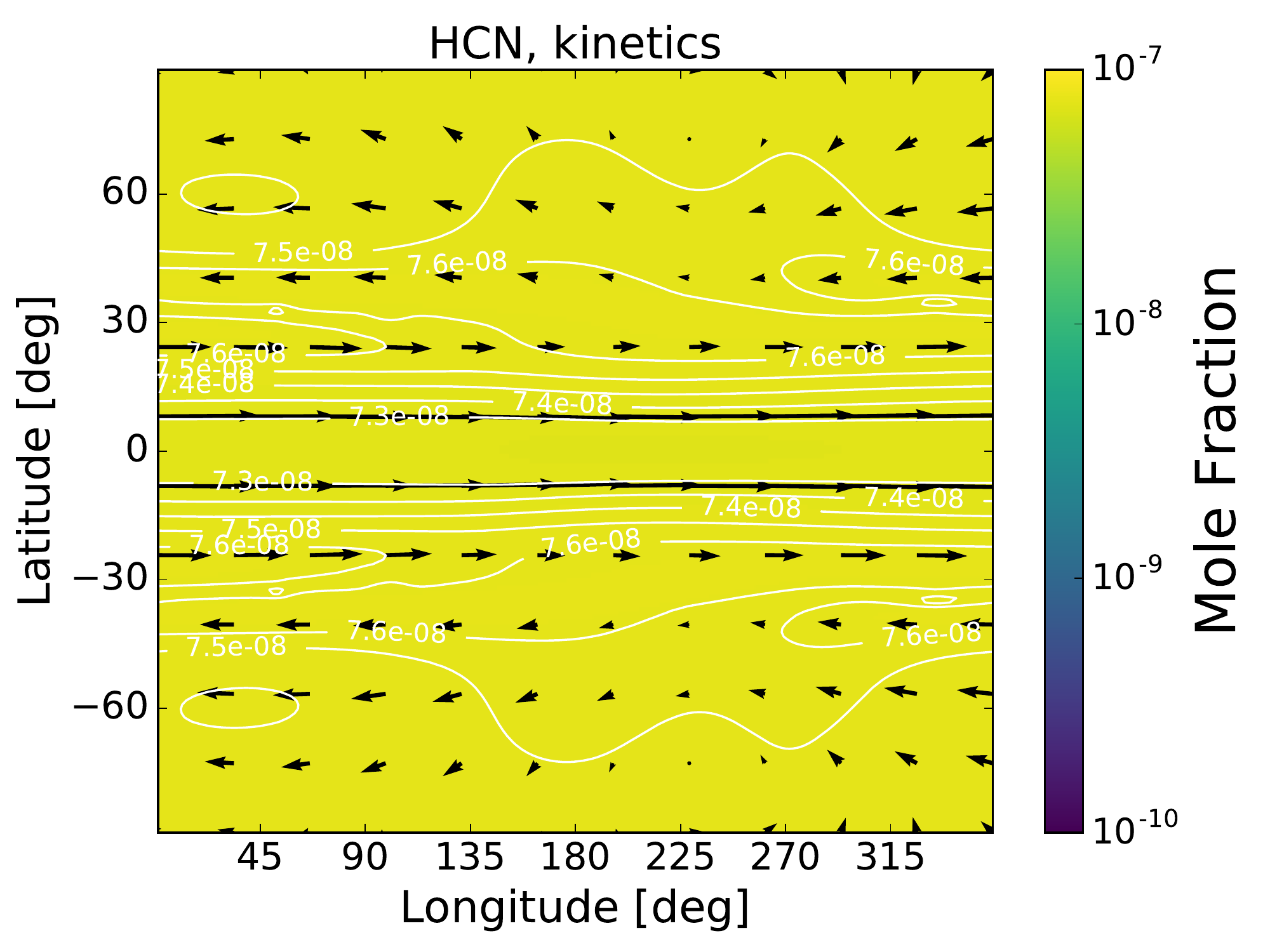} \\
  \end{center}
\caption{}
\end{figure*}

\cref{figure:hd189_mf_eq} (right column) shows the mole fractions of CH$_4$, CO, H$_2$O, CO$_2$, NH$_3$, and HCN on the $P=10^4$~Pa isobar for the kinetics simulation of HD~189733b. Here we see that for most molecules the effect of 3D advection is to largely remove the horizontal abundance gradients that are present in the chemical equilibrium simulation. For CH$_4$, CO, H$_2$O, HCN, and NH$_3$ the remaining horizontal abundance gradients are very small. However, it is clear that CO$_2$ does not change significantly from its chemical equilibrium abundance on the dayside. This means that temperature-dependent kinetics is still important for this molecule on the warm dayside and advection is not the dominant process in determining its abundance. Larger differences in the CO$_2$ abundance between the equilibrium and kinetics simulations are found on the cooler nightside. For CH$_4$, CO, and H$_2$O our results qualitatively agree with our earlier work using a more simplified chemical scheme \citep{DruMM18a}, as we also find that these molecules are efficiently horizontally mixed.

\begin{figure*}
  \begin{center}
    \includegraphics[width=0.45\textwidth]{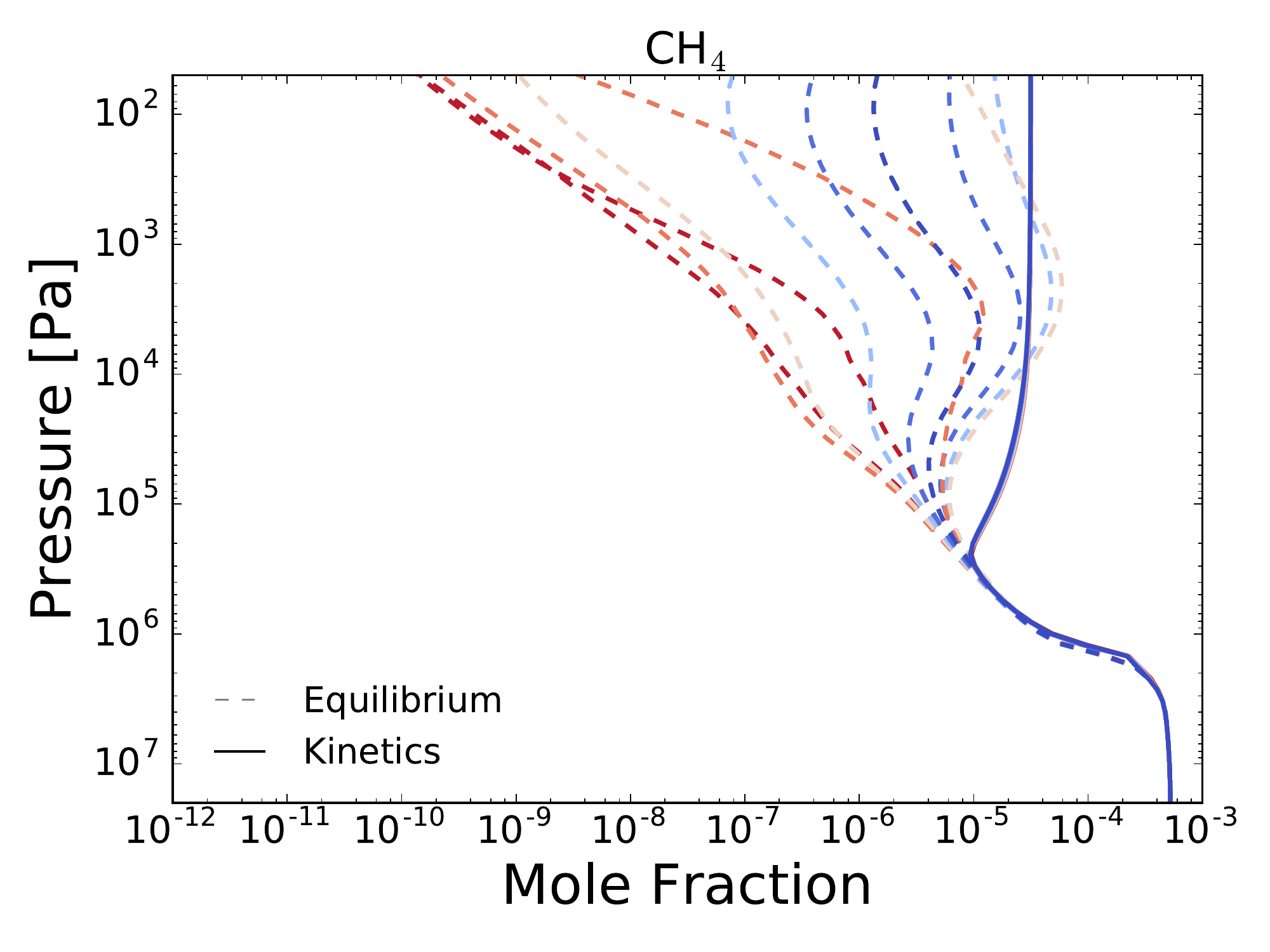}
     \includegraphics[width=0.45\textwidth]{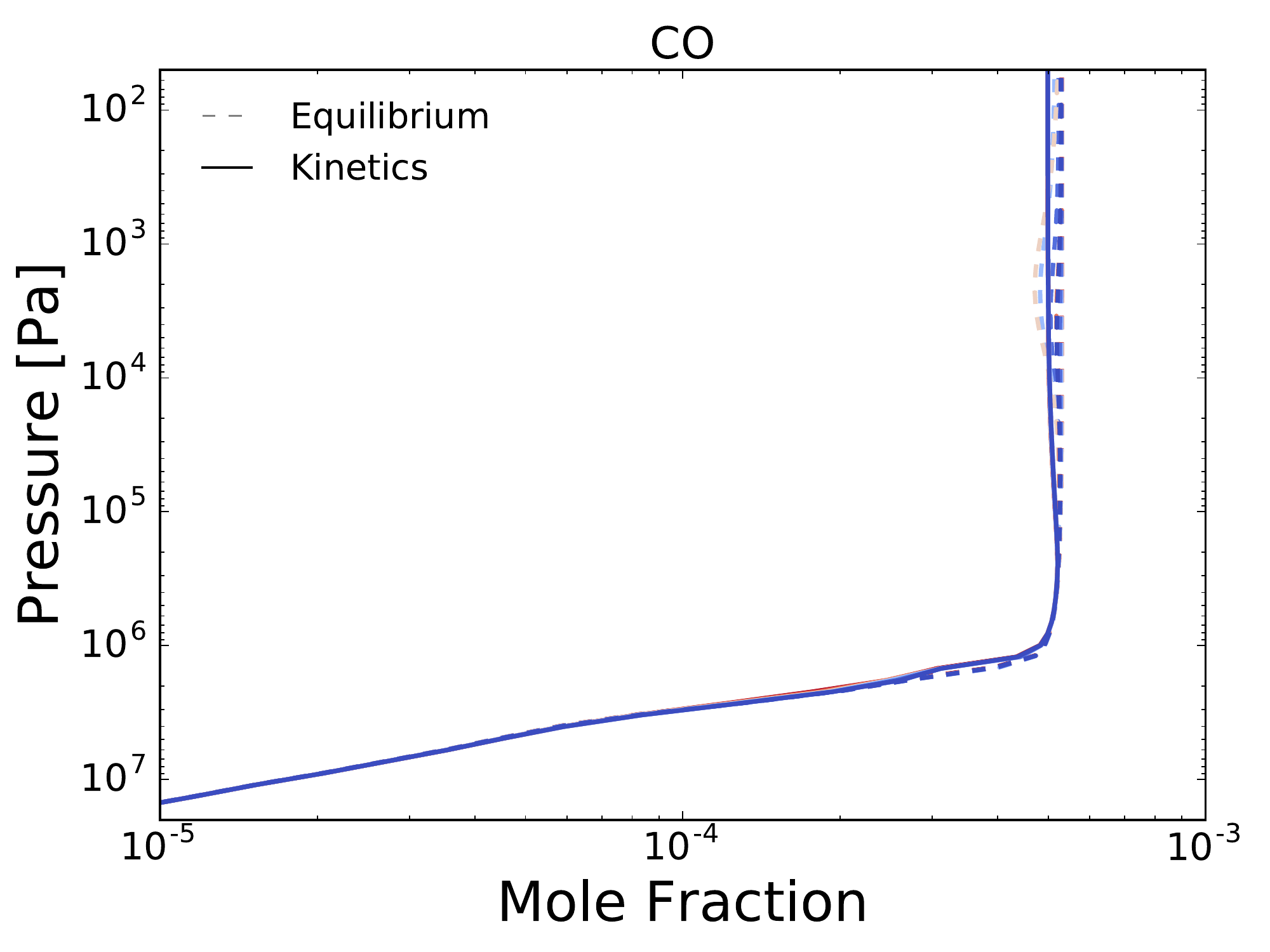} \\
     \includegraphics[width=0.45\textwidth]{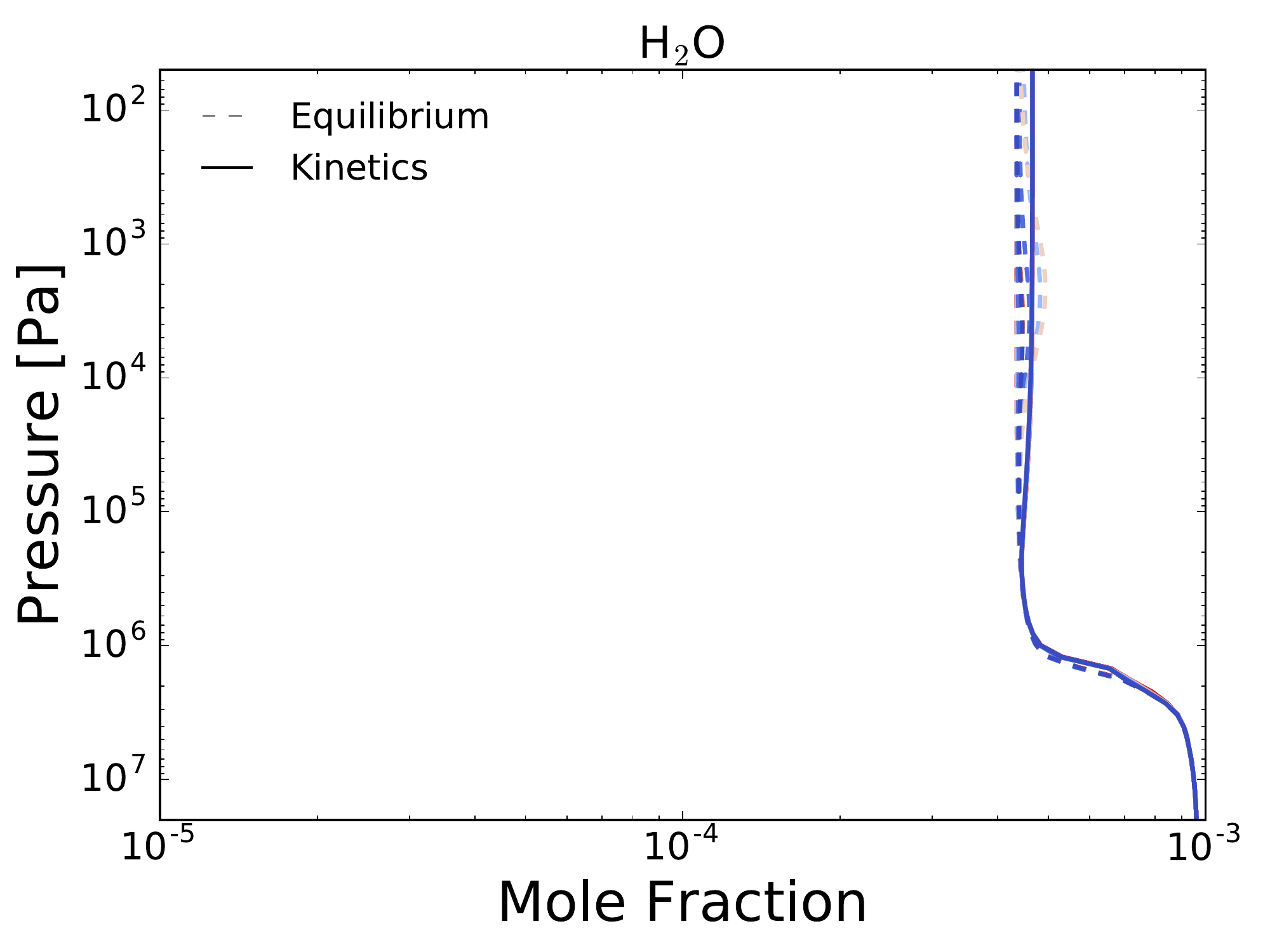}
    \includegraphics[width=0.45\textwidth]{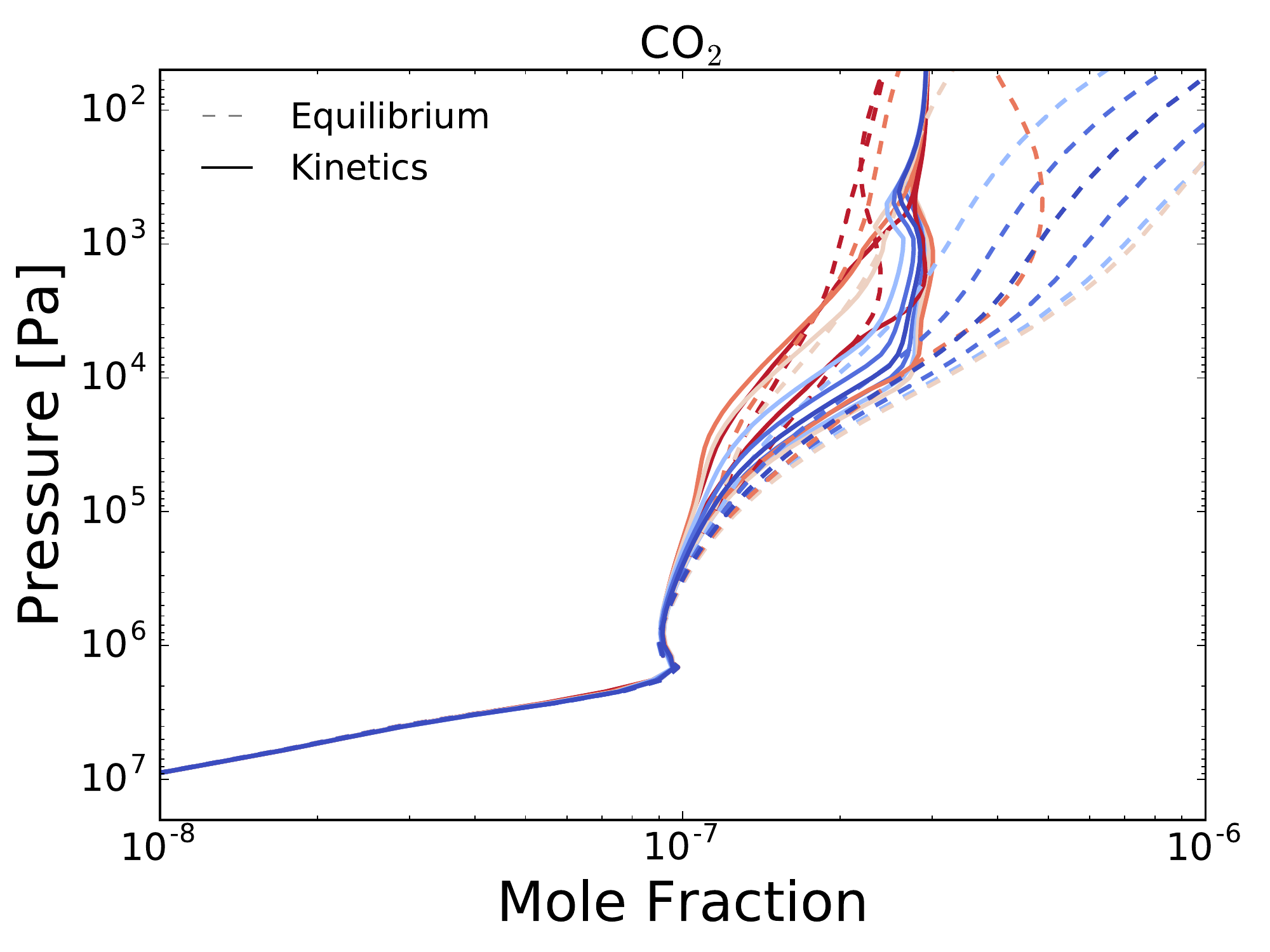}\\
     \includegraphics[width=0.45\textwidth]{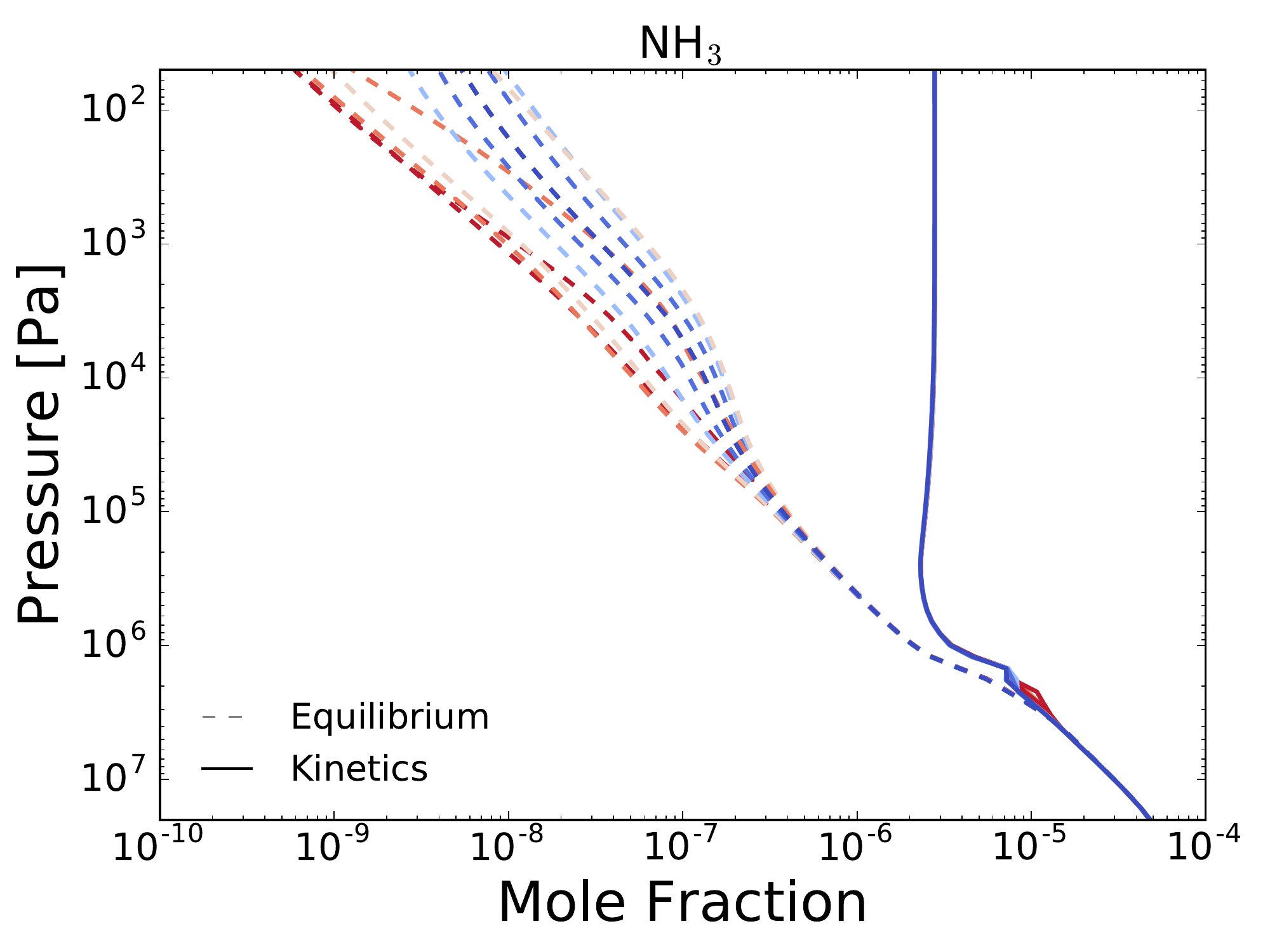}
 \includegraphics[width=0.45\textwidth]{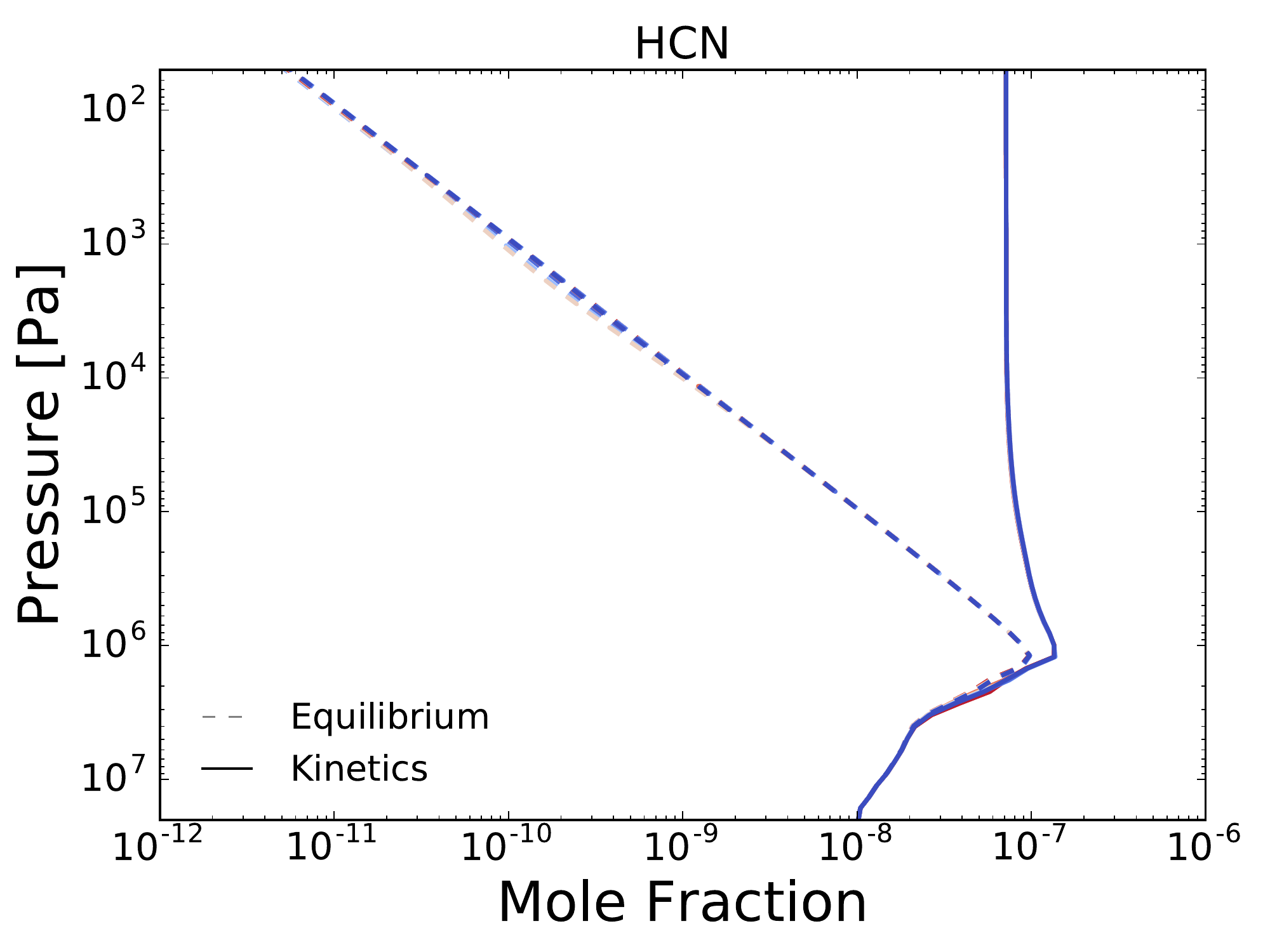}\\
  \end{center}
\caption{As \cref{figure:hd209_mf_prof}, but for the equilibrium simulation (dashed lines) and kinetics simulation (solid lines) of HD~189733b.}
\label{figure:hd189_mf_prof}
\end{figure*}

\cref{figure:hd189_mf_prof} shows both the equilibrium and kinetics abundances of CH$_4$, CO, H$_2$O, CO$_2$, NH$_3$, and HCN, as vertical (pressure) mole fraction profiles around the equator. This also shows that most of the molecules are zonally and vertically well-mixed over a large pressure range at the equator. Compared with HD~209458b (see \cref{figure:hd209_mf_prof}), the molecules typically quench at higher pressures, which is not surprising since the temperature is lower and the chemical timescale is longer.

The most obvious difference between our results for HD~189733b and HD~209458b are in the vertical profiles of CH$_4$. For HD~209458b, over a significant pressure range the abundance of CH$_4$ appears to be zonally but not vertically quenched, with vertical mixing becoming important only for pressures less than $10^4$~Pa. For HD~189733b on the other hand we find that CH$_4$ becomes vertically and horizontally well-mixed (i.e. quenched) much deeper, approximately at the base of the equatorial jet (\cref{figure:zonal_wind}). In the pressure range $10^5$~Pa to $10^4$~Pa the equatorial abundance of CH$_4$ becomes larger than any of its equatorial equilibrium abundances, significantly increasing the relative abundance of CH$_4$ for this region of the atmosphere. This structure is very similar to that found previously using a more simplified chemical scheme \citep{DruMM18a,DruMM18b}. In those earlier works we argued that this feature is due to meridional transport of material from the cooler mid-latitudes, where CH$_4$ is more abundant at local chemical equilibrium, to the equatorial region. 

In \citet{DruMM18b} we conducted a simple tracer experiment that demonstrated that mass can be transported from the mid-latitudes into the equatorial region on a timescale of several hundred days. The tracer was initialised to an abundance of zero everywhere, except in a `source' region at high-pressure and high-latitudes. After several hundred days of model integration, the tracer had increased in abundance significantly at the equator with a clear equator-ward transport occurring between 10$^5$~Pa and 10$^3$~Pa. Once the material is transported into the equatorial region, the material is rapidly distributed zonally and vertically by the fast equatorial jet and large dayside upwelling. Since the circulation for both HD~209458b and HD~189733b is broadly similar between this work and that tracer experiment, we expect the same transport process to be occuring. The difference between the two cases is in the chemical timescale, with a faster chemical timescale leading to a lower pressure quenching in the case of HD~209458b, above the region where the meridional transport is important.

The abundance profiles of CO$_2$ are quite similar to those found for HD~209458b. CO$_2$ remains close to its chemical equilibrium value to much lower pressures than other molecules shown. For $P>10^3$~Pa, we attribute the departure of the abundances of CO$_2$ from chemical equilibrium as being due to its pseudo-equilibrium with CO and H$_2$O. But for $P\lesssim10^3$~Pa, where CO$_2$ becomes approximately zonally uniform we expect this to be due to zonal quenching of CO$_2$ itself.

\begin{figure}
  \begin{center}
    \includegraphics[width=0.45\textwidth]{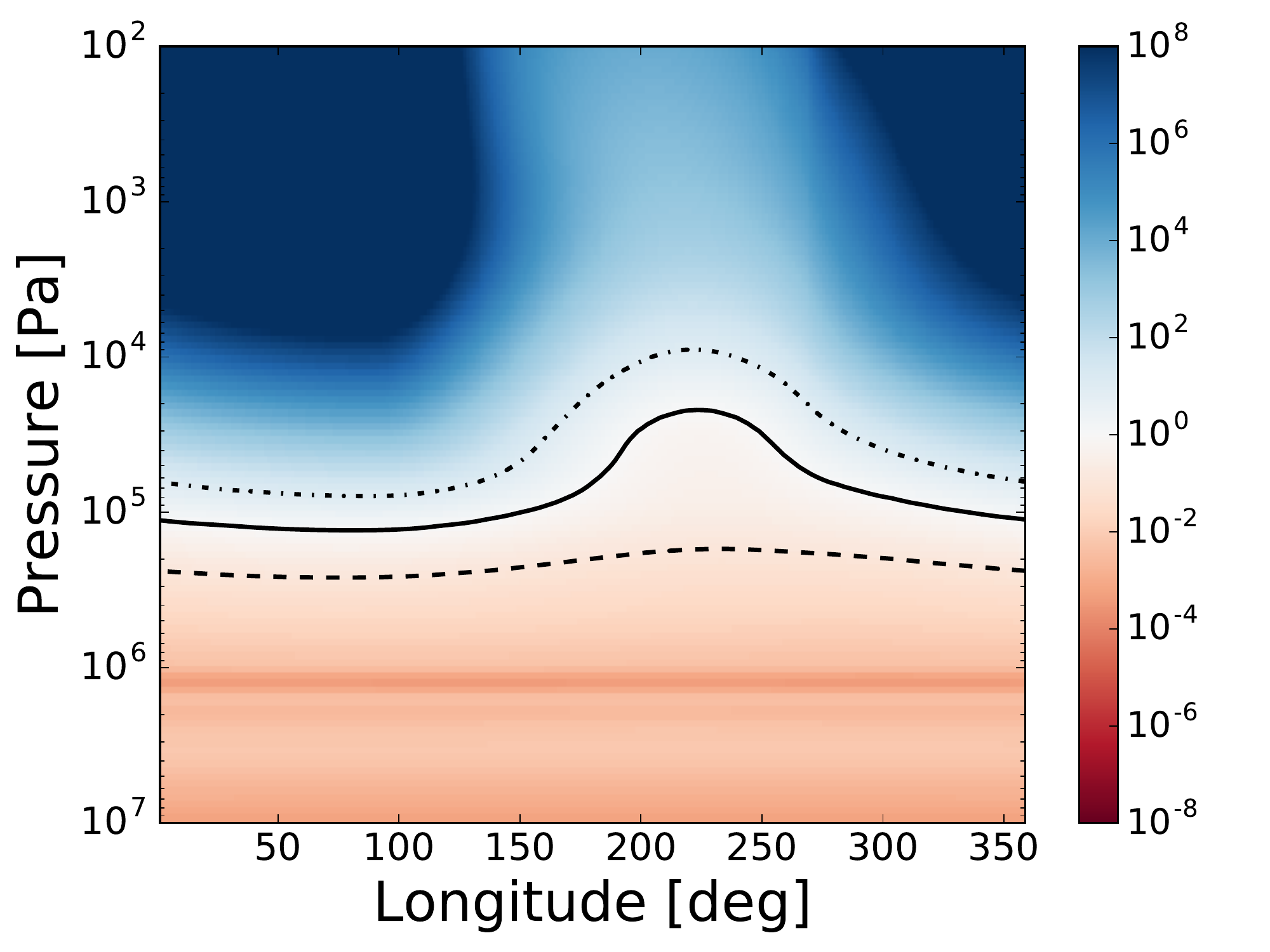} \\
    \includegraphics[width=0.45\textwidth]{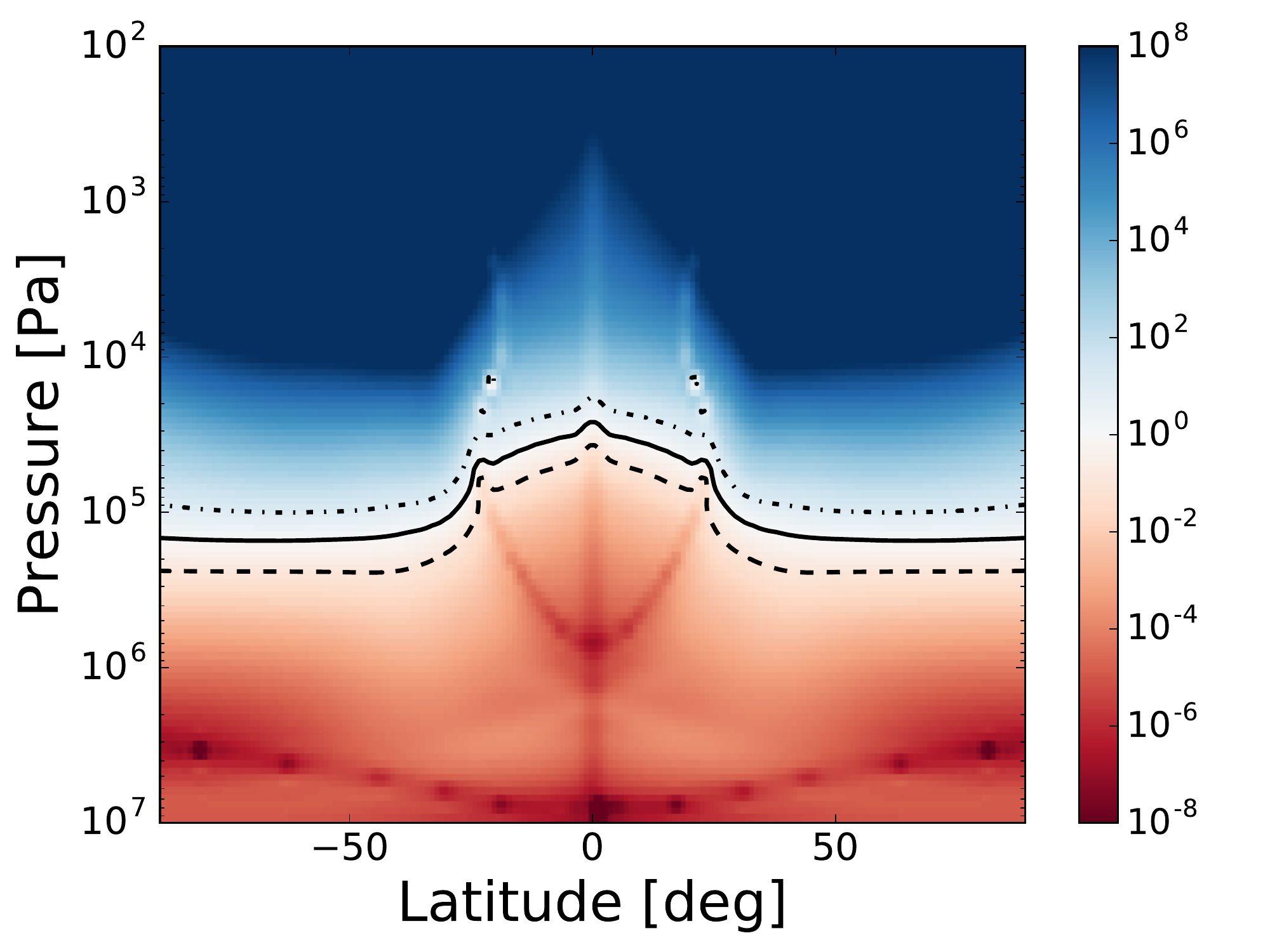} \\
    \includegraphics[width=0.45\textwidth]{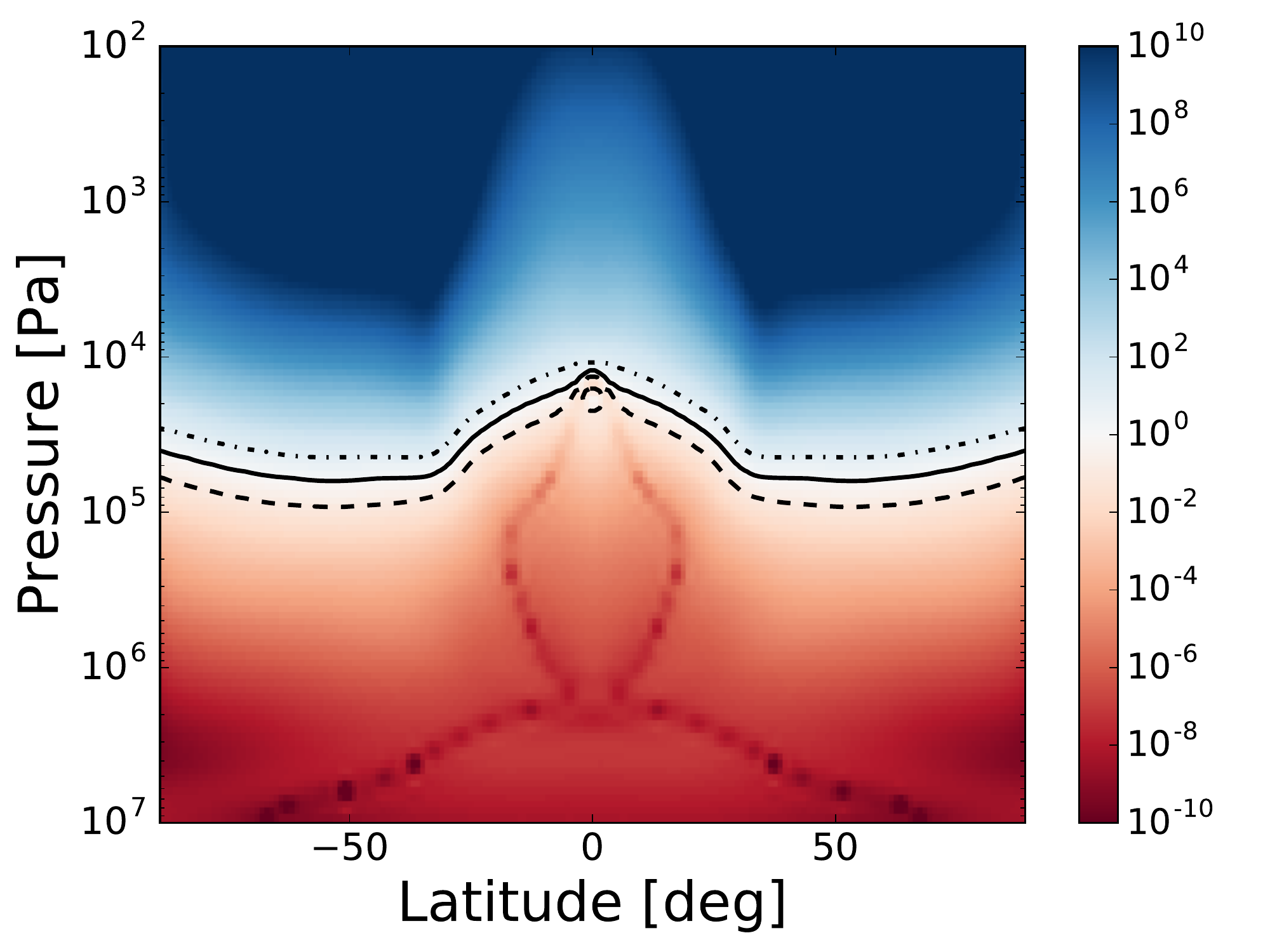} \\
  \caption{As \cref{figure:hd209_ts}, but for the kinetics simulation of HD~189733b.}
  \label{figure:hd189_ts}
  \end{center}
\end{figure}

\cref{figure:hd189_ts} shows the ratio of the chemical to advection timescales for CH$_4$ for the kinetics simulation of HD~189733b. Generally, the point at which the chemical and advection timescales become comparable (black line) lies at higher pressures than for HD~209458b. In addition, the pressure level at which the advection and chemical timescales become comparable is much more similar between the three components (zonal, meridional and vertical) compared with for HD~209458b, with the transition occurring at $P\sim10^5$~Pa. Together this means that we expect the molecules to quench deeper, and for zonal, meridional and vertical quenching to act in approximately the same regions.  This explains why we see such different behaviour for CH$_4$ between HD~209458b and HD~189733b. 

\subsection{Chemical-radiative feedback}
\label{subsection:Chemical-radiative-feedback}

The advected chemical abundances of CH$_4$, CO, CO$_2$, H$_2$O, NH$_3$, and HCN are used to calculate the total opacity in each grid cell which is then used to calculate the radiative heating rates. This allows for a chemical-radiative feedback process as changes in the chemical composition, due to the transport-induced quenching, can affect the thermal structure and circulation. Of course, the chemistry itself is strongly dependent on the temperature via the temperature-dependent rate coefficients.

Feedback between transport-induced quenching of chemical species and the temperature profile has previously been investigated in 1D considering only vertical diffusion \citep{DruTB16}, showing temperature changes of up to 100~K. In 3D, \citet{DruMM18b} found that advection of CH$_4$ can lead to changes in the temperature by up to 5--10\% for HD~189733b, compared with the chemical equilibrium case. \citet{StePS18} artificially varied the CO/CH$_4$ ratio in their 3D model and found that significant temperature responses can result, compared with the assumption of local chemical equilibrium.

For HD~209458b we find negligible differences ($<1\%$) in the atmospheric temperature and wind velocities between the equilibrium and kinetics simulations. This is in agreement with our earlier work using a simplified chemical relaxation scheme \citep{DruMM18a}. While there are significant changes in the relative abundances of several molecules, as shown in the previous section, the abundances of these molecules remain relatively small and therefore make only a minor contribution to the total opacity.

On the other hand, for HD~189733b we find more significant differences ($5-10\%$) in the temperatures and wind velocities between the equilibrium and kinetics simulations. The results are very similar to those presented in \citet{DruMM18b} for the same planet but using a simplified chemical relaxation scheme (only considering CH$_4$, CO, and H$_2$O) and we therefore do not show the results here. The fact that the results are so similar is not overly surprising since in \citet{DruMM18b} we found that the changes in the temperature and winds were mainly due to the changes in the CH$_4$ abundance, and the CH$_4$ abundance profiles  for HD~189733b are very similar between this work and \citet{DruMM18b}. This suggests that quenching of other chemical species considered in this work is not important for the radiative heating rates.

\subsection{Spectra and phase curves}

In this section we present synthetic spectra calculated directly from our 3D atmosphere simulations. We calculate both transmission spectra and secondary eclipse emission spectra, and search for signatures of the chemical changes between the equilibrium and kinetics simulations. We also present simulated JWST and ARIEL observations to assess the observational significance of any differences in the model spectra from the perspective of these up-coming space-based instruments.

PandExo \citep{Batalha2017} was used to simulate JWST observations of the emission and transmission spectra. The simulated instruments used were the Near InfraRed Spectrograph (NIRSpec) and the Mid-Infrared Instrument (MIRI), using their G395H and LRS modes respectively. The stellar parameters were sourced from the TEPCat database \citep{Southworth2011}. The results were computed assuming a single orbit/transit, and an equal amount of observing time in-and-out of transit, corresponding to a total observation time of 6.096 hours for HD~209458 and 3.614 hours for HD~189733. The ARIEL Noise Simulator will be described in a future work \citep{mugnai_Arielrad}. For the ARIEL simulated observations five eclipses where used for the emission spectra and ten transits for transmission spectra.

\subsubsection{HD~209458b}

\begin{figure*}
  \begin{center}
  \includegraphics[width=0.45\textwidth]{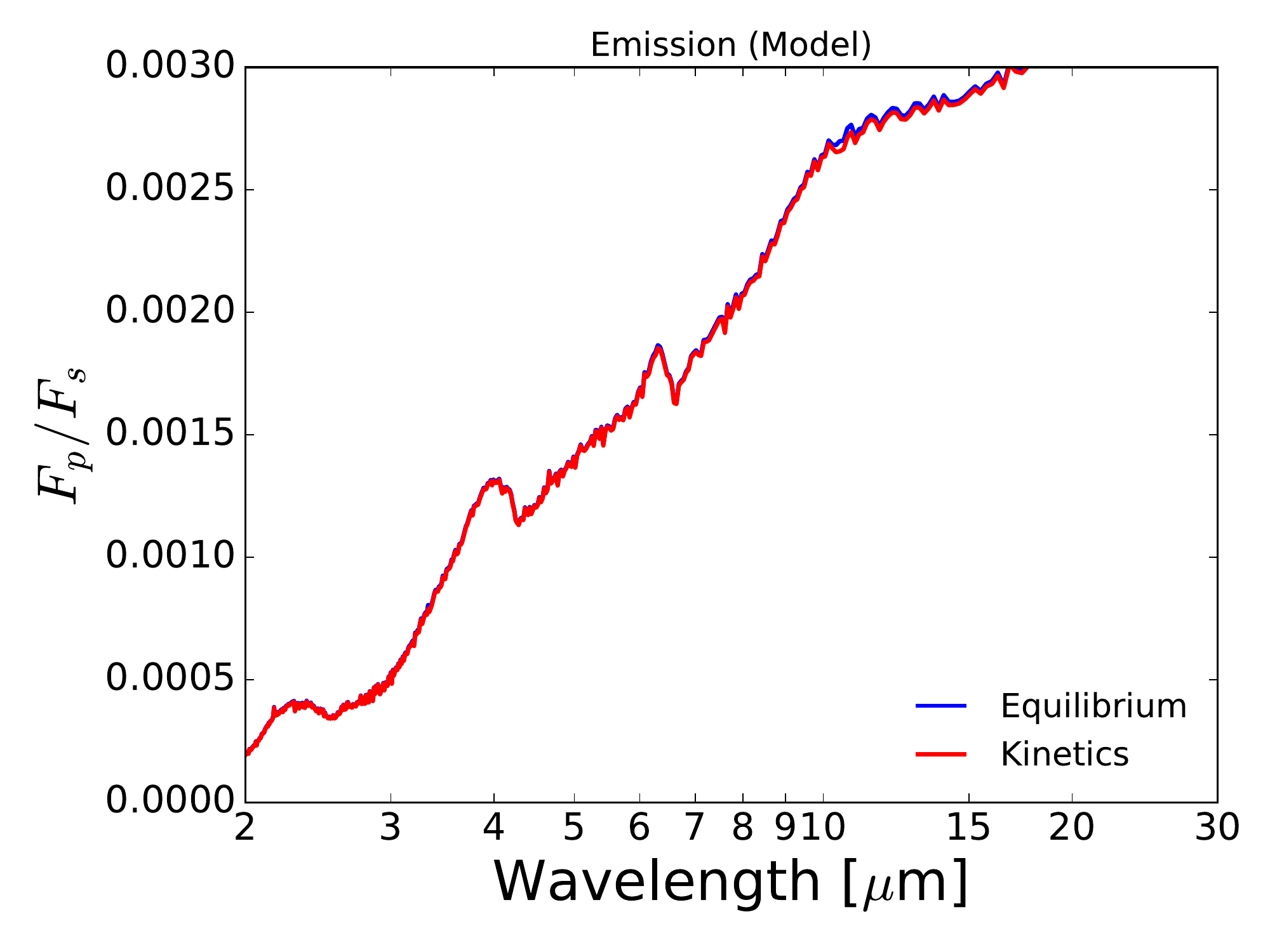} 
  \includegraphics[width=0.45\textwidth]{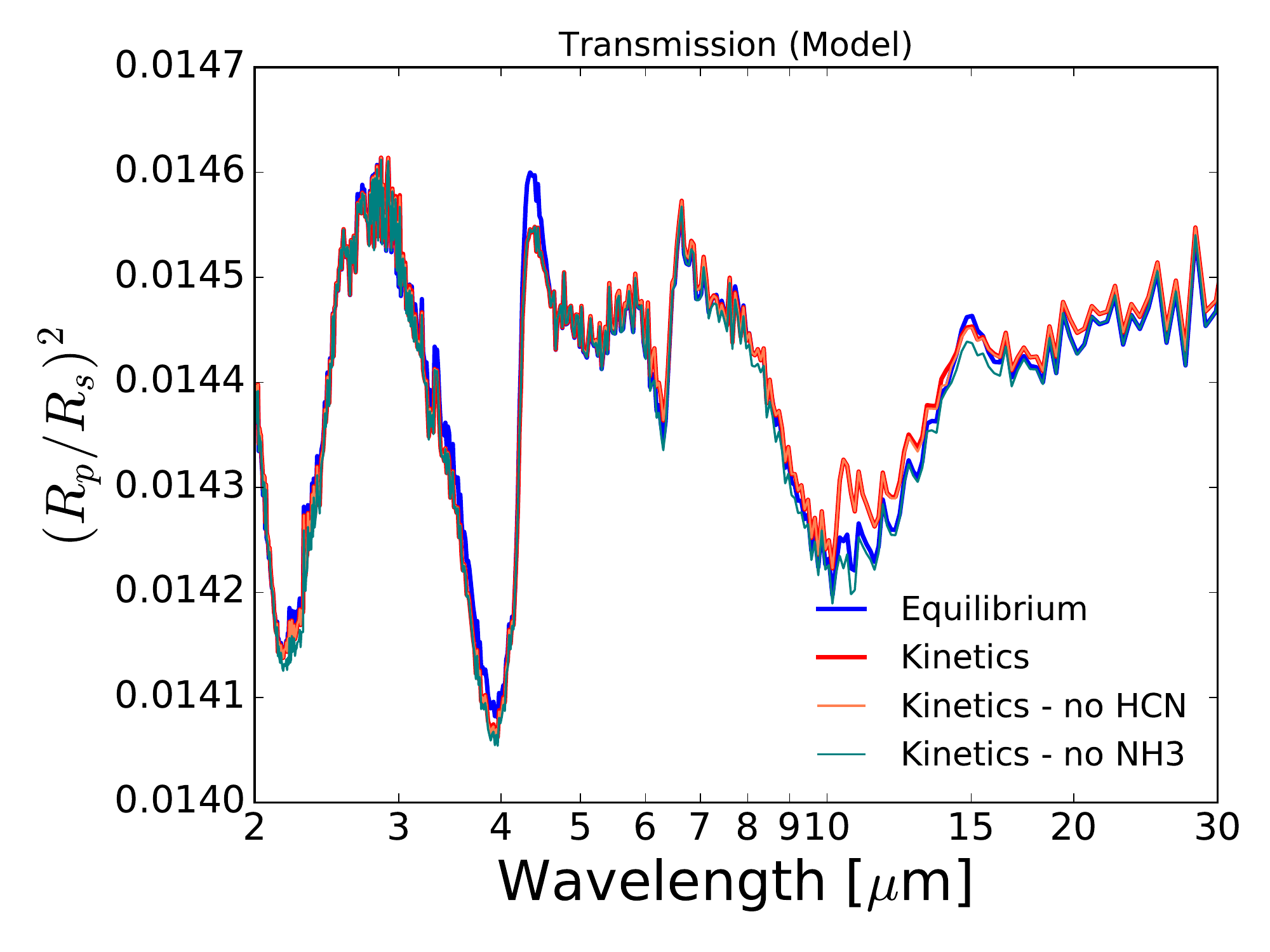} \\
   \includegraphics[width=0.45\textwidth]{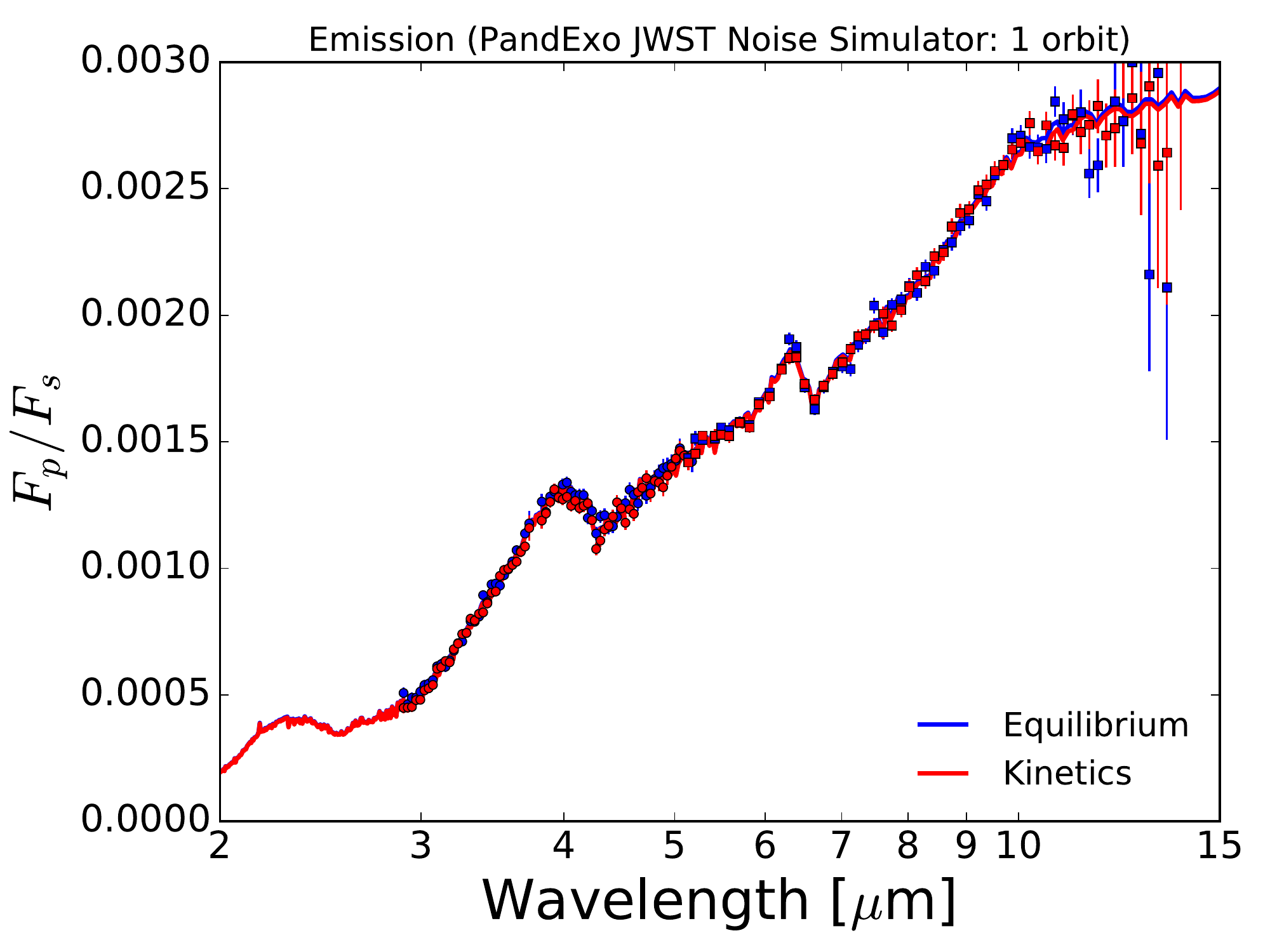} 
  \includegraphics[width=0.45\textwidth]{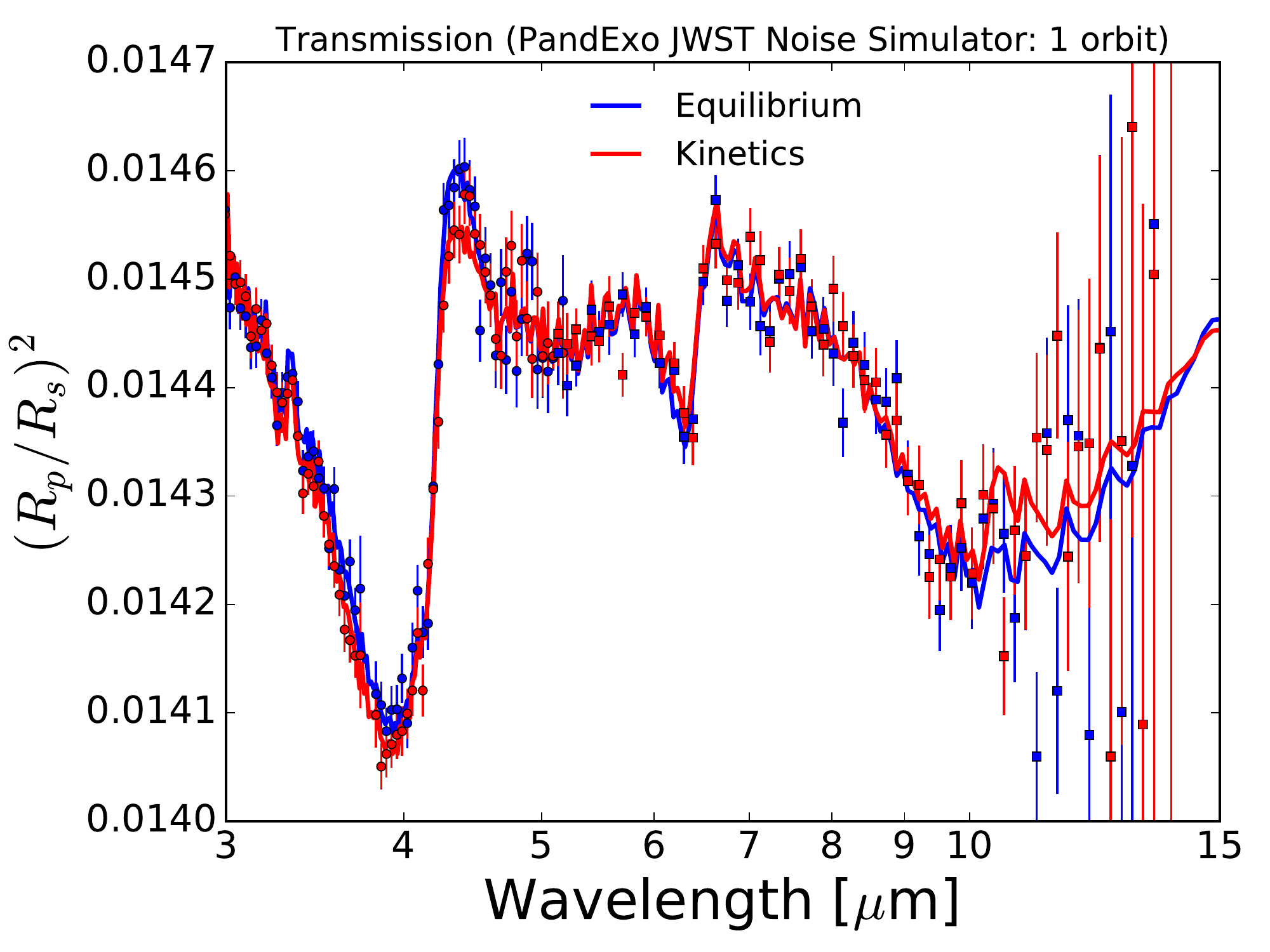} \\
   \includegraphics[width=0.45\textwidth]{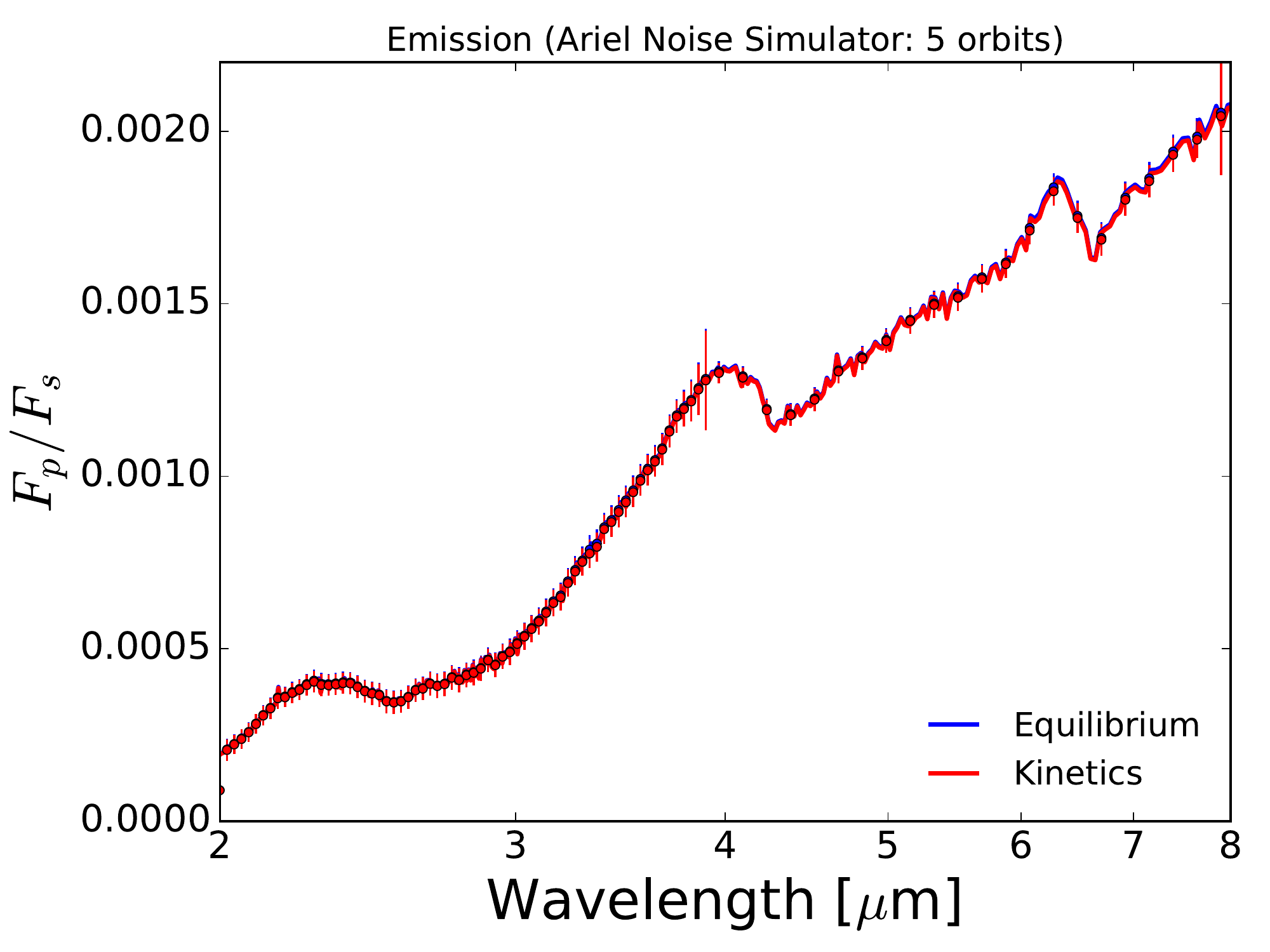} 
  \includegraphics[width=0.45\textwidth]{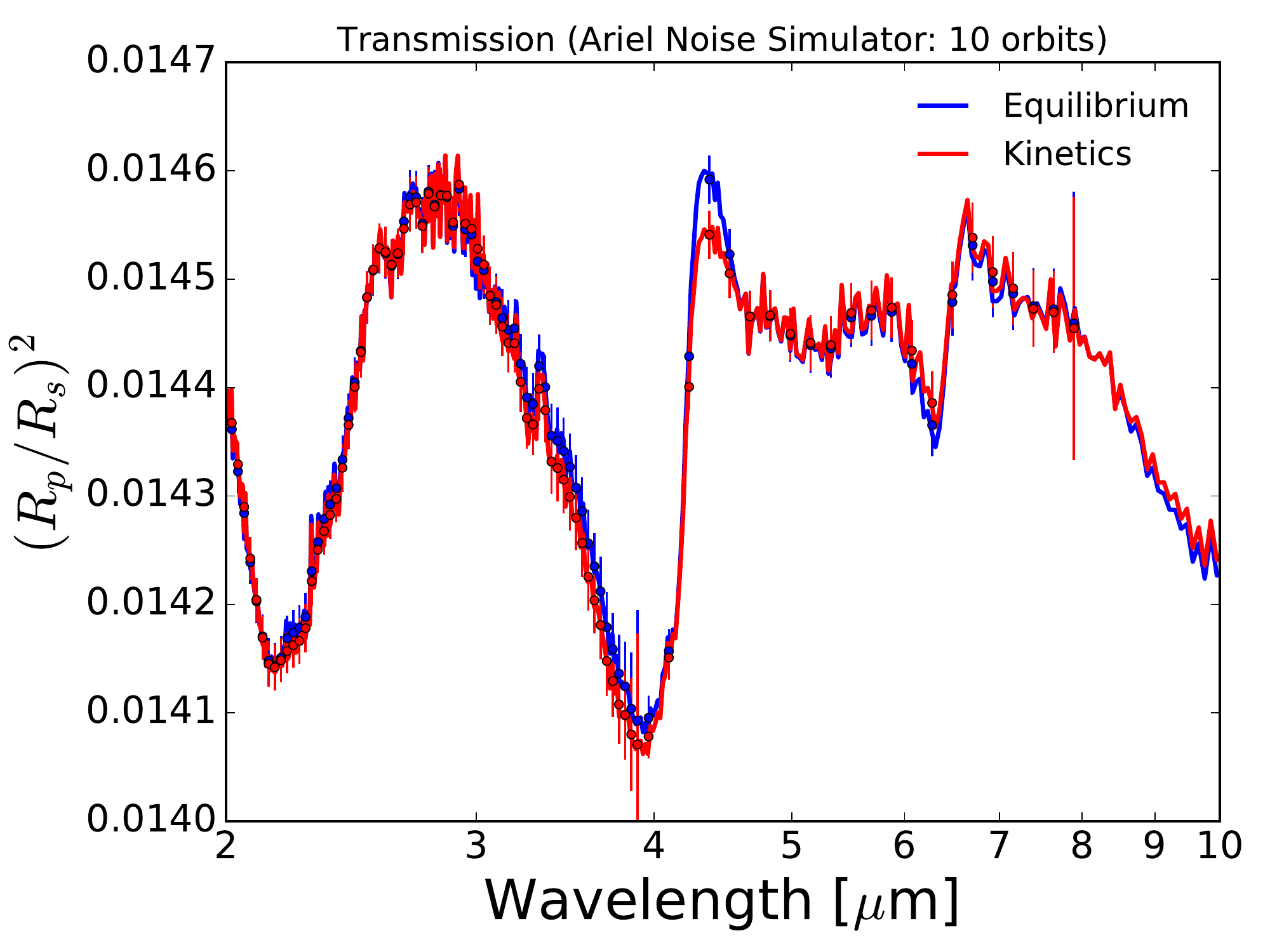} \\
  \end{center}
  \caption{Transmission (left) and emission (right) spectra for the equilibrium (blue) and kinetics (red) simulations of HD~209458b. The top row shows the model spectra, the middle row overlays simulated JWST observations and the bottom row overlays simulated ARIEL observations.}
  \label{figure:hd209_spectra}
\end{figure*}

\cref{figure:hd209_spectra} shows the emission and transmission spectra calculated for both the equilibrium and kinetics simulations of HD~209458b. In emission it is clear that there are negligible differences between the two cases. This is not surprising since the most abundant absorbing species (CO and H$_2$O) show relatively little difference in abundance between the equilibrium and kinetics simulations. Whilst the abundances of CH$_4$, NH$_3$, and HCN change significantly, their relative abundances are typically below $\sim10^{-7}$ and therefore make only a small contribution to the total opacity. CO$_2$ remains close to its equilibrium abundance on the dayside, which the secondary eclipse is probing. The emission spectrum is also very sensitive to the temperature structure, though there are only very minor ($<1\%$) changes in the temperature. 

In transmission, however, there are two clear spectral regions where the equilibrium and kinetics simulations show differences. Firstly, at $\sim4.5$ microns the kinetics simulation shows a reduced transit depth compared with the equilibrium simulation. This is due to a reduction in the abundance of CO$_2$ in the terminator regions (see \cref{figure:hd209_mf_prof}) which decreases the opacity. In addition, there is an increase in the transit depth between $10-12$ microns, which we find is due to enhanced NH$_3$ absorption. The NH$_3$ abundance is increased significantly in the kinetics simulation compared with the equilibrium simulation, which increases the transit depth in this region where NH$_3$ absorbs.

The middle and bottom rows of \cref{figure:hd209_spectra} overlay the simulated JWST and ARIEL observations, respectively. The JWST simulated observations assume a single transit/eclipse while the ARIEL simulated observations include five transits for transmission and ten eclipses for emission. For a single transit the NIRSpec G395H instrument is able to resolve the feature at 4.5 microns, with additional eclipses expected to improve this. Ten transits with ARIEL are also enough to resolve this feature. The feature due to NH$_3$ at longer wavelengths is not well resolved by the MIRI LRS simulated observations, and is not included in the spectral coverage of ARIEL. 

\begin{figure}
    \centering
    \includegraphics[width=0.45\textwidth]{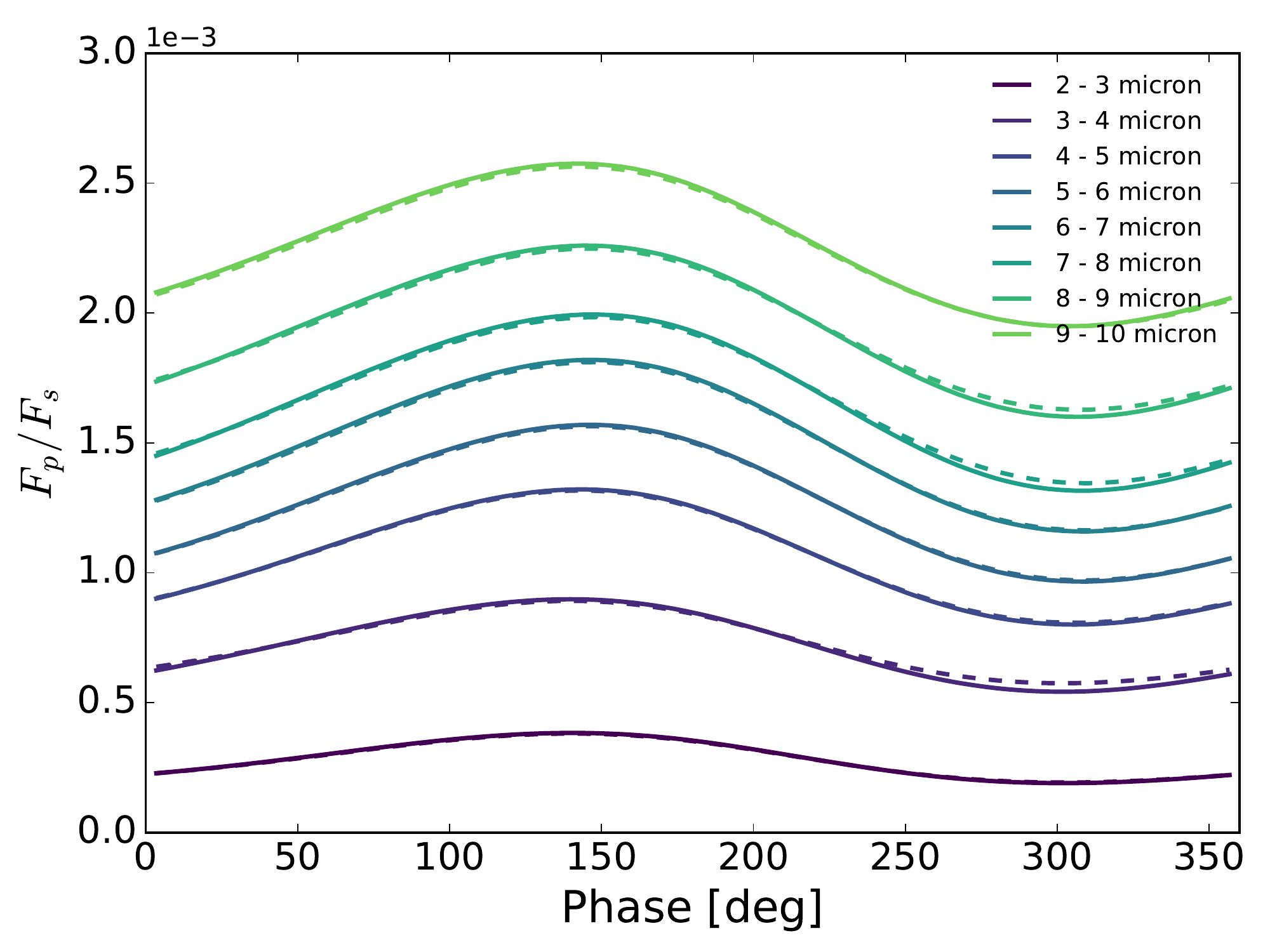}
    \caption{Emission phase curves for the equilibrium (solid) and kinetics (dashed) simulations of HD~209458b in a series of spectral bands. The y-axis is the ratio of the planet to stellar flux.}
    \label{figure:hd209_phase}
\end{figure}

\cref{figure:hd209_phase} shows the simulated emission phase curves for HD~209458b. Considering the secondary eclipse emission spectra previously discussed, it is unsurprising that there is relatively small difference between the equilibrium and kinetics simulations. The most notable difference between the two cases occurs when the nightside of the planet is in view, with the kinetics simulation showing an increased flux ratio compared with the equilibrium simulation. This occurs in the 3-4, 7-8 and 8-9 micron bands, where CH$_4$ is a prominent absorber and is almost certainly due to the decreased nightside abundance of CH$_4$, due to zonal quenching (\cref{figure:hd209_mf_prof}). The decreased CH$_4$ abundance shifts the photosphere to higher pressures and temperatures, resulting in a larger emission flux.

\subsubsection{HD~189733b}

\begin{figure*}
  \begin{center}
  \includegraphics[width=0.45\textwidth]{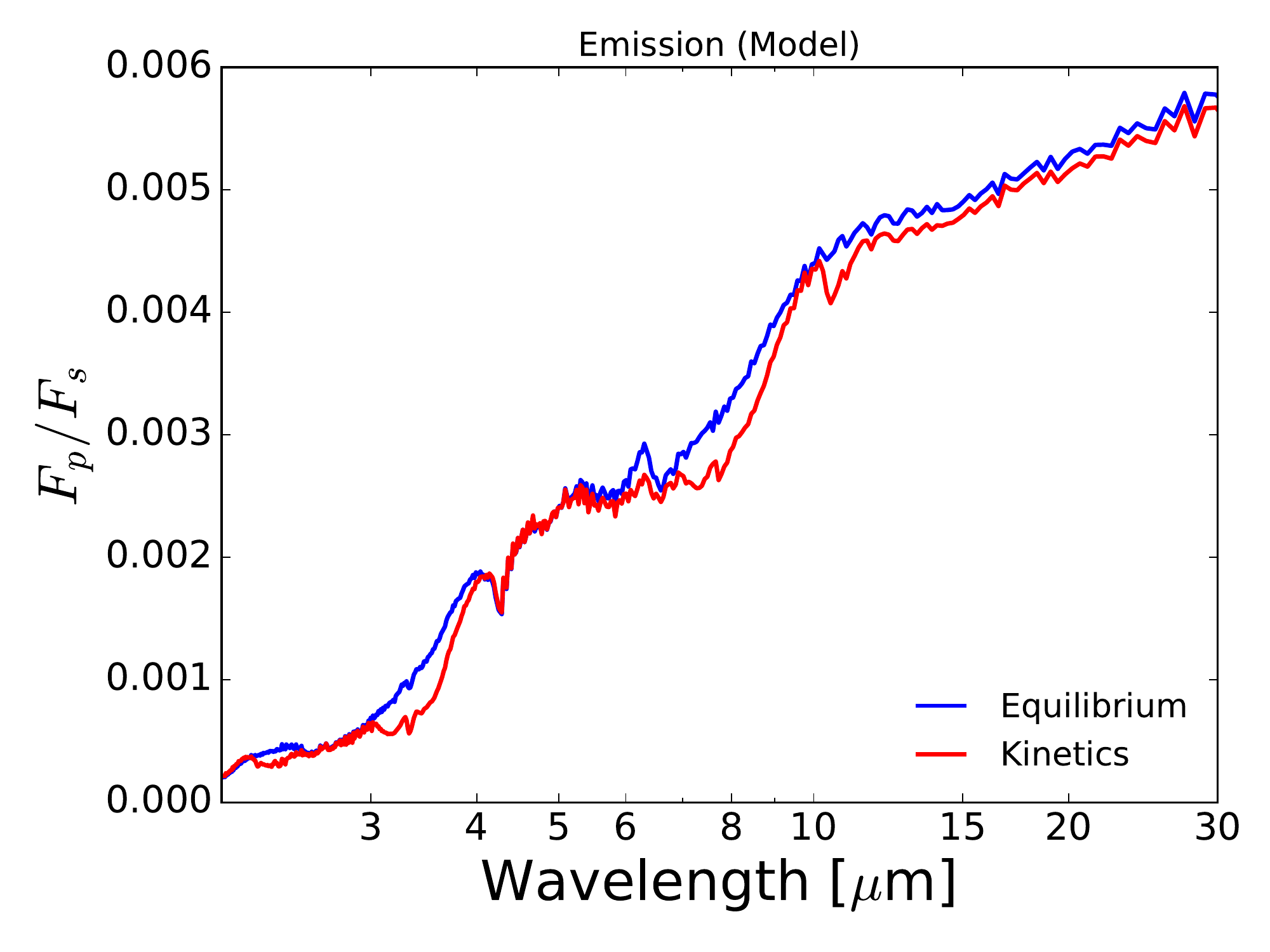} 
  \includegraphics[width=0.45\textwidth]{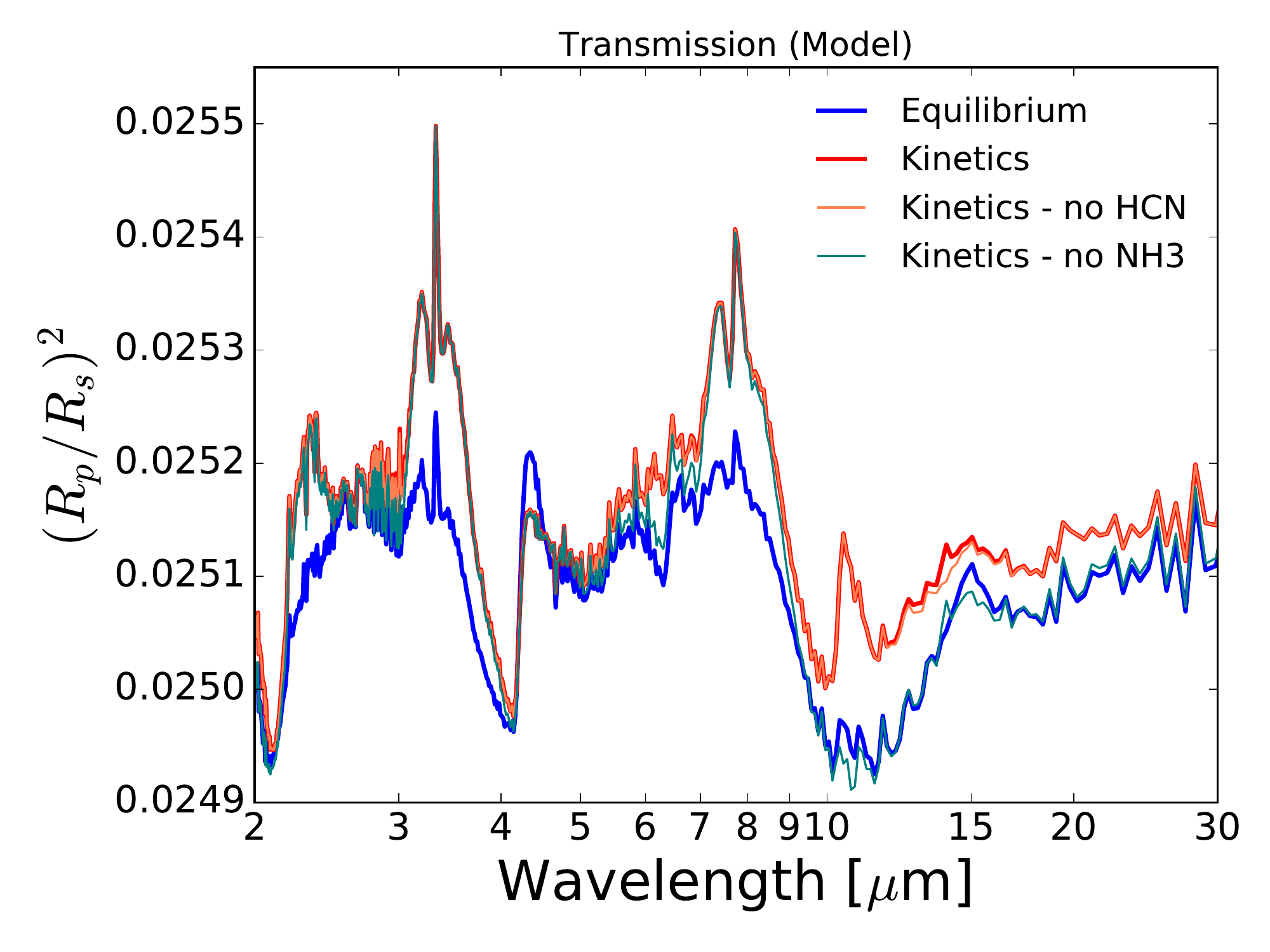} \\
  \includegraphics[width=0.45\textwidth]{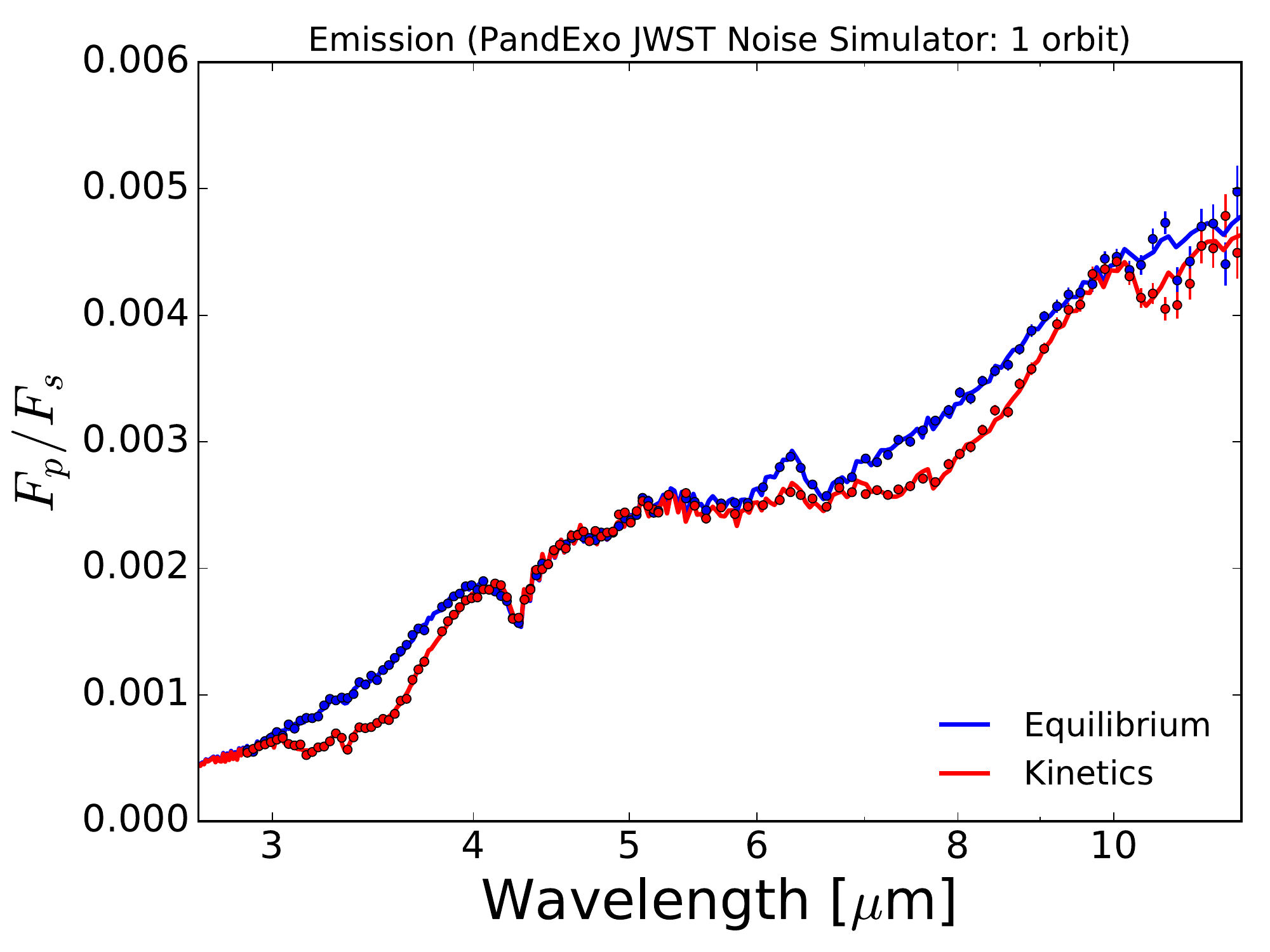} 
  \includegraphics[width=0.45\textwidth]{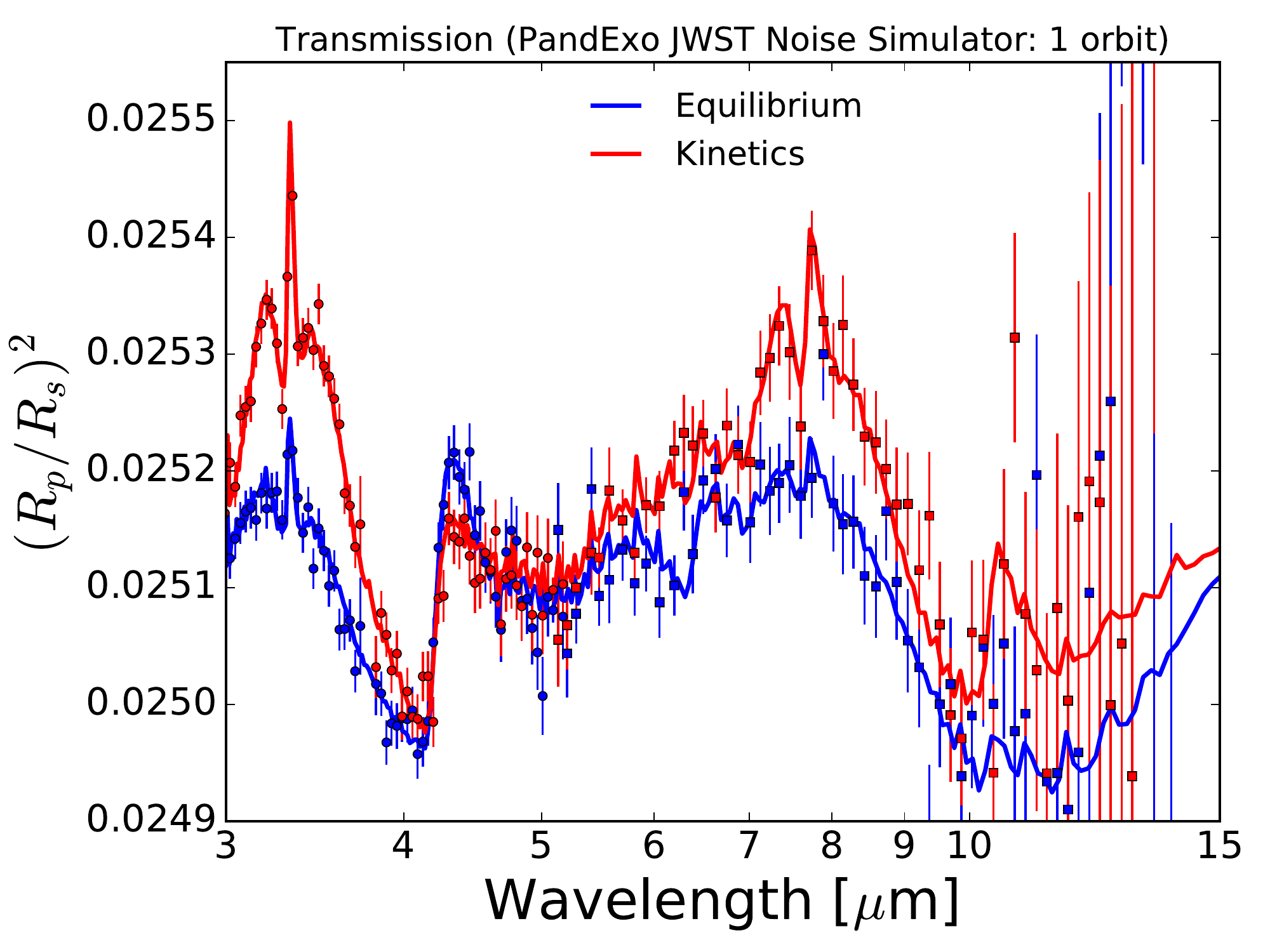} \\
  \includegraphics[width=0.45\textwidth]{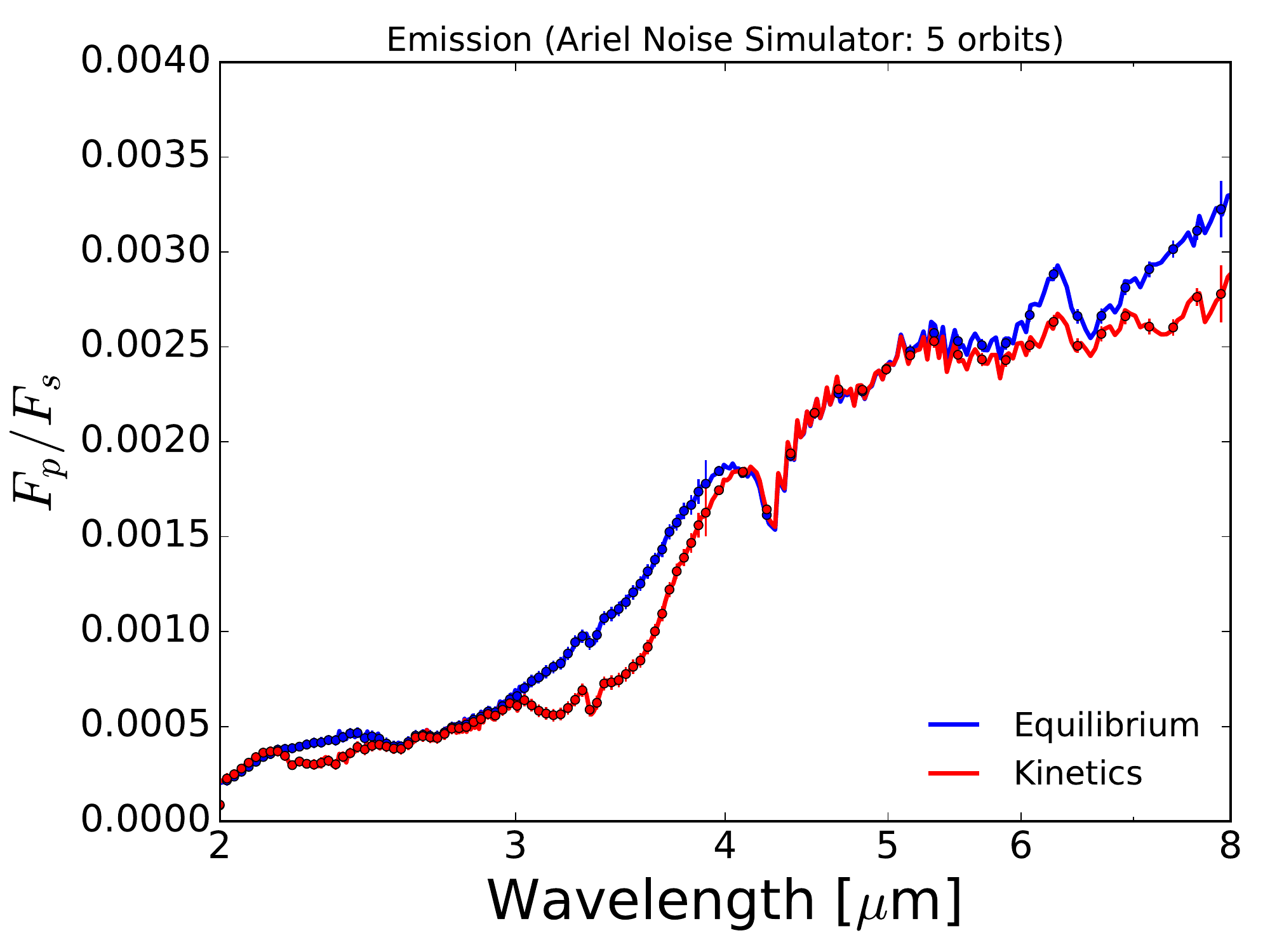} 
  \includegraphics[width=0.45\textwidth]{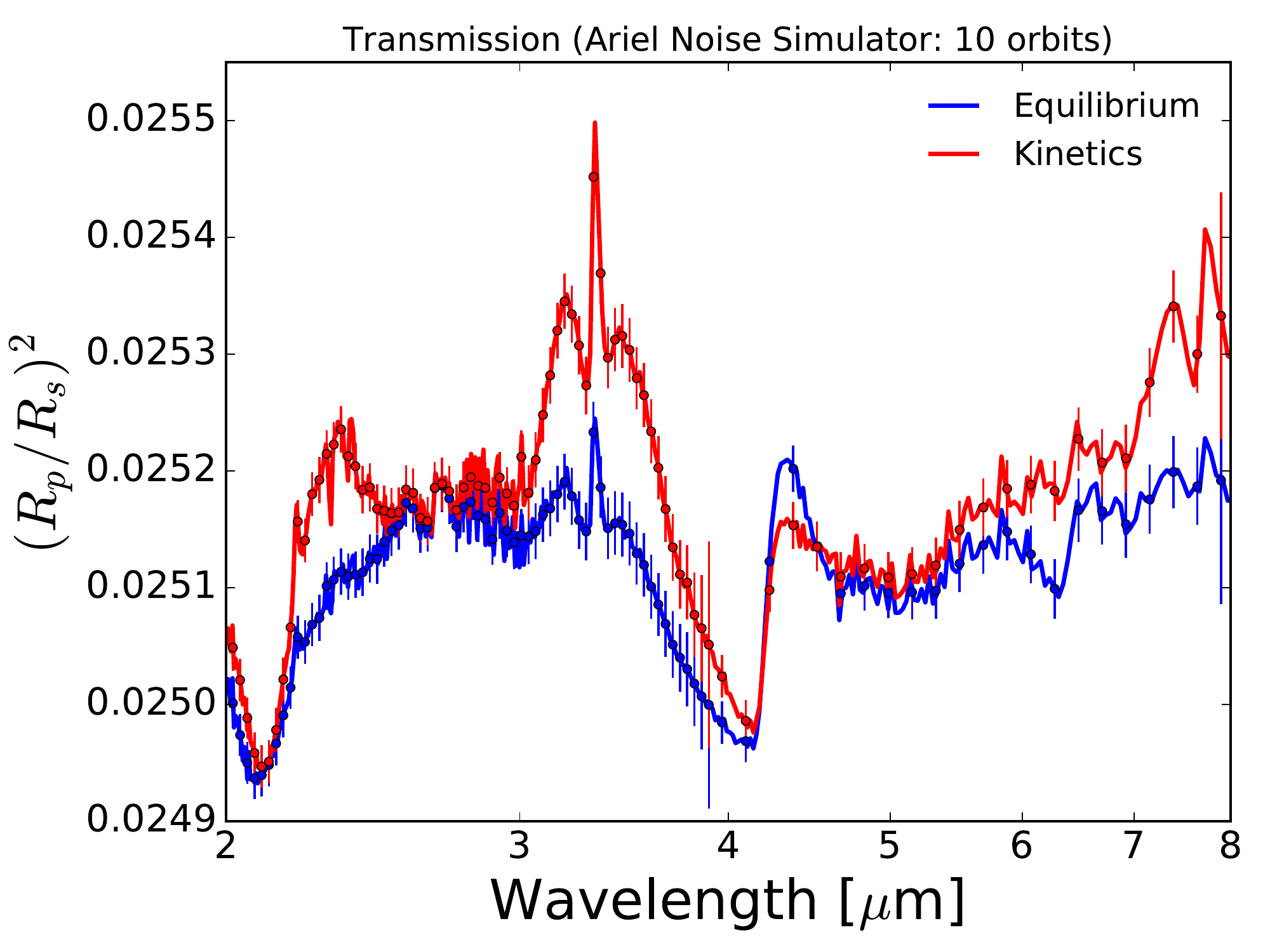} \\
  \end{center}
  \caption{As \cref{figure:hd209_spectra} but for HD~189733b.}
  \label{figure:hd189_spectra}
\end{figure*}

\cref{figure:hd189_spectra} shows the emission and transmission spectra calculated for both the equilibrium and kinetics simulations of HD~189733b. In emission there are much more obvious differences between the equilibrium and kinetics simulations for HD~189733b, compared with for HD~209458b. This is primarily due to an enhanced absorption by CH$_4$ which is increased in abundance by several orders of magnitude in the kinetics simulation. Additionally there are features due to NH$_3$ at $\sim6.5$ microns and $10-12$ microns, where increased abundances lead to increased opacity and therefore a reduced eclipse depth. In the middle and lower panels of \cref{figure:hd189_spectra} it is clear that a single eclipse observation with JWST and five eclipses with ARIEL are sufficient to clearly differentiate between the two models. 

In transmission there are very significant differences between the equilibrium and kinetics simulations across a wide range of wavelengths. The most prominent features are due to CH$_4$ between $3-4$ microns and $7-10$ microns, where an increased CH$_4$ abundance leads to an enhanced transit depth. There is also a significant contribution from NH$_3$ which increases the transit depth at $\sim6.5$ microns and $10-15$ microns. The pale orange line in \cref{figure:hd189_spectra} (top right) shows a calculation that excludes absorption due to NH$_3$ in the spectrum calculation only (it is still included in the main model integration) for the kinetics simulation. Comparing this with the spectra for the standard kinetics simulation clearly shows the importance of NH$_3$ in shaping the transmission spectrum for HD~189733b, which is greatly enhanced by 3D mixing. HCN also absorbs in this wavelength region, however a similar test that removes the contribution of HCN to the transit depth shows negligible difference with the kinetics spectrum that includes the full set of absorbers. The transit depth is decreased at 4.5 microns due to a decrease in the CO$_2$ abundance, similarly to HD~209458b.

The simulated JWST (single transit) clearly shows the enhanced transit depths in the region where CH$_4$ absorbs. However, the large features due to NH$_3$ at longer wavelengths are lost in the noise. Ten transits with ARIEL enables the clear differentiation between the two models in the regions of the enhanced CH$_4$ absorption and decreased CO$_2$ absorption. The enhanced transit depth at $\sim6.5$ microns is also apparent in the simulated ARIEL observations.

\begin{figure}
    \centering
    \includegraphics[width=0.45\textwidth]{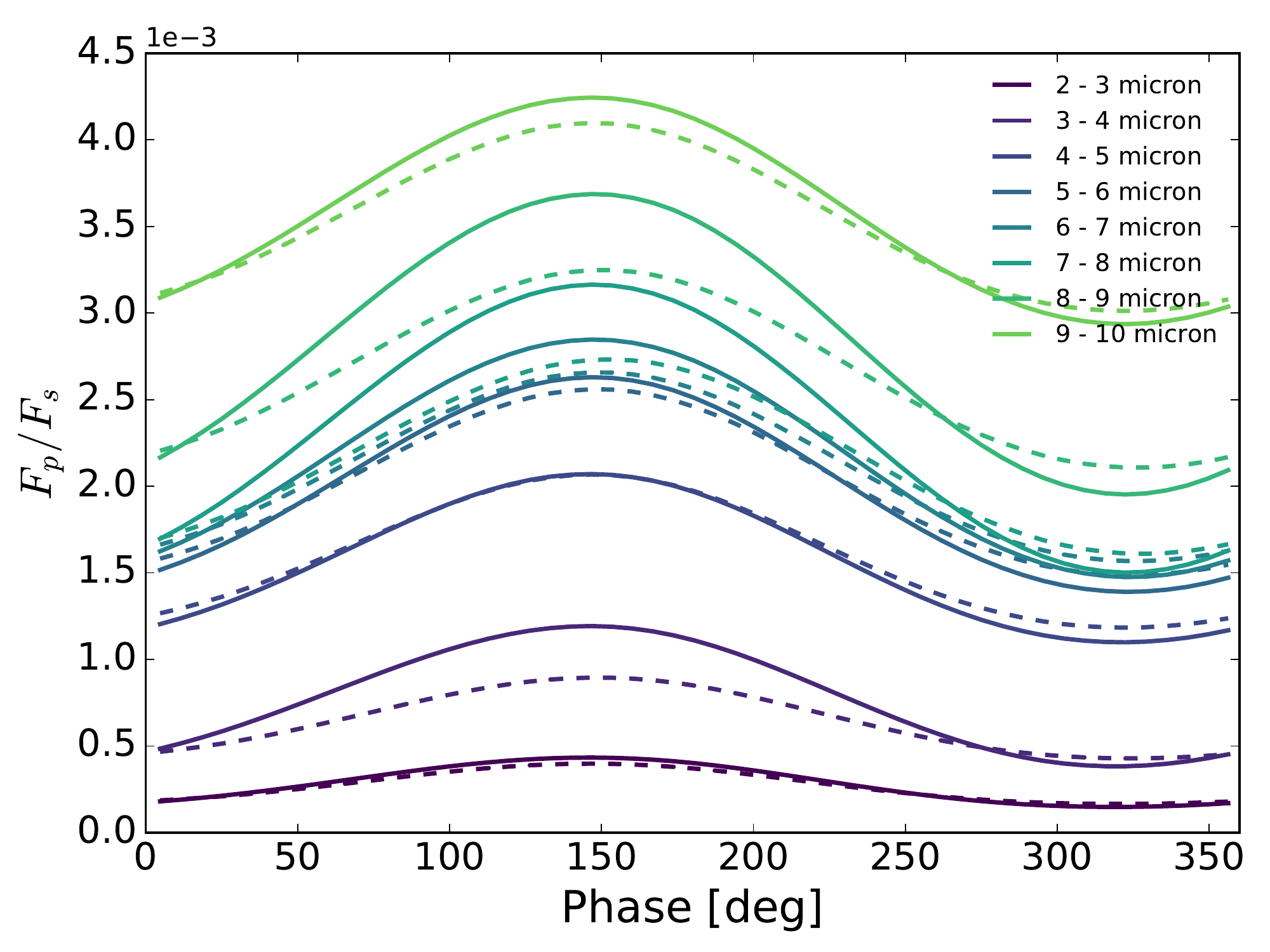}
    \caption{As \cref{figure:hd209_phase} but for HD~189733b.}
    \label{figure:hd189_phase}
\end{figure}

\cref{figure:hd189_phase}  shows the simulated emission phase curves for HD~189733b. There is a significant difference between the equilibrium and kinetics simulations for many of the wavelength bands presented. In most of the wavelength bands the simulation including chemical mixing shows a decreased dayside emission (as shown previously in the secondary eclipse emission spectra) and an increase in the nightside emission. This is primarily due to increases in the abundance of CH$_4$, which is particularly important on the dayside. However, the large increase in the NH$_3$ abundance also contributes in the 6-7 micron band. Increases in the abundances of these species increases the atmospheric opacity and leads to a lower temperature emission (since the temperature decreases with altitude). The 4-5 micron channel shows very little change in the dayside emission but a small increase in the nightside emission. This is likely due to the decrease in the nightside CO$_2$ abundance, leading to a deeper and hotter photosphere in this wavelength region.

The main effect of chemical mixing on the emission phase curve is to decrease the amplitude (the difference between the maximum dayside emission and minimum nightside emission). However, it is also apparent that there is a slight increase in the offset of the phase curve, with the peak emission being shifted further away from secondary eclipse. This is consistent with our previous simulations of HD~189733b using much simpler chemical scheme and is due to the temperature response of the atmosphere \citep{DruMM18b}. The temperature response of the atmosphere is very similar to that already presented in \citet{DruMM18b} and in Figure 5 of that paper we show that the largest temperature increase occurs east of the substellar point, which shifts the hotspot further east.

%% file: discussion.tex

\section{Discussion}
\label{section:discussion}

Here we discuss the results presented in this work in the context of both our own previous studies, and those from the literature, with particular focus on two key works \citep[][]{CooS06,AguPV14}.

\subsection{Comparison with our earlier work}
\label{section:disc_earlier}

In our earlier work \citep{DruMM18a,DruMM18b} we used a much simpler and more computationally efficient chemical scheme, by revisiting the chemical relaxation scheme developed by \citet{CooS06} where only the advection and chemical relaxation of CO is considered. The chemical timescale was based on a rate-limiting reaction that was identified in the interconversion steps from CO to CH$_4$. The mole fractions of CH$_4$ and H$_2$O were subsequently found by assuming mass balance assuming that all carbon is in the form of CO and CH$_4$ and all oxygen is in the form of CO and H$_2$O.

For HD~189733b there is a very good agreement for the abundance distributions of CH$_4$, H$_2$O, and CO found using the \citet{CooS06} chemical relaxation scheme \citep{DruMM18b} or using the coupled chemical kinetics scheme (this work). This implies that in the particular case of HD~189733b the parameterised chemical timescale developed by \citet{CooS06} gives a similar chemical evolution to the reduced chemical network of \citet{VenBD19} for CO. The results and conclusions presented in \citet{DruMM18b} are therefore validated here with the use of a more accurate chemical scheme. This important result demonstrates that, at least in some cases, simplified chemistry schemes can be used in 3D atmosphere models, greatly increasing computational efficiency while maintaining accurate results, compared with more sophisticated chemical schemes.

For HD~209458b the picture is quite different and there are significant differences in the abundance distributions of CH$_4$ between simulations using the chemical relaxation scheme and simulations using the coupled chemical kinetics scheme. Specifically, CH$_4$ remains in chemical equilibrium at lower pressures when using the coupled chemical kinetics scheme rather than the chemical relaxation scheme \citep{DruMM18a}. This has the important consequence that when CH$_4$ does undergo mixing, the quench point lies in a different region of the atmosphere where the circulation is different, resulting in a rather different CH$_4$ distribution.

\begin{figure}
  \begin{center}
  \includegraphics[width=0.45\textwidth]{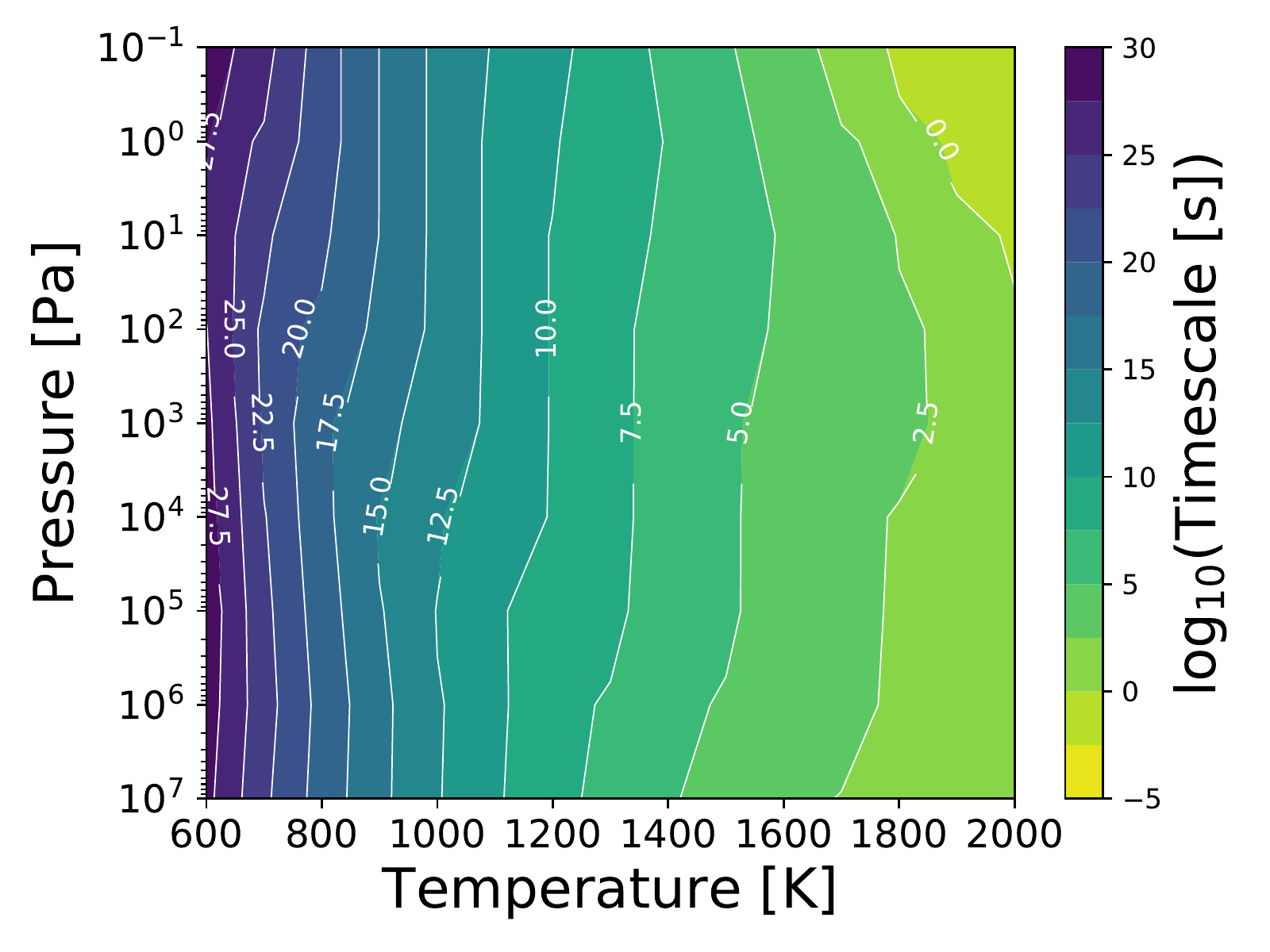} \\
  \includegraphics[width=0.45\textwidth]{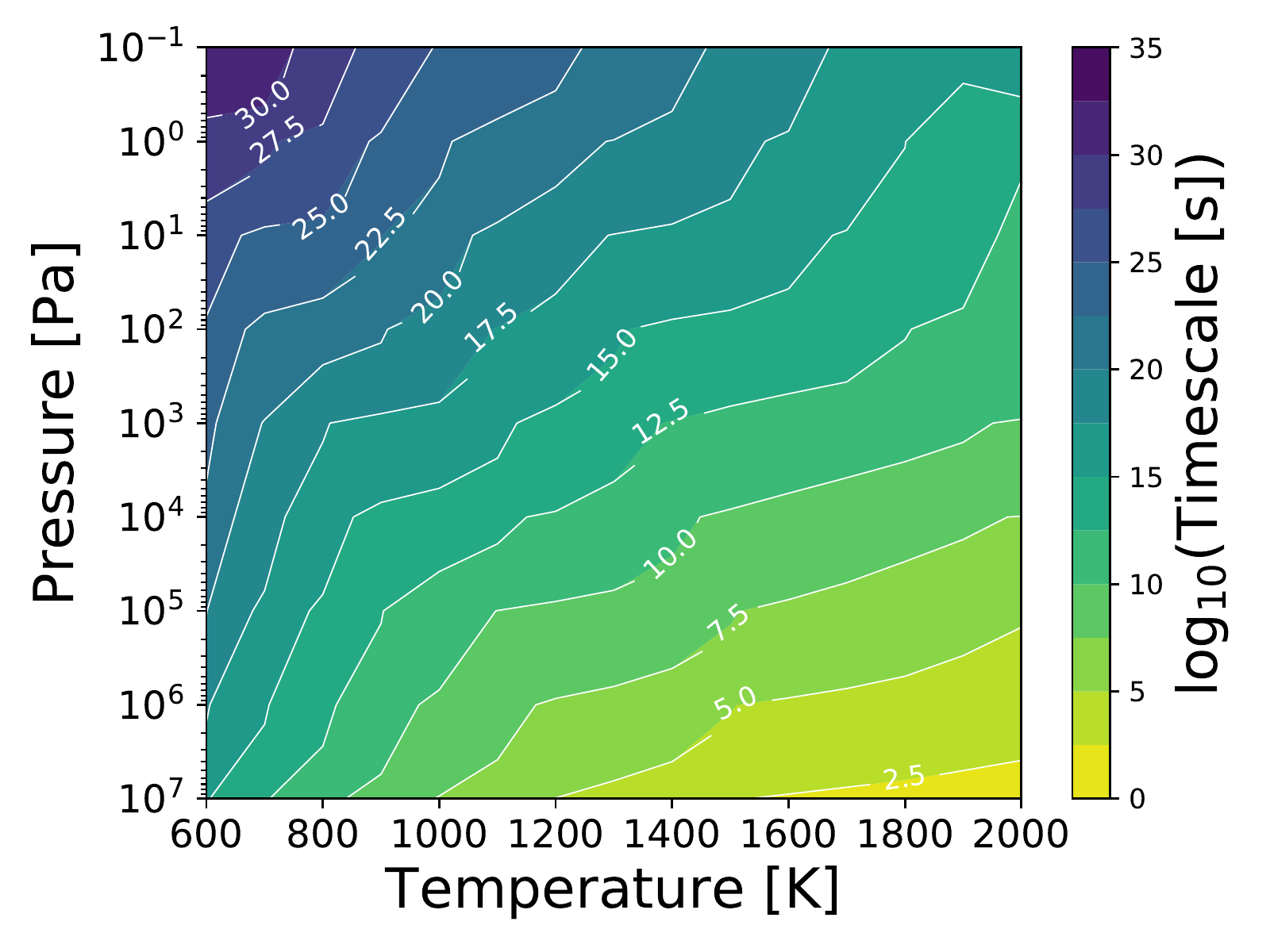} 
  \end{center}
  \caption{Chemical timescales of CH$_4$ (top) and CO (bottom) estimated for the \citet{VenBD19} chemical network as a function of pressure (in bar, y-axis) and temperature (in K, x-axis). The quantity shown is the common logarithm of the timescale in seconds. The colourscales are different between the two panels.}
  \label{figure:tau_ch4}
\end{figure}

In \citet{CooS06}, the chemical timescale represents the time for the chemical interconversion of two molecules, namely CO and CH$_4$, to take place. This, combined with the fact that CH$_4$ is found by mass balance with CO, means that both of these molecules have the same chemical timescale by definition. In practice, this entails that both molecules quench at the same location in the atmosphere. In contrast, in the chemical relaxation scheme more recently developed by \citet{TsaKL18} the chemical timescale represents the time taken for a particular species to evolve back to its local chemical equilibrium, following a perturbation. 

In \cref{figure:tau_ch4} we show the chemical timescales for CH$_4$ and CO, as a function of pressure and temperature, calculated specifically for the \citet{VenBD19} network, using the same method as described in \citet{TsaKL18}. Comparing the timescales for CH$_4$ and CO it is apparent that for some pressure-temperature regimes the timescales for these two molecules can be significantly different \citep[see also][]{TsaKL18}. For example, at $P=10^5$~Pa and $T=1200$~K the timescale for CH$_4$ is $\log_{10}(\tau[{\rm s}])\sim10$ while for CO it is $\log_{10}(\tau[{\rm s}])\sim12.5$, more than two orders of magnitude difference. This suggests that the assumption of a single timescale describing the quenching of both CO and CH$_4$ as used in previous work is not accurate \citep{CooS06,DruMM18a,DruMM18b}. \citet{CooS06} present their CO-CH$_4$ interconversion timescale in their Table 2. Comparing our CO and CH$_4$ chemical timescales to this at $P=10^5$~Pa and $T=1500$~K we find a good agreement for CO ($\sim10^7$~s) but a larger discrepancy for CH$_4$ ($\sim10^5$~s for our CH$_4$ chemical timescale compared with $\sim10^7$~s for the \citet{CooS06} CO-CH$_4$ interconversion timescale). This is expected since the \citet{CooS06} interconversion timescale is based on the reduction of CO.

To be clear, we are not suggesting that the CO-CH$_4$ interconversion timescale of \citet{CooS06} is incorrect. However, the above discussion does imply that it is not appropriate to use the CO-CH$_4$ interconversion timescale to find the quench point of CH$_4$. Finally, this explains why we find the CH$_4$ quench point to lie at lower pressures in this work compared with our previous results \citep{DruMM18a,DruMM18b} using the \citet{CooS06} interconversion timescale.

\subsection{Comparison with \citet{CooS06}}

\citet{CooS06} were the first to investigate 3D transport of chemical species in hot Jupiter atmospheres, using the simple chemical relaxation scheme described above when simulating the atmosphere of HD~209458b. The main conclusion of this work was that chemical equilibrium does not hold for pressures less than $\sim1$~bar, leading to a significant increase in CO on the cold nightside, at the expense of CH$_4$. Based on their simulations, the authors argued that it is predominantly vertical mixing that drives the disequilibrium of CO and CH$_4$. This is due to vertical quenching from a deep atmospheric level that is horizontally uniform and very abundant in CO (and therefore not very abundant in CH$_4$). This vertical quenching occurs all across the planet, leading to horizontally uniform abundances in the photosphere.

In \citet{DruMM18a} we compared our results, from a different 3D atmospheric dynamics model but using the same simple chemical relaxation scheme, with those of \citet{CooS06} and found important quantitative and qualitative differences. Firstly, we found a significantly different chemical equilibrium composition, with CO being the dominant carbon-species everywhere (except for very high pressures), as opposed to the CO-rich dayside and CH$_4$-rich nightside atmosphere found by \citet{CooS06}. We also found that horizontal (specifically, meridional) quenching, in combination with vertical quenching, is important in determining the photospheric abundances of CH$_4$. We attributed these differences to being almost entirely due to differences in the way the atmospheric heating is taken into account in the model, leading to a very different thermal and chemical structure. \citet{CooS06} used a simple temperature forcing (or Newtonian cooling) scheme whereas \citet{DruMM18a} used a coupled radiative transfer scheme to calculate the radiative heating rates consistently with the chemical composition. The temperature forcing scheme developed for HD~209458b significantly overpredicts the day-night temperature contrast with a dayside that is too hot and a nightside that is too cold, compared with the coupled radiative transfer scheme \citep[e.g.][]{Showman2009,AmuMB16}.

\subsection{Comparison with \citet{AguPV14}}

\citet{AguPV14} developed a model that includes a sophisticated chemistry scheme but makes a very crude approximation to horizontal chemical transport. The \citet{AguPV14} model is in practice a 1D vertical column model with a time-varying pressure-temperature profile sourced from previous 3D atmosphere model output for different longitudes. This so-called `pseudo-2D' model represents therefore a column of atmosphere rotating around the equator (i.e. within the equatorial zonal jet) with a solid-body rotation. Vertical diffusion was also included and parameterised with an eddy diffusion coefficient. In terms of the chemistry, \citet{AguPV14} included a chemical kinetics scheme and used the 105 species chemical network of \citet{Venot2012}. The authors also included photochemical dissociation. Overall, compared with other works investigating horizontal chemical mixing in hot Jupiter atmospheres \citep[][and this work]{CooS06,DruMM18a,DruMM18b,MenTM18} the model developed by \citet{AguPV14} includes a much more sophisticated approach to chemistry, and is so far the only one to consider photochemistry, but takes a much cruder approach to horizontal transport, and only considers the zonal direction.

The main conclusions from \citet{AguPV14} are that horizontal quenching can occur for certain molecules leading to their abundances being homogenised with longitude, at values typical of their chemical equilibrium values on the dayside. This was found to have a particularly important effect on the abundance of CH$_4$, which is significantly decreased in their pseudo-2D model on the nightside relatively its local chemical equilibrium abundance. This is due to quenching from the dayside where CH$_4$ is much lower in abundance in chemical equilibrium. This agrees with our results for HD~209458b where we find zonal quenching of CH$_4$ over the pressure range 0.1~bar to 0.01~bar, decreasing the nightside abundance. However, for HD~189733b we find that deeper meridional transport has a more important impact on the CH$_4$ abundance, with equatorward transport of CH$_4$ from mid-latitudes increasing the equatorial abundance. Such a process would not be captured by the purely zonal-vertical pseudo-2D model of \citet{AguPV14}. We note that the HD~209458b model presented by \citet{AguPV14} includes a thermal inversion which is likely to lead to differences in the chemical structure and transport, in addition to differences that result from the very different modelling approach.

\citet{AguPV14} include photochemical dissociation in their model, which is found to have some impact on molecular abundances in the upper atmosphere. One of the principle effects is the increase the abundance of HCN. However, these photochemical effects are restricted to pressures less than $\sim1$~Pa, above the upper domain of our model. Therefore, even if we were to include photochemical dissociation in our 3D model the effects on the chemical abundances in the pressure region covered by our domain are likely to be small.

Our results qualitatively agree with the conclusions of \citet{AguPV14} for the case of HD~209458b. Zonal quenching appears to `contaminate' the cooler nightside with the composition from the warmer dayside. In this case, it may be possible that a 2D model only accounting for zonal and vertical transport can be sufficient to accurately model the equatorial regions of hot Jupiters. However, for HD~189733b we find that meridional transport, occurring deeper in the atmosphere, has a more important role in setting the quenched abundances, especially for CH$_4$. In this case, a full 3D model is required to capture all of the important mixing processes.

\subsection{Comparison with other works}

Several other recent works have also tackled the issue of chemical mixing in hot Jupiter atmospheres. \citet{MenTM18} coupled a chemical relaxation scheme to the THOR general circulation model to investigate the effect of 3D mixing in the atmosphere of WASP-43b. The chemical relaxation scheme was recently developed in \citet{TsaKL18} and includes more chemical species and a more accurate chemical timescale calculation than the earlier method of \citet{CooS06}. \citet{MenTM18} conclude that zonal quenching is the dominant process in the equatorial region of WASP-43b, with vertical and meridional mixing having only a secondary effect. 

The results of \citet{MenTM18} are qualitatively consistent with our findings for HD~209458b but different to our findings for HD~189733b. For HD~209458b we find that not all horizontal abundance gradients are removed, in agreement with the conclusions of \citet{MenTM18} for WASP-43b, while for HD~189733b we find that the chemistry is very efficiently horizontally homogenised. However, given that \citet{MenTM18} simulate the atmosphere of a different planet, with a significantly different circulation and thermal structure, it is not surprising that we should find differences. WASP-43b has a similar planetary equilibrium temperature to HD~209458b ($\sim1450$~K), but is considerably warmer than HD~189733b ($\sim1200$~K). The circulation and temperature structure is also likely to be quite different due to the much higher surface gravity ($\sim55$~m~s$^{-2}$) and slower rotation rate ($\sim1.5\times10^{-5}$~s$^{-1}$), compared with both HD~209458b and HD~189733b (see \cref{table:params}). In addition to these physical differences in the planet-stellar properties, there are many differences in the model implementation, level of assumption, and included model physics between our model and that of \citet{MenTM18}. For example, \citet{MenTM18} use a relatively simple 2-band radiative transfer scheme compared with our non-grey radiative transfer scheme. Also, \citet{MenTM18} include the effect of additional cloud opacity on the nightside, while we choose to assume a cloud-free atmosphere. The very large number of differences, both in the details of the numerical model and also in the properties of the planetary system under investigation, means that a meaningful comparison is not possible.

One of the more interesting features discussed in \citet{MenTM18} is the presence of chemically distinct polar vortexes. We do not find similar features in our simulations, however this may again be simply due to the use of different planetary and stellar parameters associated with different planets. A more controlled comparison of our model with that of \citet{MenTM18}, using model parameters and a setup that are as close as feasibly possible would be an interesting future exercise.

\citet{StePS18} recently investigated the feedback effect of non-equilibrium chemical abundances on the 3D circulation and thermal structure of HD~189733b. Assuming globally uniform chemical abundances (i.e. no interactive chemistry scheme) the authors varied the ratio of CO and CH$_4$. The assumption of globally uniform abundances for HD~189733b is likely a good one based on our simulation results presented here and in \citet{DruMM18a}, with both of these molecules having roughly uniform abundances for pressures less than $\sim10^5$~Pa. 

\citet{StePS18} found that in cases where CO remains the dominant carbon-species, varying the CO/CH$_4$ ratio can alter the temperature by 50-100~K. In cases where CH$_4$ is more abundant than CO (a less likely scenario) a larger temperature response (200-400~K) is found for changes in the CO/CH$_4$ ratio. In the more likely CO-dominated regime, our results using a more sophisticated approach to the chemistry support the conclusions of \citet{StePS18} that disequilibrium chemical abundances can alter the thermal structure by up to 10\% compared to abundances predicted assuming local chemical equilibrium. For the equatorial region of HD~189733b, our simulations predict that HD~189733b remains in the CO-dominated regime, but that 3D vertical mixing can decrease the CO/CH$_4$ ratio by several orders of magnitude, due to an increase the equatorial abundance of CH$_4$. This leads to temperature changes of 10\% compared with the local chemical equilibrium simulation \citep{DruMM18b}.

%% file: conclusions.tex

\section{Conclusions}
\label{section:conclusions}

In this paper we presented results from a model that consistently couples a chemical kinetics scheme to a 3D radiation-hydrodynamics atmosphere model. We use the recently released 30 species carbon-oxygen-nitrogen-hydrogen chemical network of \citet{VenBD19}. We investigated the effect of 3D advection on the chemical structure of hot Jupiter atmospheres, using the particular cases of HD~209458b and HD~189733b. Focusing on six important absorbing species (CH$_4$, CO, H$_2$O,  CO$_2$, NH$_3$, and HCN) we found that 3D advection can lead to significant changes in the chemical composition compared with a simulation that assumes local chemical equilibrium. These chemical differences lead to important signatures in simulated emission and transmission spectra, primarily due to CH$_4$, CO$_2$, and NH$_3$. The model presented in this work represents an improvement on our earlier studies \citep{DruMM18a,DruMM18b} which used a much simpler chemical relaxation scheme \citep[that of][]{CooS06}, while using essentially the same radiative transfer and hydrodynamics setup.

For HD~189733b, interestingly, we find qualitatively very similar results to our earlier work \citep{DruMM18b}. The abundance profiles of CH$_4$, CO, and H$_2$O (the only three molecules considered in that earlier work) are very similar. The most important consequence of 3D mixing is that the abundance of CH$_4$ is significantly increased, particularly in the low pressure regions of the atmosphere, in comparison to the chemical equilibrium case, leading to a decrease in the eclipse depth and an increase in the transit depth, where CH$_4$ absorbs. Our new results that include more chemical species show that CO$_2$ and NH$_3$ also make important contributions to both the emission and transmission spectra, and that the abundance profiles of these species are strongly effected by 3D mixing, particularly NH$_3$. We therefore conclude that CH$_4$ and NH$_3$ are important targets for future observations of HD~189733b as they may help to constrain the chemical-dynamical interactions occurring in HD~189733b-like atmospheres.

For HD~209458b we find qualitatively quite different results when using the newly coupled chemical kinetics scheme, compared with our earlier results using a much simpler chemical relaxation scheme \citep{DruMM18a}. In \citet{DruMM18a} we found that CH$_4$ undergoes a quenching behaviour similar to that which we found for HD~189733b and which significantly increases the abundance of CH$_4$, at low pressures, in comparison to the chemical equilibrium case. Here, however, we find that CH$_4$ remains in equilibrium to lower pressures before quenching occurs, leading to a much smaller quenched abundance of CH$_4$. This has the result that we no longer find the significant effects of increased CH$_4$ absorption in the emission and transmission spectra as previously found in \citet{DruMM18a}.

An important conclusion of this work is that understanding the chemical structure of hot Jupiter atmospheres is not trivial, and differences in the winds or the temperature structure can have large impacts on the predicted chemical structure. There is likely to be a huge diversity of exoplanetary atmospheres, showing much more variety in the atmospheric circulation patterns and temperature structures than captured by the study of just two specific hot Jupiters, as featured in this work. The consequences of this on the predicted chemical composition would require a large parameter space study, which would be difficult to achieve with such computationally expensive models, requiring a step--change in the efficiency of the model framework.

Another important conclusion from this work is that the advection processes that determine the chemical structure are strongly three-dimensional, and these processes cannot be captured in a 1D, or even 2D, model. For HD~209458b our kinetics simulation shows that in some cases the chemistry can become efficiently zonally mixed but is not strongly effected by vertical mixing. This occurs because the horizontal advection timescale is typically a few orders of magnitude faster than the vertical advection timescale. At lower pressures, when vertical mixing eventually becomes important, the vertically quenched abundance is effectively set by the zonally quenched abundance that is determined at higher pressures. The note of caution, however, is that precise absolute wind velocities can not be easily predicted by such models \citep[see discussion in][]{heng_2011}, and require observational constraint \citep[e.g.][]{Snedd10,Louden2015}.

The quantitative wind velocities, and therefore the advection timescales, are partly determined by the dissipation included in the model (required for the numerical stability) the magnitude of which is currently unconstrained by observations \citet{heng_2011}. Adjusting this model free parameter could lead to changes in the quenching behaviour of chemical species, as the balance of advection and chemistry will change. However, we do not expect that changes of a factor of a few in the wind velocities (and therefore also in the advection timescales) will cause a very large change in the quenching behaviour since the chemical timescale that depends on the temperature varies much more steeply spatially in the atmosphere. Differences could also arise between models based on the chosen chemical network. Different choices of reactions or their associated rate constants can lead to a variety of quenching points and quenched abundances for a given circulation \citep[e.g.][]{Venot2012,Moses2014}. The precise chemical structure may vary from one model to another, with different quench points and different quenched abundances, due to differences in the predicted wind and temperature fields and also the use of different chemical networks. An intercomparison of models will be required in future to validate our results, once a family of models with similar capability are available in the community.

Recently, \citet{sainsbury_martinez_2019} confirmed, using a 3D dynamical atmosphere model, that the deep atmosphere is much hotter than what is predicted by 1D radiative-convective equilibrium models, due to vertical advection of potential temperature \citep{TreCM17}. Integration times of much longer than 1000 (Earth) days, as used here, are required to achieve a steady-state in the deep atmosphere, when initialised from a `cold' initial profile. However, \citet{sainsbury_martinez_2019} demonstrated that initialising the model with a hotter initial profile can reduce the integration time required for steady-state to be reached. \citet{LinMB18}, who coupled a cloud kinetics formation scheme to a 3D radiation-hydrodynamics atmosphere model, has shown that initialising the atmosphere with a `hot' initial profile can lead to significantly different cloud structures in the upper atmosphere, due to differences in the temperature profile at $P>10^5$~Pa. We performed a similar test for our HD~209458b case (see \cref{section:hot_initial}) but found no significant differences in the atmospheric structure or composition between the hot and cold initial profile models. This is likely because most species quench at pressures $P\lesssim10^5$~Pa while the temperature profile is only different for pressures greater than this. Since the chemistry quenches deeper in the case of HD~189733b it is possible that considering the hot deep atmosphere may lead to more significant differences in the upper atmosphere chemistry, if the quench points are deep enough. If the deep temperature profile has not yet reached a steady-state at the pressure level of quenching, then a wrong quenched abundance will be found, affecting the abundance at lower pressures. This potential issue should be considered when interpreting our results, and should be investigated in more detail in future by running the model for a much longer integration time and/or initialising with a hotter initial thermal profile \citet{sainsbury_martinez_2019}.

The model described in this paper represents a major step forward in the ability to study the interactions of chemistry, radiation and fluid dynamics in a 3D exoplanet atmosphere. However, there are a number of important physical elements that are clearly missing that are likely to have a significant impact. Firstly, though the model takes into account 3D transport of chemical species by the resolved winds we do not yet account for photochemical processes, as included routinely in 1D chemistry models \citep[e.g.][]{Moses2011,Venot2012}. Photodissociation and photoionisation have the potential to significantly change the atmospheric composition, particularly at high altitudes (low pressures). This is not only due to the direct effect of additional loss terms for molecules and atoms via dissociation and ionisation. The products of photodissociation and photoionisation are often very reactive species themselves, which in turn impact the chemical balance even further. Photochemical products formed on the irradiated dayside may well be efficiently transported to the cooler nightside, exciting additional chemical pathways. However, 1D models typically show that photochemical effects only become important for pressures less than around $~1$~Pa \citep[e.g.][]{Moses2011,DruTB16} which is typically the pressure of the upper boundary of 3D simulations. In order to study photochemical effects in a 3D atmosphere, the model may first have to be extended to lower pressures which is would require significant development and testing. 

A further major limitation of this work is the absence of condensates and clouds in the simulations. Here we refer to condensate clouds rather than photochemical hazes. In this work we consider a gas-phase only composition. Condensates and clouds of various compositions are expected to form in hot Jupiter atmospheres and there is a wide-variety of approaches with varying complexity to model them in both 1D \citep[e.g.][]{barstow_2017,morley_2017,goyal_2018,Powell2019} and 3D models \citep[e.g.][]{charnay_2015,LeeDH16,LinMB18,LinMM18,LinMM19}. The presence, distribution and composition of cloud can strongly effect the radiation field, which determines the thermal structure and observable properties \citep{LinMM18,LinMM19}. The formation of cloud can also have an important effect on the gas-phase composition, as the gas becomes depleted in the elements that form the cloud condensates themselves \citep[see discussion in][]{Lee_2015,WoiHH18,LinMB18}. In this work we focus on understanding the gas-phase chemistry which, fundamentally, is the necessary precursor to any cloud formation. However, fully understanding the composition of hot Jupiter atmospheres likely requires coupling gas-phase chemical kinetics and cloud kinetics formation schemes within the framework of a 3D model. This represents another very significant scientific and technical challenge.

%% file: appendix.tex
\section{Model evolution and steady-state}
\label{section:appendix_evo}
\begin{figure}
  \begin{center}
    \includegraphics[width=0.4\textwidth]{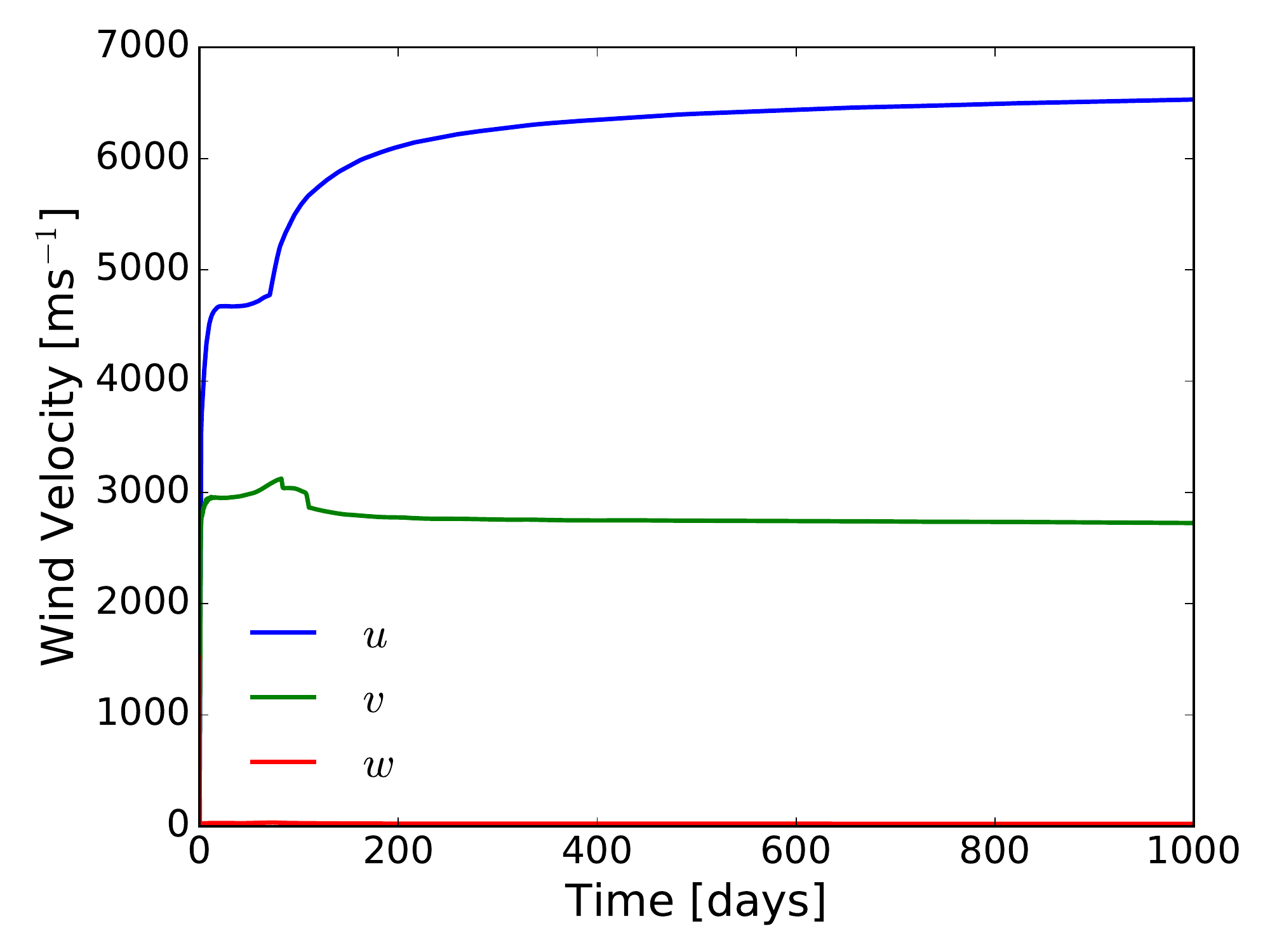}\\
    \includegraphics[width=0.4\textwidth]{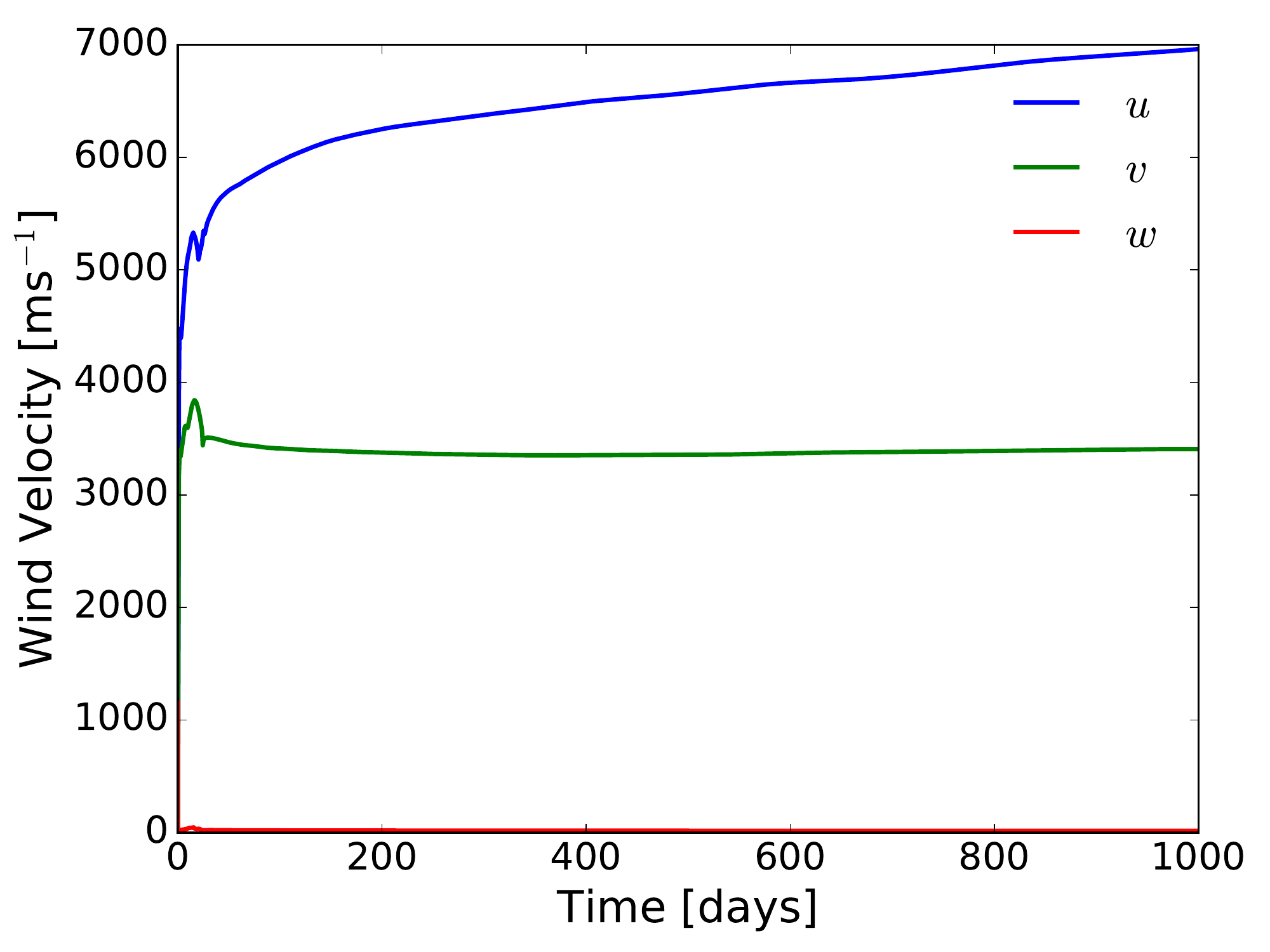} 
  \end{center}
\caption{Evolution of the global maximum for the zonal (blue), meridional (green) and vertical (red) components of the wind velocities for the kinetics simulation of HD~209458b (top) and HD~189733b (bottom).}
\label{figure:windmax}
\end{figure}

Here we assess the model evolution in order to judge whether a steady-state has been reached. \cref{figure:windmax} shows the evolution of the global maximum in each of the three components (zonal, meridional and vertical) of the wind velocities. After starting from rest (0~km~s$^{-1}$) the winds quickly spin up to several km~s$^{-1}$ over a timescale of a few hundred days. By 1000 days, the maximum integration time of the simulations presented here, the acceleration of the global maximum winds is very small. The zonal wind component for HD~189733b shows a small gradient at 1000 days. 

\begin{figure}
  \begin{center}
    \includegraphics[width=0.35\textwidth]{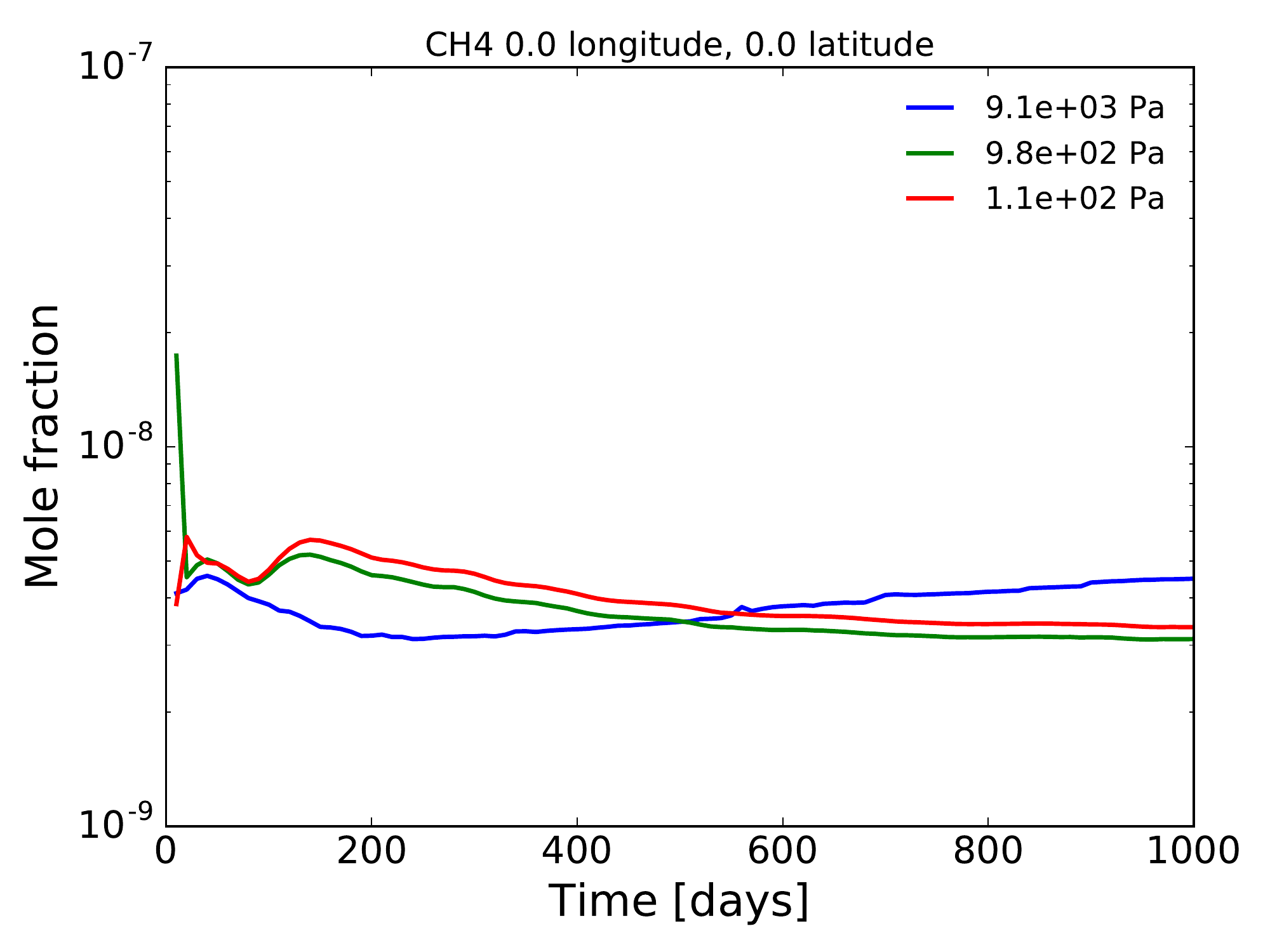} \\
    \includegraphics[width=0.35\textwidth]{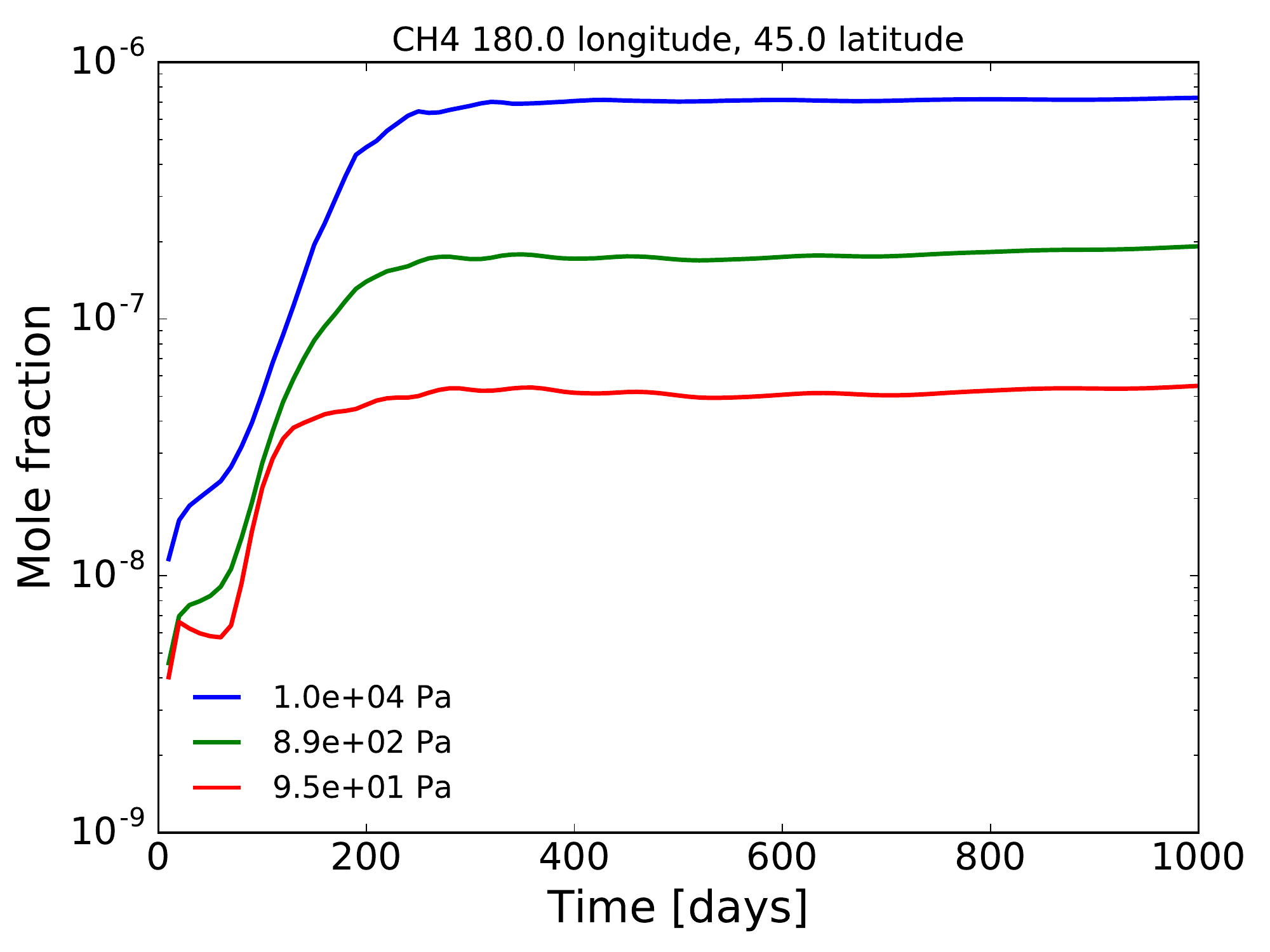} \\
     \includegraphics[width=0.35\textwidth]{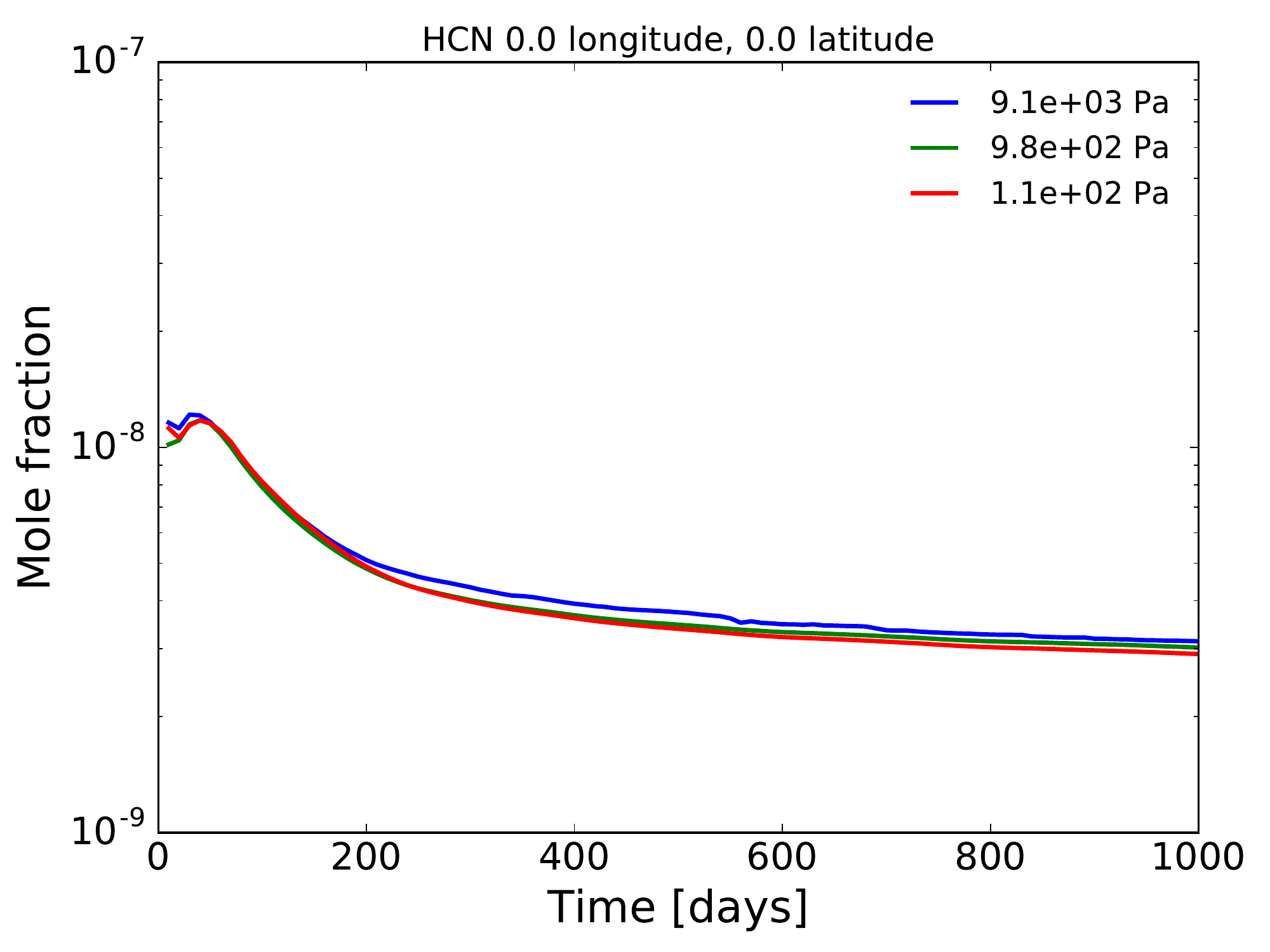} \\
    \includegraphics[width=0.35\textwidth]{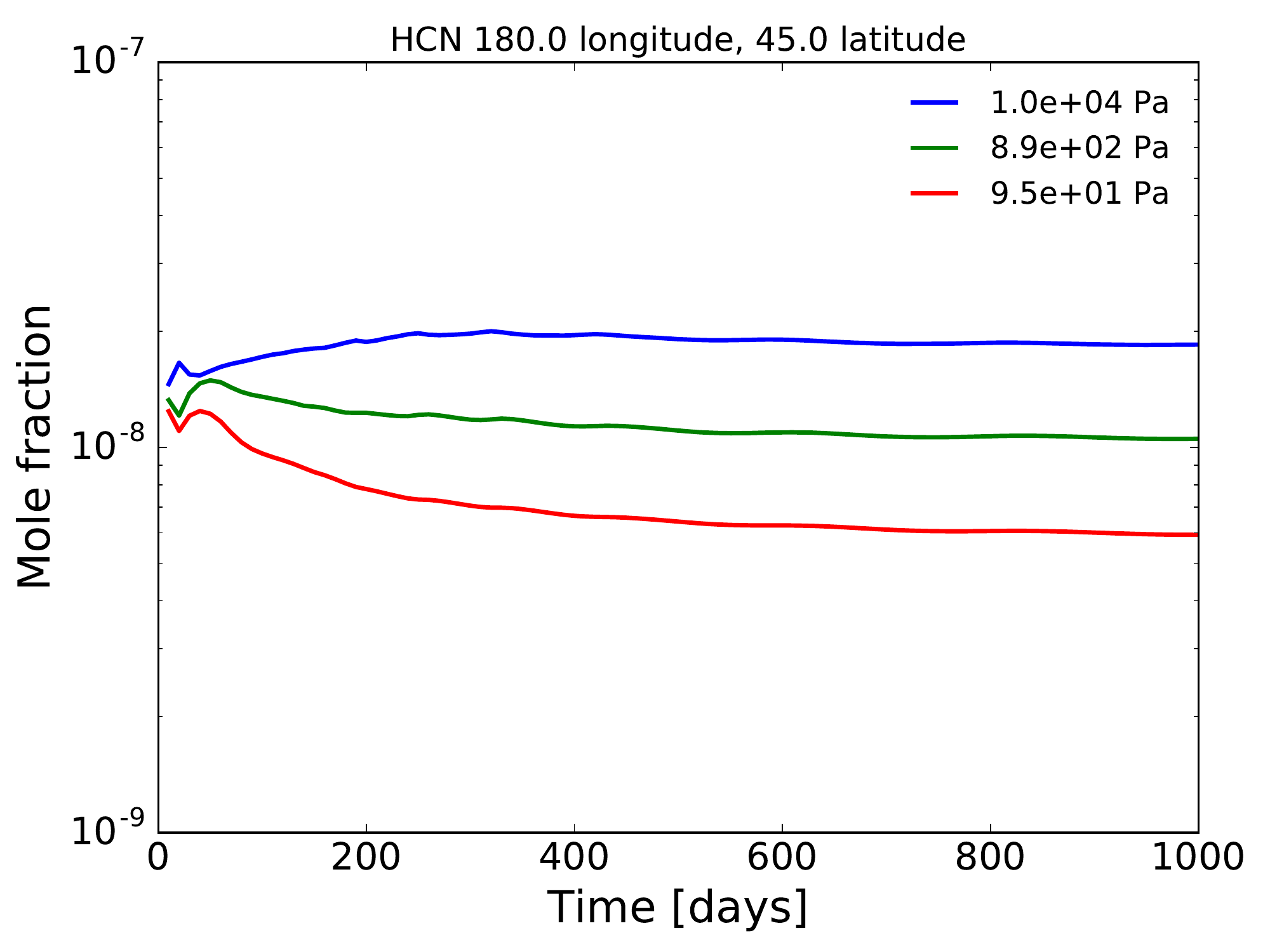}
  \end{center}
\caption{Evolution of the mole fractions of CH$_4$ (top) and HCN (bottom) on three particular model vertical (altitude) levels ($\sim10^2$~Pa, $\sim10^3$~Pa, and $\sim10^4$~Pa) at two arbitrary coordinates ($\lambda=\phi=0^{\circ}$ and $\lambda=180^{\circ}$, $\phi=45^{\circ}$) for the kinetics simulation of HD~209458b.}
\label{figure:evohd209}
\end{figure}

\begin{figure}
  \begin{center}
    \includegraphics[width=0.35\textwidth]{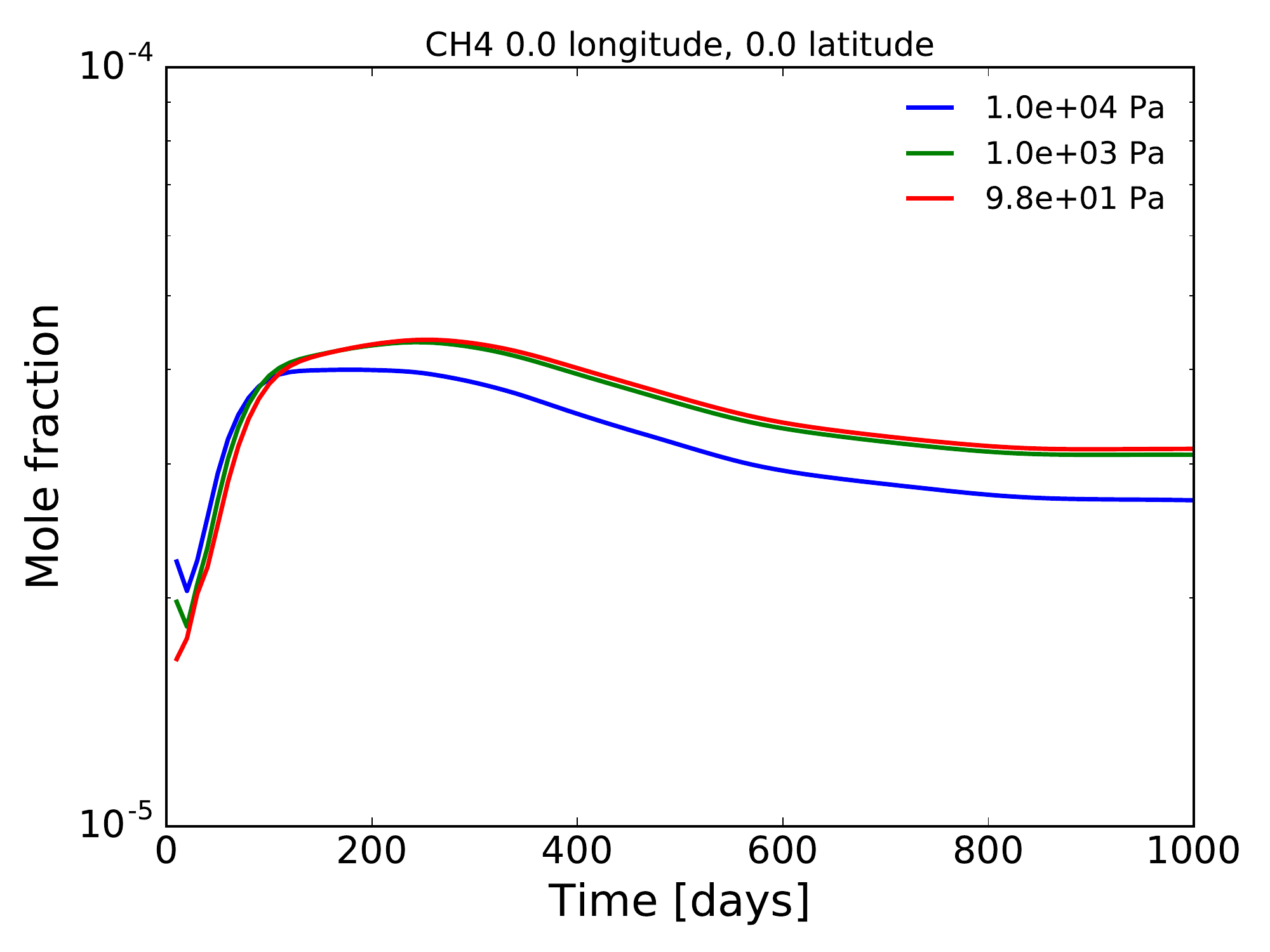} \\
    \includegraphics[width=0.35\textwidth]{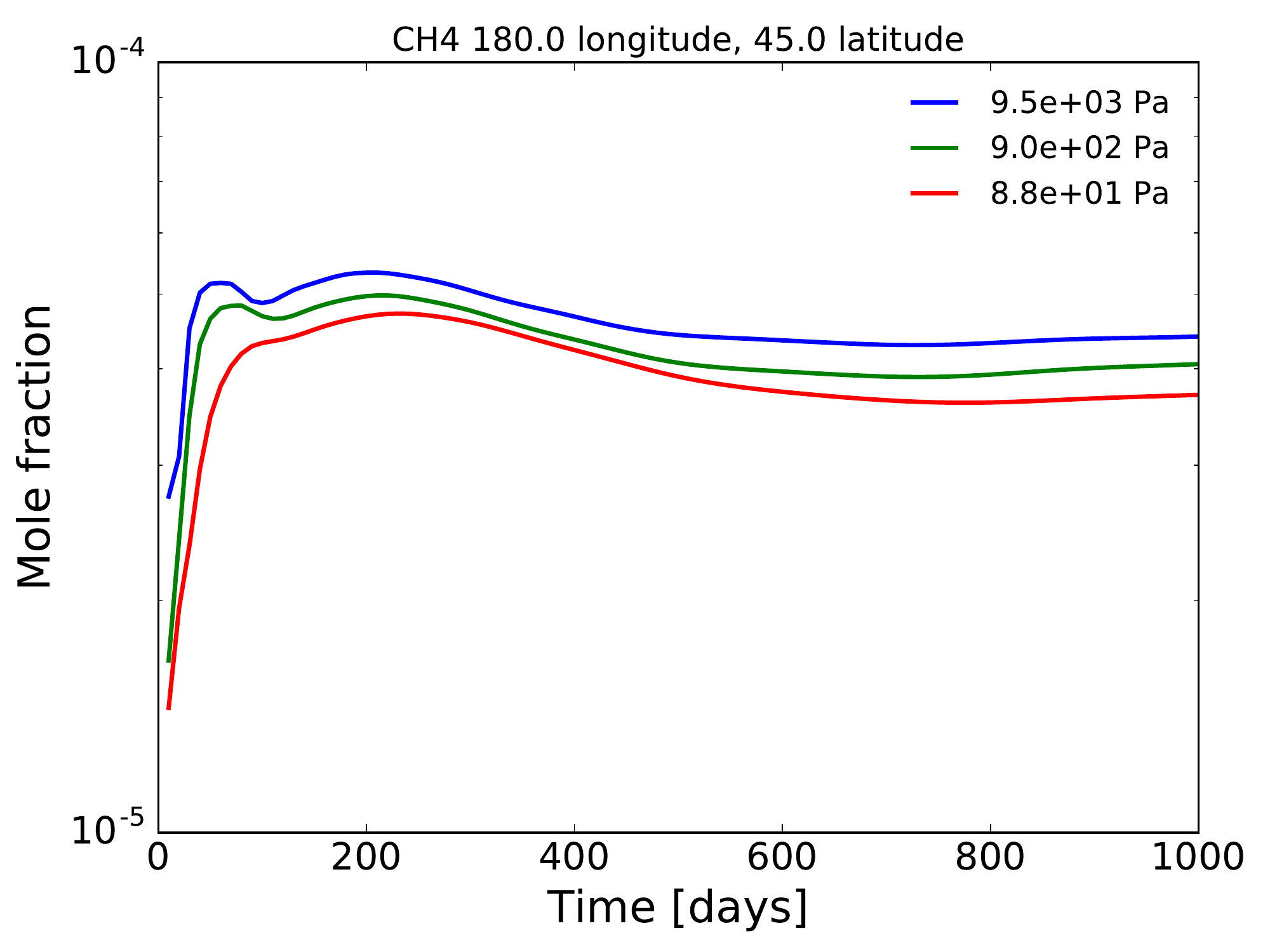} \\
     \includegraphics[width=0.35\textwidth]{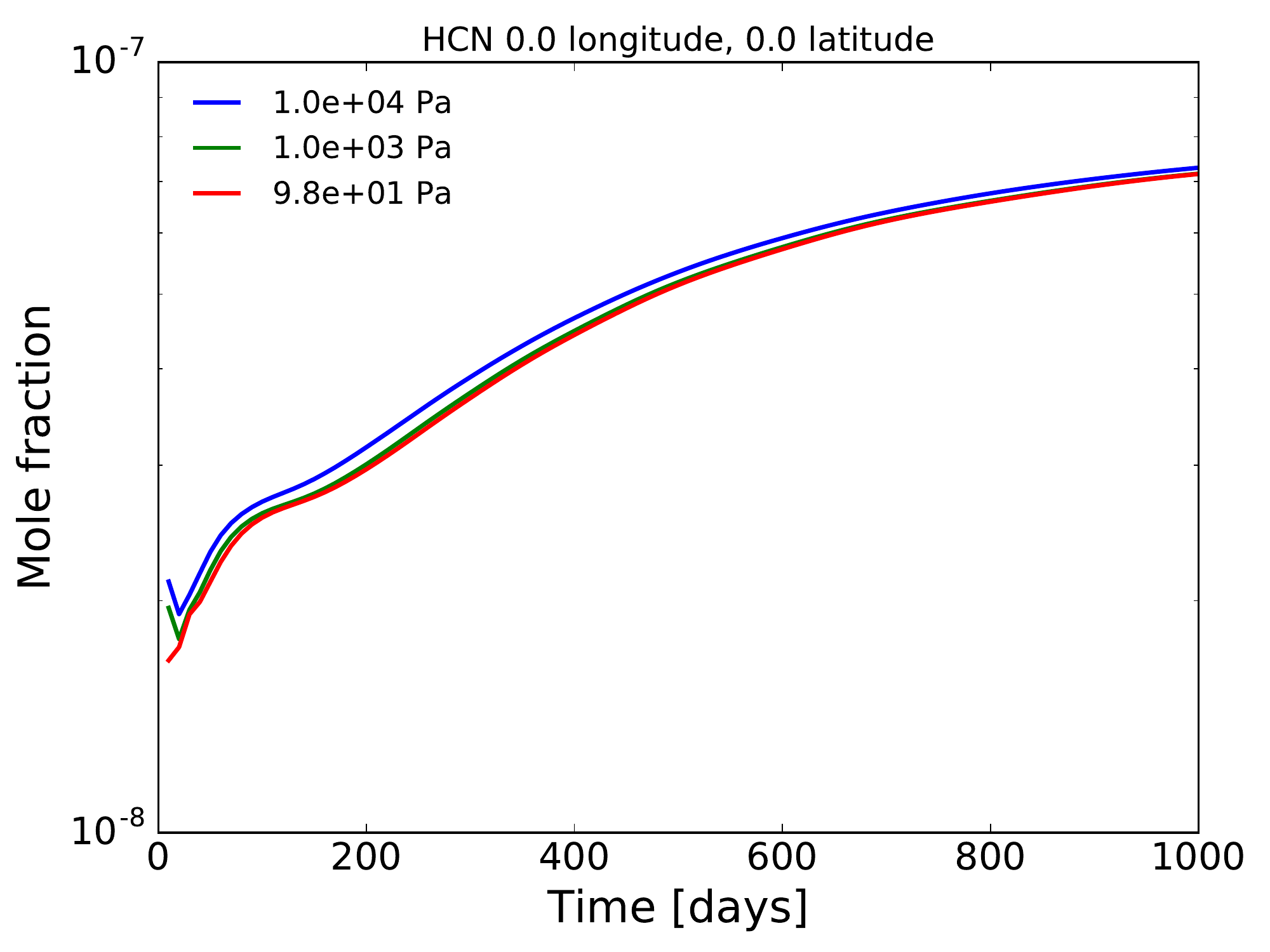} \\
    \includegraphics[width=0.35\textwidth]{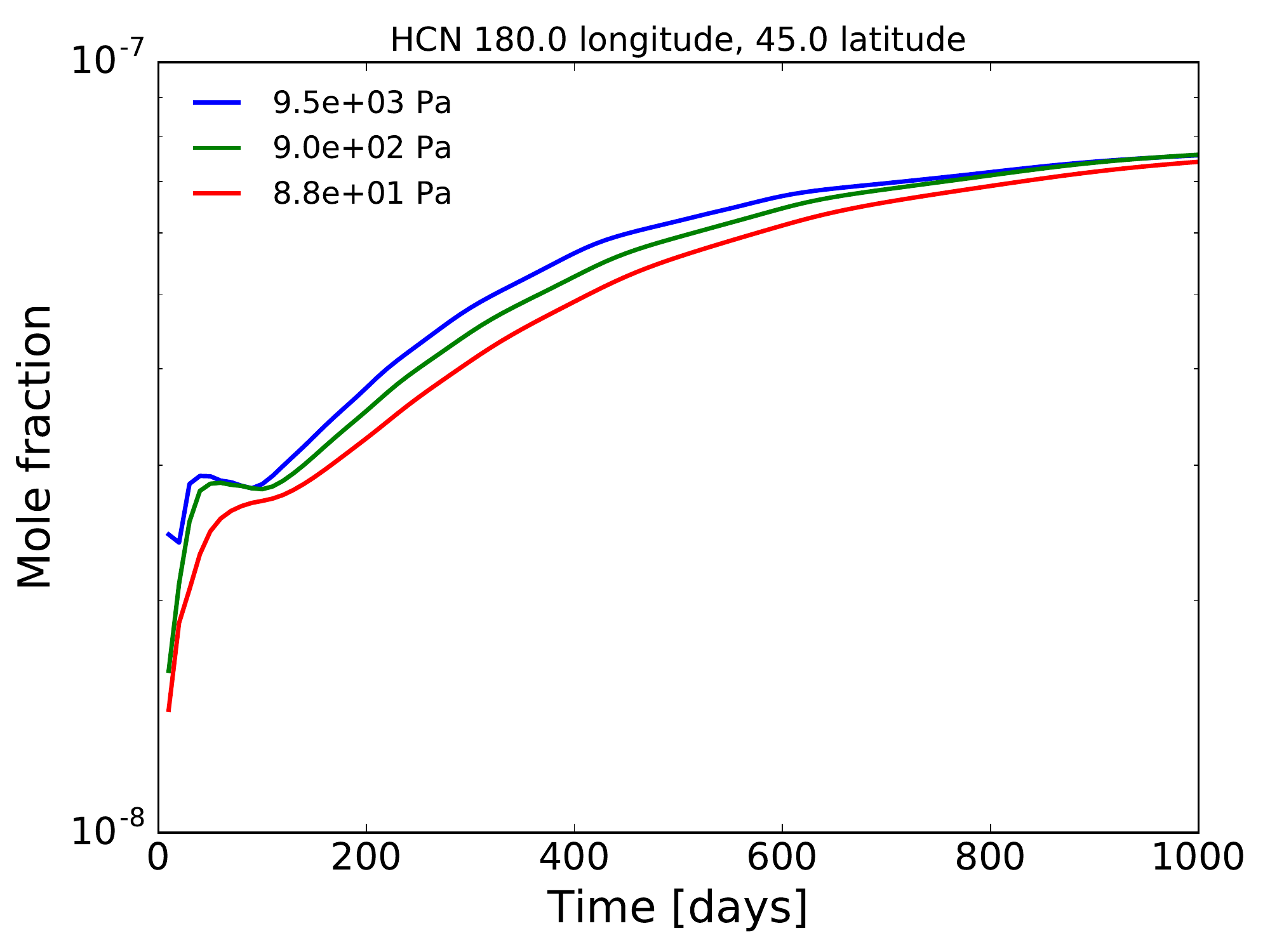}
  \end{center}
\caption{As \cref{figure:evohd209} but for the kinetics simulation of HD~189733b.}
\label{figure:evohd189}
\end{figure}

\cref{figure:evohd209} shows the evolution of the mole fractions of CH$_4$ and HCN in a series of arbitrarily chosen grid cells for the kinetics simulation of HD~209458b. The grid cells correspond to three model vertical (altitude) levels ($\sim10^2$~Pa, $\sim10^3$~Pa, and $\sim10^4$~Pa) at two arbitrary coordinates ($\lambda=\phi=0^{\circ}$ and $\lambda=180^{\circ}$, $\phi=45^{\circ}$). The legends refer to the pressures of the corresponding grid cells at the end of the simulation (1000 days). Clearly, most of the changes in the mole fractions, from their initial values, occurs within the first few hundred days, reflecting somewhat the evolution of the global maximum wind velocities (\cref{figure:windmax}). At the end of the simulation the gradients are approximately flat. \cref{figure:evohd189} shows the same information but for the kinetics simulation of HD~189733b. The evolution profile shapes are quite different. However, as for HD~209458b, the gradients of the chemical evolution are approximately flat by 1000 days. 

Based on the evolution of the global maximum wind velocities and the mole fractions of CH$_4$ and HCN, all of which show very small gradients by 1000 days, we conclude that our simulations have reached a pseudo-steady state (for $P>10^5$~Pa) by 1000 days. However, we cannot rule out the possibility of long-term chemical evolution that might become apparent for much longer integration times. The challenge is that it is very difficult to run these complex models for very long integration times.

\section{Conservation}
\label{section:appendix_con}

The UM does not explicitly enforce global axial angular momentum in the numerical scheme. However, all simulations presented conserve initial global axial angular momentum to within 99\%.

\begin{figure}
  \begin{center}
    \includegraphics[width=0.45\textwidth]{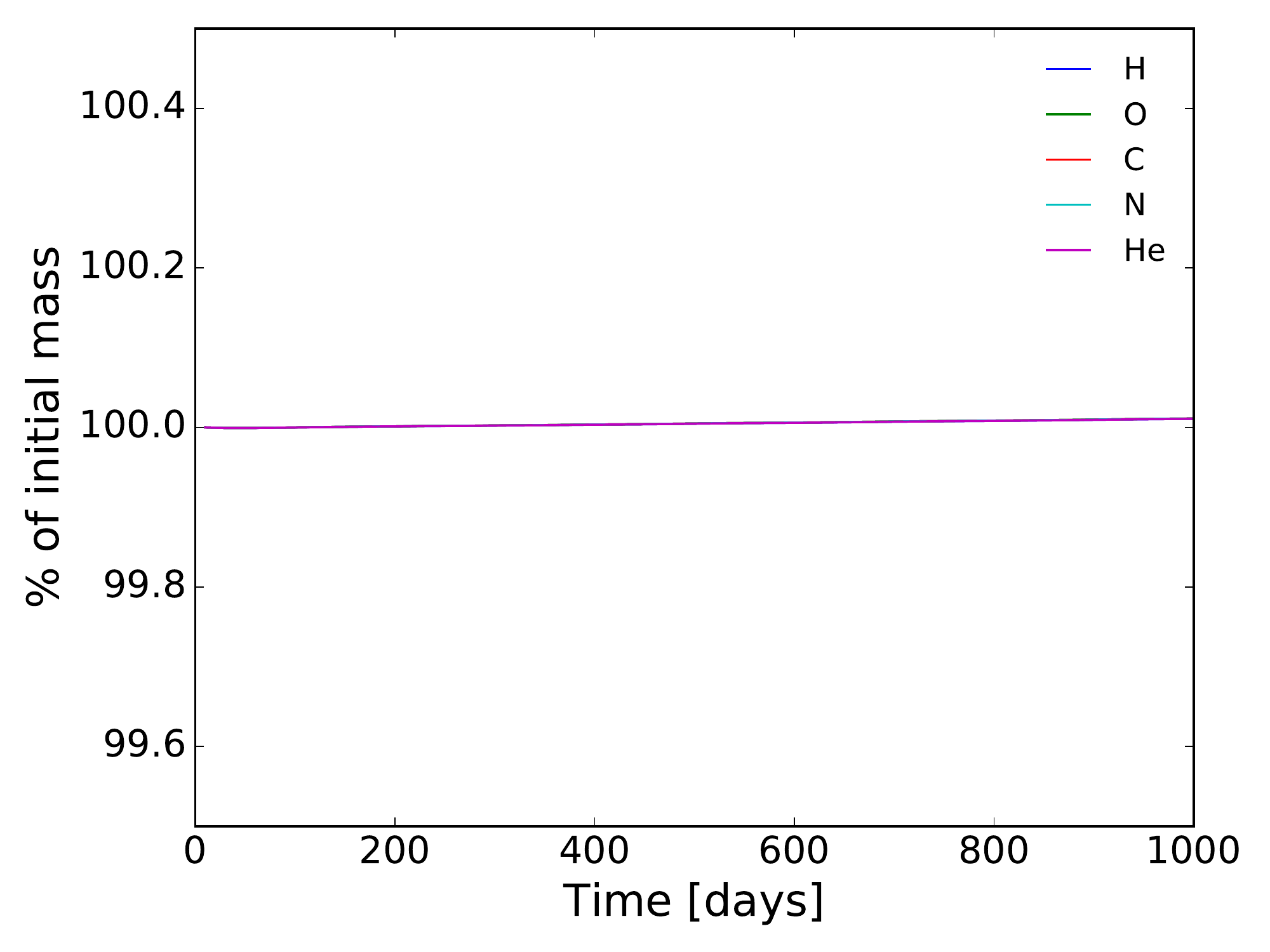} \\
    \includegraphics[width=0.45\textwidth]{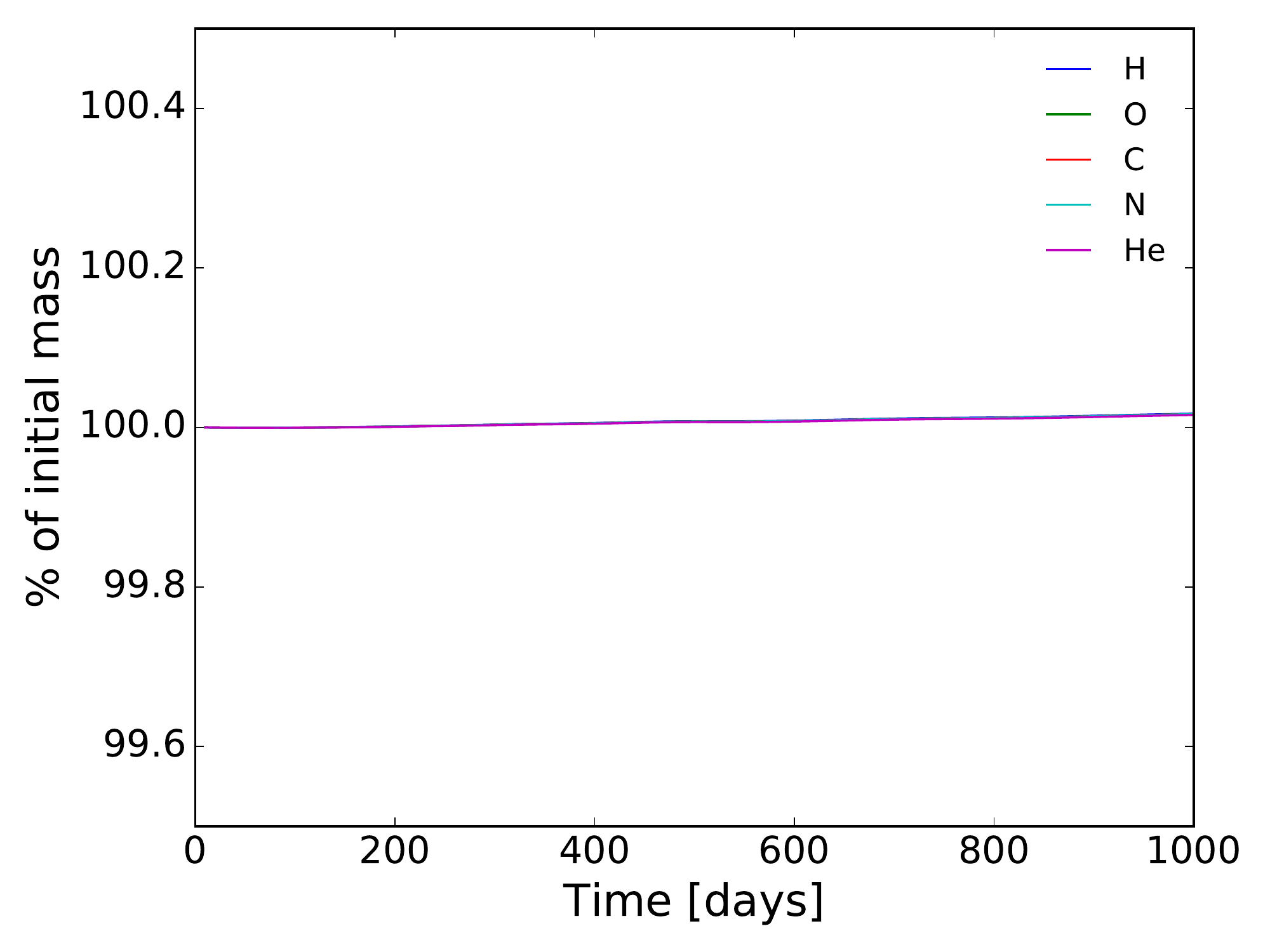}
  \end{center}
\caption{Conservation of global element mass shown as fraction of the initial mass for the kinetics simulation of HD~209458b (top) and HD~189733b (bottom). The global element mass is conserved to within better than 99.9\% for all elements.}
\label{figure:element_conservation}
\end{figure}

An important quantity to conserve is the global (i.e. summed over all grid cells) mass of each element: hydrogen, helium, carbon, oxygen, and nitrogen. The global mass of each chemical species is not a conserved quantity due to chemical transformation. We calculate the global mass of each element from our simulations using Equation E10 presented in \citet[][]{DruMM18b} and show it as a function of model integration time in \cref{figure:element_conservation} for both planets. The mass of each element shows a very small increase over the course of the simulation but importantly is conserved to within 99.9\% at 1000 days, the end of the simulation.

\section{Hot initial profile}
\label{section:hot_initial}

\begin{figure}
    \centering
    \includegraphics[width=0.45\textwidth]{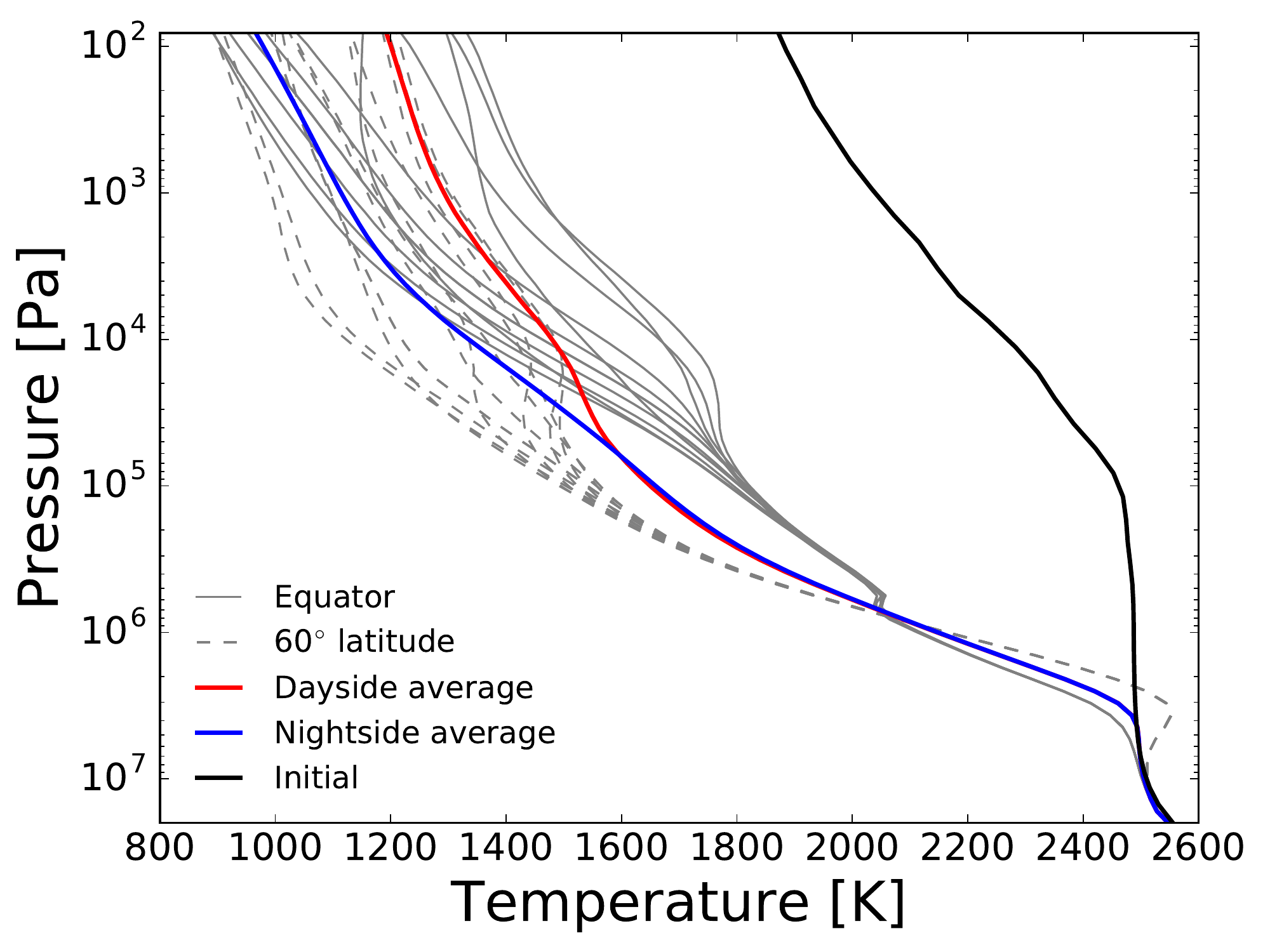}
    \caption{As \cref{figure:temp_prof} but for the simulation of HD~209458b initialised with a hot thermal profile.}
    \label{figure:hd209_hot_pt}
\end{figure}

\cref{figure:hd209_hot_pt} shows the pressure-temperature profiles from the chemical equilibrium simulation of HD~209458b initialised with a hotter thermal profile. Specifically the initial profile simply has 800~K added to the temperature at every pressure level, compared with the nominal case. The temperature profiles in \cref{figure:hd209_hot_pt} can be directly compared with the temperature profiles for the nominal HD~209459b simulation in \cref{figure:temp_prof}. The only significant difference is in the deep atmosphere ($P\gtrsim10^6$~Pa) where the nominal simulation shows a much lower temperature as the model has not been evolved for long enough to reach a steady-state. At lower pressures differences in the temperature profiles between the simulations are minimal. There are also negligible differences in the zonal-mean zonal wind (not shown).

\begin{figure*}
  \begin{center}
    \includegraphics[width=0.45\textwidth]{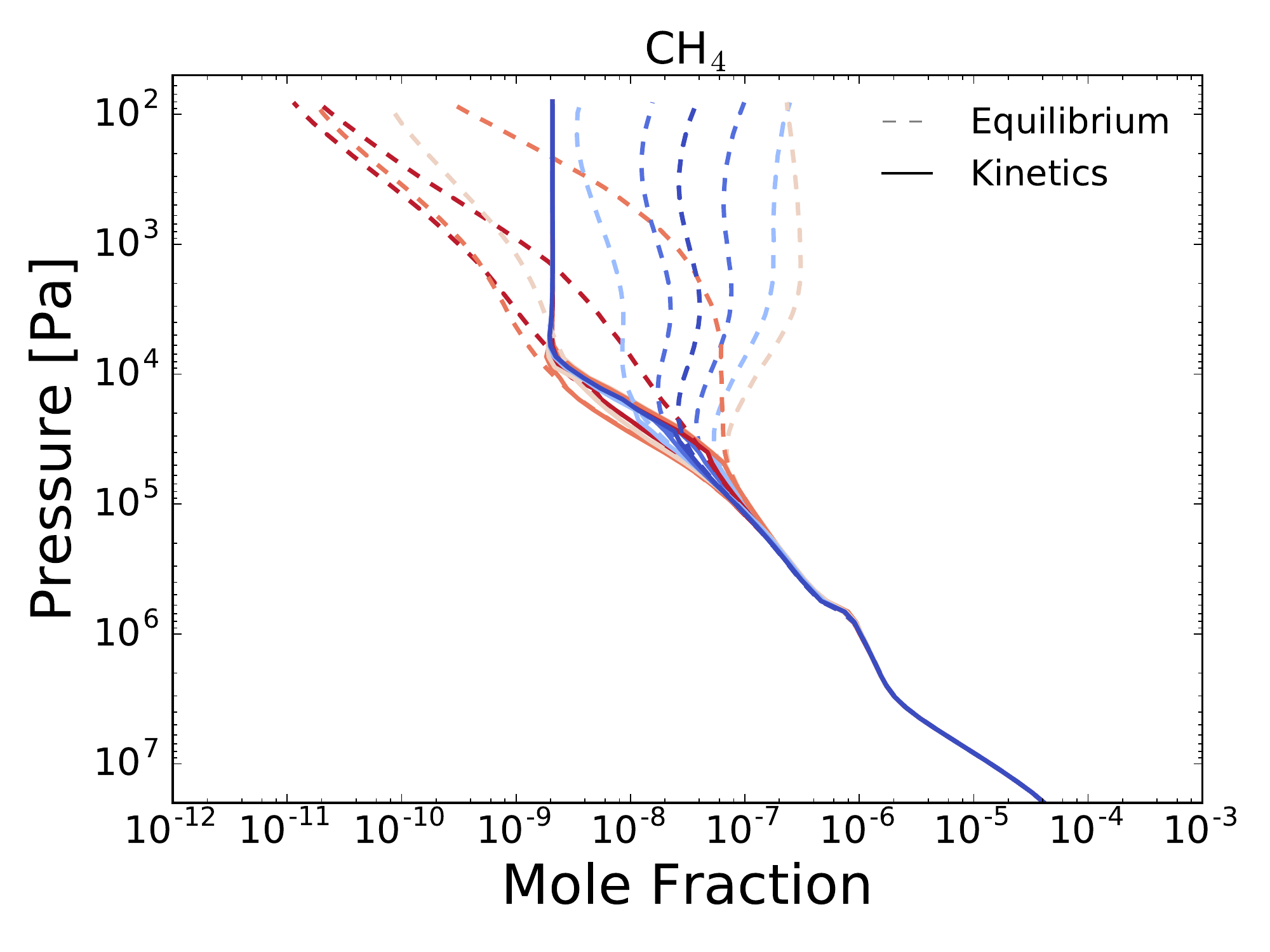}
     \includegraphics[width=0.45\textwidth]{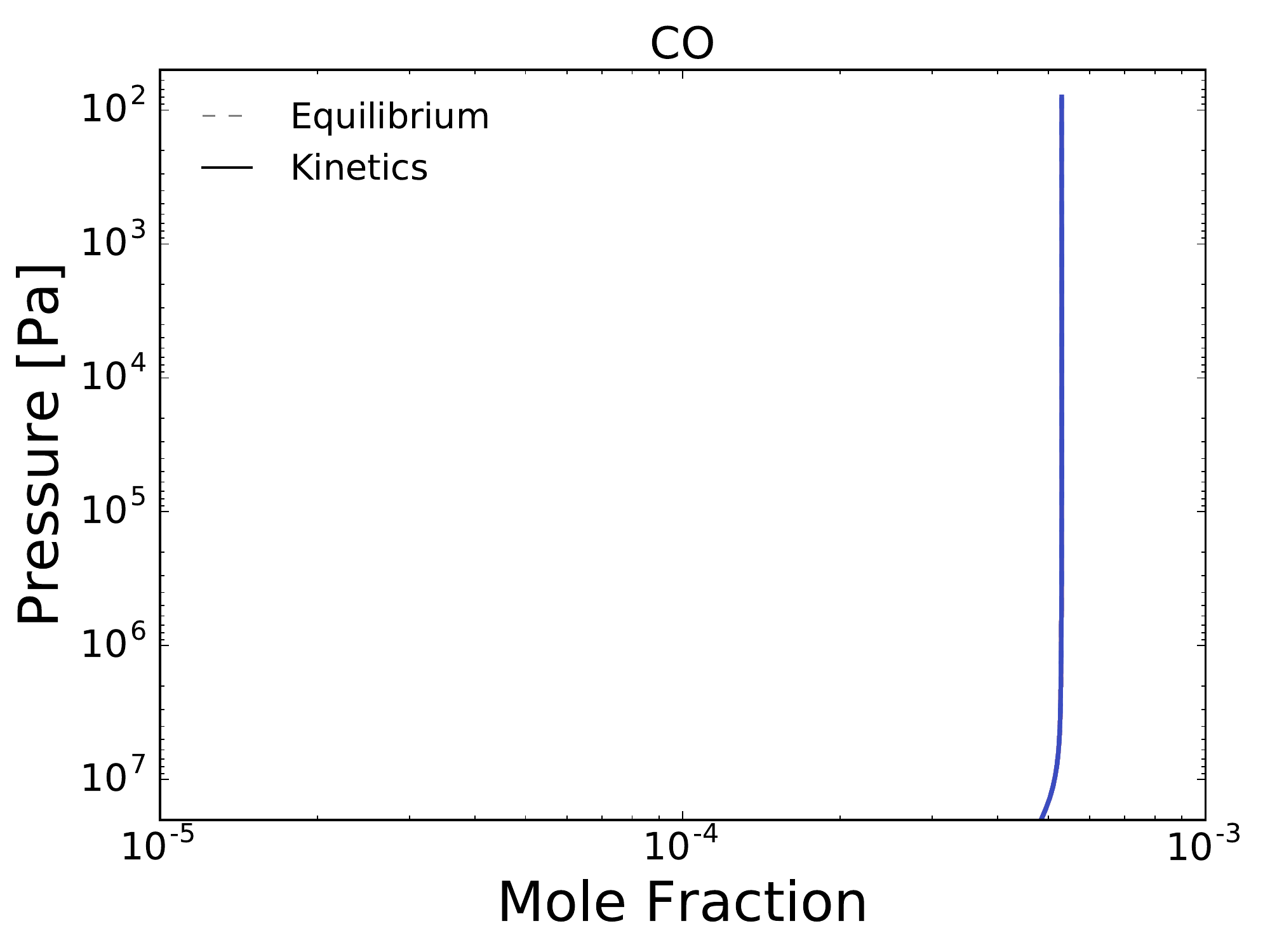} \\
     \includegraphics[width=0.45\textwidth]{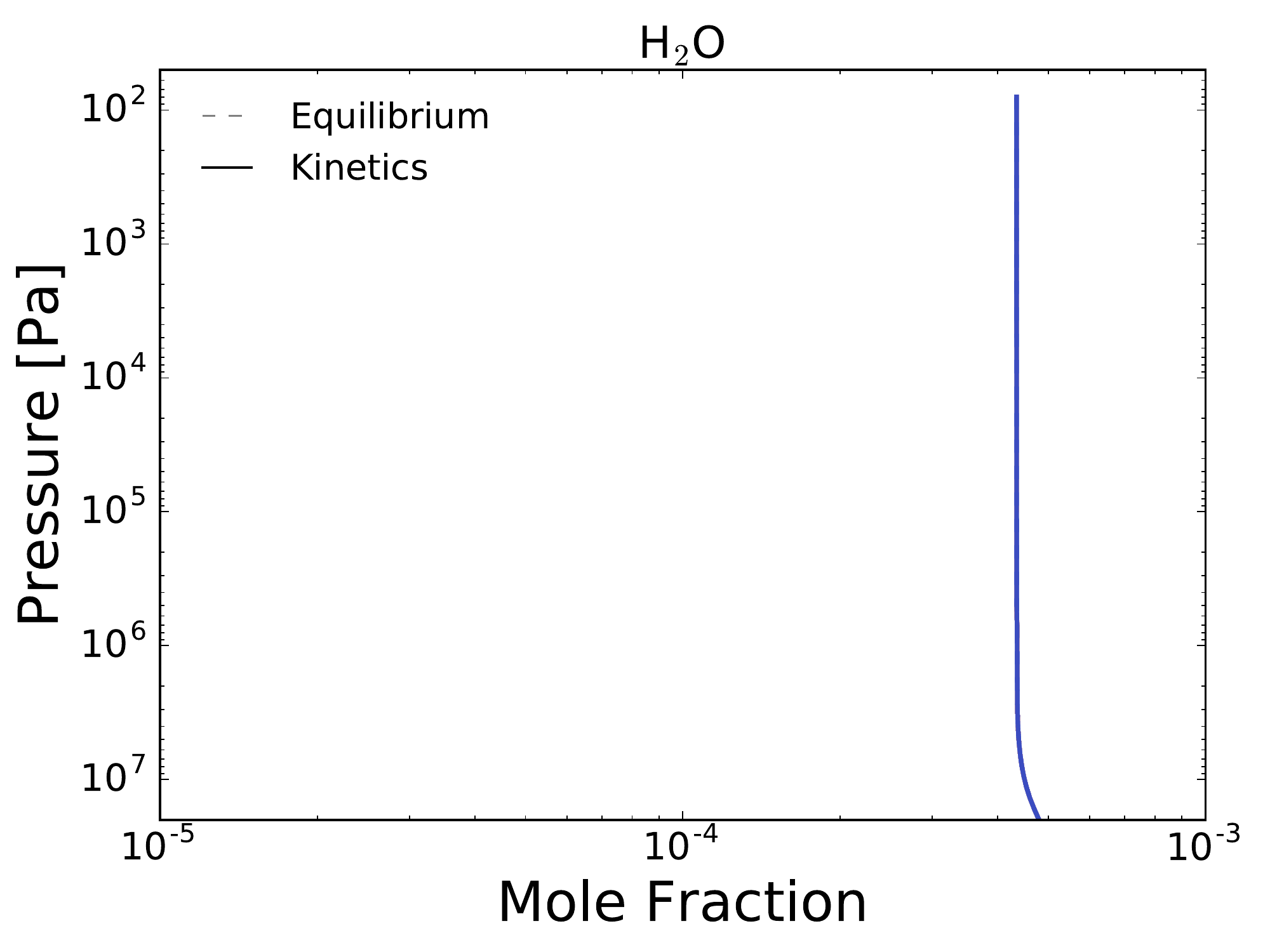}
    \includegraphics[width=0.45\textwidth]{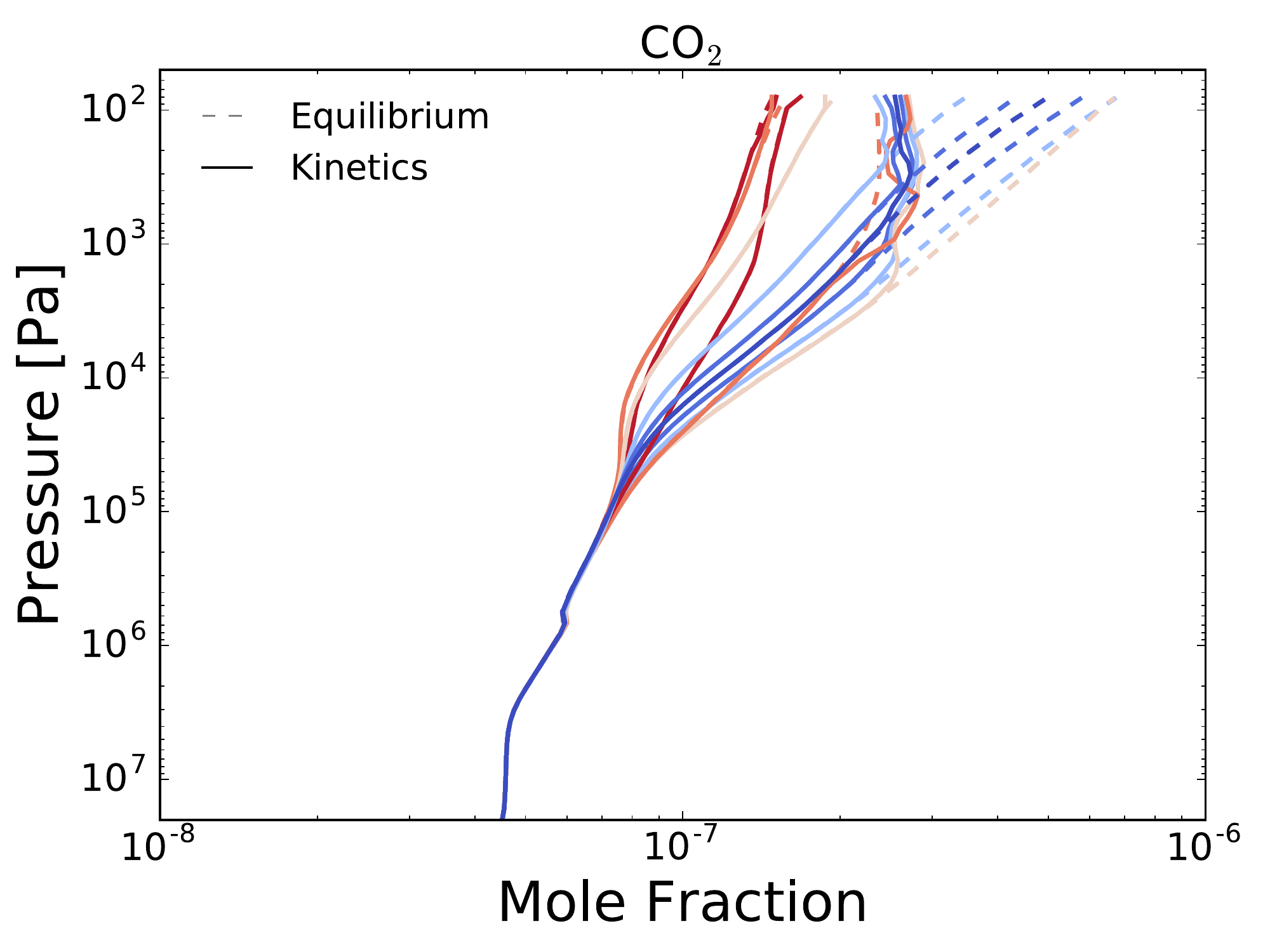}\\
     \includegraphics[width=0.45\textwidth]{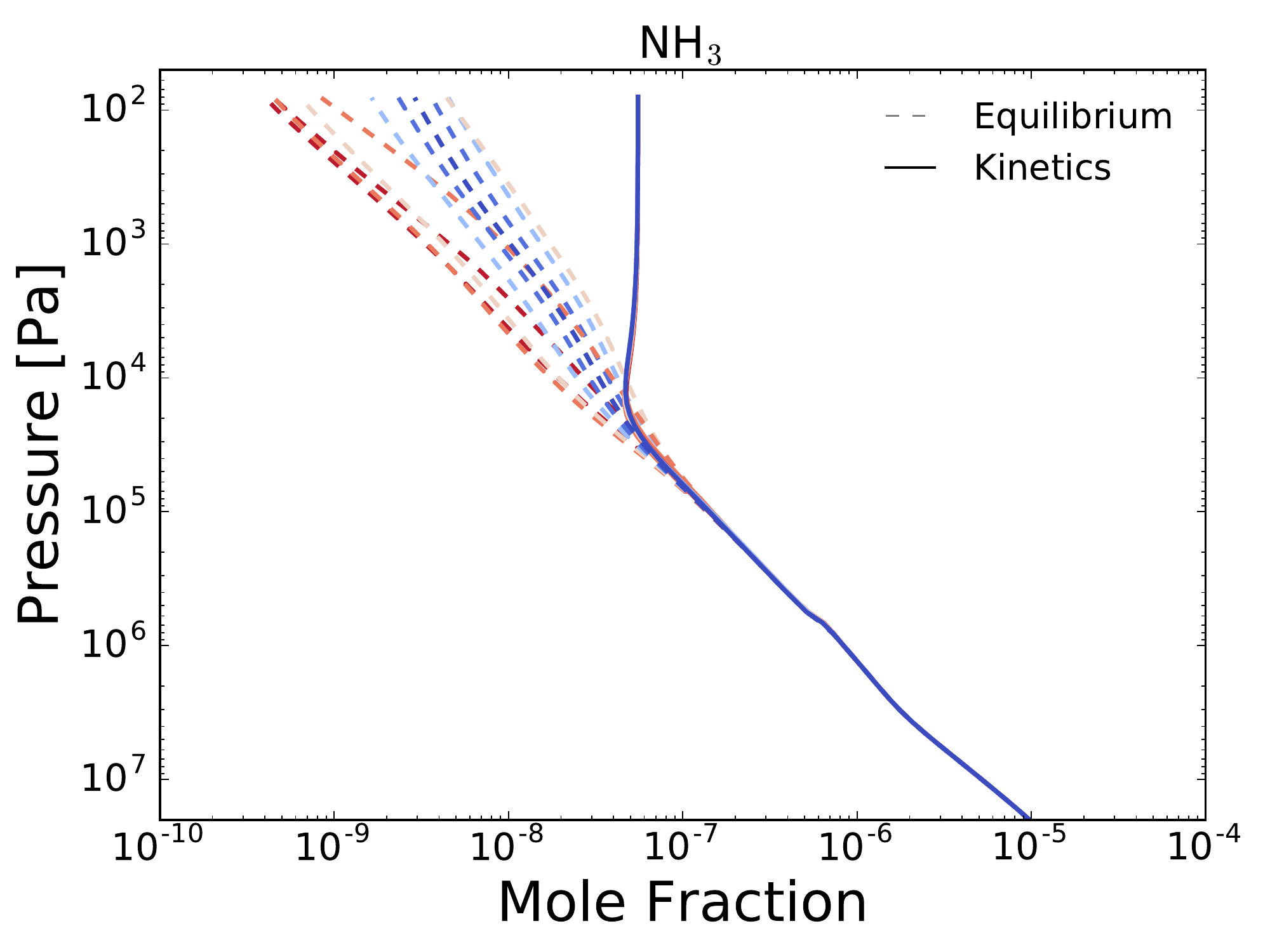}
 \includegraphics[width=0.45\textwidth]{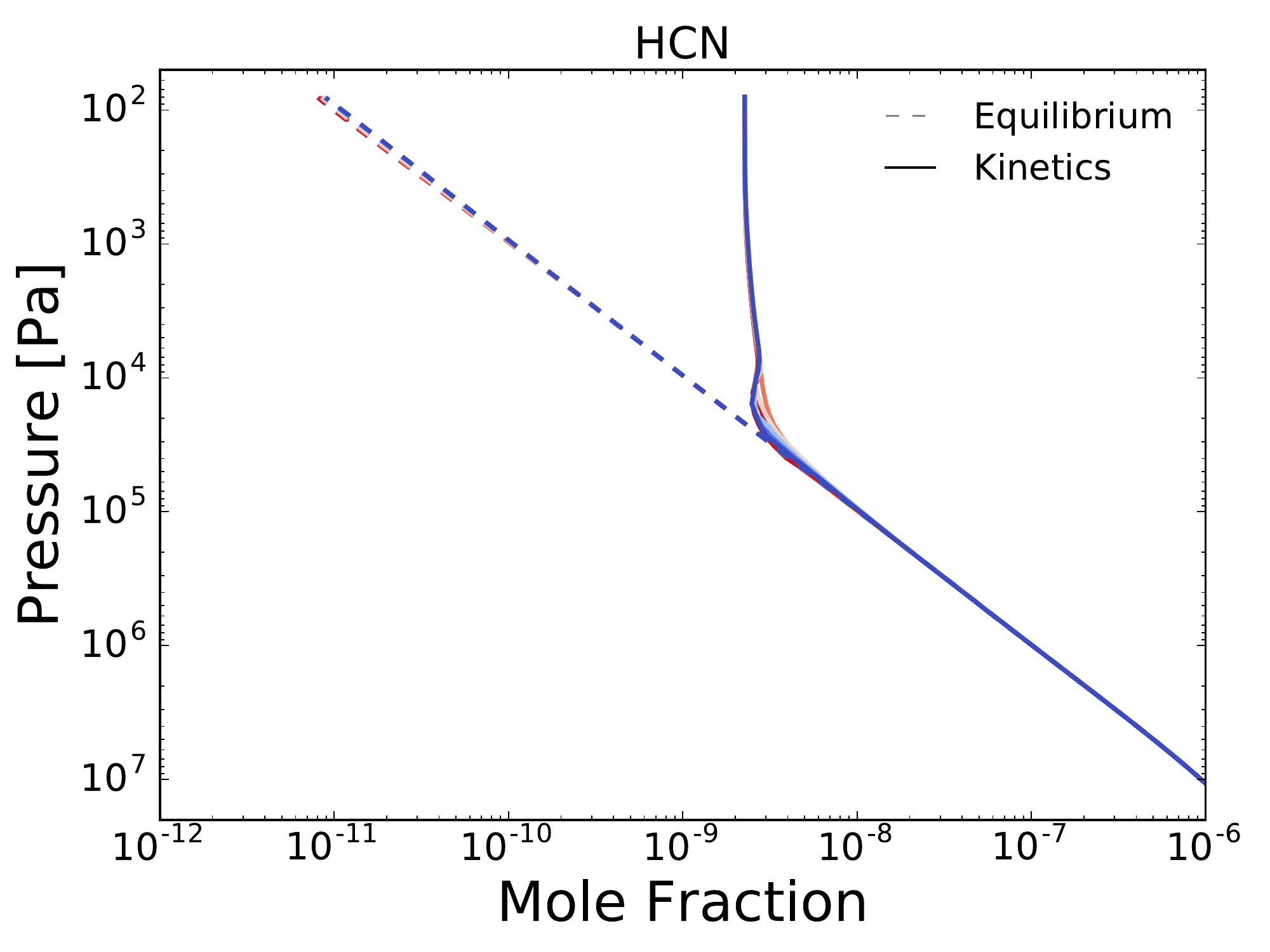}\\
  \end{center}
\caption{Vertical mole fraction profiles for different longitude points around the equator, for the equilibrium simulation (dashed lines) and kinetics simulation (solid lines) of HD~209458b initialised with a hot temperature profile.}
\label{figure:hd209_hot_mf_prof}
\end{figure*}

\cref{figure:hd209_hot_mf_prof} shows abundance profiles of the six key molecules considered in the main body of the paper, for both the equilibrium and kinetics simulations with a hot initial profile. Comparing this figure with abundance profiles for the nominal HD~209459b simulations (\cref{figure:hd209_mf_prof}) clearly shows that there are negligible differences between the simulations with a `hot' and `cold' initial profile for most of the model domain. Differences are apparent for $P>10^6$~Pa, where all six species are in equilibrium, which is simply a result of the aforementioned temperature difference. Importantly, the deep atmosphere does not effect the lower atmosphere in this particular case, as the quench point for all six molecules lies above (towards lower pressures) the region where the atmosphere has not reached a steady-state.

A further outcome of this experiment is that it shows the insensitivity of the model result to the initial conditions with respect to the chemical abundances. The initial chemical abundances correspond to the chemical equilibrium at the start of the simulation, which is defined by the initial thermal profile. Since the initial profile is uniformly 800~K hotter in this test, compared with the nominal case, this corresponds to a significantly different chemical equilibrium composition. Despite this, the simulations show a very similar result (except in the deep atmosphere) after 1000~days of integration. This suggests that the results are not strongly sensitive to the initial conditions.

\section{Abundance profiles at higher latitudes}
\label{section:midlat}

\begin{figure*}
  \begin{center}
    \includegraphics[width=0.45\textwidth]{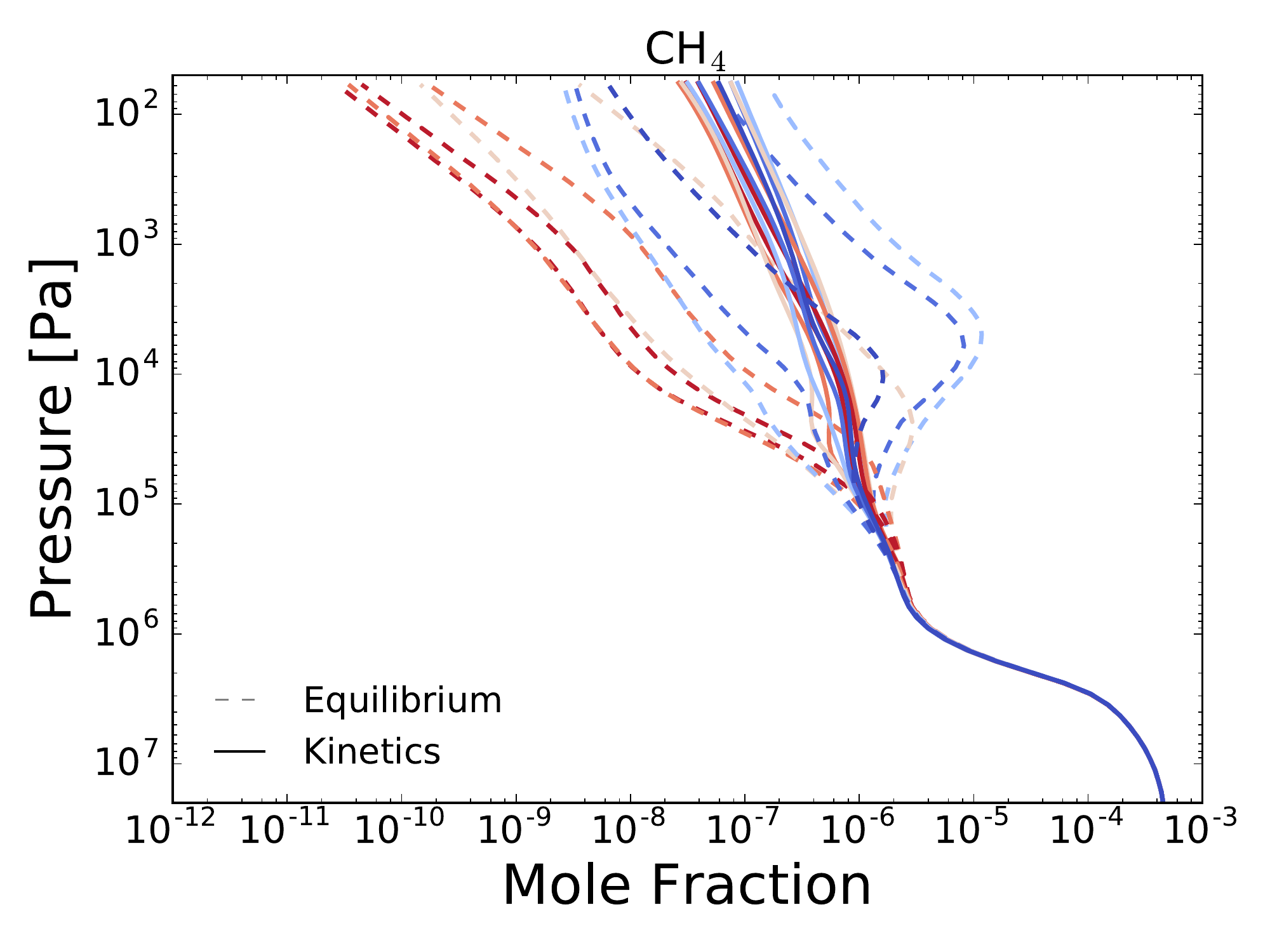}
     \includegraphics[width=0.45\textwidth]{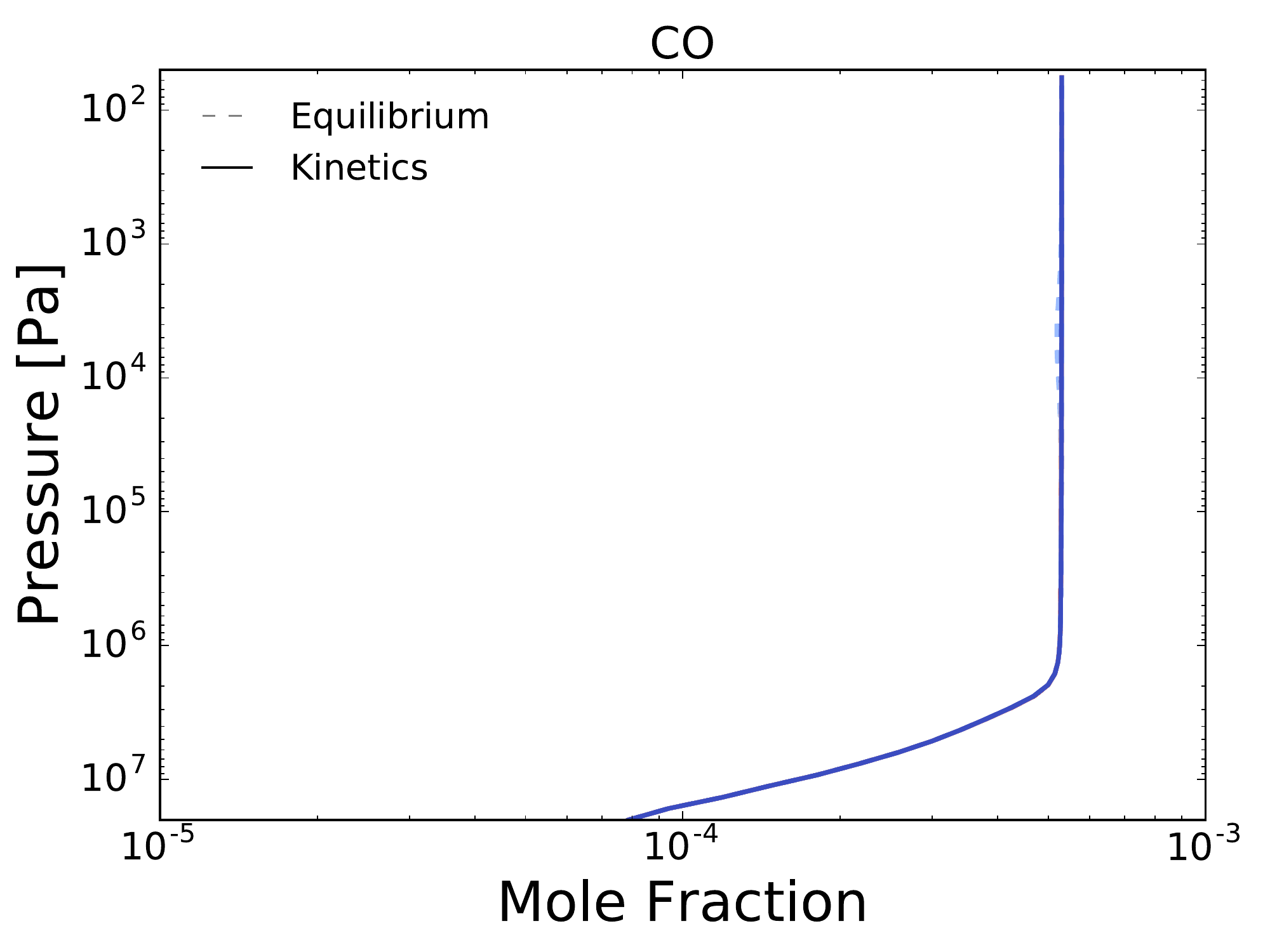} \\
     \includegraphics[width=0.45\textwidth]{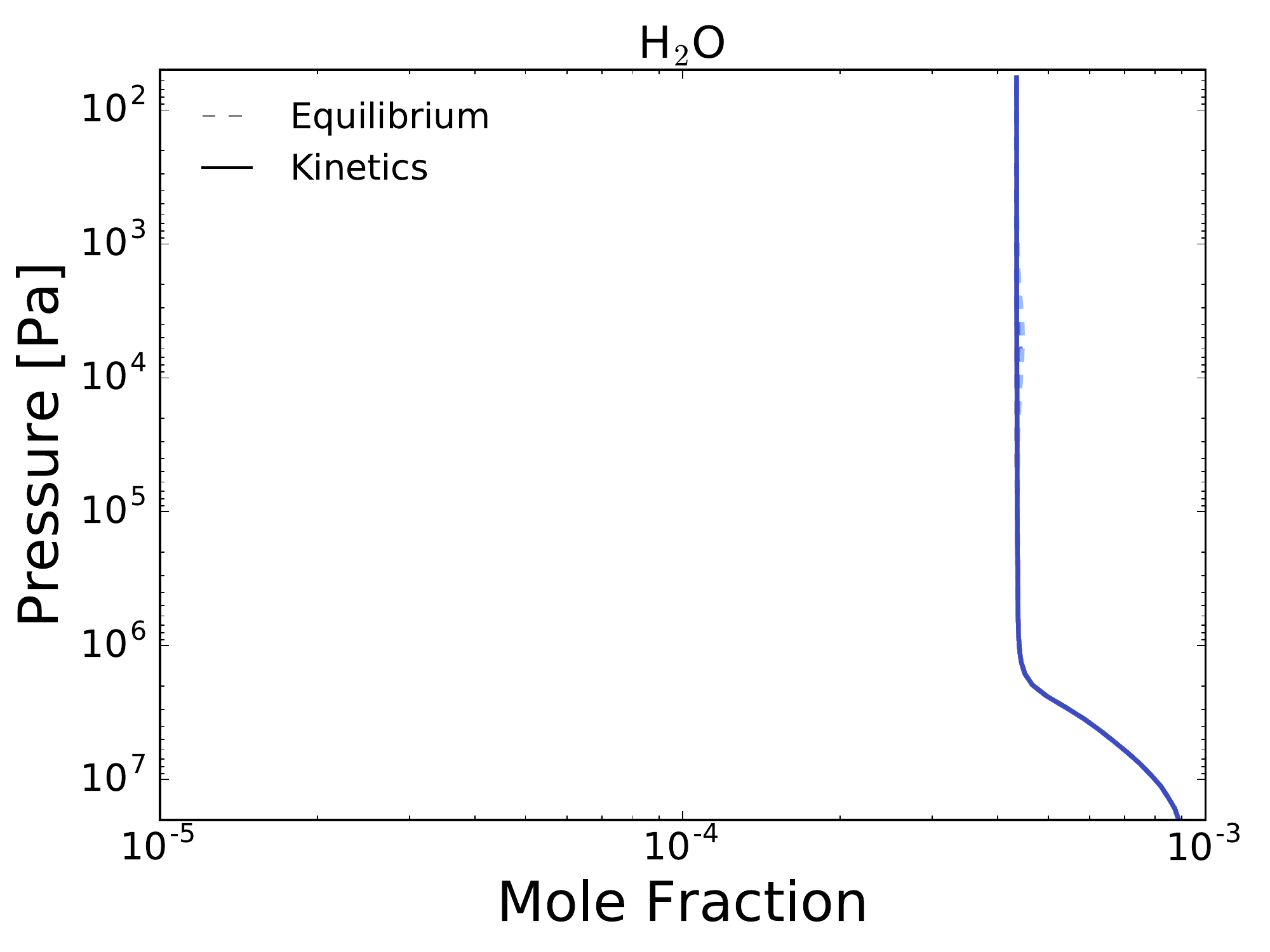}
    \includegraphics[width=0.45\textwidth]{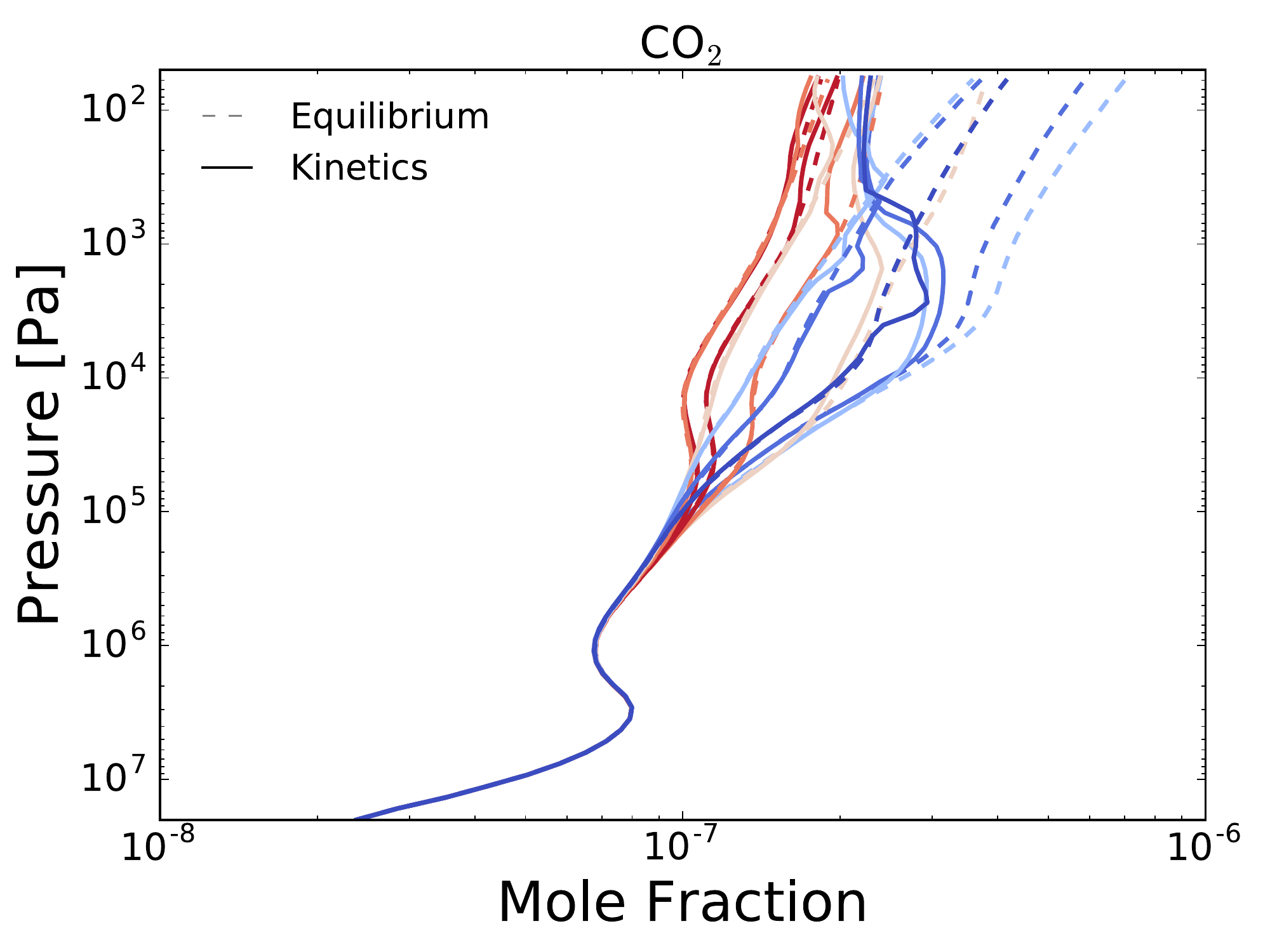}\\
     \includegraphics[width=0.45\textwidth]{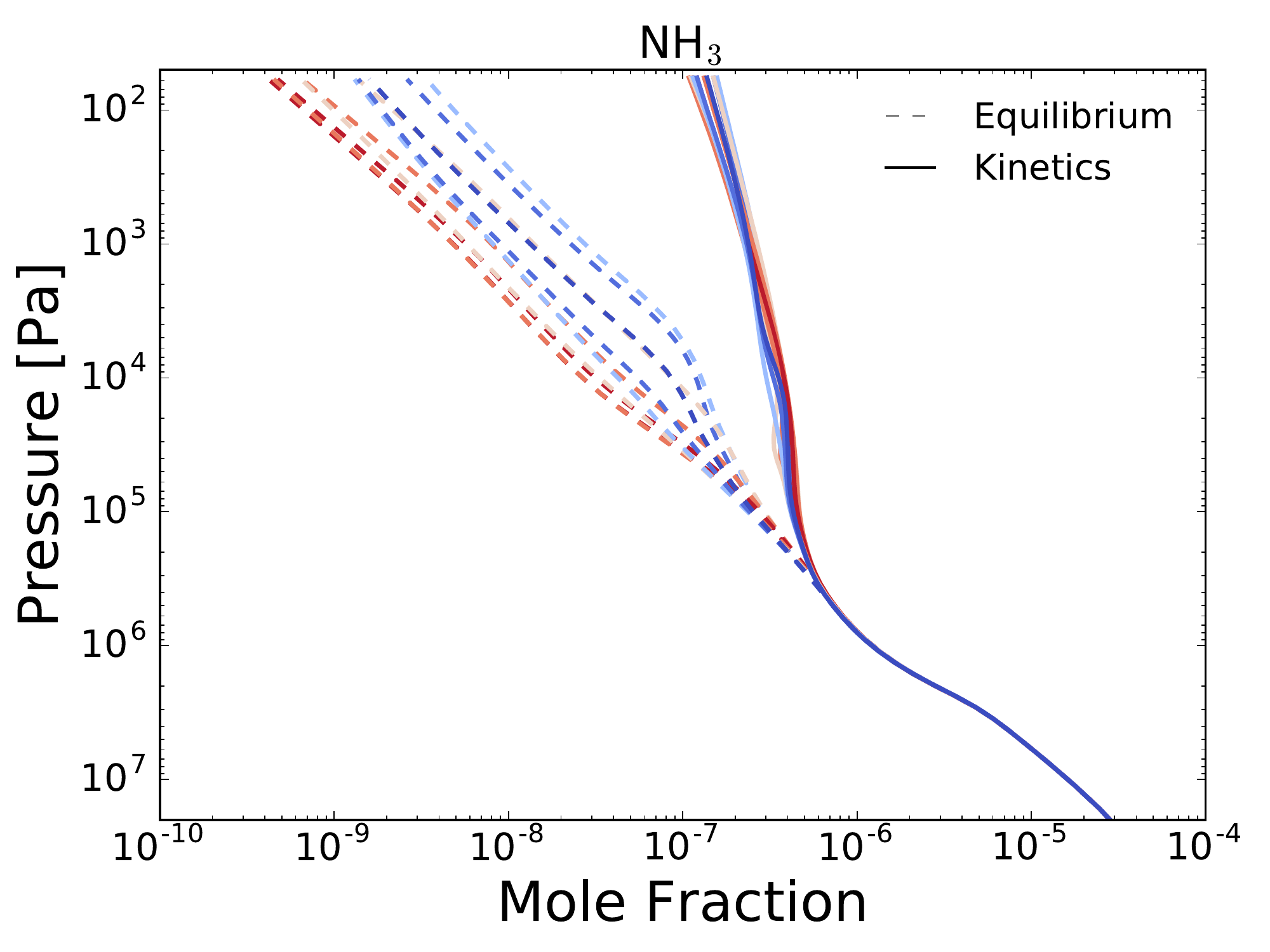}
 \includegraphics[width=0.45\textwidth]{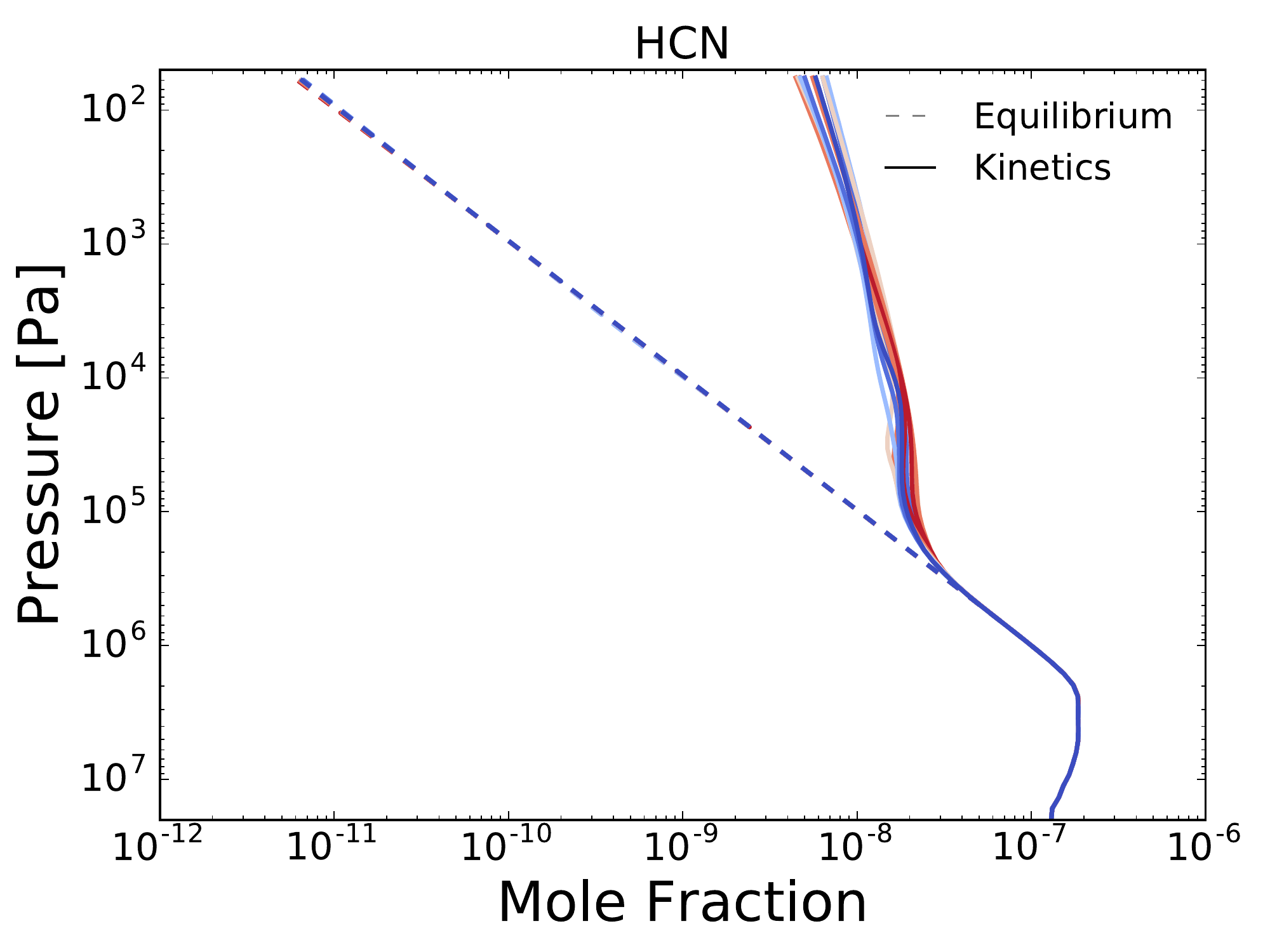}\\
  \end{center}
\caption{Vertical mole fraction profiles for different longitude points around the latitude of 45$^{\circ}$, for the equilibrium simulation (dashed lines) and kinetics simulation (solid lines) of HD~209458b.}
\label{figure:hd209_mf_prof_midlat}
\end{figure*}

\begin{figure*}
  \begin{center}
    \includegraphics[width=0.45\textwidth]{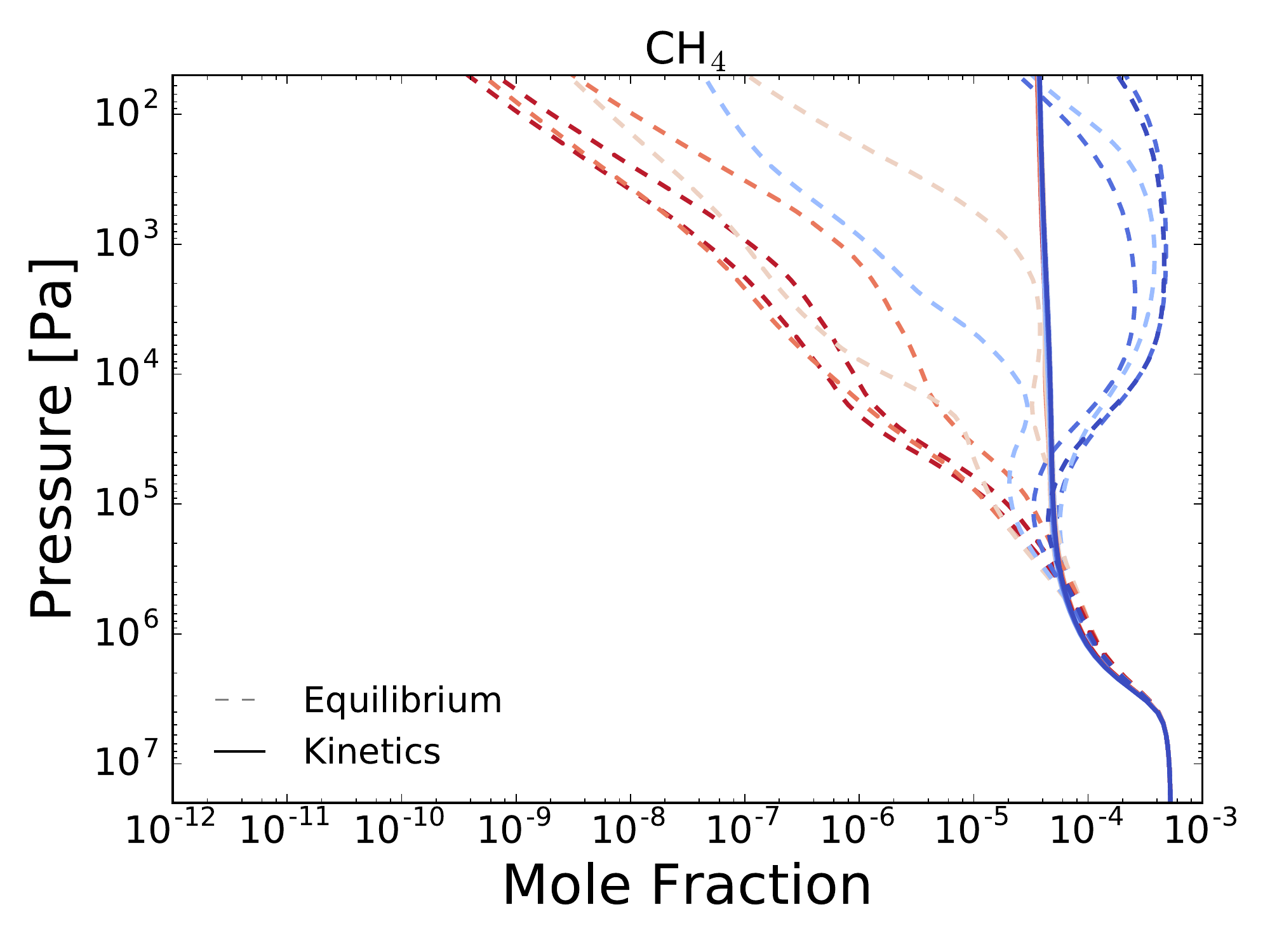}
     \includegraphics[width=0.45\textwidth]{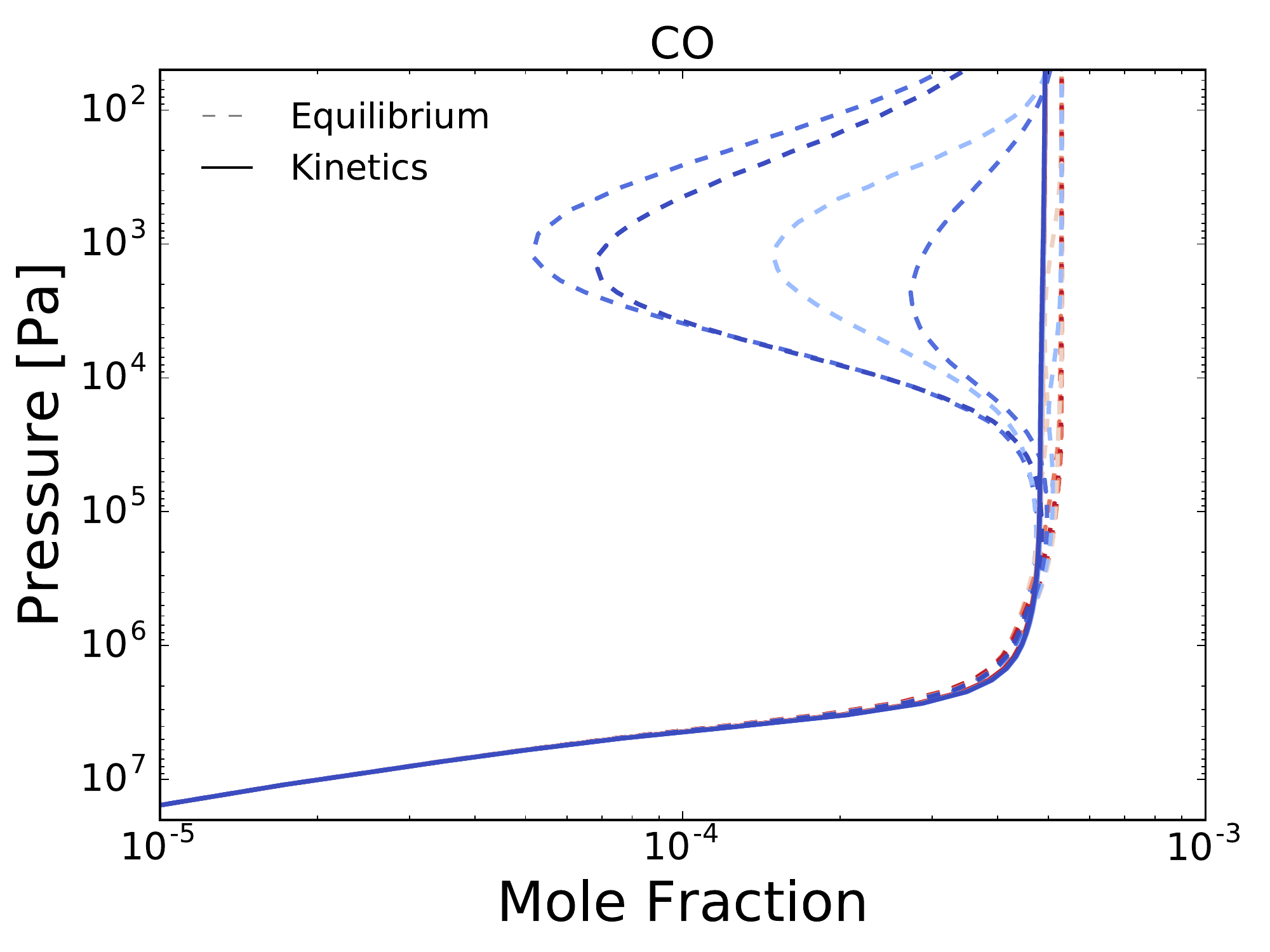} \\
     \includegraphics[width=0.45\textwidth]{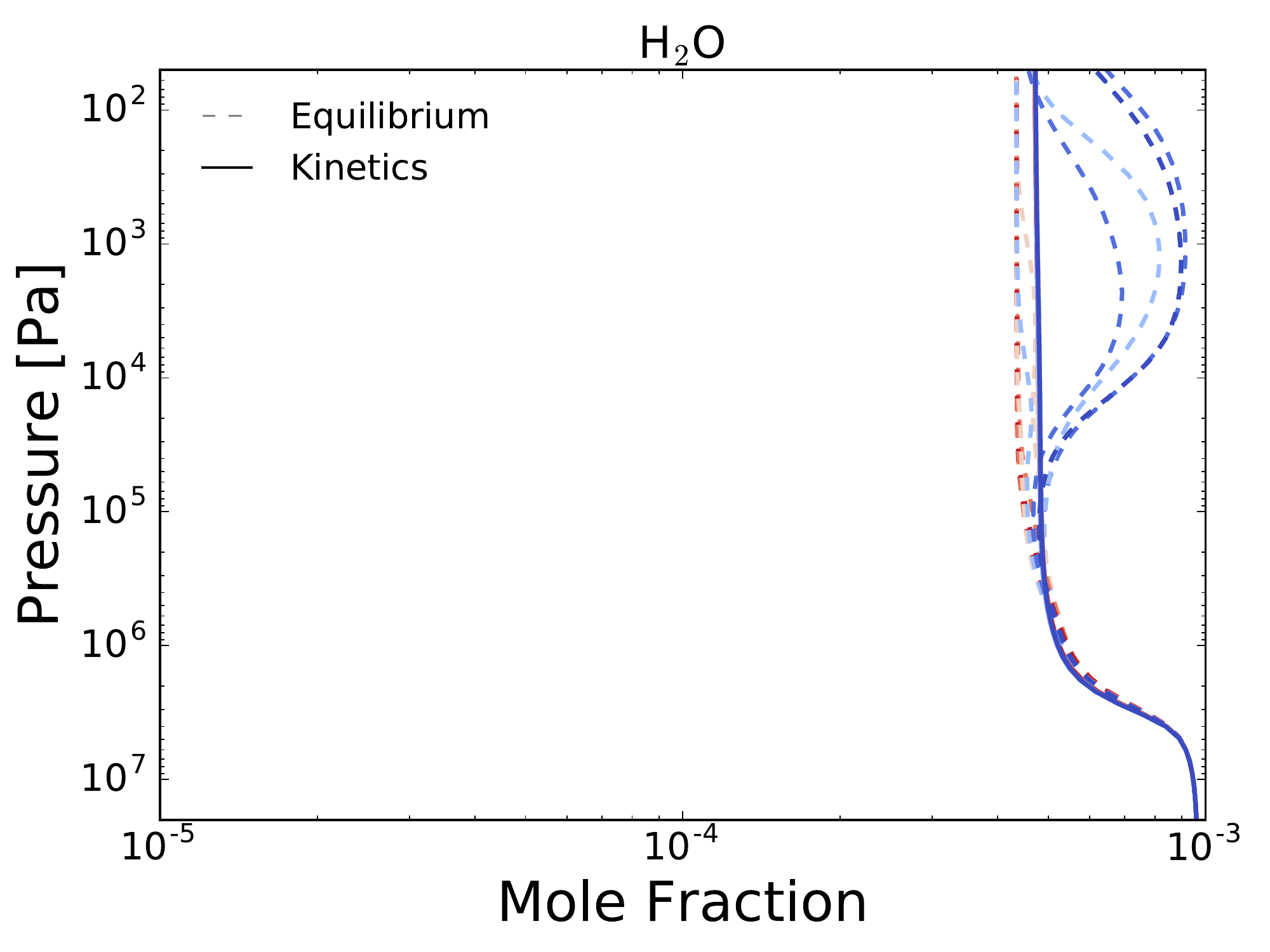}
    \includegraphics[width=0.45\textwidth]{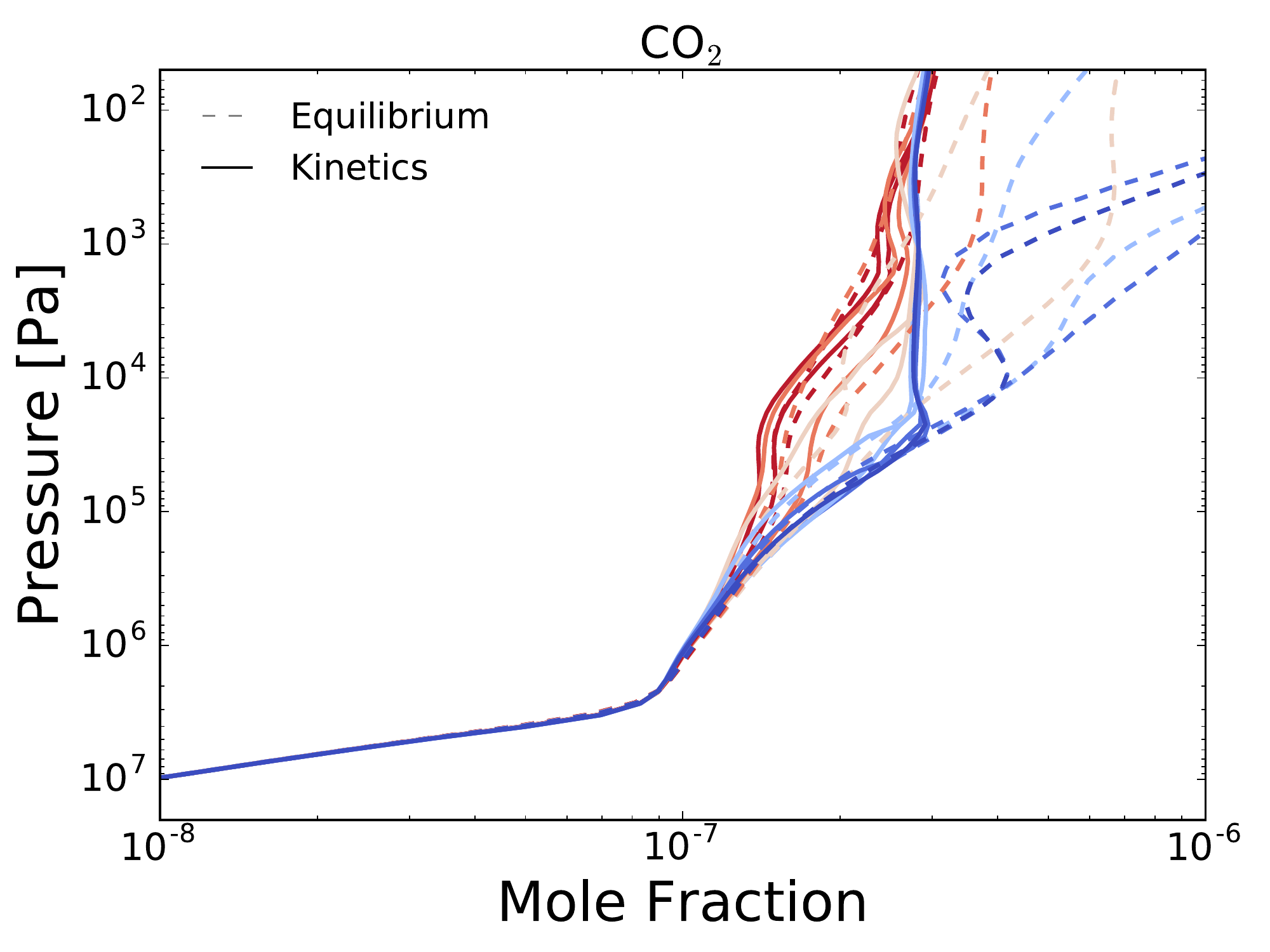}\\
     \includegraphics[width=0.45\textwidth]{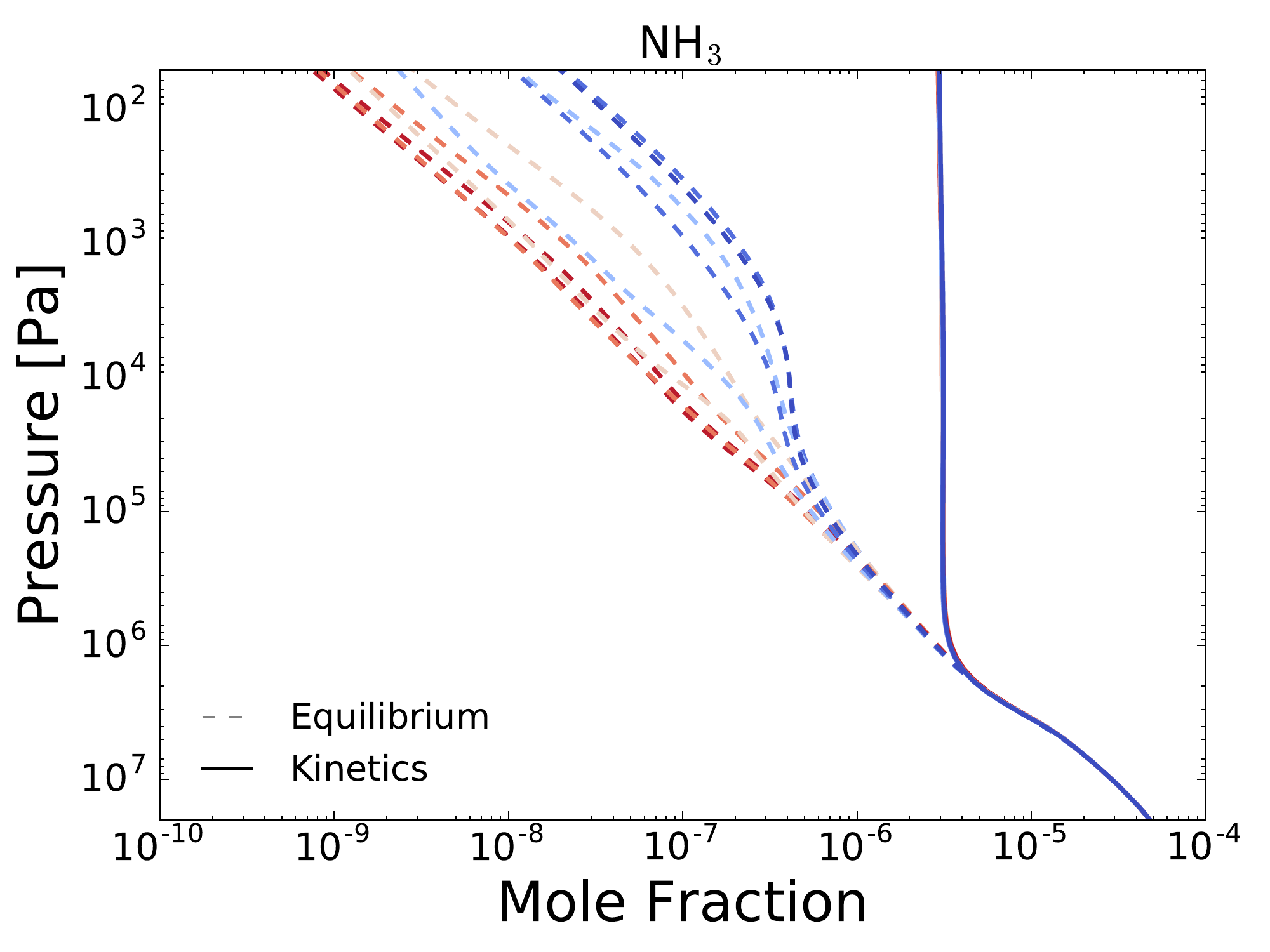}
 \includegraphics[width=0.45\textwidth]{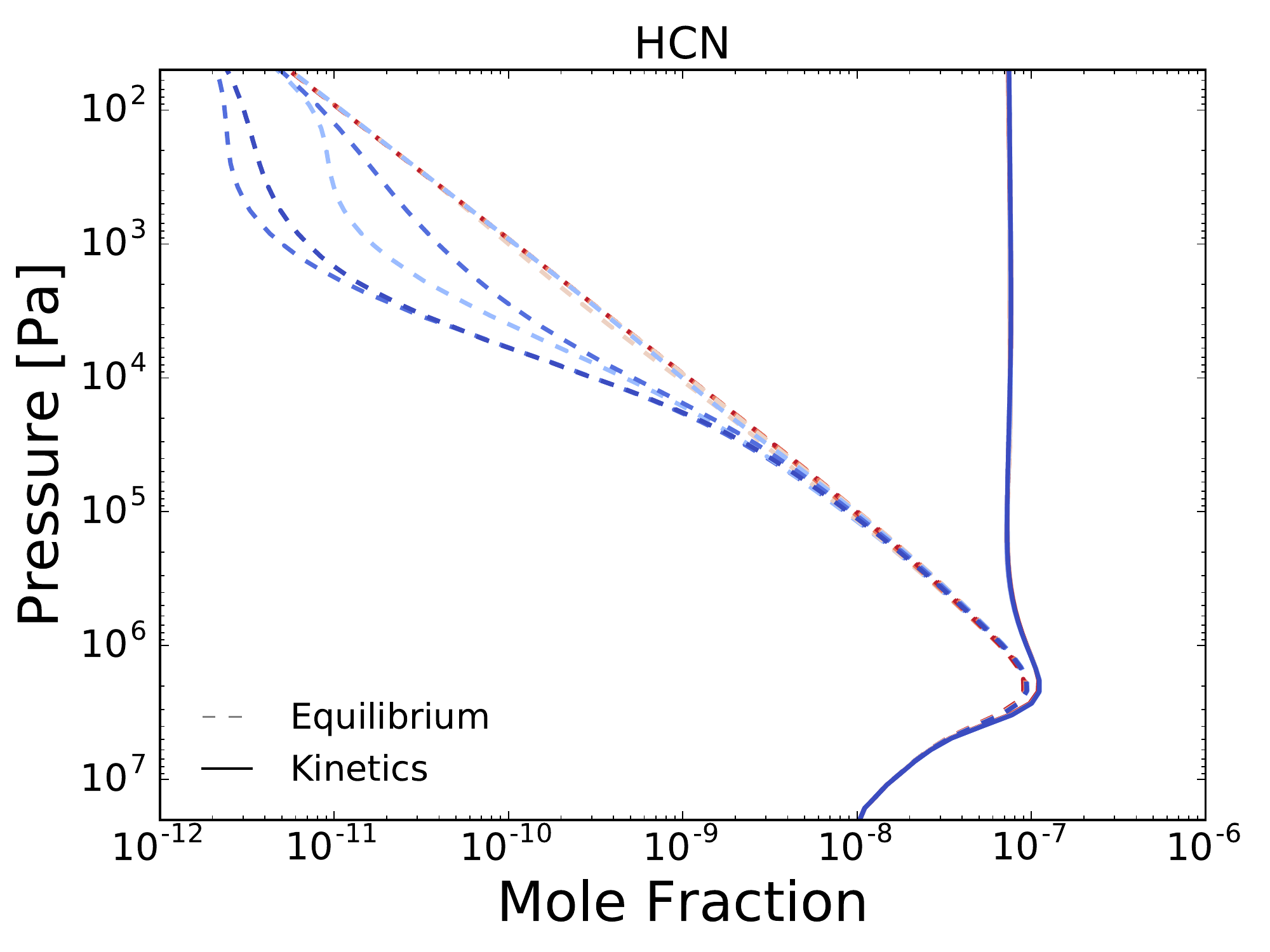}\\
  \end{center}
\caption{Vertical mole fraction profiles for different longitude points around the latitude of 45$^{\circ}$, for the equilibrium simulation (dashed lines) and kinetics simulation (solid lines) of HD~189733b.}
\label{figure:hd189_mf_prof_midlat}
\end{figure*}

In this section we present abundance profiles of the six species of interest at a latitude of 45$^{\circ}$. These complement the abundance profiles at the equator shown in the main body of the text. The abundance profiles at 45$^{\circ}$ latitude are shown in \cref{figure:hd209_mf_prof_midlat} and \cref{figure:hd189_mf_prof_midlat} for HD~209458b and HD~189733b, respectively.

\section{Timescale profiles}

In this section we present profiles of the advection and chemical timescales (for CH$_4$ and CO) at specific points around the planet. The advection timescales are estimated using the pressure profile of the wind velocity (no spatial averaging) for each component, using the equations presented in \cref{results:209}. The chemical timescales are estimated for the \citet{VenBD19} network using the method described in \citet{TsaKL18}.

\begin{figure*}
    \centering
    \includegraphics[width=0.39\textwidth]{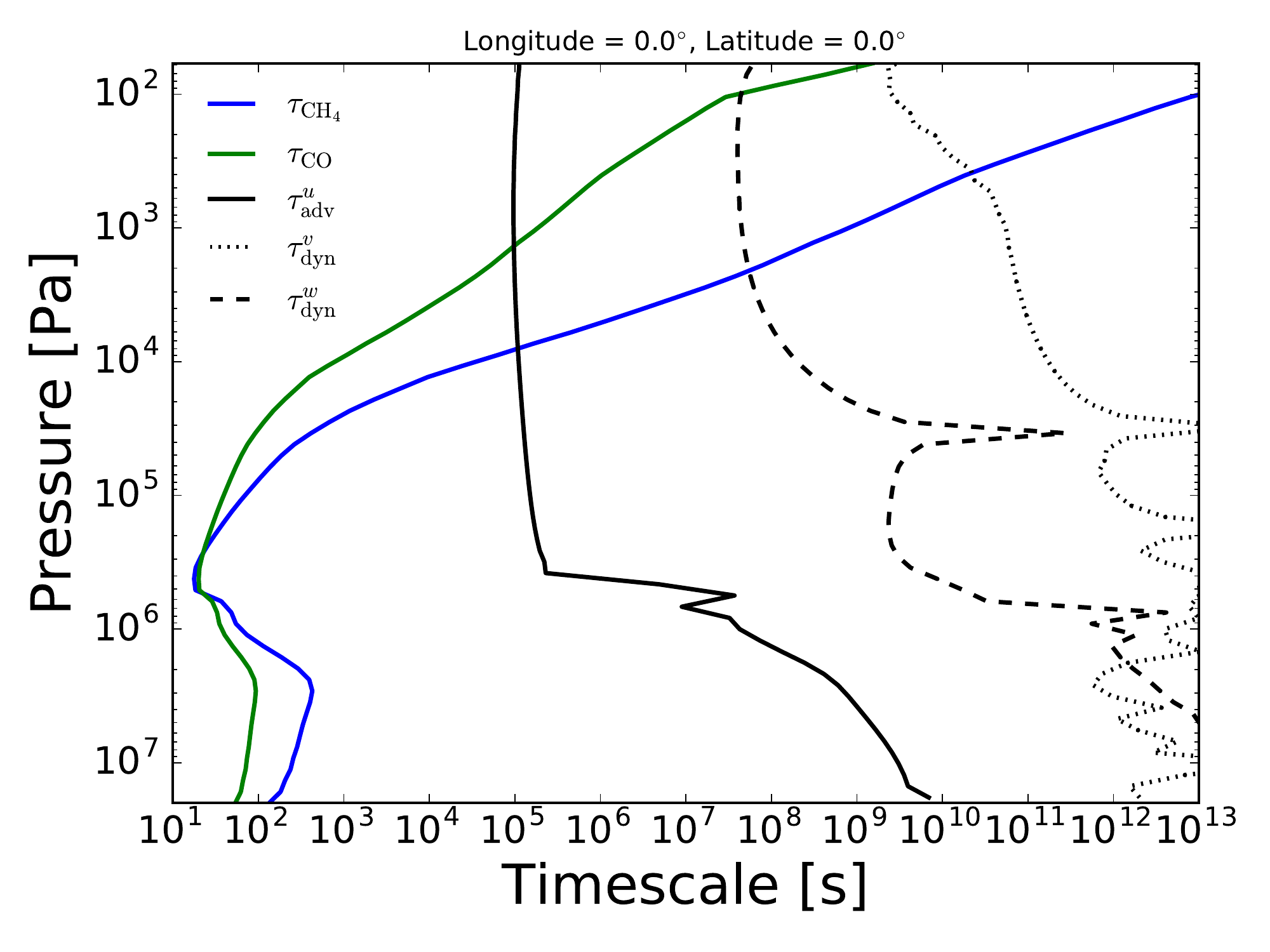} 
    \includegraphics[width=0.39\textwidth]{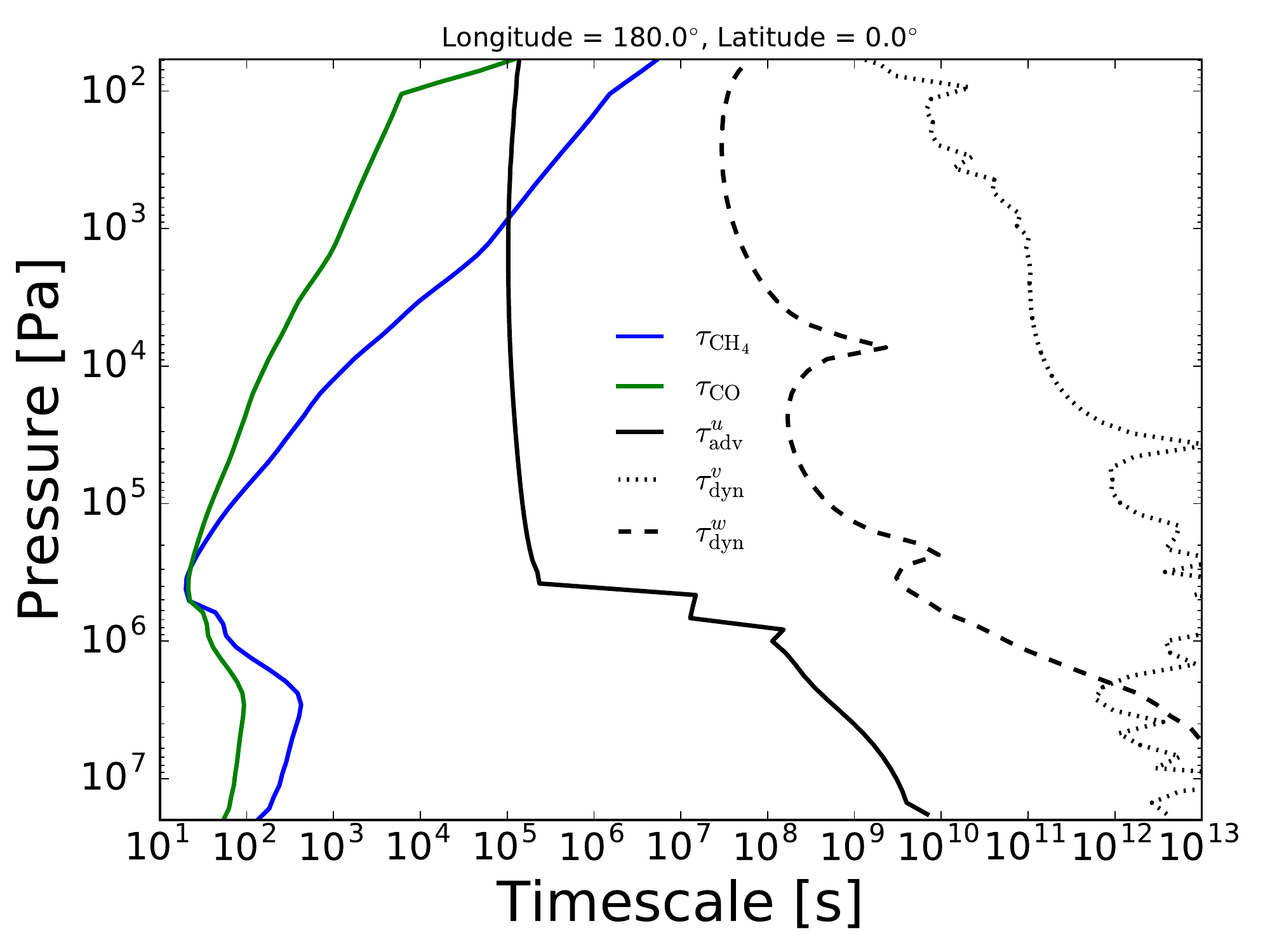} \\
    \includegraphics[width=0.39\textwidth]{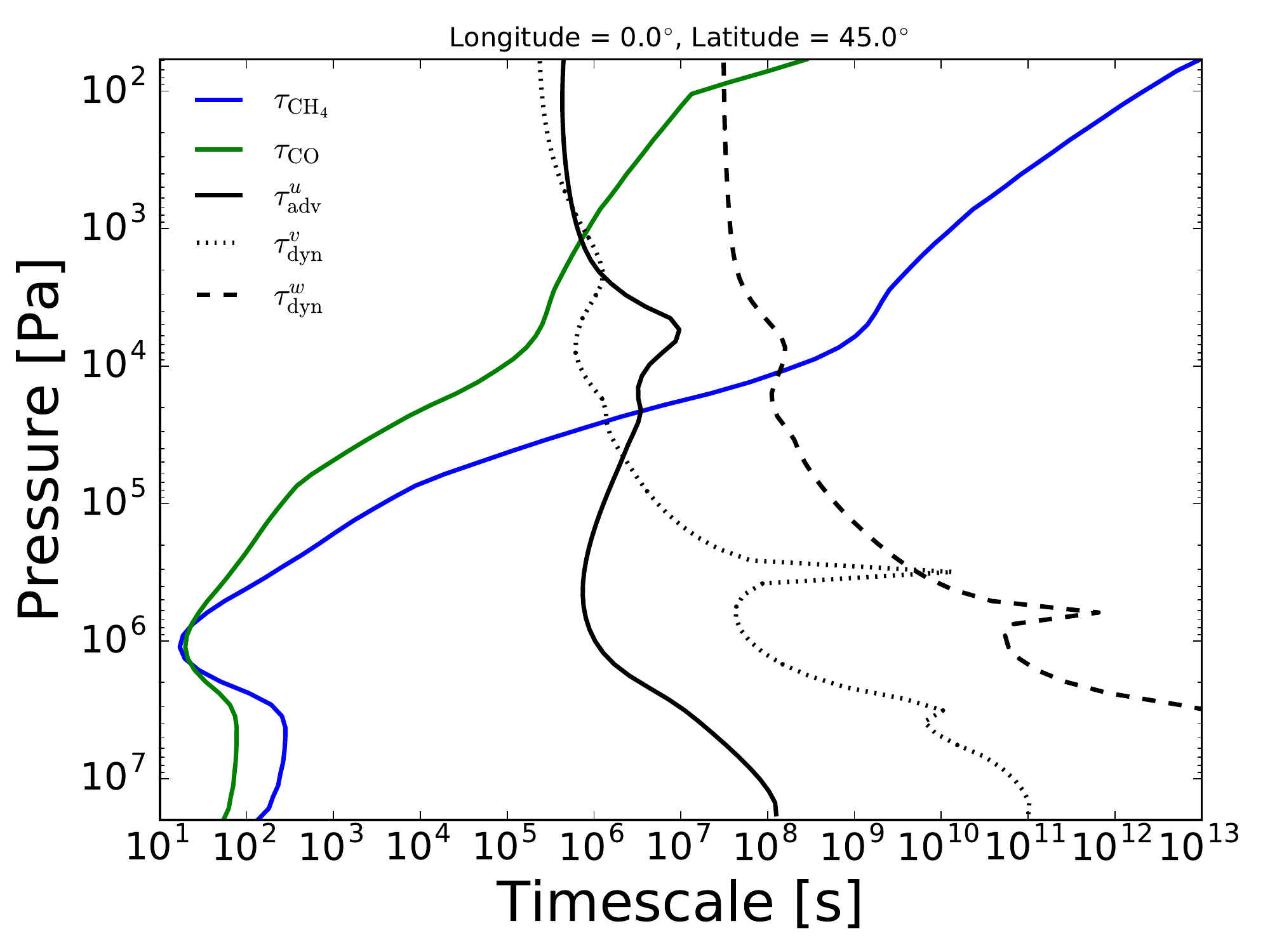} 
    \includegraphics[width=0.39\textwidth]{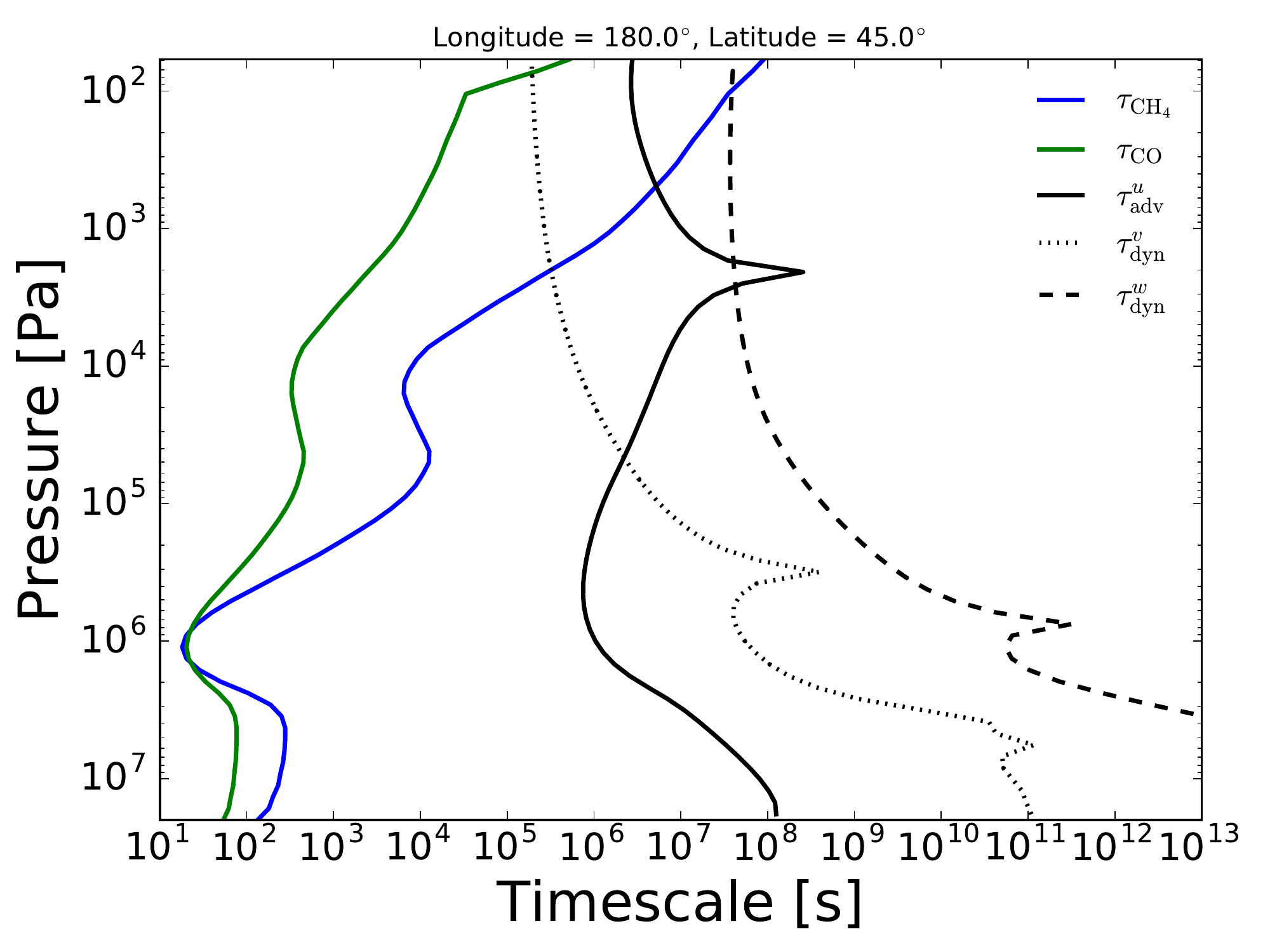} 
    \caption{Pressure profiles of the advection timescale (showing seperately the zonal ($u$), meridional ($v$), and vertical ($w$) components) and the chemical timescale for CH$_4$ and CO for the HD~209458b simulation for different columns around the planet.}
    \label{figure:hd209_ts_profiles}
\end{figure*}

\begin{figure*}
    \centering
    \includegraphics[width=0.39\textwidth]{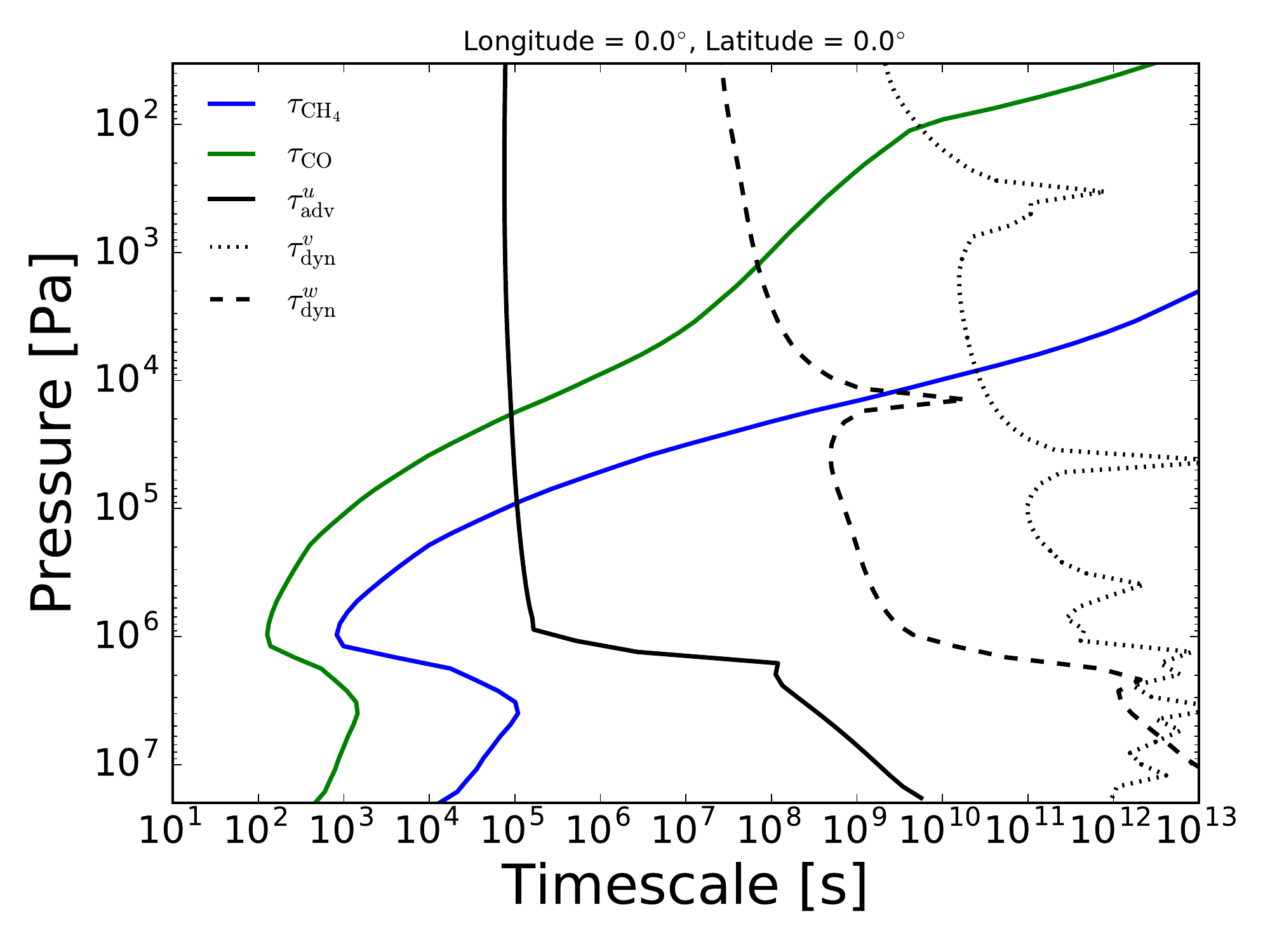} 
    \includegraphics[width=0.39\textwidth]{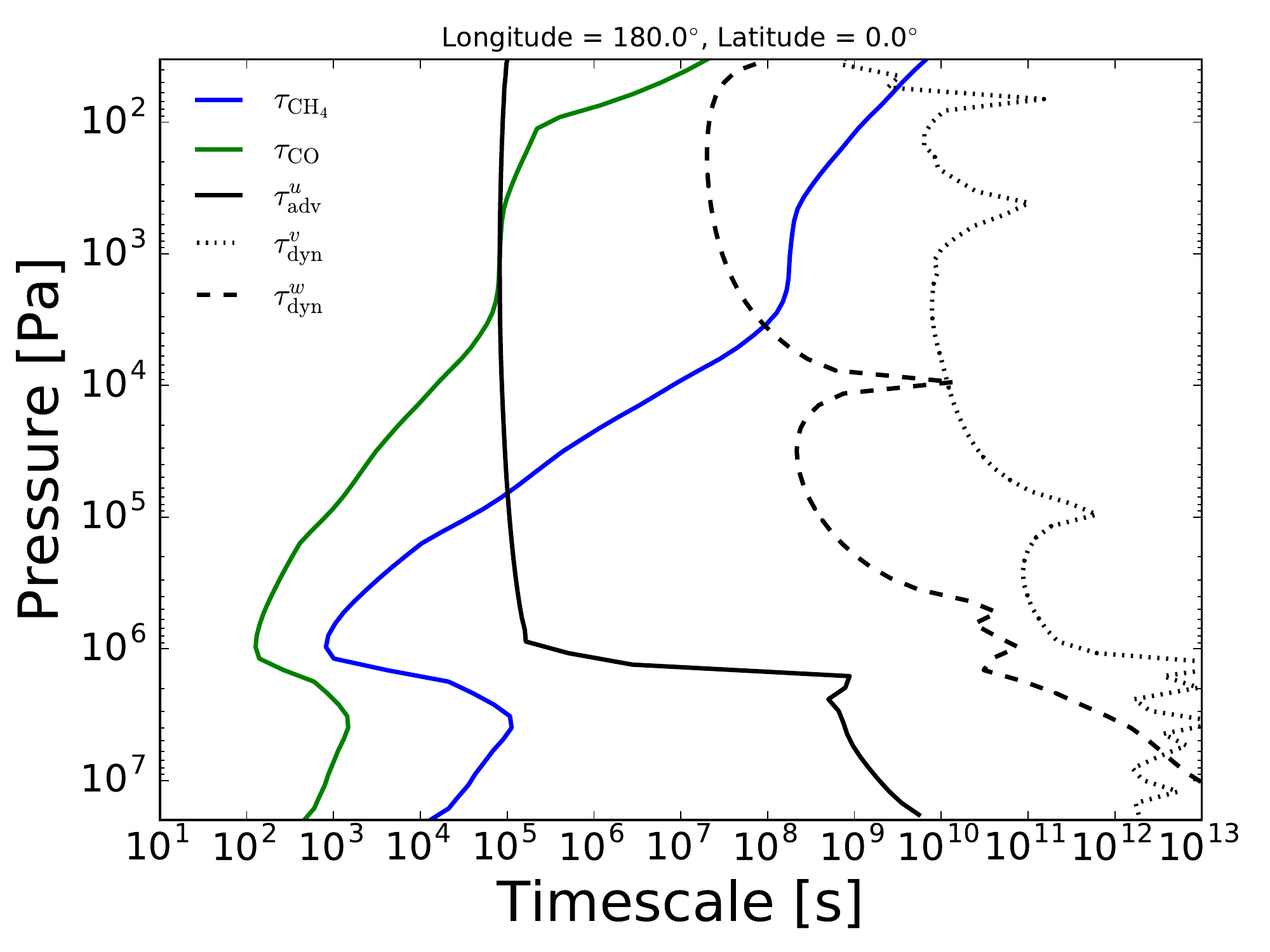} \\
    \includegraphics[width=0.39\textwidth]{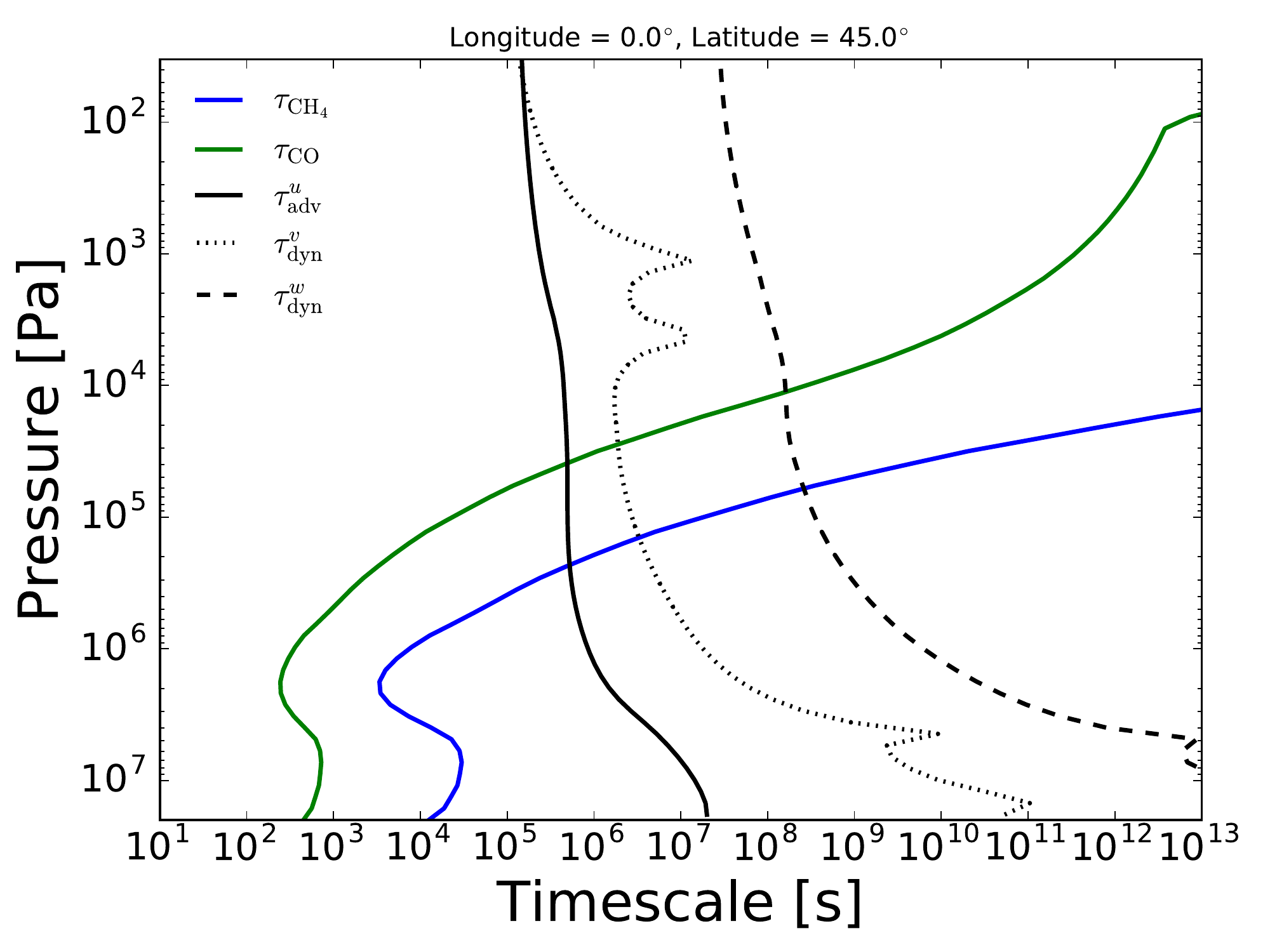} 
    \includegraphics[width=0.39\textwidth]{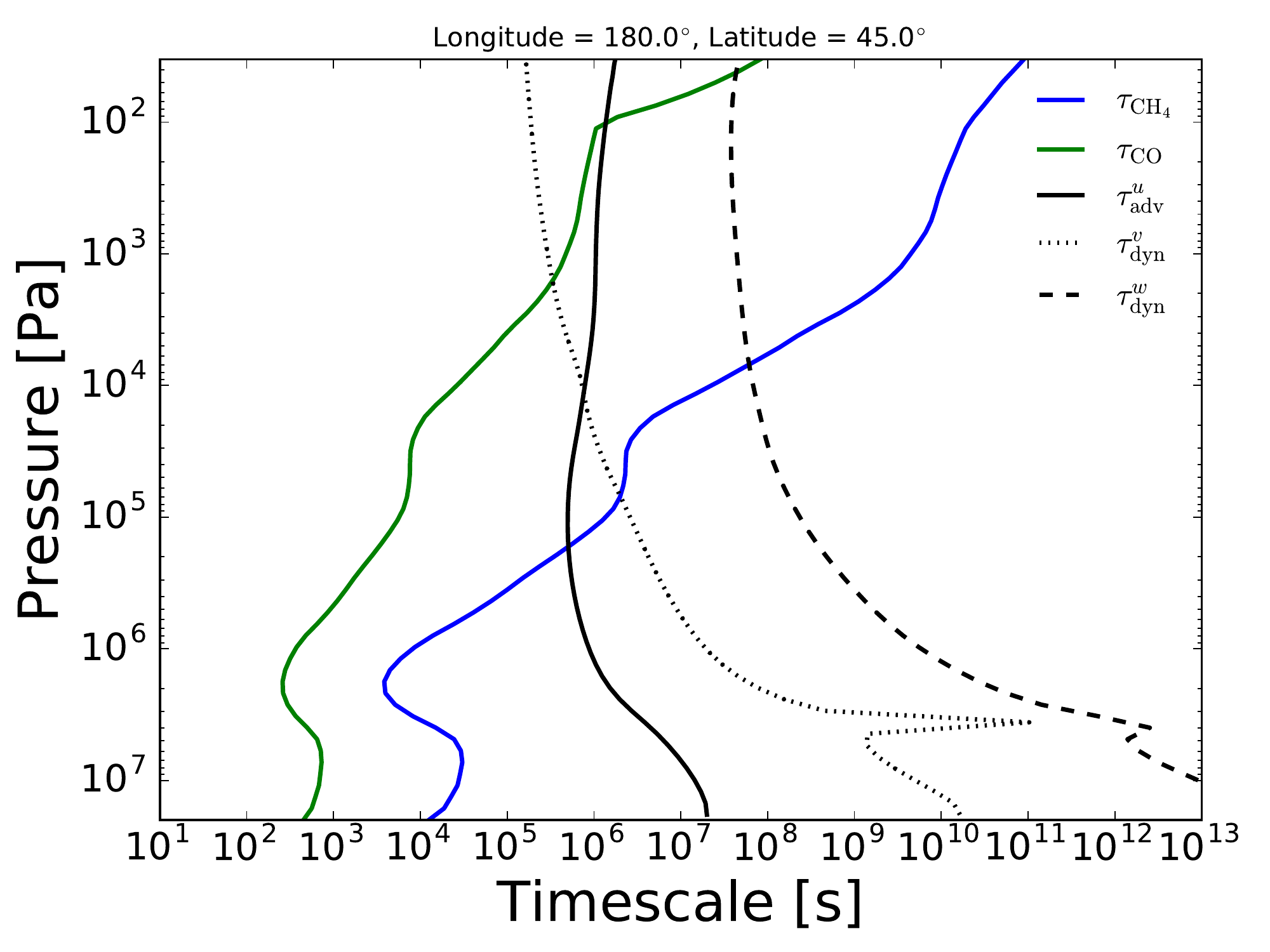} 
    \caption{As \cref{figure:hd209_ts_profiles} but for the HD~189733b simulation.}
    \label{figure:hd189_ts_profiles}
\end{figure*}